\pdfoutput=1
\documentclass[
12pt, % Default font size is 10pt, can alternatively be 11pt or 12pt
a4paper, % Alternatively letterpaper for US letter
onecolumn,
twoside, %oneside,%twoside,% Alternatively onecolumn portrait % Alternatively portrait
hidelinks,%Remove boxes from hyperlink
table]{book}

\usepackage[columnsep=2cm,
 left=1.575in,
 right=0.78in,
 top=1.18in,
 bottom=.88in]{geometry} % Margins includeheadfoot, includefoot,

\usepackage[T1]{fontenc}
\usepackage[utf8]{inputenc}

\usepackage{sectsty}
\usepackage[titles]{tocloft} % subfigure option only if using subfigure package

\usepackage{textcomp}
\usepackage{gensymb}
\usepackage{enumitem}
\usepackage{graphicx}
\usepackage[colorinlistoftodos]{todonotes}
\usepackage{fancyhdr}
\usepackage{setspace}
\usepackage{url}
\usepackage{verbatim}
\usepackage[bottom]{footmisc}
\usepackage[english]{babel}
\usepackage{multirow}

\usepackage[nottoc, numbib]{tocbibind}

\usepackage{amsmath}
\usepackage{amsfonts}
\usepackage{amsthm}
\usepackage{amssymb}
\usepackage{longtable,tabularx}
\usepackage{csquotes}
\usepackage{siunitx}
\usepackage{booktabs}

\usepackage{moreverb}
\usepackage{bm}
\usepackage{algpseudocode}
\usepackage{algorithm}
\usepackage{subfig}
\usepackage{minitoc}

\usepackage{multicol}
\usepackage{hyperref}
\usepackage{cleveref}
\usepackage{bookmark}
\usepackage{xcolor}
\usepackage{pdfpages}
\usepackage{tikz}
\usepackage[titletoc]{appendix}
\usepackage{mathtools}
\usepackage{etoolbox}
\usepackage[htt]{hyphenat}
\usepackage{soul}

\usepackage[acronym, nogroupskip, nopostdot]{glossaries} 
\newglossary[slg]{symbol}{sld}{sdn}{Nomenclature}
\makeglossaries
\glstoctrue
\usepackage{afterpage}

\usepackage{emptypage}
    
\let\Algorithm\algorithm
\renewcommand\algorithm[1][]{\Algorithm[#1]\setstretch{1.4}}

\usepackage{lscape}
\usepackage{adjustbox}
\usepackage{array}
\usepackage{multirow}
\usepackage{rotating}
\PassOptionsToPackage{table,xcdraw}{xcolor}
\usepackage{makecell}

\renewcommand{\texttt}[1]{%
	\begingroup
	\ttfamily
	\begingroup\lccode`~=`/\lowercase{\endgroup\def~}{/\discretionary{}{}{}}%
	\begingroup\lccode`~=`[\lowercase{\endgroup\def~}{[\discretionary{}{}{}}%
	\begingroup\lccode`~=`.\lowercase{\endgroup\def~}{.\discretionary{}{}{}}%
	\catcode`/=\active\catcode`[=\active\catcode`.=\active
	\scantokens{#1\noexpand}%
	\endgroup
}

\newboolean{rev}
\setboolean{rev}{false}
\newcommand{\setrevtrue}{\color{blue}}
\newcommand{\setrevfalse}{}
\newcommand{\rev}{\ifthenelse{\boolean{rev}}{\setrevtrue}             
	{\setrevfalse}}

\usepackage[backend=biber,
 giveninits=true,
 isbn=true,
 url=true,
 doi=true,
 style=ieee,
 citestyle=numeric-comp,
 maxnames = 5]{biblatex}
{%
  \fancyhf{}% clear all header and footer fields
  \fancyhead[R]{\nouppercase{\leftmark}}
  \fancyfoot[C]{\thepage} % except the center
  \renewcommand{\headrulewidth}{0pt}
  \renewcommand{\footrulewidth}{0pt}
}

\newcommand{\devEcho}{Amazon Echo}
\newcommand{\devAugust}{August doorbell}
\newcommand{\devAwair}{Awair air quality}
\newcommand{\devBelkinMotion}{Belkin motion}
\newcommand{\devBelkinSwitch}{Belkin switch}
\newcommand{\devBelkinCam}{Belkin cam}
\newcommand{\devBlipcareBP}{Blipcare BP}
\newcommand{\devCanaryCam}{Canary cam}
\newcommand{\devDlinkCam}{Dlink cam}
\newcommand{\devDropcam}{Dropcam}
\newcommand{\devHelloBarbie}{Hello Barbie}
\newcommand{\devHPprinter}{HP printer}
\newcommand{\devLifx}{LiFX bulb}
\newcommand{\devNestSmoke}{NEST smoke}
\newcommand{\devNetatmoWeather}{Netatmo weat.}
\newcommand{\devNetatmoCam}{Netatmo cam}
\newcommand{\devHuebulb}{Hue bulb}
\newcommand{\deviHome}{iHome}
\newcommand{\devPixstar}{Pixstar photo}
\newcommand{\devRingdoor}{Ring doorbell}
\newcommand{\devSamsungcam}{Samsung cam}
\newcommand{\devSmartThings}{SmartThings}
\newcommand{\devTPswitch}{TPlink plug}
\newcommand{\devTPcam}{TPlink cam}
\newcommand{\devChromeCast}{Chromecast}
\newcommand{\devTriby}{Triby speaker}
\newcommand{\devWithingsBaby}{Withings baby mon.}
\newcommand{\devWithingsSleep}{Withings sleep}
\newcommand{\devWithingsScale}{Withings scale}
\newcommand{\devNonIoT}{Non IoT}

\usepackage{hhline}
\definecolor{GrayCell}{HTML}{DDDDDD}
\definecolor{LightGrayCell}{HTML}{EEEEEE}
\definecolor{RedCell}{HTML}{FF0000}
\definecolor{GreenCell}{HTML}{32CB00}
\definecolor{YellowCell}{HTML}{F8FF00}
\definecolor{WhiteCell}{HTML}{FFFFFF}
\newcolumntype{L}[1]{>{\raggedright\let\newline\\\arraybackslash\hspace{0pt}}m{#1}}
\newcolumntype{C}[1]{>{\centering\let\newline\\\arraybackslash\hspace{0pt}}m{#1}}
\newcolumntype{R}[1]{>{\raggedleft\let\newline\\\arraybackslash\hspace{0pt}}m{#1}}

\newcolumntype{M}[1]{%
	>{\begin{turn}{90}\begin{minipage}{#1}%
				\raggedright\hspace{0pt}}l%
			<{\end{minipage}\end{turn}}}

\newcommand{\bordersForColoredTable}{\setlength\arrayrulewidth{0.6pt}}
\newcommand{\myverb}{\fontsize{10}{48}\usefont{OT1}{lmtt}{b}{n}\noindent }

\newcommand{\ie}{{\em i.e., }}
\newcommand{\eg}{{\em e.g., }}

\newcommand\TP{\mathit{TP}}

\newcommand\FP{\mathit{FP}}
\newcommand\FN{\mathit{FN}}

\newacronym{iot}{IoT}{Internet of Things}
\newacronym{iiot}{IIoT}{Industrial Internet of Things}  
\newacronym{ip}{IP}{Internet Protocol}
\newacronym{mac}{MAC}{Media Access Control}
\newacronym{cop}{CoP}{Code of Practice}
\newacronym{hids}{HIDS}{Host-based Intrusion Detection System}
\newacronym{nids}{NIDS}{Network-based Intrusion Detection System}
\newacronym{ietf}{IETF}{Internet Engineering Task Force}
\newacronym{mud}{MUD}{Manufacturer Usage Description}
\newacronym{ids}{IDS}{Intrusion Detection System}
\newacronym{cmmpp}{CMMPP}{Coupled Markov Modulated Poisson Processes}
\newacronym{pca}{PCA}{Principal Component Analysis}
\newacronym{svm}{SVM}{Support Vector Machines}
\newacronym{knn}{KNN}{K-Nearest Neighbor}
\newacronym{mlp}{MLP}{Multilayer Perceptron}
\newacronym{plc}{PLC}{Programmable Logic Controller}
\newacronym{ics}{ICS}{Industrial Control System}
\newacronym{scada}{SCADA}{Supervisory Control And Data Acquisition}
\newacronym{icn}{ICN}{Industrial Control Networks}
\newacronym{ifttt}{IFTTT}{If This Then That}
\newacronym{ldap}{LDAP}{Lightweight Directory Access Protocol}
\newacronym{gps}{SDN}{Software Defined Networking}
\newacronym{ipp}{IPP}{Internet Printing Protocol}
\newacronym{nfc}{NFC}{Near Field communication}
\newacronym{cdf}{CDF}{Cumulative Distribution Function}
\newacronym{sip}{SIP}{Session Initiation Protocol}
\newacronym{tcp}{TCP}{Transmission Control Protocol}
\newacronym{udp}{UDP}{User Datagram Protocol}
\begin{document}
\frontmatter
\thispagestyle{empty}
\pagenumbering{alph}
\begin{titlepage}
\thispagestyle{empty}
% \pdfbookmark[chapter]{Titlepage: Computationally Efficient Non-linear Kalman Filters for On-board Space Vehicle Navigation}{Thesis}
\begin{center}

% Upper part of the page

% Title

\Huge \textbf{IoT Behavioral Monitoring\\via Network Traffic Analysis}\\
\vspace{1.5cm}
\huge \textbf{Arunan Sivanathan}\\
\vspace{2cm}
\large A dissertation submitted in fulfillment\\of the requirements for the degree of\\
\vspace{0.5cm}
\large \textbf{Doctor of Philosophy}\\
\vspace{2cm}
\includegraphics[width=0.4\columnwidth]{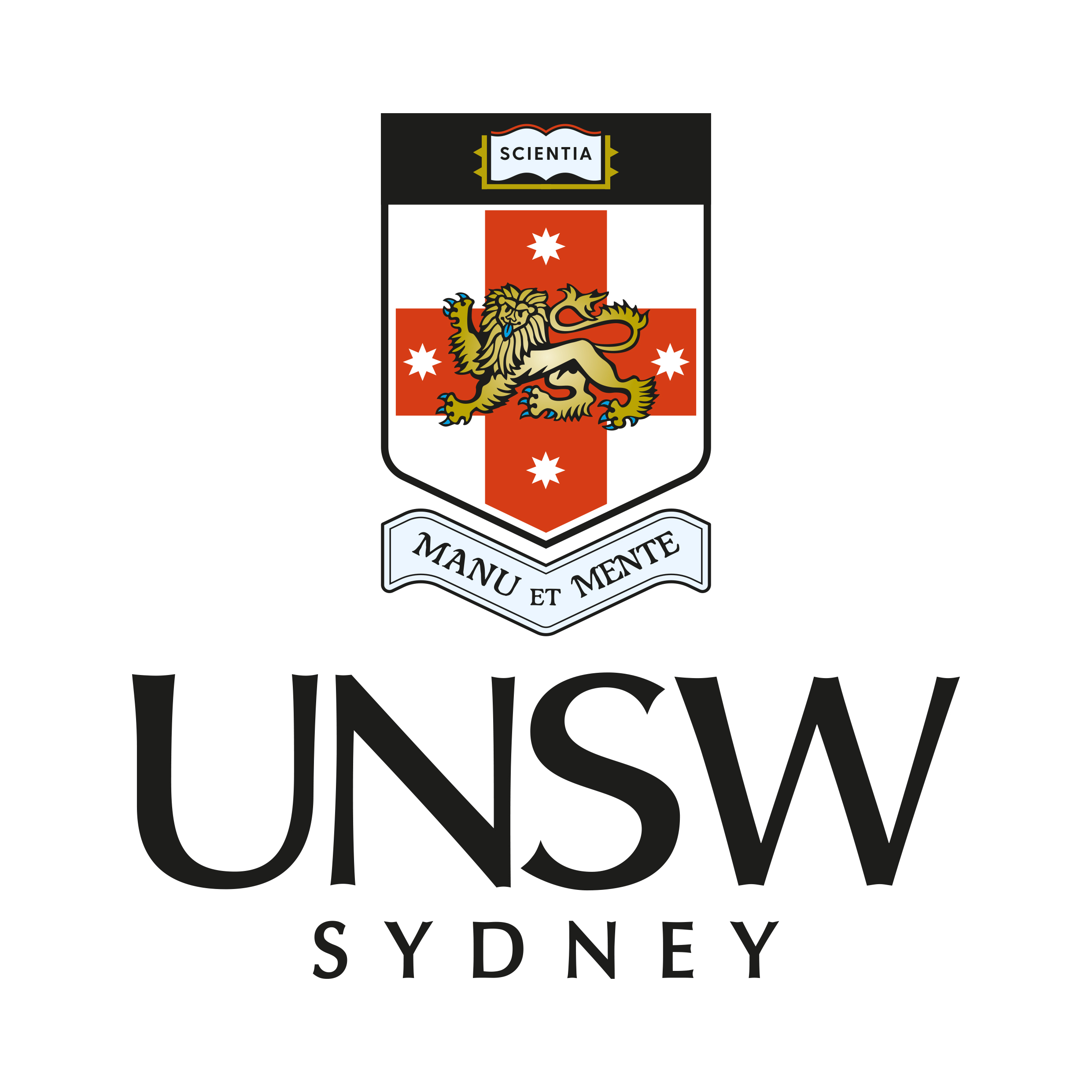}\\
\vspace{0.5cm}
\large School of Electrical Engineering and Telecommunications\\
\vspace{0.5cm}
\large The University of New South Wales\\
\vspace{2cm}
\large September 2019
\end{center}

\end{titlepage}

\pagenumbering{roman}
\setstretch{1.18}
\chapter*{Abstract}
%\mtcaddchapter[Abstract]
\bookmarksetupnext{level=section}
\addcontentsline{toc}{chapter}{Abstract}
\onehalfspace
\bookmarksetupnext{level=section}
\thispagestyle{plain}
\adjustmtc
\vspace{-1cm}

The Internet of Things (IoT) is being hailed as the next wave revolutionizing our society. Smart homes, enterprises, and cities are increasingly being equipped with a plethora of IoTs, ranging from smart-lights to smoke alarms and security cameras. While IoT networks have the potential to benefit society and our lives, they create privacy and security challenges not seen with traditional IT networks. The unprecedented scale and heterogeneity of IoT devices make today's security measures inapplicable to IoT networks. Due to the lack of tools for real-time visibility into IoT network activity, operators of such smart environments are not often aware of their IoT assets, let alone whether each IoT device is functioning properly safe from cyber-attacks. This thesis is the culmination of our efforts to develop techniques to profile the network behavioral pattern of IoTs, automate IoT identification and classification, deduce their operating context, and detect anomalous behavior indicative of cyber-attacks.

We begin this thesis by surveying IoT market-segments, security risks, and stakeholder roles, while reviewing current approaches to vulnerability assessments, intrusion detection, and behavioral monitoring. For our first contribution, we collect traffic traces and characterize the network behavior of IoT devices via attributes such as activity cycles and signaling patterns. We develop a robust machine learning-based inference engine trained with these attributes and demonstrate real-time classification of 28 off-the-shelf IoT devices in the lab with over 99\% accuracy. Our second contribution enhances the classification by reducing the cost of attribute extraction (via flow-level telemetry at multiple timescales) while also identifying IoT device states (bootup, user-interaction, and idle). Prototype implementation and evaluation demonstrate the ability of our supervised machine learning method to detect behavioral changes (including firmware updates) for five IoT devices. Our third and final contribution develops a modularized unsupervised inference engine that dynamically accommodates the addition of new IoT devices and/or updates to existing ones, without requiring system-wide retraining of the model. We demonstrate via experiments that our model can automatically detect attacks (\ie direct, spoofing, and reflection) and firmware changes in ten IoT devices with over 94\% accuracy.
\cleardoublepage

%\setstretch{1.5}
\chapter*{List of Publications} %
%\mtcaddchapter[List of Publications]
\bookmarksetupnext{level=section}
\addcontentsline{toc}{chapter}{List of Publications}
%\onehalfspace
\bookmarksetupnext{level=section}
\thispagestyle{plain}

\vspace{-5mm}
During the course of this thesis project, a number of publications have been made based on the work presented here and are listed below for reference.

\vspace{5mm}
\noindent{\underline{\large Journal Publications}}
\begin{enumerate}

\item \textbf{A. Sivanathan}, H. Habibi Gharakheili, and V. Sivaraman , ``Detecting IoT Behavioral Changes Using Clustering-Based Network Traffic Modeling'', (Under review at IEEE Internet of Things Journal)

\item \textbf{A. Sivanathan}, H. Habibi Gharakheili, and V. Sivaraman , ``Managing IoT Cyber-Security using Programmable Telemetry and Machine Learning'', (Under going revision at IEEE Transactions on Network and Service Management).

\item H. Habibi Gharakheili, \textbf{A. Sivanathan}, A. Hamza, and V. Sivaraman, ``Network-Level Security for the Internet of Things: Opportunities and Challenges'', \emph{Computer},vol. 52(8):58-62, Aug. 2019.

\item \textbf{A. Sivanathan}, H. Habibi Gharakheili, F. Loi, A. Radford, C. Wijenayake, A. Vishwanath, and V. Sivaraman, ``Classifying IoT Devices in Smart Environments Using Network Traffic Characteristics'', \emph{IEEE Transactions on Mobile Computing},18(8):1745-1759, Aug 2019.

\end{enumerate}

\vspace{5mm}
\noindent{\underline{\large Conference Publications}}
\begin{enumerate}
\setcounter{enumi}{4}

\item \textbf{A. Sivanathan}, H. Habibi Gharakheili and V. Sivaraman, ``Inferring IoT Device Types from Network Behavior Using Unsupervised Clustering'', \emph{IEEE LCN}, Osnabruck, Germany, Oct 2019.

\item S. Madanapalli, \textbf{A. Sivanathan}, H. Habibi Gharakheili, V. Sivaraman, S. Patil and B. Pularikkal, ``Modeling and Monitoring Wi-Fi Calling Traffic in Enterprise Networks Using Machine Learning'', \emph{IEEE LCN}, Osnabruck, Germany, Oct 2019.

\item \textbf{A. Sivanathan}, H. Habibi Gharakheili and V. Sivaraman, ``Can We Classify an IoT Device Using TCP Port Scan?'', \emph{IEEE ICIAfS}, Colombo, Sri Lanka, Dec 2018.

\item \textbf{A. Sivanathan}, F. Loi, H. Habibi Gharakheili and V. Sivaraman, ``Experimental Evaluation of Cybersecurity Threats to the Smart-Home'', \emph{IEEE ANTS}, Bhubaneswar, India, Dec 2017.

\item F. Loi, \textbf{A. Sivanathan}, H. Habibi Gharakheili, A. Radford and V. Sivaraman, ``Systematically Evaluating Security and Privacy for Consumer IoT Devices'', \emph{Workshop on Internet of Things Security and Privacy (IoT S\&P)}, Dallas, Texas, USA, Nov 2017.

\item M. Lyu, D. Sherratt, \textbf{A. Sivanathan}, H. Habibi Gharakheili, A. Radford and V. Sivaraman, ``Quantifying the Reflective DDoS Attack Capability of Household IoT Devices'', \emph{ACM WiSec}, Boston, MA, USA, Jul 2017.

\item \textbf{A. Sivanathan}, D. Sherratt, H. Habibi Gharakheili, A. Radford, C. Wijenayake, A. Vishwanath and V. Sivaraman, ``Characterizing and Classifying IoT Traffic in Smart Cities and Campuses'', \emph{IEEE Infocom SmartCity17 Workshop on Smart Cities and Urban Computing}, Atlanta, GA, USA, May 2017.

\item \textbf{A. Sivanathan}, D. Sherratt, H. Habibi Gharakheili, V. Sivaraman and A. Vishwanath, ``Low-Cost Flow-Based Security Solutions for Smart-Home IoT Devices'', \emph{IEEE ANTS}, Bangalore, India, Nov 2016.

\end{enumerate}

\vspace{5mm}
\noindent{\underline{\large Patent}}
\begin{enumerate}
	\setcounter{enumi}{12}
	\item ``An IoT Device Classification Apparatus and Process'', \textbf{A. Sivanathan}, H. Habibi Gharakheili, V. Sivaraman, Australian Provisional Patent, Application No. 2018904759, filed Dec 2018.
\end{enumerate}
\cleardoublepage

\chapter*{Acknowledgment}
	\bookmarksetupnext{level=section}
	\addcontentsline{toc}{chapter}{Acknowledgment}
	\onehalfspace
	\bookmarksetupnext{level=section}

	First and foremost, I would like to express my sincere gratitude to my primary supervisor, Prof. Vijay Sivaraman.  Prof. Vijay, thank you for your guidance and support throughout my Ph.D. research. It has been an absolute privilege to work with you, and this work would not have been possible without your contagious enthusiasm for research. I am equally grateful to my joint supervisor, Dr. Hassan Habibi Gharakheili. Dr. Hassan, thank you for giving me valuable pointers and ideas, shaping my research, and carefully reviewing all the manuscripts I produced during my Ph.D. I owe you a big debt of gratitude for your time, careful attention to detail, and challenging questions.
	\vspace{0.5em}
	
	I would like to express my sincere thanks to Prof. Eliathamby Ambikairajah, who offered me this Ph.D. candidature at UNSW. Prof. Ambi, it wouldn't be possible to pursue this opportunity without your recognition. I am very thankful for the scholarship and your keen monitoring on my progress throughout my carrier. My special thanks also go to Dr. Tharmarajah Thiruvaran, who recognized and recommended me for this position.
	\vspace{0.5em}

	This thesis is the result of three and a half years of a wonderful collaboration. The development and execution of the ideas presented here simply would not have been possible without the hard work, deep discussions, and shared excitement of all my co-authors, Vijay Sivaraman, Hassan Habibi Gharakheili, Chamith Wijenayake, Adam Radford, Arun Vishwanath, Daniel Sherratt, Franco Loi, Minzhao Lyu, Sharat Chandra Madanapalli, and Mohammed Ayyoob Hamza. I deeply appreciate your contributions.
	\vspace{0.5em}

	I graciously acknowledge and appreciate the collaboration with the teams at Cisco Systems and the Australian Communications Consumer Action Network (ACCAN). Their involvement and the input provided at various stages of this work were inevitably productive.
	\vspace{0.5em}

	I have been extremely fortunate to meet and interact with several talented, interesting, and fun fellow research colleagues, Mohammad Hossein, Ayyoob, Sharat, Iresha, Jawad, Minzhao, and Thanchanok (Tara), who spared their time to help me with both the happy and boring parts of the Ph.D.
	\vspace{0.5em}

	I cannot forget friends who went through hard times together, cheered me on, and celebrated each accomplishment. Anu Raghavi, Arunkumar, Gajan, Kaavya, Kawsihen, Navaroshan, Sirojan, and Tharshini -- thank you for keeping my life nourished and fun- filled.
	\vspace{0.5em}

	Words cannot express my gratitude to my parents and family, Sivanathan, Sasikala, Abhayan, Aparajithan, and Dhanya for constantly encouraging me to pursue a doctoral degree. My deepest appreciation is dedicated to my wife, Anusuya. You have been extremely caring and supportive of me throughout this entire process and have made countless sacrifices to help me get to this point whilst carrying your own burdens. Also, I would like to thank Anusuya's parents and family members for their constant moral support during this journey. Aekan and Pavinesh, because  of you, I laughed harder and smiled more.
	\vspace{0.5em}

	Finally, I'm grateful to the people of Australia, who as a country has generously offered me all the necessary rights, respect, peace and safe environment, some of which were even refused in my country of origin.

	%It would not have been possible to read a doctoral degree without the help and support from all of you; I would like to say a \textbf{`Big Thanks'} to all of you.

\onehalfspace

\vspace{-8mm}

\onehalfspace
\cleardoublepage
%\doublespace
\dominitoc
%\onehalfspace
\clearpage
\tableofcontents
\adjustmtc
\cleardoublepage

\clearpage
\listoffigures
\adjustmtc
%\mtcaddchapter[List of Figures]
\cleardoublepage

\clearpage
\listoftables
%\adjustmtc
%\mtcaddchapter[List of Tables]
\doublespacing
\adjustmtc
\glsaddall

\clearpage
%\printglossary[style=super, type= symbol , nonumberlist]
\printglossary[style=super, type=\acronymtype, nonumberlist]

\onehalfspace
\raggedbottom

%\afterpage{\blankpage}
%\afterpage{\blankpage}

\mainmatter
\doublespacing
\pagenumbering{arabic}
\pagestyle{fancy}{%
    \fancyhf{}
    \fancyhead[L,C]{}
    \fancyhead[R]{\nouppercase{\leftmark}}
    \fancyfoot[L]{}
    \fancyfoot[C]{\thepage}
    \fancyfoot[R]{}
    \renewcommand{\headrulewidth}{0pt}
    \renewcommand{\footrulewidth}{0pt}
}

\thispagestyle{fancy}
%\setstretch{1.5}

\chapter{Introduction}
	\adjustmtc[2]
	\mtcsetfeature{minitoc}{open}{\vspace{1em}}
	\minitoc

	The number of devices connecting to the Internet is rapidly increasing, signalling the beginning of the era of ``Internet of Things'' (IoT). IoT refers to the tens of billions of low--cost devices autonomously communicating with each other and with remote servers on the Internet. They include everyday objects such as lights, cameras, motion sensors, door locks, thermostats, fitness trackers, power switches and household appliances. With shipments projected to reach nearly 20 billion by 2020~\cite{IoTShipments}, thousands of IoT devices are expected to find their way in homes, enterprises, campuses, and cities of the near future, engendering ``smart'' environments benefiting our society and our lives.
	\vspace{-0.6em}

	While the benefits of IoT devices are well understood, they have unfortunately also become weapons of destruction in the hands of cyber-attackers. Recent incidents show that the consequences of exploiting IoT vulnerabilities can be high: an eavesdropper can illegitimately snoop into family activities, an attacker can take control or shut down a power grid, and household devices can become launch-pads to attack popular web-services. Securing IoT devices from attack remains a formidable challenge. The large scale and heterogeneity in IoT devices, each with its own hardware, firmware, and software, makes the security vulnerabilities diverse and attack vectors complex. The reasons for such vulnerabilities can be manifold, for example, devices do not have any host protection because of the resource constraints, device integrators obtain device parts from various suppliers without conducting any systematic security testing, and device manufacturers have low motivation to embed  security in consumer IoT devices, as they are dissuaded by low margins, time-to-market pressure, and limited skills.
	\vspace{-0.7em}

	Network operators are not always fully aware of their IoT assets and they lack tools that provide visibility into device operational behavior. Obtaining visibility in a timely manner is paramount to network operators so as to ensure that devices are in appropriate network security segments, and device behavioral changes indicative of cyber-attacks are able to be detected, so they can be quarantined rapidly when a breach is identified.
	\vspace{-0.7em}

	This thesis begins by surveying the security challenges in the IoT ecosystem by categorizing connected devices based on market segments, assessing the risks and challenges compared to traditional IT networks, and recognizing the major stakeholders and their responsibilities. We also develop a systematical approach to evaluate the security of smart home devices, validate it empirically in the lab using many consumer IoT devices, and compare it against threats and solutions used in traditional IT security. 
	\vspace{-0.8em}

	The observations above emphasize the need to develop a deep understanding of the behavior of network traffic to/from IoT devices. The objective in this thesis is therefore to develop behavioral models of IoT devices, which allows for feature extraction, automated classification, and anomaly detection using machine learning algorithms. Equally important to this thesis is the empirical validation of the models using data collected from IoT devices in the lab and in the wild. In this context, the major contributions of this thesis are as follows:
	
	\section{Thesis Contributions}		
		\begin{enumerate}
			\item Our first contribution is to learn the unique traffic behavioral characteristics of various IoT devices. We build an IoT experimental testbed instrumented with 28 consumer IoT devices and five non-IoT devices. The traffic characteristics of each device are analyzed via attributes comprising activity patterns (\eg distribution of flow volume, flow duration, traffic rate, and device sleep time) and signalling patterns (\eg server ports, domain names, cipher suites, DNS, and NTP queries) extracted from the network traffic traces. These attributes are used to develop an inference engine to classify the IoT device types using machine learning techniques. The proposed approach is trained and validated using the data collected over a six months period from our lab testbed, and demonstrated to achieve over 99\% accuracy. 
			
			\item Our second contribution is to develop techniques to extract flow-based attributes at multiple timescales using a programmable telemetry architecture and to minimize the cost of attribute extraction. We then develop an inference engine by using these extracted flow level attributes along with a multi-stage supervised learning architecture to detect the behavioral changes of IoT devices, distinguish IoT traffic from non-IoTs, classify individual types and identify states (\eg bootup, user-interaction, and idle) during its normal operations. We then quantify the trade-off between performance and cost of our solution, and demonstrate how our monitoring scheme can be used in real-time operation for detecting behavioral changes (\ie firmware upgrade or cyber-attacks).
			
			\item For our third contribution, we invent unsupervised, modularized machines to achieve a per-device type classification. This allows us to dynamically accommodate changes (\eg firmware upgrade or addition of a new type) in an IoT network without requiring a system-wide retraining. The machines are sensitive to minor deviations in traffic characteristics, allowing us to identify changes arising from low-rate cyber-attacks. We validate this by launching multi-rate attacks including port scanning, ARP spoofing, smurf, fraggle, TCP SYN flooding and UDP/TCP/ICMP reflections in our testbed. We have shown that our machine is able detect attacks with a high true positive rate (TPR).			
		\end{enumerate}
	
	\section{Thesis Organization}
        The rest of this thesis is organized as follows. \hyperref[chap:survey]{Chapter~\ref*{chap:survey}} published in \cite{LoiIoTSP,ANTS2017,ICIAfS2018} surveys the landscape of the IoT ecosystem and highlights related work and contributions made in recent years. In \hyperref[chap:characterization]{Chapter~\ref*{chap:characterization}}, published in \cite{infocom17,TMC18}, we characterize the IoT traffic based on the network activity and signalling pattern, and classify the device types using machine learning-based inference engines. \hyperref[chap:telemetry]{Chapter~\ref*{chap:telemetry}} published in \cite{ANTS16,TNSM19} presents a low-cost attribute extraction architecture using the software defined networking (SDN) paradigm and enhances the inference engine to recognize IoT devices along with the device types and states. In \hyperref[chap:anomaly]{Chapter~\ref*{chap:anomaly}}, published in \cite{LCN19,IOTJ19}, we enhance the inference to an unsupervised modular device classification architecture to accommodate behavioral changes of the devices while also developing a methodology to monitor the consistency of the classifier and validate the classification framework using benign and attack traffic. \hyperref[chap:conclusion]{Chapter~\ref*{chap:conclusion}} concludes the thesis with pointers to direction for future work.	

\chapter{Survey on IoT Ecosystems} \label{chap:survey}
	\minitoc

	The IoT ecosystem is in its early stages, and concerns about security and privacy threats have been getting increasing attention. In this chapter, we summarize the IoT ecosystem and challenges involved in protecting the device from cyber-attacks. Our first contribution explores the IoT ecosystem in terms of the market segments, risks, and challenges, as well as the expected responsibilities of major stakeholders in securing the devices. The second contribution proposes a systematical approach to evaluate the security of the devices by exploring aspects of confidentiality, integrity, access control and the possibility to launch reflection attacks. Then we review the existing IT security solutions and the challenges in adapting them to the IoT domain. Finally, we study existing techniques for characterization, classification, and anomaly detection in IoT network traffics. Parts of this chapter have been published in~\cite{LoiIoTSP},~\cite{ANTS2017} and~\cite{ICIAfS2018}.

\section{Introduction}
	The phrase ``Internet of Things'' introduced by Kevin Ashton in 1999, refers to the Internet-connected cyber-physical systems which sense and control the physical environment without much human interventions unlike traditional general purpose computers and smartphones~\cite{Ashton2009}. Internet-connected devices have already started to create a profound effect on us by offering the promise of unparalleled freedom and flexibility, be it for fitness, health, efficiency, safety, or entertainment~\cite{Gubbi2013}. The number of IoT devices in use is exponentially growing -- 20.4 billion Internet-connected devices will be instrumented across the globe by 2020 as per~\cite{Gartner20bn}. Australia's largest telco operator  Telstra predicts that an average Australian household which had 13 Internet-connected devices in 2017 will reach 30 by 2020~\cite{Testra}.  
	
\section{IoT Market Segments}
	\vspace{-1em}
	According to IDC's forecast, spending on IoT technology is expected to surpass US\$ 1.2 trillion in 2022 with the compound annual growth of 13.6\%~\cite{idc2022}. It covers a wide range of application domains such as smart home, wearables, automobiles, industrial IoT, smart cities, smart agriculture, intelligent retail, energy management, and health care. The IoT market can be folded into three main categories: 1) Industrial IoT; 2) Consumer IoT; and 3) Enterprises / Commercial IoT. This section provides a brief overview of these segments.
	
	\vspace{-1em}
	\subsection{Industrial IoT (IIoT)}
		\vspace{-1em}
		Industrial IoT (IIoT) brings the fourth industrial revolution (Industry 4.0) into reality by facilitating remote monitoring and automation in value chains of the manufacturing industries (\eg transportation, oil-and-gas, mining, energy/utilities, aviation, and logistics). IIoT includes interconnected industrial sensors, controllers, and actuators to maximize the efficiency and reliability in  mission-critical infrastructures of the industries. IIoT devices are specifically designed to tolerate rugged environment and operate for the long-term. They are mainly instrumented in new or legacy systems to perform relatively simple tasks like measuring the fuel levels or monitoring the product quality with the assistance of sensors~\cite{Yonomi2019,Kelltontech}. 

		Typically, IIoT devices are deployed in large scale Industrial Control Networks (ICN) which are transparent as opposed to a corporate IT network. To maintain interoperability and management through single management systems (\eg ICS, SCADA), industries tend to use homogeneous devices and similar protocols (\eg MQTT, DDS, CoAP) within a specific network. Also, industries mostly have an in-house ability to maintain life cycles (\eg updating firmware or patching vulnerabilities) of the devices. Due to the transparency of ICNs and predictable behavior of the IIoT devices, it is relatively easy to barricade IIoT using simple security policies~\cite{Cisco2017}.

    \subsection{Consumer IoT}
		Consumer IoT refers to the connected gadgets built for personal use, ranging from wearable health monitors, smart bulbs, smoke-alarms, and webcams to smart home appliances such as fridges. These devices come in different form factors to perform heterogeneous functionalities. Unlike IIoT, consumer IoT manufacturers prioritize low cost, advanced functionalities, user convenience, and elegant interfaces over performance, reliability, and long-term support of the devices. Typically, these devices collect and deal with a lot of private and sensitive user information since they work in an environment very close to users.

\begin{figure}[t]
			\centering
			\includegraphics[width=0.6\textwidth]{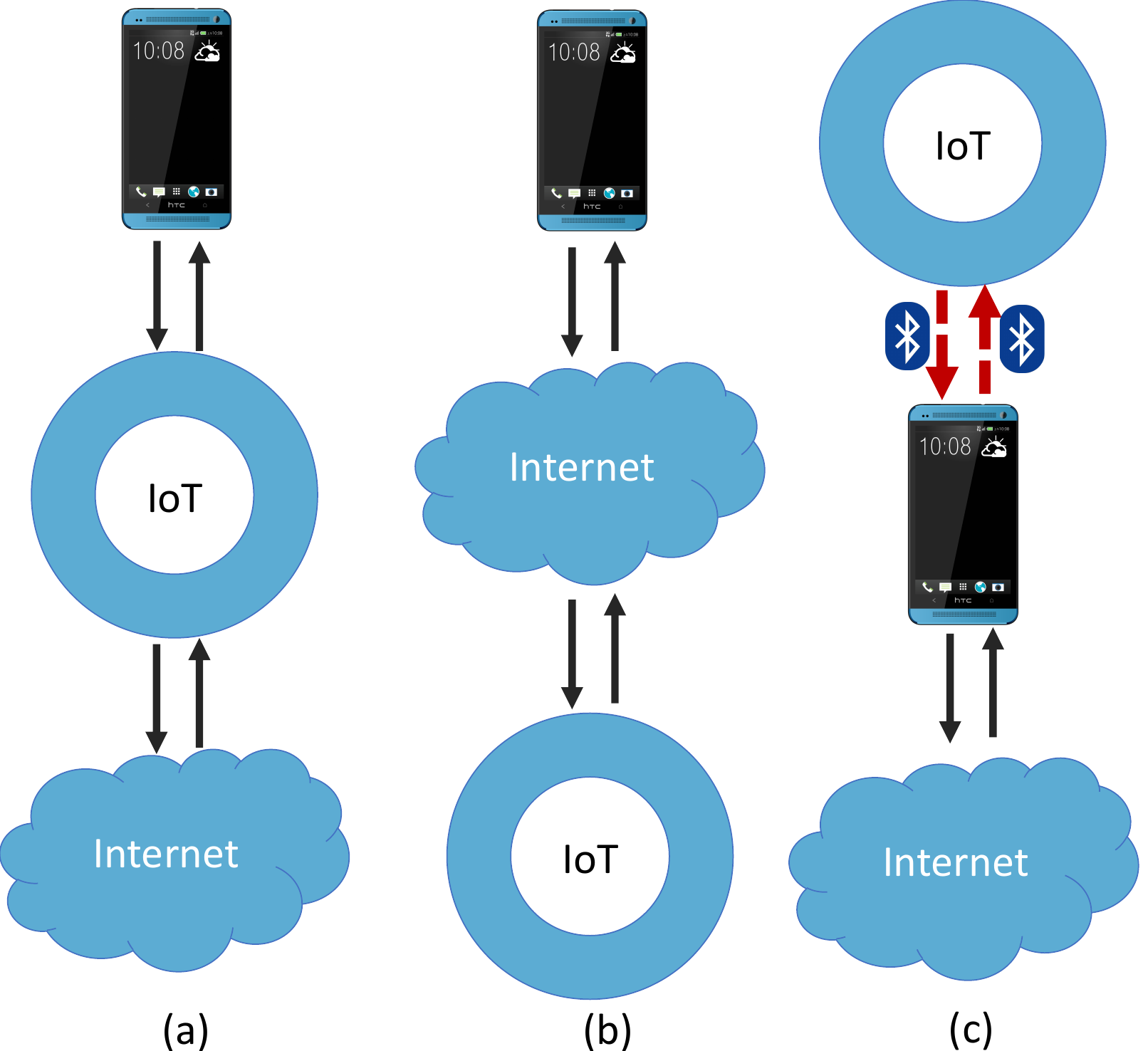}
			
			\caption{Communication model: (a) Direct, (b) External, (c) Transit}
			
			\label{fig:c2_commodels}
		\end{figure}

		The consumer IoT devices are mostly connected to small scale networks similar to a home network where they function independently. The devices from different vendors offer different management portals or mobile apps to control the IoT devices~\cite{Ray2017}. Fig~\ref{fig:c2_commodels} shows the main three communication models used by the devices to exchange the data with users and the cloud-based services: 1) Direct access model -- the devices not only communicate directly with cloud services but also, can be controlled directly through the management portals or smartphone Apps (\eg LiFX bulb, HP Envy Printer); 2) External access model -- the device directly communicates with the cloud services only. It doesn't provide any direct interface or API to control by users. However, users can get updates via connecting with the cloud servers (\eg Awair air quality monitor, Nest smoke sensor); and 3) Transit model -- this method is mostly used in low power devices which do not have the direct Internet connectivity. These devices communicate with smartphones using low powered communication mediums such as Bluetooth or Near Field Communication (NFC). Then the smartphone relays the data to a vendor cloud server over the Internet (\eg Fitbit, Tile Bluetooth tracker)~\cite{M2Msec14}.
		
		The interoperability of consumer IoTs is inherently limited since each manufacturer uses dissimilar protocols for authorization and communication. Still, consumer IoTs use commonly accepted protocols (\eg UPnP, Bonjour, REST, mDNS) to discover, control, and communicate with other devices. On the other hand, integration platforms such as IFTTT, voice-activated assistances (\eg Google Home, Amazon Echo) and dedicated IoT hubs (\eg Samsung SmartThings) enable somewhat interoperability using local or cloud to cloud APIs.

		The level of inbuilt security mechanisms of the consumer IoTs varies between vendor to vendor. Also, consumer IoT platforms usually use third-party tools and services, which adds another level of complexity in security. On the other hand, the heterogeneous behavior of consumer IoT devices makes it difficult to protect using simple security policies.

	\subsection{Enterprise / Commercial IoT}
		Enterprise IoT (also referred to commercial IoT) can be found in large organizational networks such as smart buildings, retail spaces, and smart cities (\eg smart lighting, connected HVAC system). The characteristics of enterprise IoT overlaps the behavior of industrial IoTs and consumer IoTs~\cite{EnterpriseIoT}. For instance, enterprise IoT, like smart lighting system, offers a user-friendly interface to consumers, whereby they optimize the power usage of the whole organization by controlling thousands of light bulbs.

		The enterprise IoT mostly support automation protocols, namely BACnet and Modbus, to exchange data with other devices. Although typical enterprise IoTs are managed by the private cloud services, they may update the information to the public cloud APIs as well (\eg a public transportation company which equipped GPS trackers on their buses may update the location of buses to users through common mapping services similar to Google map).

		Industrial IoT is typically managed by a centralized controller in an industry; nevertheless, a single organizational network may contain several independent enterprise IoT systems and be controlled by different departments. For example, a smart city may have street lights which are monitored by councils and traffic signals which are controlled by city police. Due to this complexity, the authentication and access permissions are sophisticated compared to consumer or pure Industrial IoTs, which may require support with hierarchical access control and directory services (\eg LDAP, Active Directory). Despite the fact that enterprise IoTs are commonly connected behind relatively secured corporate networks, the involvement of large people, departments, in the IoT ecosystem, may lead to insecure settings. 
		
		\vspace{1em}
	\section{IoT Security Risks}
		{\rev
			IoTs are being rapidly adopted as they give us the opportunity to enjoy incredible experiences in our life. Nevertheless, they are susceptible to attack by those wishing to harm us. Many Internet-connected devices have poor in-built security measures that make them vulnerable, and these flaws have the potential to reveal private data and information that may further hurt or alarm us. A typical smart home with many IoT devices is under significant risk of cyberattack. This vulnerability compromises data and threatens our safety. The frequency and severity of cyber-attacks has been escalating in recent years. As each month brings new consumer IoT devices to the market and millions of deployments in households worldwide, new security and privacy attack vectors open up that can be exploited at a scale never seen before.
		} 
		
		Furthermore, search engines, such as Shodan~\cite{Shodan} and Inseccam~\cite{insecam}, are discovering vulnerable IoT devices exposed to the Internet; openly available lists of IoT default username-password combinations~\cite{DPEProject}; as well as the publicly available botnet codes similar to Mirai~\cite{Mirai16,Antonakakis2017}, which make it an effortless task to launch a cyberattack. 

		The attacks on IoT devices have already started to show the impact on the economy of the companies as well as the privacy of users. One-fourth of the companies, which are rapidly moving towards IoT, reported at least \$34 million security-related losses in only the last year~\cite{DigicertInc2018}. Studies show that 91.5\% of data generated by IoT devices are exposed as plain text -- readable by anybody snooping the transaction~\cite{Greene}. It includes private and sensitive information such as medical records~\cite{Wood2017}.

		\subsection{Threats in IoT Network Compare to IT Network}    
			The attacks currently targeting the IoT devices are not completely brand new in cyberspace. Although they have been encountered in traditional IT network over the decades, they pose new dimensions of challenges in the IoT ecosystem~\cite{Suo2012}. First of all, the scale of IoT and the number of exposed endpoints create a massive exploitable threat surface which has never been seen in the traditional networks. The traditional IT networks are built upon a very limited number of platforms (\ie applications, operating systems, and device vendors) which are well grown and constantly undergo security evaluations by experts. However, every year hundreds of new IoT devices are introduced to the market by newly emerging startups -- mostly who do not have expertise in security. This makes the situation worse.

			Almost 90\% of IoT devices closely monitor and collect some form of personal data like location, health, habits, interests, etc~\cite{HewlettPackard2014}. Also, they interact with the physical environment without any intervention from human users. These factors, as a result of the security breach, possibly create severe consequences as it results in a loss of privacy or infrastructure damage, or worse it results in safety hazards to users~\cite{Zhang2015}. In 2015,  security evaluators demonstrated this by hijacking a jeep remotely~\cite{killjeep}. Furthermore, the IoT behind the secured enterprise networks becomes an easy infiltration point for attackers~\cite{Schneier2014} and this can then jeopardize other devices in the network. The compromised fish tank sensors used to hack a casino is a good example of this scenario~\cite{fishtank}.

		\subsection{Drawbacks of Traditional Security Measures}
			The present IT ecosystem is mainly protected by host-based threat protection mechanisms (\eg Anti-virus) and network perimeter defenses (\eg firewalls and instruction detection systems (IDS)). Unlike the general-purpose computers or smartphones in the IT ecosystem, IoT devices come with a very limited computational power. Therefore they cannot run in-built protection mechanisms like anti-virus software as well as key certificate exchanges and state of the art encrypted algorithms during communications. Moreover, IoTs do not have enough resources like storage, battery, and computational power to support automatic firmware upgrading and security patching mechanism~\cite{Schneier2014} like our smartphones.

			On the other hand, the scale and heterogeneity of the IoT ecosystem make it difficult to protect using the existing network-level defense systems. For example, the traditional networks can be protected for a certain degree by allowing well-known protocols (\ie port numbers) only through firewalls. However, the IoT ecosystem makes it impossible because of the diverse amount of protocols and standards used in the devices~\cite{Jing2014}. Meanwhile, the classical attack signature identification using IDS is also not an easy task since the attack and response patterns vary between devices. The detailed analysis of the existing IDS will be discussed in \S\ref{existsec}.

			IoT devices are vulnerable to indirect attacks as well. For example, in a smart home, door locks might be configured to unlock automatically while the smoke sensor alarm is triggered. An attacker may exploit this cross-device communication to open the door by compromising under secured smoke sensor~\cite{Vyas2015}. This kind of attack can be preventable only if security systems distinguish the context of devices in the network~\cite{Yamauchi2019}.

		{\rev
		\subsection{Types of Cyber Attacks on IoT}
			Cyber attacks on IoT can be categorized into 4 different criteria:
			1) attacks that affect the confidentiality of the data communication of them;
			2) attacks that affect the integrity and authentication of connections they establish with other entities (local or external);
			3) attacks that affect the access control and availability to make the connection with legitimate devices; and
			4) attacks that use the IoT devices as reflectors to attack new devices.
			Following attacks recorded in recent history can explain the severity of each types. In November 2015, hackers compromised a Hello Barbie doll and gained access to user accounts and encrypted audios~\cite{Barbie15} (Confidentiality violation).
			In 2016, a large scale attack used the Zigbee protocol in the Phillips Hue lightbulb to spread a worm to control other lightbulbs~\cite{Phillips16} (integrity violation).
			In 2017, a hacker with the name ``Stackoverflowin'' gained illegitimate access to 150,000 printers by exploiting the Internet printing protocol (IPP), and was able to send out rogue print jobs~\cite{Printer17}(access control violation).
			In the same year, a university became a victim of DDoS attack from its campus lamp posts and vending machines~\cite{vendingmachine2017} and one of the largest distributed denial of service (DDoS) attack to dates, recorded in 2016, used an army of compromised IoT devices to bring down the Dyn DNS server. It affected many popular websites~\cite{DDoS16} (availability violation)).
			Although we have not seen any mass scale reflection attacks using IoT devices yet, the researches suggest that present IoT devices are highly capable to reflect the attacks\cite{Wisec17}. We develop a systematic approach to evaluate these 4 types of vulnerabilities in \S\ref{ch2:syseval}. Later, in Chapter~\ref{chap:anomaly} we will validate the efficacy of our anomaly detection engine by launching DDoS (Type 3) and Reflection attacks (Type 4) which are commonly used for IoT devices at scale. 
		}
	 	
	\section{Perspectives and Roles of Stakeholders}
		It is a well-known fact that there is no one-off solution to secure all IoT devices. It requires a lot of responsibilities and actions to be taken by various entities related to the IoT ecosystem~\cite{ACCAN}. This section identifies the main players of the IoT ecosystem and discuss their role and responsibilities in the security domain.

		\subsection{Consumers}
			Mostly IoT consumers do not scrutinize the security of devices during their purchase. There are two main reasons for that: 1) they don't seem to be aware of how detailed and sensitive data are being collected by their devices nor are they aware the consequences if that data get compromised; and 2) they don't have the knowledge to rate the safety of a device. Also, many IoT users do not follow good security practices such as applying strong passwords -- 10 out of 100 devices have never been changed from their default username and passwords (\eg <admin,admin>, <admin,password>)~\cite{threatscape2017}. They assume that device manufacturers or service providers apply the software updates and patches until the lifetime of the device  -- actually, this is a myth. Nevertheless, it is not reasonable to expect them to be tech-savvy enough to patch the devices manually as well. Experts say, although consumers do not have the capacity to understand the technical terms related to the security, they can be indicated using a rating scheme similar to ``energy efficiency star rating'' in electrical appliances~\cite{Stilgherrian}.

		\subsection{Manufacturers}
			The peak demand for IoT leads the manufacturers to focus on rushing the device to the market rather than prioritizing the security. Furthermore, manufacturers hesitate to provide long term supports to devices, especially due to development costs. They are more likely to release a new version with the improved functionalities to get more profits than supporting the previous version.  Even though some of the manufacturers are aware that their devices support mass scale DDoS, they don't give close attention towards fixing the issues. The reason is those kinds of attacks don't impact the customers directly~\cite{Fernandes2015}. 

		\subsection{Regulators}
			As security and privacy concern rises about the IoTs, there have been calls for government regulations in the IoT ecosystem. These regulations are expected to urge manufacturers to build devices with minimum security standards. The main implication in this process is, according to the current settings, different domains of the IoT may fall under the different departments' regulations. For example, devices related to health and medical come under the rules and regulations of the Therapeutic Goods Administration within the Department of Health. Meanwhile, services and technologies such as telecommunications, broadcasting, radio communications, and the Internet are regulated by the Australian Communications and Media Authority. In the scenario of Medical grade IoT, it may require the attention of both departments~\cite{ACCAN}. On the other hand, some people believe strict government regulations may create additional bureaucracy and stifle the innovation and agility in the IoT development~\cite{Richardson2017}. Several governments have already started to propose very basic level legislations to handle this trade-off.

			\subsubsection{United States of America}
				The United States Congress introduced a bill ``Internet of Things (IoT) Cybersecurity Improvement Act of 2019'' to set minimum standards to procure and use the IoTs for government agencies. It requested the recommendations from the National Institute of Standards and Technology (NIST) to propose minimum standards on IoT development, identity management, patching, and configuration management~\cite{USsecurityLaw2019}. However, this bill does not consider consumer or business use cases.

			\subsubsection{UK}
				In 2019, the UK proposed a basic set of code of practice (CoP) to be followed on IoT design and development~\cite{James2019}. It includes: 1) unique factory reset settings for each device -- cannot have universal default passwords for all devices; 2) a public point of contact has to be provided by manufacturers to disclose the vulnerabilities; 3) requirement to explicitly state the minimum duration that a device will continue to receive security updates or patches; and 4) a labeling system to determine the level of security -- similar to health star rating on foods or energy rating on electrical appliances. Currently, the officials say the government is planning to impose the first three practices as mandatory `Secure by Design' rules and the labeling system as a voluntary scheme to improve the knowledge of  consumers about the basic security standards of devices.
			
			\subsubsection{Japan}
				The National Institute of Information and Communications Technology (NICT) of Japan has announced a scanning over the nationwide Internet-connected devices to identify the vulnerabilities. This project has been estimated to continue until 2022. During the experiment, agencies especially probe the devices using the list of default usernames and passwords without the concern of citizens and businesses to identify the IoTs with easily guessable credentials. The owners of the devices will be notified if the scan finds any potential security issues. Although this search can help to uncover a portion of vulnerable devices, fixing them might have implications. For example, owners may not have enough knowledge to fix the vulnerabilities~\cite{JPsecurityLaw2019}.

			\subsubsection{Europe}
				`ETSI TS 103 645' is a new cybersecurity standard for consumer Internet of Things devices released by the European Telecommunications Standards Institute (ETSI) in February 2019~\cite{Eecke2019}. It proposes 13 best practices to support manufacturers: no default passwords; keeping software updated; manage vulnerability reports; securely store security-sensitive data; communicate securely; minimize attack surfaces; ensure software integrity; protect personal data; be resilient to outages; make use of telemetry data; allow users to delete personal data; make installation and maintenance easy; and validate input data. It claims that these standards allow flexibility for innovation rather than being rigid rules.    
			
			\subsubsection{Australia}
				Compared to other countries discussed earlier, Australia is still lagging in imposing the legislation for the protection of IoT. In the past, the federal government has proposed the idea of a rating logo for Internet-connected devices which is named as ``Cyber Kangaroo''~\cite{Mikolic-Torreira2017} -- assuring a basic level of quality for consumers.
			
				However, it received criticism from the experts for various reasons. The resilience to attack of the devices cannot be expressed by a static rating logo. The security weaknesses of the devices may unveil over time, but the rating logo on the packaging cannot reflect those changes. Also, the security rating may reflect different meaning on different domains. For example, the consequences of an attack on a connected car are different from the breach on connected Barbie dolls -- these cannot be rated by a single rating scheme. The security labels also make a false impression to users that these devices are always secure. Thus, users tend to negate the best practices, such as changing default passwords / updating the devices immediately after the patch available~\cite{Chapman2017}.

		\subsection{Insurers}
			In spite of the precautions taken to secure the IoT, there is the possibility to still be affected by cyberattacks along the line. It may cause harm to users and affect the reputation of the manufacturers. To mitigate the impact and avoid  bankruptcy, they may invest in cyber-insurance. The increasing premium for these manufacturers who are more likely to be vulnerable will also force them to bring security to the top of their priority list. It is claimed that the global market size of Cyber Insurance is estimated to grow from US\$2.9 billion in 2019 to US\$16.7 billion by 2024~\cite{ARC2019}.

\section{Systematical Evaluation of Cybersecurity Threats}\label{ch2:syseval}

	Emerging research work \cite{M2Msec14, Dhanjani2015, SmartThings16, Vyas2015,Morgner2019,Notra2014,Wisec17} has focused on understanding and identifying potential security and privacy threats for IoT. However, there is little research into a systematic way for identifying security flaws in existing and emerging IoT devices. We believe our work is the first to develop a systematic methodology for profiling the security posture of consumer IoT devices, which can lead to a security-star rating that can inform consumers, regulators, and insurance bodies of the associated risks.

	With regards to this scenario, we develop a suite of security tests categorized under four criteria: {\em confidentiality} of data sent/received by the IoT device; {\em integrity and authentication} of connections the IoT device establishes with other (local or external) entities; the {\em access control and availability} of the IoT device to connection requests; and the capability of the IoT device to participate in {\em reflection attacks}. Next, we apply our automated security test suite to 20 IoT devices available in the market today, chosen to cover a range of applications including home security (cameras and motion sensor), health (weighing scale, blood-pressure monitor and air-quality sensors), energy management (light-bulbs and power-switch), and entertainment (photo frame, printer and speaker). Finally, using the outputs of our automated test suite, we assign a color-coded security score to each of the devices under each of the four criteria, thereby giving an intuitive visual representation of the device's security posture. 

	\subsection{Security Test Suite}
		In this section, we develop a suite of security tests to categorize threats that exploit security/privacy vulnerabilities in IoT devices under four dimensions namely confidentiality, integrity, access control and availability, and reflection.   
	
		\subsubsection{Confidentiality}\label{syseval:method:conf}
			Confidentiality involves ensuring the exchanged data between endpoints cannot be understood by unwanted snoopers. We evaluate the confidentiality of exchanged data using three measures, whether it is \textit{plaintext}, \textit{encoded}, or \textit{encrypted}. We assess all communication channels of a given IoT device -- between: device and cloud server; device and user App; user App and cloud server. We therefore wrote a Python script that performs ARP spoofing inside the home network to intercept all traffic to/from the IoT device as well as the user's smartphone. 
			
			\textbf{Encryption protocol}: We use this test to determine the security protocol being used for a particular communication channel. The security protocol is obtained  by checking the \verb|protocol| field of the packet capture on Wireshark to see if it is identifiable.
			
			\textbf{Plaintext}: After inspecting the protocol field, we analyze the \verb|data| field (i.e. payload) to check if it contains any human-readable text. This test determines whether the data is in plaintext or not, but it does not differentiate between encoded data and encrypted data as both are not human-readable. 
			
			\textbf{Entropy}: Since the above tests cannot always evaluate the confidentiality of data, we use the entropy test to verify whether a certain communication is encrypted, encoded or in plaintext. Entropy can not only be used to determine whether data is encrypted, but also to assess the strength of encryption. The better the level of encryption the higher the entropy as it will contain more information.
			
			We wrote a Python script that is fed raw data from captured packets to compute the Shannon entropy of the data one byte at a time (i.e. a value between 0 and 8) -- we look at the data in bytes.
			In order to have an accurate entropy value, we use at least 100 KB worth of packets. Our entropy test verifies whether the data is encrypted in conjunction with the encryption protocol test and confirms the plaintext test. {\rev We note that the entropy test may fail to distinguish encrypted from encoded communications specially when it is applied to traffic of compressed video generated by cameras -- video compression yields a high entropy value though it is unencrypted.}
		
		\vspace{-0.5em}
		\subsubsection{Integrity} \label{syseval:method:int}
		\vspace{-0.5em}
			Integrity assessment ensures a given IoT device performs its intended functions without any manipulation and no message to/from the device is modified without detection. We therefore test the following: 		
			
			\textbf{Replay attack}: We feed captured packets sent from the user App to the IoT device (using the technique mentioned in Confidentiality) into our Python script which will then replay them to the IoT device. The attack is successful if the device performs a certain function specified in the packet. 
			Furthermore, if packets are in plaintext (or encoded), we modify certain fields inside the packets and replay them to check whether the device responds to tampered packets. {\rev Replay attacks are launched on a third of IoT devices which use plaintext or encoded packets only. We note that other devices which communicate encrypted traffic using TLS/SSL are protected against replay attacks..}
			
			\textbf{DNS security}: We also test whether the device attempts to connect to an illegitimate server. Inspecting the DNS queries and responses, we assess whether devices uses DNSSEC. {\rev We note that DNSSEC is only offered by authoritative servers, not recursive resolvers. Authoritative servers contacted by IoT devices are managed by their respective manufacturers. In order to determine whether a device validates DNSSEC certificate records or not, we spoofed the response of DNS queries made by that device. Accepting spoofed responses by the device and attempting to connect to the illegitimate server indicate that the device does not validate DNSSEC records. 
			
			If the device is vulnerable to DNS spoofing, we use a python script to perform DNS spoofing redirecting traffic to a fake server. If the device attempts to connect to this fake server, the system integrity is violated. In addition, if it sends information to the fake server it indicates the device does not conduct any form of authentication.
			}
		
		\vspace{-0.5em}
		\subsubsection{Access control and availability}
			We consider the access control and availability of an IoT device to identify how easily an attacker can gain access/control to/of the device and determine whether it is susceptible to a denial of service (DoS) attack. % which can have serious implications for users (e.g. home security and health monitoring devices). 
			We start our test by scanning for ports that are open on the device using command \verb|nmap| \verb|-sS -sU -p 0- 65535 [deviceIP]|. We then attempt to gain access via Telnet, SSH and HTTP using a list of known weak login credentials -- these ports were exploited recently by the Mirai botnet that resulted in one of the largest DDoS attacks from IoTs over the Internet \cite{Incapsula16}. 
			
			{\rev	
				\textbf{Denial of Service}: We assess the ease of launching a DoS attack using the following experiment. 
				We determine how much incoming traffic the IoT device can handle before it completely loses its expected functionality. We flood the device with ICMP \verb|ping| requests as well as UDP packets, and determine the amount of data that is required to stop the operation of the IoT. 
				We conduct these two tests using the \verb|hping3| tool by issuing the command: \verb|hping3 -d 1000 -1 (1 for ICMP and 2| \verb|for UDP) -p (port) (deviceIP)|. We also use another python script to measure the maximum number of concurrent TCP connections the device can handle before it crashes --  by flooding the device with TCP SYN packets to initiate connections to the list of open ports on the device.
		  	}
		
		\subsubsection{Reflection attacks}
			Following the public announcement of the large DDoS attack fueled by IoTs in 2016 \cite{Incapsula16} many manufacturers have consequently closed their remote access ports, or strengthened their default login credentials. We have shown that  IoT devices can still be employed to launch DDoS attacks by exploiting various protocols using source-spoofed traffic \cite{Wisec17}. Evaluating the reflection capability of device protocols is important since IoT devices are increasingly contributing to DDoS attacks to popular services providers across the Internet \cite{Ryba2015,Kuhrer2014,Rossow2014,Kuhrer2014Impact}.
			{\rev
				We experiment the reflection attacks on three standard protocols namely ICMP, SSDP and SNMP. 
				We write a python script that crafts malformed packets (with spoofed source IP address) and sends; (a) ICMP messages, (b) SSDP broadcasts, and (c) SNMP requests to a given IoT device. For the SNMP, we further check if the device supports the SNMP public community string that can potentially generate a larger volume of responses. If successful, we issue a \verb|getBulk| SNMP request that sends multiple \verb|getNext| requests at once. Responding to each of these protocols reveals that the device can be used to launch a reflection attack.
			}

	\subsection{Security Posture of IoTs}
		We now validate our assessment methodology by applying it to 20 IoT devices that have been recently introduced to the consumer market, ranging from cameras and lightbulbs to power switches and health monitoring devices.
		We verify our methodology on some devices with known security flaws \cite{Dhanjani2015} and also evaluate the security and privacy posture of other IoT devices with security vulnerabilities that are unknown to us.

		\subsubsection{Confidentiality}
			\bgroup
			\def\arraystretch{1.5}
			\bordersForColoredTable
			\begin{table}[t!]
				\rowcolors{1}{WhiteCell}{LightGrayCell}
				\centering
				
				\caption{Posture of confidentiality}
				
				\label{tab:confidEval}
				\begin{adjustbox}{max width=\textwidth}
					\begin{tabular}{|l|ccc|ccc|ccc|}
						\hline
						& \multicolumn{3}{c|}{\textbf{Device to Server}}                                                                         & \multicolumn{3}{c|}{\textbf{Device to User-app}}                                                                    & \multicolumn{3}{c|}{\textbf{User-App to Server}}                                  \\
						\rowcolor{WhiteCell}
						\multicolumn{1}{|c|}{\textbf{Devices}}   & \textbf{Plaintext}    & \textbf{Protocol}                     & \textbf{Entropy}                     & \textbf{Plaintext}                     & \textbf{Protocol}                     & \textbf{Entropy}     & \textbf{Plaintext}   & \textbf{Protocol}    & \textbf{Entropy}           \\ \hline
						\devHuebulb             & No                                     & TLSv1.2                                   & 7.70                                 & Yes                                    & None                                  & 5.48                 &                      &                      &                            \\
						\devBelkinSwitch        & Partially                              & Unknown                               & 7.74                                 & Yes                                    & None                                  & 5.16                 &                      &                      &                            \\
						\devSamsungcam          & No                                     & Unknown                               & 7.99                                 &                                        &                                       &                      & No                   & Unknown              & 7.91                       \\
						\devBelkinCam           & No                                     & Unknown                               & 7.06                                 & No                                     & SSL                                   & 7.95                 & No                   & SSL                  & 7.48                       \\
						\devAwair               & No                                     & SSL                                   & 7.89                                 &                                        &                                       &                      & No                   & SSL                  & 7.90                       \\
						\devHPprinter           &                                        &                                       &                                      & Yes                                    & None                                  & 5.38                 &                      &                      &                            \\
						\devLifx                &                                        &                                       &                                      & No                                     & Unknown                               & 4.66                 & No                   & SSL                  & 7.64                       \\
						\devCanaryCam           & No                                     & TLSv1.2                               & 7.96                                 &                                        &                                       &                      & No                   & TLSv1.2              & 7.46                       \\
						\devTPswitch            & No                                     & Unknown                               & 7.95                                 & No                                     & Unknown                               & 5.33                 & No                   & SSL                  & 7.63                       \\
						\devEcho                & No                                     & TLSv1.2                               & 7.98                                 &                                        &                                       &                      & No                   & TLSv1.2              & 7.91                       \\
						\devSmartThings         & No                                     & TLSv1.2                               & 7.69                                 &                                        &                                       &                      & No                   & TLSv1.2              & 7.80                       \\
						\devPixstar             & No                                     & TLSv1.2                               & 7.87                                 &                                        &                                       &                      &                      &                      &                            \\
						\devTPcam               & No                                     & Unknown                               & 7.97                                 & Yes                                    & None                                  & 7.51                 & No                   & TLSv1.2              & 7.73                       \\
						\devBelkinMotion        &                                        &                                       &                                      & Yes                                    & None                                  & 5.16                 &                      &                      &                            \\
						\devNestSmoke           & No                                     & Unknown                               & 7.25                                 &                                        &                                       &                      & No                   & TLSv1.2              & 7.54                       \\
						\devNetatmoCam          & No                                     & IPsec                                 & 8.00                                 & Partially                              & HTTP                                  & 7.97                 & No                   & TLSv1.2              & 7.98                       \\
						\devDlinkCam            & Yes                                    & None                                  & 5.40                                 &                                        &                                       &                      &                      &                      &                            \\
						\devHelloBarbie         & No                                     & TLSv1.2                               & 7.99                                 &                                        &                                       &                      &                      &                      &                            \\
						\devWithingsSleep       & No                                     & Unknown                               & 7.84                                 &                                        &                                       &                      & No                   & TLSv1.2              & 7.63                       \\
						\devDropcam             & No                                     & TLSv1.2                               & 7.99                                 &                                        &                                       &                      & No                   & TLSv1.2              & 7.94                       \\

						\hline
					\end{tabular}
				\end{adjustbox}
			\end{table}
			\egroup		
				
			\begin{figure}[t!]
				\centering
				\includegraphics[width=0.6\textwidth]{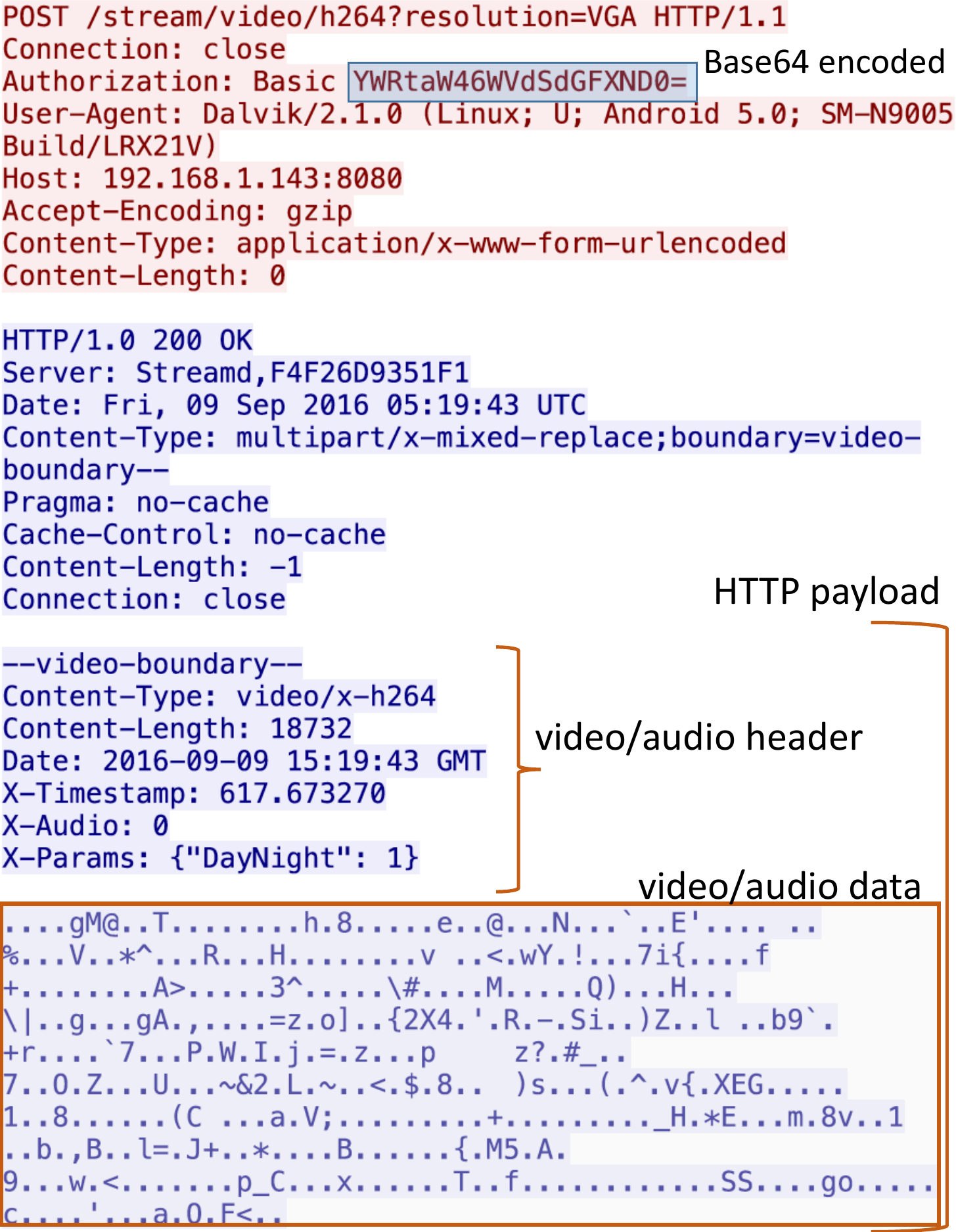}
				\caption{TPLink camera POST message}
				\label{fig:tplinkcam}
			\end{figure}
		
			\begin{figure}[t]
				\centering
				\includegraphics[width=0.8\textwidth]{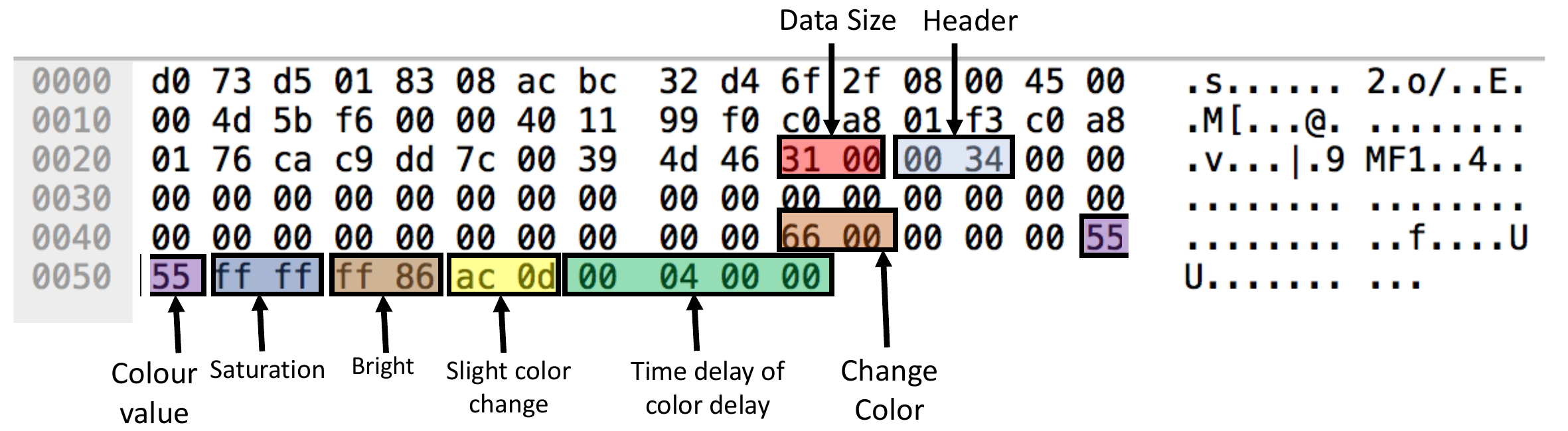}
				\caption{Control bit pattern of LiFX lightbulb}
				\label{fig:lifx}
			\end{figure}
		
			Our confidentiality assessment results are shown in Table~\ref{tab:confidEval} by three measures over three communication channels (as discussed in \S\ref{syseval:method:conf}).
			{\rev
				It can be seen that most of the devices have fairly secure communication in two channels namely device-to-server and mobile-app-to-server since they use secure protocols like TLS/SSL most of the time. However, a majority of the vulnerabilities arise when the device communicates with the mobile-App -- five devices send in plaintext, only one device uses SSL with fairly lower entropy values. 
			}
			Note that for some devices (e.g. Belkin Switch, Samsung Smart Cam), the security protocol is not identified but together with plaintext and entropy tests, we can evaluate the confidentiality of a given channel. Considering the user privacy, we see quite a few devices such as Phillips Hue lightbulb, Belkin power switch, HP Envy printer,  TPLink camera, and Belkin motion sensor, communicate in plaintext (some of them were discussed in \cite{M2Msec14}), -- revealing private information, for example, whether the Belkin power switch is on/off,  or when the Phillips Hue lightbulb was last used. 

			Our results also enable us to discover new vulnerabilities in some devices such as the TPLink camera.
			Fig.~\ref{fig:tplinkcam}, which depicts a detailed insight into packets captured from the TPLink camera (i.e. a \verb|POST| request packet payload in red text followed by the \verb|HTTP| response packet in blue text). The video/audio stream is sent in plaintext (the video/audio header is human-readable even though its data doesn't seem human-readable). This data can be sniffed by an attacker and then used to reassemble the video/audio data.
			Surprisingly, it reveals not only the video/audio data but also the authentication password required for logging in to the device. This password is exposed in the basic authentication field of the packet shown in Fig.~\ref{fig:tplinkcam} (i.e. \verb|YWRtaW46WvdSdGFND0=|)  -- this is a Base64 version of ``admin''. Given the password, we are able to log into the device by simply guessing the user-name as ``admin'' which is a common default credential used in many IoT devices. 
			\vspace{0.5em}
			
\begin{figure}[t!]
				\centering
				\includegraphics[width=0.8\textwidth]{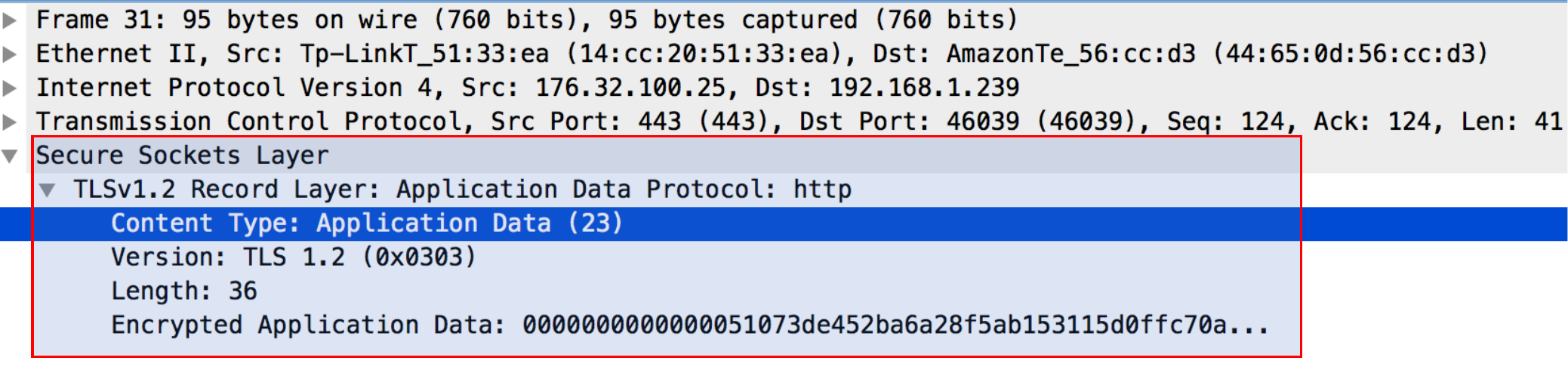}
				\caption{Wireshark capture of Amazon Echo}
				\label{fig:amazon}
			\end{figure}

\begin{figure}[b!]
				\centering
				\includegraphics[width=0.8\textwidth]{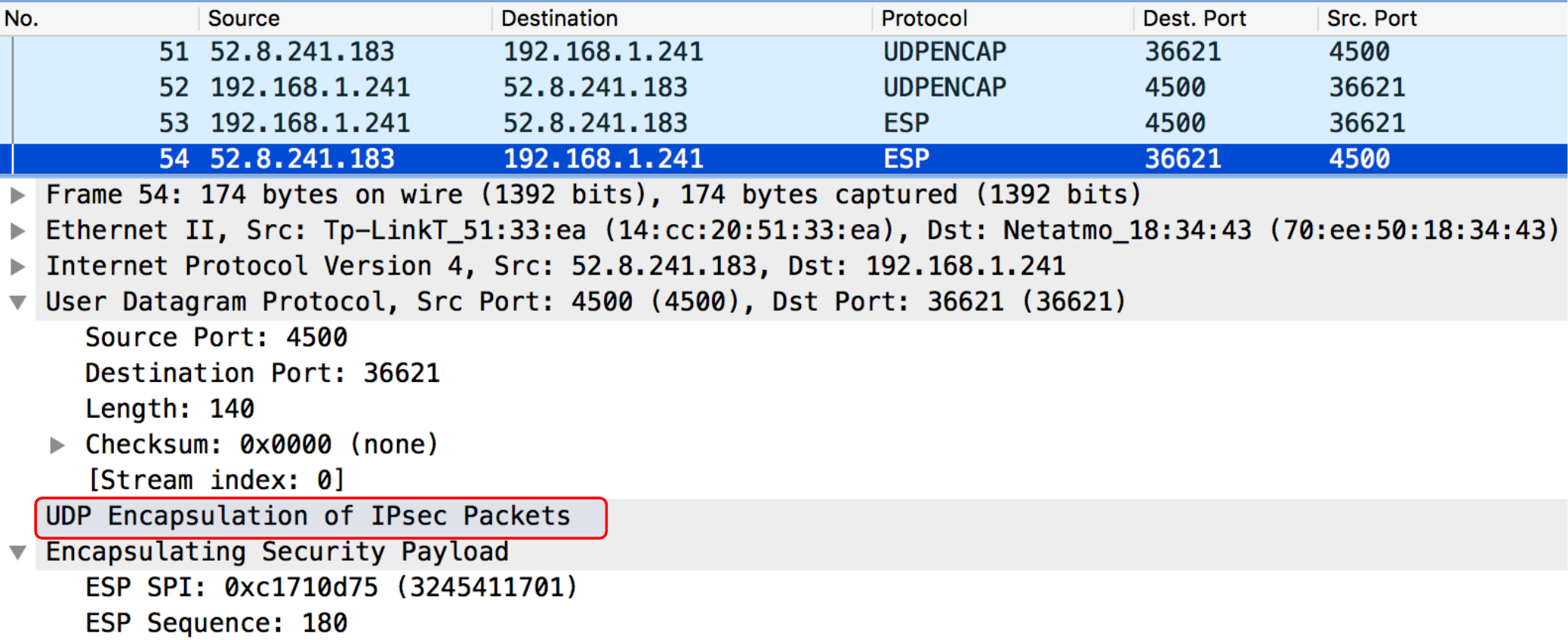}
				\caption{Wireshark capture of Netatmo camera}
				\label{fig:netatmocam}
			\end{figure}

			The efficacy of our entropy measure can be seen in the LiFX lighbulb. Our plaintext test for this device shows that the LiFX bulb is not communicating in a human-readable format, whereas its traffic data has a low entropy value of $4.56$. When taking a closer look into the LiFX packets, we are able to discover that packets associated with certain commands (from the user App) are identical and certain bits represent specific functions of the device, meaning that the data is just encoded as shown in Fig.~\ref{fig:lifx}. Similarly in the TPLink power switch, we see that the data is not in plaintext but the entropy value is $5.33$, suggesting that it could possibly be encoded or poorly encrypted. By guessing that the data is sent in JSON format (i.e. \verb|{data}|), we attempt to XOR the first byte with the character ``\{'' to obtain the single byte key. We then apply the key to the encrypted message and are able to extract the message in plaintext. This indicates a weak encryption is used in the TPLink power switch. Note that some devices employ stronger encryption protocols. For example, Amazon Echo uses TLSv1.2 for all traffic it communicates (shown in Fig.~\ref{fig:amazon}), or Netamo camera implements IPsec, protecting the IP address of endpoints from potential attackers (shown in Fig.~\ref{fig:netatmocam}).
			

\begin{figure}[t!]
				\centering
				\includegraphics[width=0.6\textwidth]{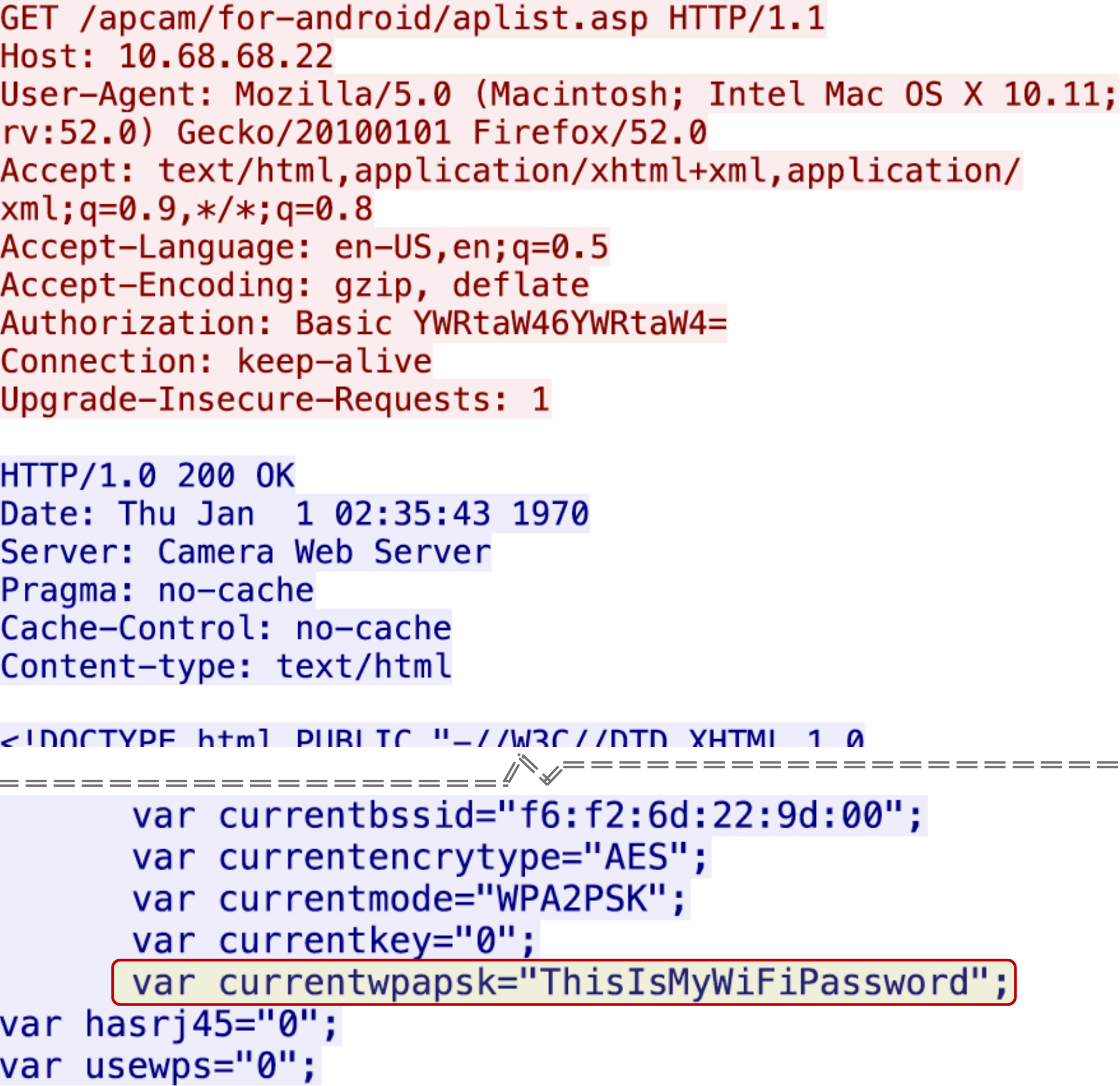}
				\caption{Belkin camera is pairing with user app}
				%\vspace{2mm}
				\label{fig:belkincam}
			\end{figure}

			Lastly, we evaluate the confidentiality of devices' communication after their initial setup phase is complete. There are, however, some devices that communicate in an insecure manner when they initially pair with the user App. For example, Fig.~\ref{fig:belkincam} shows that Belkin camera exposes the password of the local WiFi network in plaintext (i.e. \verb|ThisIsMyWiFiPassword| in Fig.~\ref{fig:belkincam}) when responding to a \verb|GET| request.
			
		\subsubsection{Integrity and Authentication}

			Our assessment results for the posture of integrity and authentication in twenty IoT devices are shown in Table~\ref{tab:integEval}. Considering the test for replay attacks, five of our IoT devices are susceptible such as the Philips Hue light bulb, Belkin power switch, HP Envy printer, LiFX light bulb, and TPLink switch. Some of these exploits have been already reported. For example, the Belkin switch was evaluated to be insecure against replay attacks due to the lack of authentication \cite{M2Msec14} or the LiFX lightbulb that communicates encoded messages with the user App \cite{LiFXpkt15}. An attacker can turn on/off the Belkin switch with a well-crafted fresh packet, or change the color/brightness of the LiFX bulb using the control bit pattern shown in Fig.~\ref{fig:lifx}. On the other hand, those IoT devices that employ secure protocols (e.g. SSL) are protected against replay attacks such as the Awair air monitor and Amazon Echo.

			\bgroup
			\def\arraystretch{1.5}
			\bordersForColoredTable
			\begin{table}[t!]
				\rowcolors{1}{WhiteCell}{LightGrayCell}
				\centering
				\caption{Posture of integrity and authentication}
				\label{tab:integEval}
				\begin{adjustbox}{max width=0.6\textwidth}
					{\rev
					\begin{tabular}{|l|c|c|c|}
						\hline
						%					\rowcolor{WhiteCell}
						\multicolumn{1}{|c|}{\textbf{Devices}} & \textbf{Replay Attack} & \textbf{DNS spoofing} & \textbf{Fake Server} \\ \hline
						\devHuebulb             & Yes                   & Yes                   & HTTP                 \\
						\devBelkinSwitch        & Yes                   & Yes                   & Fail SSL             \\
						\devSamsungcam          & No                    & Yes                   & Fail SSL             \\
						\devBelkinCam           & No                    & Yes                   & Fail SSL             \\
						\devAwair               & No                    & Yes                   & Fail SSL             \\
						\devHPprinter           & Yes                   & Yes                   & Fail SSL             \\
						\devLifx                & Yes                   & Yes                   & Plaintext            \\
						\devCanaryCam           & No                    & Yes                   & Fail SSL             \\
						\devTPswitch            & Yes                   & Yes                   & Fail SSL             \\
						\devEcho                & No                    & Yes                   & Fail SSL             \\
						\devSmartThings         & No                    & Yes                   & Fail SSL             \\
						\devPixstar             & No                    & Yes                   & Fail SSL             \\
						\devTPcam               & No                    & Yes                   & Fail SSL             \\
						\devBelkinMotion        & Yes                   & Yes                   & Plaintext            \\
						\devNestSmoke           & No                    & Yes                   & Fail SSL             \\
						\devNetatmoCam          & No                    & Yes                   & Fail Ipsec           \\
						\devDlinkCam            & Yes                   & Yes                   & Plaintext            \\
						\devHelloBarbie         & No                    & Yes                   & Fail SSL             \\
						\devWithingsSleep       & No                    & Yes                   & Fail SSL             \\
						\devDropcam             & No                    & Yes                   & Fail SSL             \\ \hline                                                                                                                           
					\end{tabular}
							}
				\end{adjustbox}
			\end{table}
			\egroup

			Our DNS security test results show that none of 20 IoT devices implements DNSSEC protocol that is primarily designed to prevent DNS spoofing attacks. This vulnerability enables attackers to hijack the DNS query and possibly impersonate the legitimate server to the IoT device.  Even if DNS spoofing is successful, the victim IoT device may protect itself by some form of authentication.  According to the last column of Table~\ref{tab:integEval}, some devices such as the Phillips Hue lightbulb and LiFX bulb do communicate with the fake server, after a successful DNS spoofing. The Phillips Hue lightbulb sends an HTTP message to the fake server that is listening on the same port as the real server, while the LiFX bulb sends data to our fake server which appears to be in its own unique data format (as shown in Fig.~\ref{fig:lifx}).
			\vspace{1em}

		\subsubsection{Access Control and Availability}
		
			\bgroup
			\def\arraystretch{1.3}
			\bordersForColoredTable
			\begin{table}[t!]
				\rowcolors{1}{WhiteCell}{LightGrayCell}
				\centering
				\caption{Posture of access control and availability}
				\label{tab:accessEval}
				\begin{adjustbox}{max width=\textwidth}
					\begin{tabular}{|l|p{2.5cm}|p{5cm}|C{2cm}|C{2cm}|C{2cm}|C{2cm}|C{2cm}|}
						\hline		
						\multicolumn{1}{|c|}{\textbf{\shortstack[c]{Devices}}} & \multicolumn{1}{c|}{\textbf{\shortstack[c]{Open Ports\\(TCP)}}} & \multicolumn{1}{c|}{\textbf{\shortstack[c]{Open Ports\\(UDP)}}} & \multicolumn{1}{c|}{\textbf{\shortstack[c]{Vulnerable\\Ports}}}  & \multicolumn{1}{c|}{\textbf{\shortstack[c]{Weak\\Passwords}}} & \multicolumn{1}{c|}{\textbf{\shortstack[c]{ICMP\\DoS}}} & \multicolumn{1}{c|}{\textbf{\shortstack[c]{UDP\\DoS}}} & \multicolumn{1}{c|}{\textbf{\shortstack[c]{No. of\\TCP Con.}}}\\ \hline
						\devHuebulb             & 80, 8080                                          & 1900, 5353                                                                                                                                                                                                             & 80                                            & No                                          & Protected                              & Protected                    & 112                      \\
						\devBelkinSwitch        & 53, 49155                                         & 53, 1900, 3111, 7638, 13965, 14675, 17143, 19422, 22894, 23835, 26011, 27047, 38849, 40014, 41970, 42518, 43403, 47836, 53121, 53330, 55353, 65484                                                                     & None                                          &                                             & 23Mbps                                 & 6.3Mbps                      & 97                       \\
						\devSamsungcam          & 80, 443, 554, 943, 4520, 49152                    & 161, 5353                                                                                                                                                                                                              & 80                                            & No                                          & 90Mbps                                 & 4.1Mbps                      & 17                       \\
						\devBelkinCam           & 80, 81, 443, 9964, 49153                          & 1900, 10000, 13105, 19827, 26854, 28971, 32596, 32435, 33435, 35042, 35316, 35056, 36500, 36943, 38587, 38606, 39632, 39714, 43588, 43834, 47709, 48190, 44179, 49156, 49201, 49360, 52042, 52144, 52603, 55254, 56284 & 80                                            & No                                          & 7.7Mbps                                & 74Kbps                       & 256                      \\
						\devAwair               & Filtered                                          & Filtered                                                                                                                                                                                                               &                                               &                                             & 36Mbps                                 & 7.2Mbps                      &                          \\
						\devHPprinter           & 80, 443, 631, 3910, 3911, 8080, 9100, 9220, 53048 & 137, 161, 543, 3702, 5353, 5355, 7235, 53592, 56693, 56723                                                                                                                                                             & 80, All ports allow telnet                    & No                                          &                                        &                              & 1                        \\
						\devLifx                & Closed                                            & Filtered                                                                                                                                                                                                               &                                               &                                             & 6Mbps                                  & 82Kbps                       &                          \\
						\devCanaryCam           & Closed                                            & Closed                                                                                                                                                                                                                 &                                               &                                             & 6.4Mbps                                &                              &                          \\
						\devTPswitch            & 80, 9999                                          & 1040                                                                                                                                                                                                                   & 80                                            & No                                          & 5.5Mbps                                & 25Mbps                       & 15                       \\
						\devEcho                & 4070                                              & 5353                                                                                                                                                                                                                   & None                                          &                                             & Protected                              & 9.2Mbps                      & 258                      \\
						\devSmartThings         & 23, 39500                                         & Filtered                                                                                                                                                                                                               & 23                                            & No                                          & 130Mbps                                & 8.8Mbps                      & 1                        \\
						\devPixstar             & Closed                                            & 137                                                                                                                                                                                                                    &                                               &                                             & Protected                              & Protected                    &                          \\
						\devTPcam               & 80, 554, 1935, 2020, 8080                         & 1068, 3702, 5353, 42941                                                                                                                                                                                                & 80                                            & Yes                                         & 48Mbps                                 & 870Kbps                      & 130                      \\
						\devBelkinMotion        & 53, 49152                                         & 53, 1900, 3080, 3081, 3082, 3179, 3229, 3236, 3619, 4050, 4052, 4053, 4054, 4055, 4289, 4996, 4997, 4998, 14675                                                                                                        & None                                          &                                             & 11.3Mbps                               & 350Kbps                      & 109                      \\
						\devNestSmoke           & Closed and filtered                               & 17395, 17466, 17471, 18184, 18234, 18455, 18721, 18916, 19090, 19112, 19217, 19458, 19581                                                                                                                              &                                               &                                             & Protected                              & Protected                    &                          \\
						\devNetatmoCam          & 80, 5555                                          & 654, 7242, 26082, 29110, 31574, 35826, 39408, 46721, 48080, 56943                                                                                                                                                      & 80                                            & No                                          & 8.2 Mbps                               & 45Kbps                       & 256                      \\
						\devDlinkCam            & 21, 23, 5001, 5004, 16119                         & 1900, 5002, 5003, 10000                                                                                                                                                                                                & 5004                                          & No Password                                 & 49Mbps                                 & 292Kbps                      & 20                       \\
						\devHelloBarbie         & Closed                                            & Closed                                                                                                                                                                                                                 &                                               &                                             & 10Mbps                                 &                              &                          \\
						\devWithingsSleep       & 22, 7685, 7888                                    & 5353                                                                                                                                                                                                                   & 22                                            & No                                          & Protected                              & Protected                    & 22                       \\
						\devDropcam             & Closed                                            & Closed of filtered                                                                                                                                                                                                     &                                               &                                             & 4Mbps                                  &                              &                          \\ \hline
					\end{tabular}	
				\end{adjustbox}
			\end{table}
			\egroup
			
			{\rev
				
				Our access control and availability evaluation shown in Table~\ref{tab:accessEval} assesses the state of ports as: ``open'' indicating that a service is actively accepting TCP connections and/or UDP datagrams, ``closed'' indicating that the port receives and responds to probe packets, but there is no service listening on it, and ``filtered'' indicating a port scanner cannot determine whether the port is open or closed (a filtering method prevents probes to reach the port).
			}
			The results shown in Table~\ref{tab:accessEval} indicate that almost all devices have some form of vulnerabilities in terms of open ports which enable intruders to communicate with or access into the device. 
			For example, the Belkin smart camera exposes a large number of ports, 5 TCP and 31 UDP. 
			Another vulnerable device is the HP printer with 9 open TCP ports and 10 open UDP ports. Among all these open ports, we note that the HP printer responds on a special TCP port 9100 that is used for printing with no authorization -- this vulnerability was recently exploited to attack more than 150000 printers \cite{stackoverflowin17}. 
			{ \rev
				On the other hand, a device like the Awair air monitor has all ports closed, and hence is protected against common attacks such as SYN flooding.
			}
			
			We note that some IoT devices allow remote access via SSH (port 22), Telnet (port 23), or HTTP(port 80). Until recently, many IoT devices had weak credentials (from a list of about 60 common defaults) that Mirai malware \cite{Mirai16} exploited to hijack hundreds of thousands of IoTs, launching a major DDoS attack on the Internet. 
			None of these 60 defaults were valid when we used them for our 20 IoT devices. 
			Surprisingly, we have two devices with no protection for remote access: HP printer allows Telnet without asking for a password; and the DLink camera asks for no credentials during SSH access -- some manufacturers seemingly open remote access ports for testing/debugging purposes.
				
			From the DoS attack test results shown in Table~\ref{tab:accessEval}, it can be seen that most devices are susceptible to at least one form of DoS attacks, either of ICMP-, UDP- or TCP-based. 
			We note that the required traffic rate to cause a device to stop functioning is not significant in many cases especially when UDP is used (i.e. less than 1 Mbps for Belkin SmartCam, LiFX lightbulb or TPLink camera). For Samsung Smart camera, it can handle ICMP traffic rate up to $90$ Mbps, however it stops functioning (the camera will not be able to transmit live video stream to the user App), if it is bombarded by UDP-based traffic at a rate more than $4.1$ Mbps.
		
		\subsubsection{Reflection Attacks}
						\bgroup
			\def\arraystretch{1.5}
			\bordersForColoredTable
			\begin{table*}[t!]
				\rowcolors{1}{WhiteCell}{LightGrayCell}
				\centering
				
				\caption{Posture of reflection attacks}
				
				\label{tab:reflectionEval}
				\begin{adjustbox}{max width=0.7\textwidth}
					\begin{tabular}{|l|c|c|c|}
						\hline		
						\multicolumn{1}{|c|}{\textbf{\shortstack[c]{Devices}}} & \multicolumn{1}{c|}{\textbf{\shortstack[c]{ICMP Reflection}}} & \multicolumn{1}{c|}{\textbf{\shortstack[c]{SSDP Reflection}}} & \multicolumn{1}{c|}{\textbf{\shortstack[c]{SNMP Reflection}}} \\ \hline
						\devHuebulb             & Yes                      & Yes                      & No                       \\
						\devBelkinSwitch        & Yes                      & Yes                      & No                       \\
						\devSamsungcam          & Yes                      & No                       & v2c                      \\
						\devBelkinCam           & Yes                      & Yes                      & No                       \\
						\devAwair               & Yes                      & No                       & No                       \\
						\devHPprinter           & Yes                      & No                       & v1                       \\
						\devLifx                & No                       & No                       & No                       \\
						\devCanaryCam           & Yes                      & No                       & No                       \\
						\devTPswitch            & Yes                      & No                       & No                       \\
						\devEcho                & Yes                      & No                       & No                       \\
						\devSmartThings         & Yes                      & No                       & No                       \\
						\devPixstar             & Yes                      & No                       & No                       \\
						\devTPcam               & Yes                      & No                       & No                       \\
						\devBelkinMotion        & Yes                      & Yes                      & No                       \\
						\devNestSmoke           & Yes                      & No                       & No                       \\
						\devNetatmoCam          & Yes                      & No                       & No                       \\
						\devDlinkCam            & Yes                      & Yes                      & No                       \\
						\devHelloBarbie         & Yes                      & No                       & No                       \\
						\devWithingsSleep       & Yes                      & No                       & No                       \\
						\devDropcam             & Yes                      & No                       & No                       \\ \hline
					\end{tabular}	
				\end{adjustbox}
			\end{table*}

			Lastly, we consider ICMP, SSDP and SNMP protocols by checking if a given device reflects traffic of these types. Our results are shown in Table~\ref{tab:reflectionEval}. We can see that all devices, except the LiFX lightbulb, are reflecting ICMP traffic. 
			We then test the SSDP protocol which is commonly enabled in many IoT devices for ease of discovery. When we use SSDP, the reflected traffic (i.e. response) is amplified by a large factor since it contains service and presence information of the IoT device -- this makes it an attractive protocol for DDoS attackers. We observe that five of our devices are vulnerable to SSDP reflection attacks -- the rest of them do not use SSDP for discovery. Lastly, we examine SNMP protocol which is not widely used by IoT devices. Furthermore, with SNMP v2c (and v3), it is possible to use public community strings such that the amplification factor is significantly high. The SNMP v2c is only available in the Samsung Smart camera. Sending a \verb|getBulk| request to the camera, it will iterate the \verb|getNext| request multiple times, and hence a larger amount of traffic is generated.

		{\rev 
		\subsection{Security Rating of IoT Devices}\label{sec:ch2_rating}
		
		\bgroup
		\def\arraystretch{1.5}
		\begin{table*}[!t]
			\centering
			\vspace{-3mm}	
			\caption{Security rating}
			\vspace{-3mm}	
			\label{tab:ch2_securityrating}
			\begin{adjustbox}{max width=\textwidth}
				\begin{tabular}{|>{\raggedright\arraybackslash}p{4cm}|>{\centering\arraybackslash}p{1cm}|>{\centering\arraybackslash}p{1cm}|>{\centering\arraybackslash}p{1cm}|>{\centering\arraybackslash}p{1cm}|>{\centering\arraybackslash}p{1cm}|>{\centering\arraybackslash}p{1cm}|>{\centering\arraybackslash}p{1cm}|>{\centering\arraybackslash}p{1cm}|>{\centering\arraybackslash}p{1cm}|>{\centering\arraybackslash}p{1cm}|>{\centering\arraybackslash}p{1cm}|>{\centering\arraybackslash}p{1cm}|>{\centering\arraybackslash}p{1cm}|>{\centering\arraybackslash}p{1cm}|>{\centering\arraybackslash}p{1cm}|>{\centering\arraybackslash}p{1cm}|>{\centering\arraybackslash}p{1cm}|>{\centering\arraybackslash}p{1cm}|>{\centering\arraybackslash}p{1cm}|>{\centering\arraybackslash}p{1cm}|>{\centering\arraybackslash}p{1cm}|>{\centering\arraybackslash}p{1cm}|>{\centering\arraybackslash}p{1cm}|>{\centering\arraybackslash}p{1cm}|>{\centering\arraybackslash}p{1cm}|>{\centering\arraybackslash}p{1cm}|>{\centering\arraybackslash}p{1cm}|>{\centering\arraybackslash}p{1cm}|>{\centering\arraybackslash}p{1cm}|>{\centering\arraybackslash}p{1cm}|>{\centering\arraybackslash}p{1cm}|>{\centering\arraybackslash}p{1cm}|>{\centering\arraybackslash}p{1cm}|>{\centering\arraybackslash}p{1cm}|>{\centering\arraybackslash}p{5cm}|}
					\hline
					& \multicolumn{10}{c|}{\textbf{Confidentiality}}                                                                                                   & \multicolumn{3}{>{\centering\arraybackslash}p{3.6cm}|}{\textbf{Integrity and Authentication}}                                                                        & \multicolumn{7}{c|}{\textbf{Access Control}}                                                                                                                                                                                                              & \multicolumn{4}{c|}{\textbf{Reflection Attacks}}                                                                                                                  \\ \cline{2-25}
					& \multicolumn{3}{c|}{\textbf{\shortstack{ \\Device\\to\\Server}}} & \multicolumn{3}{c|}{\textbf{\shortstack{ \\Device\\to\\Application}}} & \multicolumn{3}{c|}{\textbf{\shortstack{ \\Application\\to\\Server}}} & \textbf{All}     & \multirow{2}{*}{ \rotatebox{90}{\parbox{4cm}{\textbf{Replay \newline Attack}}}} & \multirow{2}{*}{\rotatebox{90}{\parbox{4cm}{\textbf{DNS \newline Spoofing}}}} & \multirow{2}{*}{\rotatebox{90}{\parbox{4cm}{\textbf{Fake Server}}}}& \multirow{2}{*}{\rotatebox{90}{\parbox{4cm}{\textbf{Open Ports \newline (TCP)}}}} & \multirow{2}{*}{\rotatebox{90}{\parbox{4cm}{\textbf{Open Ports \newline (UDP)}}}} & \multirow{2}{*}{\rotatebox{90}{\parbox{4cm}{\textbf{Vulnerable Ports}}}} & \multirow{2}{*}{\rotatebox{90}{\parbox{4cm}{\textbf{Weak\newline Passwords}}}} & \multirow{2}{*}{\rotatebox{90}{\parbox{4cm}{\textbf{ICMP DoS}}}} & \multirow{2}{*}{\rotatebox{90}{\parbox{4cm}{\textbf{UDP DoS}}}} & \multirow{2}{*}{\rotatebox{90}{\parbox{4cm}{\textbf{No. of TCP \newline Connections}}}} & \multirow{2}{*}{\rotatebox{90}{\parbox{4cm}{\textbf{ICMP \newline Reflection}}}} & \multirow{2}{*}{\rotatebox{90}{\parbox{4cm}{\textbf{SSDP \newline Reflection}}}} & \multirow{2}{*}{\rotatebox{90}{\parbox{4cm}{\textbf{SNMP \newline Reflection}}}} & \multirow{2}{*}{\rotatebox{90}{\parbox{4cm}{\textbf{SNMP Public \newline Community String}}}} \\ \cline{2-11} 
					\multicolumn{1}{|c|}{\textbf{Devices}}                 & \rotatebox{90}{\parbox{3.2cm}{\textbf{Plaintext}}}&\begin{turn}{90}\textbf{Protocol}\end{turn}&\begin{turn}{90}\textbf{Entropy}\end{turn}&\begin{turn}{90}\textbf{Plaintext}\end{turn}&\begin{turn}{90}\textbf{Protocol}\end{turn}&\begin{turn}{90}\textbf{Entropy}\end{turn}&\begin{turn}{90}\textbf{Plaintext}        \end{turn}&\begin{turn}{90}\textbf{Protocol}\end{turn}&\begin{turn}{90}\textbf{Entropy}\end{turn}&\begin{turn}{90}\textbf{Privacy}\end{turn}&&&&&&&&&&&&&&\\ \hline
					Phillip Hue lightbulb   &\cellcolor{GreenCell}A &\cellcolor{GreenCell}A &\cellcolor{GreenCell}A &\cellcolor{RedCell}C   &\cellcolor{RedCell}C   &\cellcolor{RedCell}C   &\cellcolor{GreenCell}A &\cellcolor{GreenCell}A &\cellcolor{GreenCell}A &\cellcolor{RedCell}C   &\cellcolor{RedCell}C   &\cellcolor{RedCell}C   &\cellcolor{RedCell}C   &\cellcolor{RedCell}C   &\cellcolor{RedCell}C   &\cellcolor{RedCell}C   &\cellcolor{GreenCell}A &\cellcolor{YellowCell}B&\cellcolor{RedCell}C   &\cellcolor{RedCell}C   &\cellcolor{RedCell}C   &\cellcolor{RedCell}C   &\cellcolor{GreenCell}A &\cellcolor{GreenCell}A \\ \hline
					Belkin Switch           &\cellcolor{YellowCell}B&\cellcolor{GrayCell}   &\cellcolor{GreenCell}A &\cellcolor{RedCell}C   &\cellcolor{RedCell}C   &\cellcolor{RedCell}C   &\cellcolor{GreenCell}A &\cellcolor{GreenCell}A &\cellcolor{GreenCell}A &\cellcolor{RedCell}C   &\cellcolor{RedCell}C   &\cellcolor{RedCell}C   &\cellcolor{GreenCell}A &\cellcolor{RedCell}C   &\cellcolor{RedCell}C   &\cellcolor{GreenCell}A &\cellcolor{GreenCell}A &\cellcolor{RedCell}C   &\cellcolor{RedCell}C   &\cellcolor{RedCell}C   &\cellcolor{RedCell}C   &\cellcolor{RedCell}C   &\cellcolor{GreenCell}A &\cellcolor{GreenCell}A \\ \hline
					Samsung Smart Cam       &\cellcolor{GreenCell}A &\cellcolor{GrayCell}   &\cellcolor{GreenCell}A &\cellcolor{GreenCell}A &\cellcolor{GreenCell}A &\cellcolor{GreenCell}A &\cellcolor{GreenCell}A &\cellcolor{GreenCell}A &\cellcolor{GreenCell}A &\cellcolor{GreenCell}A &\cellcolor{GreenCell}A &\cellcolor{RedCell}C   &\cellcolor{GreenCell}A &\cellcolor{RedCell}C   &\cellcolor{RedCell}C   &\cellcolor{RedCell}C   &\cellcolor{GreenCell}A &\cellcolor{RedCell}C   &\cellcolor{RedCell}C   &\cellcolor{RedCell}C   &\cellcolor{RedCell}C   &\cellcolor{GreenCell}A &\cellcolor{RedCell}C   &\cellcolor{RedCell}C   \\ \hline
					Belkin Smart Cam        &\cellcolor{GreenCell}A &\cellcolor{GrayCell}   &\cellcolor{GreenCell}A &\cellcolor{GreenCell}A &\cellcolor{GreenCell}A &\cellcolor{GreenCell}A &\cellcolor{GreenCell}A &\cellcolor{GreenCell}A &\cellcolor{GreenCell}A &\cellcolor{GreenCell}A &\cellcolor{GreenCell}A &\cellcolor{RedCell}C   &\cellcolor{GreenCell}A &\cellcolor{RedCell}C   &\cellcolor{RedCell}C   &\cellcolor{RedCell}C   &\cellcolor{GreenCell}A &\cellcolor{RedCell}C   &\cellcolor{YellowCell}B&\cellcolor{RedCell}C   &\cellcolor{RedCell}C   &\cellcolor{RedCell}C   &\cellcolor{GreenCell}A &\cellcolor{GreenCell}A \\ \hline
					Awair air monitor       &\cellcolor{GreenCell}A &\cellcolor{GreenCell}A &\cellcolor{GreenCell}A &\cellcolor{GreenCell}A &\cellcolor{GreenCell}A &\cellcolor{GreenCell}A &\cellcolor{GreenCell}A &\cellcolor{GreenCell}A &\cellcolor{GreenCell}A &\cellcolor{GreenCell}A &\cellcolor{GreenCell}A &\cellcolor{RedCell}C   &\cellcolor{GreenCell}A &\cellcolor{YellowCell}B&\cellcolor{YellowCell}B&\cellcolor{GreenCell}A &\cellcolor{GreenCell}A &\cellcolor{RedCell}C   &\cellcolor{RedCell}C   &\cellcolor{GreenCell}A &\cellcolor{RedCell}C   &\cellcolor{GreenCell}A &\cellcolor{GreenCell}A &\cellcolor{GreenCell}A \\ \hline
					HP Envy Printer         &\cellcolor{GreenCell}A &\cellcolor{GreenCell}A &\cellcolor{GreenCell}A &\cellcolor{RedCell}C   &\cellcolor{RedCell}C   &\cellcolor{RedCell}C   &\cellcolor{GreenCell}A &\cellcolor{GreenCell}A &\cellcolor{GreenCell}A &\cellcolor{RedCell}C   &\cellcolor{RedCell}C   &\cellcolor{RedCell}C   &\cellcolor{GreenCell}A &\cellcolor{RedCell}C   &\cellcolor{RedCell}C   &\cellcolor{RedCell}C   &\cellcolor{GreenCell}A &\cellcolor{GreenCell}A &\cellcolor{GreenCell}A &\cellcolor{RedCell}C   &\cellcolor{RedCell}C   &\cellcolor{GreenCell}A &\cellcolor{RedCell}C   &\cellcolor{GreenCell}A \\ \hline
					LiFX lightbulb          &\cellcolor{GreenCell}A &\cellcolor{GreenCell}A &\cellcolor{GreenCell}A &\cellcolor{GreenCell}A &\cellcolor{GrayCell}   &\cellcolor{RedCell}C   &\cellcolor{GreenCell}A &\cellcolor{GreenCell}A &\cellcolor{GreenCell}A &\cellcolor{GreenCell}A &\cellcolor{RedCell}C   &\cellcolor{RedCell}C   &\cellcolor{RedCell}C   &\cellcolor{GreenCell}A &\cellcolor{YellowCell}B&\cellcolor{GreenCell}A &\cellcolor{GreenCell}A &\cellcolor{RedCell}C   &\cellcolor{YellowCell}B&\cellcolor{GreenCell}A &\cellcolor{GreenCell}A &\cellcolor{GreenCell}A &\cellcolor{GreenCell}A &\cellcolor{GreenCell}A \\ \hline
					Canary Camera           &\cellcolor{GreenCell}A &\cellcolor{GreenCell}A &\cellcolor{GreenCell}A &\cellcolor{GreenCell}A &\cellcolor{GreenCell}A &\cellcolor{GreenCell}A &\cellcolor{GreenCell}A &\cellcolor{GreenCell}A &\cellcolor{GreenCell}A &\cellcolor{GreenCell}A &\cellcolor{GreenCell}A &\cellcolor{RedCell}C   &\cellcolor{GreenCell}A &\cellcolor{GreenCell}A &\cellcolor{GreenCell}A &\cellcolor{GreenCell}A &\cellcolor{GreenCell}A &\cellcolor{RedCell}C   &\cellcolor{GreenCell}A &\cellcolor{GreenCell}A &\cellcolor{RedCell}C   &\cellcolor{GreenCell}A &\cellcolor{GreenCell}A &\cellcolor{GreenCell}A \\ \hline
					TPLink Switch           &\cellcolor{GreenCell}A &\cellcolor{GrayCell}   &\cellcolor{GreenCell}A &\cellcolor{GreenCell}A &\cellcolor{GrayCell}   &\cellcolor{RedCell}C   &\cellcolor{GreenCell}A &\cellcolor{GreenCell}A &\cellcolor{GreenCell}A &\cellcolor{GreenCell}A &\cellcolor{RedCell}C   &\cellcolor{RedCell}C   &\cellcolor{GreenCell}A &\cellcolor{RedCell}C   &\cellcolor{RedCell}C   &\cellcolor{RedCell}C   &\cellcolor{GreenCell}A &\cellcolor{RedCell}C   &\cellcolor{RedCell}C   &\cellcolor{RedCell}C   &\cellcolor{RedCell}C   &\cellcolor{GreenCell}A &\cellcolor{GreenCell}A &\cellcolor{GreenCell}A \\ \hline
					Amazon Echo             &\cellcolor{GreenCell}A &\cellcolor{GreenCell}A &\cellcolor{GreenCell}A &\cellcolor{GreenCell}A &\cellcolor{GreenCell}A &\cellcolor{GreenCell}A &\cellcolor{GreenCell}A &\cellcolor{GreenCell}A &\cellcolor{GreenCell}A &\cellcolor{GreenCell}A &\cellcolor{GreenCell}A &\cellcolor{RedCell}C   &\cellcolor{GreenCell}A &\cellcolor{RedCell}C   &\cellcolor{RedCell}C   &\cellcolor{GreenCell}A &\cellcolor{GreenCell}A &\cellcolor{YellowCell}B&\cellcolor{RedCell}C   &\cellcolor{RedCell}C   &\cellcolor{RedCell}C   &\cellcolor{GreenCell}A &\cellcolor{GreenCell}A &\cellcolor{GreenCell}A \\ \hline
					Samsung Smart Things    &\cellcolor{GreenCell}A &\cellcolor{GreenCell}A &\cellcolor{GreenCell}A &\cellcolor{GreenCell}A &\cellcolor{GreenCell}A &\cellcolor{GreenCell}A &\cellcolor{GreenCell}A &\cellcolor{GreenCell}A &\cellcolor{GreenCell}A &\cellcolor{GreenCell}A &\cellcolor{GreenCell}A &\cellcolor{RedCell}C   &\cellcolor{GreenCell}A &\cellcolor{RedCell}C   &\cellcolor{YellowCell}B&\cellcolor{RedCell}C   &\cellcolor{GreenCell}A &\cellcolor{RedCell}C   &\cellcolor{RedCell}C   &\cellcolor{RedCell}C   &\cellcolor{RedCell}C   &\cellcolor{GreenCell}A &\cellcolor{GreenCell}A &\cellcolor{GreenCell}A \\ \hline
					Pixstar Photo Frame     &\cellcolor{GreenCell}A &\cellcolor{GreenCell}A &\cellcolor{GreenCell}A &\cellcolor{GreenCell}A &\cellcolor{GreenCell}A &\cellcolor{GreenCell}A &\cellcolor{GreenCell}A &\cellcolor{GreenCell}A &\cellcolor{GreenCell}A &\cellcolor{GreenCell}A &\cellcolor{GreenCell}A &\cellcolor{RedCell}C   &\cellcolor{GreenCell}A &\cellcolor{GreenCell}A &\cellcolor{RedCell}C   &\cellcolor{GreenCell}A &\cellcolor{GreenCell}A &\cellcolor{GrayCell}   &\cellcolor{GrayCell}   &\cellcolor{GreenCell}A &\cellcolor{RedCell}C   &\cellcolor{GreenCell}A &\cellcolor{GreenCell}A &\cellcolor{GreenCell}A \\ \hline
					TPLink Camera           &\cellcolor{GreenCell}A &\cellcolor{GrayCell}   &\cellcolor{GreenCell}A &\cellcolor{RedCell}C   &\cellcolor{RedCell}C   &\cellcolor{GreenCell}A &\cellcolor{GreenCell}A &\cellcolor{GreenCell}A &\cellcolor{GreenCell}A &\cellcolor{RedCell}C   &\cellcolor{GreenCell}A &\cellcolor{RedCell}C   &\cellcolor{GreenCell}A &\cellcolor{RedCell}C   &\cellcolor{RedCell}C   &\cellcolor{RedCell}C   &\cellcolor{RedCell}C   &\cellcolor{RedCell}C   &\cellcolor{YellowCell}B&\cellcolor{RedCell}C   &\cellcolor{RedCell}C   &\cellcolor{GreenCell}A &\cellcolor{GreenCell}A &\cellcolor{GreenCell}A \\ \hline
					Belkin Motion Sensor    &\cellcolor{GreenCell}A &\cellcolor{GreenCell}A &\cellcolor{GreenCell}A &\cellcolor{RedCell}C   &\cellcolor{RedCell}C   &\cellcolor{RedCell}C   &\cellcolor{GreenCell}A &\cellcolor{GreenCell}A &\cellcolor{GreenCell}A &\cellcolor{RedCell}C   &\cellcolor{GreenCell}A &\cellcolor{GrayCell}   &\cellcolor{GrayCell}   &\cellcolor{RedCell}C   &\cellcolor{RedCell}C   &\cellcolor{GreenCell}A &\cellcolor{GreenCell}A &\cellcolor{RedCell}C   &\cellcolor{YellowCell}B&\cellcolor{RedCell}C   &\cellcolor{RedCell}C   &\cellcolor{RedCell}C   &\cellcolor{GreenCell}A &\cellcolor{GreenCell}A \\ \hline
					Nest Smoke Alarm        &\cellcolor{GreenCell}A &\cellcolor{GrayCell}   &\cellcolor{GreenCell}A &\cellcolor{GreenCell}A &\cellcolor{GreenCell}A &\cellcolor{GreenCell}A &\cellcolor{GreenCell}A &\cellcolor{GreenCell}A &\cellcolor{GreenCell}A &\cellcolor{GreenCell}A &\cellcolor{GreenCell}A &\cellcolor{RedCell}C   &\cellcolor{GreenCell}A &\cellcolor{YellowCell}B&\cellcolor{RedCell}C   &\cellcolor{GreenCell}A &\cellcolor{GreenCell}A &\cellcolor{GrayCell}   &\cellcolor{GrayCell}   &\cellcolor{GreenCell}A &\cellcolor{RedCell}C   &\cellcolor{GreenCell}A &\cellcolor{GreenCell}A &\cellcolor{GreenCell}A \\ \hline
					Netamo Camera           &\cellcolor{GreenCell}A &\cellcolor{GreenCell}A &\cellcolor{GreenCell}A &\cellcolor{YellowCell}B&\cellcolor{RedCell}C   &\cellcolor{GreenCell}A &\cellcolor{GreenCell}A &\cellcolor{GreenCell}A &\cellcolor{GreenCell}A &\cellcolor{GreenCell}A &\cellcolor{GreenCell}A &\cellcolor{RedCell}C   &\cellcolor{GreenCell}A &\cellcolor{RedCell}C   &\cellcolor{RedCell}C   &\cellcolor{RedCell}C   &\cellcolor{GreenCell}A &\cellcolor{RedCell}C   &\cellcolor{YellowCell}B&\cellcolor{RedCell}C   &\cellcolor{RedCell}C   &\cellcolor{GreenCell}A &\cellcolor{GreenCell}A &\cellcolor{GreenCell}A \\ \hline
					Dlink Camera            &\cellcolor{RedCell}C   &\cellcolor{RedCell}C   &\cellcolor{RedCell}C   &\cellcolor{GreenCell}A &\cellcolor{GreenCell}A &\cellcolor{GreenCell}A &\cellcolor{GreenCell}A &\cellcolor{GreenCell}A &\cellcolor{GreenCell}A &\cellcolor{GreenCell}A &\cellcolor{GreenCell}A &\cellcolor{GrayCell}   &\cellcolor{GrayCell}   &\cellcolor{RedCell}C   &\cellcolor{RedCell}C   &\cellcolor{RedCell}C   &\cellcolor{RedCell}C   &\cellcolor{RedCell}C   &\cellcolor{YellowCell}B&\cellcolor{RedCell}C   &\cellcolor{RedCell}C   &\cellcolor{RedCell}C   &\cellcolor{GreenCell}A &\cellcolor{GreenCell}A \\ \hline
					Hello Barbie Companion  &\cellcolor{GreenCell}A &\cellcolor{GreenCell}A &\cellcolor{GreenCell}A &\cellcolor{GreenCell}A &\cellcolor{GreenCell}A &\cellcolor{GreenCell}A &\cellcolor{GreenCell}A &\cellcolor{GreenCell}A &\cellcolor{GreenCell}A &\cellcolor{GreenCell}A &\cellcolor{GreenCell}A &\cellcolor{RedCell}C   &\cellcolor{GreenCell}A &\cellcolor{GreenCell}A &\cellcolor{GreenCell}A &\cellcolor{GreenCell}A &\cellcolor{GreenCell}A &\cellcolor{RedCell}C   &\cellcolor{GreenCell}A &\cellcolor{GreenCell}A &\cellcolor{RedCell}C   &\cellcolor{GreenCell}A &\cellcolor{GreenCell}A &\cellcolor{GreenCell}A \\ \hline
					Whithings Sleep Monitor &\cellcolor{GreenCell}A &\cellcolor{GrayCell}   &\cellcolor{GreenCell}A &\cellcolor{GreenCell}A &\cellcolor{GreenCell}A &\cellcolor{GreenCell}A &\cellcolor{GreenCell}A &\cellcolor{GreenCell}A &\cellcolor{GreenCell}A &\cellcolor{GreenCell}A &\cellcolor{GreenCell}A &\cellcolor{RedCell}C   &\cellcolor{GreenCell}A &\cellcolor{RedCell}C   &\cellcolor{RedCell}C   &\cellcolor{RedCell}C   &\cellcolor{GreenCell}A &\cellcolor{GrayCell}   &\cellcolor{GrayCell}   &\cellcolor{RedCell}C   &\cellcolor{RedCell}C   &\cellcolor{GreenCell}A &\cellcolor{GreenCell}A &\cellcolor{GreenCell}A \\ \hline
					Nest Drop Camera        &\cellcolor{GreenCell}A &\cellcolor{GreenCell}A &\cellcolor{GreenCell}A &\cellcolor{GreenCell}A &\cellcolor{GreenCell}A &\cellcolor{GreenCell}A &\cellcolor{GreenCell}A &\cellcolor{GreenCell}A &\cellcolor{GreenCell}A &\cellcolor{GreenCell}A &\cellcolor{GreenCell}A &\cellcolor{RedCell}C   &\cellcolor{GreenCell}A &\cellcolor{GreenCell}A &\cellcolor{YellowCell}B&\cellcolor{GreenCell}A &\cellcolor{GreenCell}A &\cellcolor{RedCell}C   &\cellcolor{GreenCell}A &\cellcolor{GreenCell}A &\cellcolor{RedCell}C   &\cellcolor{GreenCell}A &\cellcolor{GreenCell}A &\cellcolor{GreenCell}A \\ \hline
				\end{tabular}
			\end{adjustbox}
			\vspace{-4mm}
		\end{table*}
		\egroup
		Without doubt, hundreds of consumer IoT devices are going to emerge in the years ahead, and their security/privacy vulnerabilities are going to be diverse. Our results from evaluation of the twenty devices highlight the security posture of consumer IoTs, and reveal the problems that users have to deal with. In this section we discuss how our methodology can be used for a security ratings system that is beneficial to consumers or insurance companies. We propose a three-level rating: ``A'' being secure, ``B'' being moderately secure/insecure, and ``C'' being insecure. Table~\ref{tab:ch2_securityrating} shows our attempt to rate each of IoT devices that we assessed their security posture on the four dimensions -- all ratings in this table are subjective and given based on authors perceptions. One may consolidate our table by giving weights to each dimension in the future.

%		Our results from evaluation of the twenty devices highlight the security posture of consumer IoTs, and reveal the problems that users have to deal with. % due to the rapid growth of the IoT market.  
%		In this section we discuss how our methodology can be used for a security ratings system that is beneficial to consumers or insurance companies. We propose  a three-level rating: ``A'' being secure, ``B'' being moderately secure/insecure, and ``C'' being insecure. 
%		Table~\ref{tab:ch2_securityrating} shows our attempt to rate each of IoT devices that we assessed their security posture on the four dimensions -- all ratings in this table are subjective and given based on authors perceptions. 
%		One may consolidate our table by giving weights to each dimension in the future.

		We use color codes for ease of visualization, green for A rating, yellow for B rating, and red for C rating. We also use gray color for cells where the data is not available. For example, the encryption protocol of Belkin switch is not identified on Wireshark for the device-to-server communication; DNS query is not performed in Belkin motion sensor; normal functionality of the Pixtar photo frame is not affected by a DoS attack. Using our color-coded ratings table, consumers are able to quickly visualize the security posture of individual devices. All devices display some form of vulnerability in either of integrity, access control and reflection dimensions -- this raises concerns for consumers as well as for the Internet ecosystem in general. Devices such as the Amazon Echo, Hello Barbie, Nest Dropcam, Whitings Sleep monitor seem relatively secure by the measure of confidentiality. Amazon Echo in particular is a top-rated device in security with encrypted communication channels and having almost all of its ports closed. On the other hand, devices such as Phillips Hue lightbulb and the Belkin switch seem fairly poor in security. The Phillips Hue in particular communicates in plaintext to the user App, is susceptible to replay attacks, has many open ports and can be used to launch various reflection attacks to victim servers.
		
%		We use color codes for ease of visualization, green for A rating, yellow for B rating, and red for C rating. 
%		We also use gray color for cells where the data is not available. For example, the encryption protocol of Belkin switch is not identified on Wireshark for the device-to-server communication; DNS query is not performed in Belkin motion sensor; normal functionality of the Pixtar photo frame is not affected by a DoS attack.
%		Using our color-coded ratings table, consumers are able to quickly visualize the security posture of individual devices. All devices display some form of vulnerability in either of integrity, access control and reflection dimensions - this raises concerns for consumers as well as for the Internet ecosystem in general.  Devices such as the Amazon Echo, Hello Barbie, Nest Dropcam, Whitings Sleep monitor seem relatively secure by the measure of confidentiality. Amazon Echo in particular is a top-rated device in security with encrypted communication channels and having almost all of its ports closed. On the other hand, devices such as Phillips Hue lightbulb and the Belkin switch seem fairly poor in security. The Phillips Hue in particular communicates in plaintext to the user App, is susceptible to replay attacks, has many open ports and can be used to launch various reflection attacks to victim servers. 
		
		We recognize that security is but one concern amongst many that manufacturers of IoT devices are dealing with. The surge in demand for IoT is leading many manufacturers to rush to market with their product, and increasing user appeal to gain market traction can become more paramount than ensuring fool--proof security. No matter how it evolves, consumers would eventually demand for a rating system (much like the energy rating system given to home appliances) that needs to be developed by standard bodies and tracked by regulation entities. This would protect consumers rights and incentivize manufacturers to improve the security of their device to receive an acceptable rating that can lead to a good share of the market.
		
%		We recognize that security is but one concern amongst many that manufacturers of IoT devices are dealing with. The surge in demand for IoT is leading many manufacturers to rush to market with their product, and increasing user appeal to gain market traction can become more paramount than ensuring fool-proof security. No matter how it evolves, consumers would eventually demand for a rating system (much like the energy rating system given to home appliances) that needs to be developed by standard bodies and tracked by regulation entities. This would protect consumers rights and incentivize manufacturers to improve the security of their device to receive an acceptable rating that can lead to a good share of the market.
		
		}

\section{Existing IoT Security Solutions} \label{existsec}
	Security and monitoring solutions for traditional IT network have been extensively studied~\cite{Sandhu2011, Ko1997, Mukkamala2002} in the past few decades by the research community. Those studies include both the Host-based Intrusion Detection System (HIDS) and Network-based Intrusion Detection System (NIDS).
	
	Anti-virus in computers is a good example of the implementation of HIDS. It monitors the activities of the devices based on the system log files as well as network packets on its interfaces~\cite{mell2003understanding}. The host-based security systems are unlikely to be embedded in IoT devices due to resource constraints and are not resilient enough to maintain security standards for the long term since automatically applying the security patches is hard.
	
	NIDS solution protects all the devices in a network to complement device vendor security implementation~\cite{Sivaraman2015WiMob} by analyzing the activities of devices from the network traffic. NIDS monitor the activity of the devices based on the techniques such as signature-based detection, specification-based detection, and anomaly-based detection~\cite{Yeo2017}. The advantages of NIDS over HIDS are: 1) it can be implemented using a centralized controller and hosted in the cloud environment rather than using the device resources; and 2) upgrading the system is easy and protection of all devices can be handed over to security experts rather than expecting all the users to be tech-savvy.

	\subsection{Signature-based Intrusion Detection}
		The signature-based attack identification system is the commonly used technique in the present NIDS implementations such as Bro~\cite{Paxson1999}, Snort~\cite{Roesch1999}, and commercial hardware. They compare the traffic with already known attack signatures collected from the sandbox environment and honeypots. However, the signature-based methods do not show good performance with zero-day (attacks that have never seen before) vulnerabilities. The main implication of signature-based attack detection is that it is not scalable in the IoT domain due to the heterogeneity of IoTs. Generating the attack signature for a growing number of IoT types is not a feasible solution, whereas, in a traditional network, most of the devices run on similar platforms (\eg Windows, Unix, Linux, Android). 
	
		\vspace{-0.5em}
	\subsection{Specification-based Intrusion Detection}
		\vspace{-0.5em}
		The specification-based intrusion detection mechanism is monitoring the device based on the rules (\ie specification) that define the allowed or malicious network activities~\cite{Ko1997, Uppuluri2001}. The specification-based detection has the ability to act as both: 1) learn the attack characteristics and identify the attacks that follow those specification; or 2) learn the benign behavior of the device and detect the variations when a traffic flow overrule the specification~\cite{Berthier2011, Surendar2016, Le2016, BOSTANI2017}. 
		
		\vspace{-0.7em}
		In~\cite{Amaral2014}, Amaral et al. propose a specification-based approach for a wireless sensor network and expect the network operators to generate the specification by themselves. In~\cite{Nguyen2018} Nguyen et al. develop a protection system called ``IoTSAN'' which allows the users to define specification by semantic rules. However, generating the specifications for every device is a tedious task. Internet Engineering Task Force (IETF) recently released a standard called Manufacturer Usage Description (MUD) to outline the network activities of the devices intended by the manufacturers~\cite{ietfMUD18}. The work in~\cite{IoTSnP18-mudids} proposes an IDS by automatically profiling each device using MUD at the initial stage and then identifies the attacks that violate the profiles. This work has been extended in~\cite{HamzaSOSR2019} to identify the volumetric anomalies along with the specification violations.
		
		\vspace{-0.5em}
		Although the specification based IDS make better accuracy in identifying attacks and policy violations, generating specifications for the proliferation of IoT is hard, and the standards like MUD have not yet been adopted by manufacturers. 
	
	\vspace{-0.5em}
	\subsection{Anomaly-based Instruction Detection}
	\vspace{-0.5em}
		The anomaly detection technique is to learn legitimate behavior from the normal network traffic and identify the variations from it. Since anomaly detection just inspects deviations from the benign traffic rather than the attack signatures, it has the capability to identify the zero-day attacks as well.
		
		Although there is an extensive body of literature~\cite{Vinchurkar2012, Patel2012, Portnoy2001, Mukkamala2002, Hosmer1993, Dokas2002} in anomaly detection based instruction systems, it has achieved only a very limited amount of success rate~\cite{garcia2009anomaly} in the traditional IT network. The reasons are manifold~\cite{Sommer2010}: 1)legitimate traffic shows high variability in IT networks; 2) difficult to get the ground truth in the training dataset; 3) very limited public datasets to learn the normal network behaviors; 4) both wrongly identifying a legitimate traffic as illegitimate, and illegitimate traffic as legitimate incur a high cost; and 5) practical challenges in evaluating the system.
		
		However, in the domain of IoT, these methods give some promises since the activities of IoT devices is less complicated than servers or computers and due to their limited functionality and following similar patterns, it is easy to characterize the whole behavior of the device~\cite{Nomm2018} from the network traffic. 

	\vspace{-1em}
\section{IoT Behavioral Monitoring}
	\vspace{-1em}
	Nowadays network operators lack real-time visibility into connected devices -- over 40\% of today's endpoints are unknown and unmanaged by the organizations which lead to significant infrastructure blind spots, unauthorized access, and data leaks~\cite{Cisco2017}. Based on the fact that IoT devices exhibit limited traffic patterns, we believe it is possible to identify and characterize their network behavior~\cite{Amar2018}. It enables the operator to: 1) manage the assets connected in the network~\cite{infocom17}; 2) enforce the device-specific policies~\cite{Barrera2017}; and 3) locate the vulnerable and blacklisted devices effortlessly~\cite{Aluthge2017}.

	\subsection{IoT Traffic Characterization}
		Significant research work is carried out in the existing literature to characterize the general Internet traffic~\cite{Moore2005,GraphTraffic2009, Bonfiglio2007,svmSIP2012}. These prior works largely focus on application detection (e.g. Web browsing, Video streaming, Gaming, Mail, Skype VoIP, Peer-to-Peer, etc.). However, studies focusing on characterizing IoT traffic (also referred to as machine-to-machine or M2M traffic) are still in their infancy.
	
		\subsubsection{Analysis of Empirical Traces}
			The work in~\cite{Shafiq2012} is one of the first large-scale studies to delve into the nature of M2M traffic. It is motivated by the need to understand whether M2M traffic imposes new challenges for the design and management of cellular networks. The work uses a traffic trace spanning one week from a tier-1 cellular network operator and compares M2M traffic with traditional smartphone traffic from a number of different perspectives -- temporal variations, mobility, network performance, and so on. They conclude that M2M traffic is substantially different from smartphone traffic as it tends to exhibit higher uplink to downlink traffic volume, varying diurnal patterns, and larger round-trip times. The study informs network operators to be cognizant of these factors when managing their networks. 
			
			%device based traffic model
			In~\cite{Nikaein2013}, Nikaein et al. note that the amount of traffic generated by a single M2M device is likely to be small, but the total traffic generated by hundreds or thousands of M2M devices would be substantial. These observations are to some extent corroborated by~\cite{belllabsIoT, ALUIoT}, which note that a remote patient monitoring application is expected to generate about 0.35 MB per day and smart meters roughly 0.07 MB per day.
		
		\subsubsection{Aggregated Traffic Model}
			A Coupled Markov Modulated Poisson Processes (CMMPP) framework to capture the behavior of a single machine-type communication, as well as the collective behavior of tens of thousands of M2M devices, is proposed in~\cite{Laner2013}. The complexity of the CMMPP framework is shown to grow linearly with the number of M2M devices, rendering it effective for large-scale synthesis of M2M traffic. 
			
			In~\cite{Laner2014}, Markus et al. show that it is possible to split the (traffic) state of an M2M device into three generic categories, namely periodic update, event-driven, and payload exchange, and a number of modelling strategies that use these states are developed. An illustration of model fitting is shown via a use-case in fleet management comprising 1000 trucks run by a transportation company. The fitting is based on measured M2M traffic from a 2G/3G network.
			A simple model to estimate the volume of M2M traffic generated in a wireless sensor network-enabled connected home is constructed in~\cite{Orrevad2009}. Since the behavior of sensors is very application-specific, the work identifies certain common communication patterns that can be attributed to any sensor device. 
			
			Although all the above studies do fundamental studies in the IoT traffic characterizations, they do not undertake a fine-grained characterization in consumer IoT devices. On this scenario, we present insights into the underlying network traffic characteristics using statistical attributes such as activity cycles, port numbers, signalling patterns, and cipher suites in~\cite{infocom17} and~\cite{TMC18} which is discussed in Chapter~\ref{chap:characterization}.

	\subsection{IoT Finger-Printing and Classification}
		Traffic fingerprinting and classification is widely used for various applications such as network management~\cite{Roughan2004,lippmann2003}, QoS~\cite{Zhang2013,Madanapalli2019}, and cyber-security~\cite{Lyu2019,Tegeler12,EncryptedTelemetry16,EncryptedMalware16}. However, IoT traffic fingerprinting and classification methods are still in its early stages~\cite{Lopez-Martin2017}. The recent attention on IoT security and asset management has attracted researchers to observe IoT devices using both active and passive fingerprinting techniques.
		
		The active fingerprinting comprise various techniques -- identifying vendor from OUI prefix of the MAC Address, device name from the host-name field of DHCP negotiation, services offered by the device using discovery protocols or actively parsing service banners and estimating the device types by probing the ports~\cite{ICIAfS2018}. Shodan~\cite{Shodan} is one of the examples that actively scan and classify (mainly by parsing service banners) the IoT devices. The main disadvantage of the active scanning approach is they greatly impact the network and degrade the performance of connected devices.
		
		Although the passive fingerprinting techniques are not simple and straight forward as active, they tend to provide a rich set of information about the devices and their states without generating additional congestions. These passive fingerprints can be recorded by monitoring the network traffic (\ie passive network telemetry) using the middle-boxes (\eg switches) or mirroring the traffic to a dedicated inspection engine.
	
		\subsubsection{Passive Network Telemetry}
			Network traffic measurement has been a subject of interest to academia and industry. Many different methods have been proposed and practically used ranging from traditional port-based counting using SNMP~\cite{rfcsnmp}, WiFi packet sniffing~\cite{Srinivasan2008,Acar2018,Copos2016} and packet sampling~\cite{sFlow} to flow-based telemetry~\cite{rfcNetFlow,FlowRadar2016}.
			
			{\rev 
				Traditional fingerprinting methods like sniffing wireless (\eg WiFi, Zigbee, Bluetooth) packets require special hardware~\cite{Siby2017}. Although, these methods may reveal valuable information in attack scenarios, they provide network operators with very limited insights into the operation of their network.
			}
			
			Modern telemetry methods can be categorized into: (a) packet-based~\cite{sFlow, Zhu2015,Rasley2014}; and (b) flow-based~\cite{rfcNetFlow,FlowRadar2016,McKeown2008}. sFlow~\cite{sFlow} is one of the commonly used methods that randomly samples (\ie one in N) packets from the network switches. Due to its random sampling, sFlow tends to collect packets from elephant flows (those that carry heavy traffic and are long in duration), and hence mice flows are likely to get missed which results in an inaccurate measurement. To address this issue, Everflow~\cite{Zhu2015} proposes to collect specific packets (\eg TCP SYN, FIN, and RST) using the match and mirror functionality of data-center switches. Planck~\cite{Rasley2014} estimates the throughput of flows at very tight time-scales by mirroring traffic of multiple ports to a monitoring port at which a collector performs high-rate sampling. Overall, packet-level telemetry can only provide partial visibility into network traffic flows. 
			
			Commercial switches equipped with NetFlow~\cite{rfcNetFlow} (used in Chapter~\ref{chap:characterization}) engines export flow records. However, there are two limitations: (a) they only export a flow record once it expires (not in real-time); and (b) updating and maintaining flow records result in computational cost~\cite{Hofstede2013}.  FlowRadar~\cite{FlowRadar2016} overcomes the limitations of Netflow by incorporating an encoded hash table (data structure for flow counters) with low memory overheads and exporting flows periodically (\eg 10 ms). However, FlowRadar is still not supported by commercial switches available on the market. SDN APIs~\cite{McKeown2008} which is commonly available on OpenFlow supported switches, enable us to measure traffic flows at low cost with reasonable resolutions. Thus, we propose flow-level telemetry using SDN APIs as an effective solution in Chapter~\ref{chap:telemetry}.
		
		\subsubsection{Classification and Anomaly detection}
			IoT traffic classification has been the subject of recent researches for a variety of purposes: 1) recognizing IoT from mixture of IoT and non-IoT devices~\cite{TMC18,Meidan2017ProfilIoT}; 2) classifying the type of IoT device~\cite{Miettinen2017,Guo2018}; 3) identifying the operating states~\cite{Apthorpe2017,Acar2018}; and 4) detecting the abnormal behaviors~\cite{Doshi2018,Nomm2018} of the IoT traffic. 
			
			\textbf{Attribute selection:}
			The proposed classification mechanisms in the literature rely on a wide range of attributes from as simple as, a set of IP addresses (of servers) that each device communicates~\cite{Guo2018}, to sophisticated as the entropy of payloads exchanged by the devices~\cite{Bezawada2018}. However, these attributes come in various extraction cost and different level of impact on the classification process (\eg Although the set of IP addresses is a low-cost attribute, it is not much reliable since the IP address belongs to an elastic IPv4 address allocation -- employed for dynamic cloud computing like Amazon AWS). Thus, the trade-off between the cost and robustness of attributes plays a vital role in the attribute selection.
			
			{\rev
				Work in~\cite{Miettinen2017} develops a classification system called ``IoT Sentinel'' to recognize and identify the IoT device types immediately after it connected to the network. It employs a single attribute vector with the elements of 276 (\ie first 12 packets with 23 attributes for each). The proposed 23 attributes including 16 binary attributes (indicating the use of various protocols at application, transport, network, and link layers) along with IP layer header options, remote IP address/port numbers, and size and raw byte value of packets. In order to minimize the attribute extraction cost, the system limits the classification process during the device connecting phase-only -- it is not feasible for continuous monitoring.~\cite{Bezawada2018} proposes a technique to improve the ``IoT Sentinel'' by extracting payload entropy, TCP payload length, TCP window size in addition to the subset of above mentioned attributes for every five packets. Although this method solves the limitation, it incurs a high cost in attribute extraction.
			}
			
			Work in~\cite{Meidan2017} uses over 300 attributes from each TCP session of IoT traffic to classify the device type by applying majority voting for every 20 consecutive sessions. It highlight the most important attributes as packets Time-To-Live (minimum, median, and average), the ratio of transmitted-bytes to received-bytes, and the Alexa rank of servers which the device communicates. In this method, the IoT devices which rarely use TCP sessions, or which have long TCP sessions (\eg~\cite{TMC18} mentions Google Dropcam initiates TCP connection during the boot states and keep it alive as long as it has network connectivity) may take a long duration to be classified. Some researchers~\cite{Ortiz2019} argue that traffic attributes need to be automatically learned (from a raw sequence of packet payloads in TCP flows) instead of being hand-crafted. We believe that the extraction of packet payloads makes it difficult to scale this method. Our first approach to classify IoT devices (explained in Chapter~\ref{chap:characterization}) uses attributes which can be extracted relatively easily using the network elements that are instrumented with hardware-accelerated flow-level analyzers (\eg Netflow capable devices).
			
			Although the attributes mentioned above provides a rich set of information about the IoT behavior, obtaining them require specialized hardware accelerators -- becomes more expensive, and unscalable due to the need of deep packet inspection in real-time. To tackle this, researchers in~\cite{Apthorpe2017} have proposed a model which uses traffic patterns of encrypted network flows (with a server measured outside NAT) to reveal the existence of IoT specific devices inside a home network without the need for detailed packet or flow inspection. One of our works~\cite{ANTS16} show that the IoT traffic can be: 1) channeled based on either protocols; or endpoints specific flows, and 2) monitored at low cost using SDN enabled switches. This inspired us to extract the flow level measurements (\ie byte count, packet count) corresponding to individual devices as attributes of specific flows (\eg DNS query, DNS response, NTP query, NTP response, and etc.) in Chapter~\ref{chap:telemetry}. Also, we compute the attributes in different time-granularities since the traffic attributes are better in characterizing the network behavior of IoTs in multiple time-scales~\cite{multiTime2005}.
			
			\textbf{Classification \& Anomaly Techniques:}
			Machine learning techniques do a prominent role in state-of-the-art traffic classification and anomaly detection engines~\cite{Buczak2016}. The previous studies have used a different kind of machine learning techniques which can be categorized into two-fold: 1) supervised; and 2) unsupervised. 
			
			Supervised classification algorithms such as Support Vector Machines (SVM), naive Bayes, decision trees (\eg: C4.5, random forest), k-nearest neighbor (KNN), and neural network/Multilayer Perceptron(MLP) are commonly used for the traffic classification purposes. The supervised classification techniques tend to give high performance in distinguishing the known classes due to their discriminative ability. In the literature, these algorithms are used in two different modes: 1) Multi-class classification -- classify the data between tgree or more than classes (\eg IoT device classification/state classification); and 2) Binary classification -- make a binary decision (\ie Positive or Negative).
			
			In~\cite{lippmann2003}, Lippmann et al. evaluate the performance of multiple multi-class classifier performance (\ie KNN, SVM, Binary decision Tree and MLP) in classifying the computer operating systems (\eg: Linux 2.0, Linux 2.1, MacOS9, Win9x, WinNT, WinXP) using TCP/IP header information and concludes the KNN and Binary decision tree outperforms the rest. Despite the simplicity and reasonably good performance on a low dimensional dataset, KNN may be prone to be affected by high dimensionality in the data and high computational cost in classification~\cite{Agrawal2014}. Performance of SVM is very sensitive to the selection of hyperparameters, and it becomes difficult to train the accurate models~\cite{Yue2003}. In~\cite{Lopez-Martin2017}, Lopez-Martin et al. classify the network application traffic (\eg: Google, YouTube, Office 365)  using the multi-class neural network which is proven to be effective in complex data structures, but it requires a large amount of data to train the system. Decision tree-based classifiers are commonly used since they are easy to build discriminative models with relatively small amount of data. However, they are prone to being over-fit for the training dataset. Random forest perfectly handles the over-fitting issue using ensemble decision trees. Chapter~\ref{chap:characterization} and Chapter~\ref{chap:telemetry} therefore explore the use of it for classifying the device types and their states. The main constraint of the multi-class classification is scalability -- a high number of classes makes the classifier complex and updating the classifier requires full retraining.
			
			Unlike multi-class classifiers, binary classifiers are trained only based on two different classes. In the traffic classification domain, this approach is used for various purposes like distinguishing between known attack vs. benign traffic~\cite{Kumar2019} or IoT vs. non-IoT traffic, etc. In~\cite{Doshi2018}, Doshi et al. train the binary classifier to identify the DDoS attack traffics (made by Mirai botnets) and benign traffic of the IoT devices. Although this method shows high performance in detecting the trained attacks from benign traffic, the efficacy of unknown attack detection cannot be guaranteed. The reason is supervised classifiers learn only the differences between classes rather than profiling the whole behavior. On the other hand, the work in~\cite{Meidan2017ProfilIoT} proposes to build individual binary classification model for each class in a device classification problem to eliminate the complexity issue of multi-class classification. It is achieved by individually training each device traffic with the mixture of the rest of the devices -- `one vs. rest' approach. It is a well-known fact that supervised machine learning methods may suffer due to an unbalanced dataset (unequal number of data points between classes). This makes scalability issues in `one vs. rest' binary classification approach when the proportion of the `rest' part keeps increasing.  
			
			Unsupervised machine learning techniques are generative, which can model the whole behavioral pattern of the data to detect the abnormal behavioral changes in the traffic. Over the decade, many unsupervised network traffic anomaly detection methods have been proposed for general Internet traffic~\cite{Leung2005, Zhao2017, pimentel2014}. They use various algorithms such as probabilistic (\eg Gaussian mixture models),domain-based (\eg one-class SVM), and cluster-based (\eg DBSCAN, Kmeans)~\cite{Chen2017,HamzaSOSR2019}.
			The work in~\cite{Gu2008} use unsupervised clustering approaches to distinguish the botnet C\&C communication channels from the benign traffic of traditional network traffic as well as the malicious activity during the attack mode. The study in~\cite{Zhao2017} shows that unsupervised algorithms may suffer due to the curse of high dimensional data. Therefore, it proposes to reduce the attribute dimension using Principal Component Analysis (PCA) which transform the attributes to a reduced set of uncorrelated attributes.
			
			{\rev
				In the context of IoT, ,~\cite{MeidanAutoEncoders} authors use neural network-based deep autoencoders to detect anomalies. Similarly, work in ~\cite{Aneja2019} proposed a deep-learning framework to fingerprint iPads and iPhones using packets inter-arrival time. However, these intensive packet-based approaches are computationally expensive. 
			}
			In~\cite{HamzaSOSR2019}, Hamza et al. propose a volumetric attack detection mechanism by monitoring the MUD-compliant activities which require fine-grained flow for each device. To achieve this, they use X-Means, which is a version of K-Means clustering algorithm.
			
			The main negative aspect of the unsupervised learning method is they are not discriminative as supervised machine learning algorithms -- difficult for classification purposes. One of our work~\cite{LCN19} resolves this issue by proposing a probability-based conflict resolver, which is extended to anomaly detection in Chapter~\ref{chap:anomaly}.

\section{Conclusion}
	In this chapter, we have comprehensively studied the IoT ecosystem as well as the challenges introduced by it, especially in the areas of cybersecurity and user privacy. Initially, the market segments of IoT devices are identified based on the category of industrial, consumer, and enterprises IoTs. The challenges in securing IoT devices compared to traditional general-purpose devices are investigated in order to further move on to specialized security solutions for the IoT ecosystem. We have discussed the roles and responsibilities of key players in securing the IoT ecosystem. We proposed a systematic approach to evaluate the security of IoT devices. From there we compared the existing security solutions and the challenges to be encountered when adapting it to the IoT domain. Finally, we set the foundation for characterization, classification, and anomaly detection in IoT traffic, which will be discussed deeply in the following chapters. 

\chapter{IoT Traffic Characterization and Classification}
	\label{chap:characterization}
	\vspace{-5mm}
	\minitoc
	
	Our study in the previous chapter showed the importance of real-time visibility into IoT network using behavioral monitoring. In this chapter, we study the network behavioral patterns of IoT devices using traffic characteristics obtained at the network level. Using this, we develop an inference engine to classify the device types. Our contributions are fourfold. First, we instrument a smart environment with 28 different IoT devices spanning cameras, lights, plugs, motion sensors, appliances and health-monitors. We collect and synthesize traffic traces from this infrastructure for a period of six months, a subset of which we release as open data for the community to use. Second, we present insights into the underlying network traffic characteristics using statistical attributes such as activity cycles, port numbers, signalling patterns and cipher suites. Third, we develop a multi-stage machine learning based classification algorithm and demonstrate its ability to identify specific IoT devices with over 99\% accuracy based on their network activity. Finally, we discuss the trade-offs between cost, response time, and performance involved in deploying the classification framework in real-time. Parts of this chapter have been published in~\cite{infocom17} and~\cite{TMC18}.

	\section{Introduction}\label{sec:c1_introduction}
		% growth of IoT
		The number of devices connecting to the Internet is ballooning, ushering in the era of the ``Internet of Things'' (IoT). As we mentioned in previous chapters,
		the proliferation of IoT, creates an important problem. Operators of smart environments can find it difficult to determine what IoT devices are connected to their network and further to ascertain whether each device is functioning normally. This is mainly attributed to the task of managing assets in an organization, which is typically distributed across different departments. For example, in a local council, lighting sensors may be installed by the facilities team, sewage and garbage sensors by the sanitation department and surveillance cameras by the local police division. Coordinating across various departments to obtain an inventory of IoT assets is time consuming, onerous and error-prone, making it nearly impossible to know precisely what IoT devices are operating on the network at any point in time.  Obtaining ``visibility'' into IoT devices in a timely manner is of paramount importance to the operator, who is tasked with ensuring that devices are in appropriate network security segments, are provisioned for requisite quality of service, and can be quarantined rapidly when breached. The importance of visibility is emphasized in Cisco's most recent IoT security report~\cite{Cisco2017}, and further highlighted by two recent events: sensors of a fishtank that compromised a casino in Jul 2017~\cite{fishtank}, and attacks on a University campus network from its own vending machines in Feb 2017~\cite{vendingmachine2017}. In both cases, network segmentation could have potentially prevented the attack and better visibility would have allowed rapid quarantining to limit the damage of the cyber-attack on the enterprise network.

		One would expect that devices can be identified by their MAC address and DHCP negotiation. However, this faces several challenges: (a) IoT device manufacturers typically use NICs supplied by third-party vendors, and hence the Organizationally Unique Identifier (OUI) prefix of the MAC address may not convey any information about the IoT device; (b) MAC addresses can be spoofed by malicious devices; (c) many IoT devices do not set the Host Name option in their DHCP requests \cite{rfcDHCP}; indeed we found that about half the IoT devices we studied do not reveal their host names, as shown in Table~\ref{tab:macdhcp}; (d) even when the IoT device exposes its host name it may not always be meaningful (e.g. WBP-EE4C for Withings baby monitor in Table~\ref{tab:macdhcp}); and lastly (e) these host names can be changed by the user (e.g. the HP printer can be given an arbitrary host name). For these reasons, relying on DHCP infrastructure is not a viable solution to correctly identify devices at scale.

		\begin{table}[t]
			\centering
			\caption{MAC address and DHCP host name of IoT devices used in our testbed.}
			\label{tab:macdhcp}
			\begin{adjustbox}{max width=0.9\textwidth}
				\def\arraystretch{1.2}
				\begin{tabular}{|l|l|l|l|}
					\hline
					\textbf{IoT device}           & \textbf{MAC address} & \textbf{OUI}                  & \textbf{DHCP host name} \\ \hline
					Amazon Echo               & 44:65:0d:56:cc:d3    & Amazon Technologies Inc.      &                         \\
					August Doorbell Cam       & e0:76:d0:3f:00:ae    & AMPAK Technology, Inc.        &                         \\
					Awair air quality monitor & 70:88:6b:10:0f:c6    &                               & Awair-4594              \\
					Belkin Camera             & b4:75:0e:ec:e5:a9    & Belkin International Inc.     & NetCamHD                \\
					Belkin Motion Sensor      & ec:1a:59:83:28:11    & Belkin International Inc.     &                         \\
					Belkin Switch             & ec:1a:59:79:f4:89    & Belkin International Inc.     &                         \\
					Blipcare BP Meter         & 74:6a:89:00:2e:25    & Rezolt Corporation            &                         \\
					Canary Camera             & 7c:70:bc:5d:5e:dc    & IEEE Registration Authority   & Ambarella/C100F1615229  \\
					Dropcam                   & 30:8c:fb:2f:e4:b2    & Dropcam                       &                         \\
					Google Chromecast         & 6c:ad:f8:5e:e4:61    & AzureWave Technology Inc.     & Chromecast              \\
					Hello Barbie              & 28:c2:dd:ff:a5:2d    & AzureWave Technology Inc.     & Barbie-A52D             \\
					HP Printer                & 70:5a:0f:e4:9b:c0    & Hewlett Packard               & HPE49BC0                \\
					iHome PowerPlug           & 74:c6:3b:29:d7:1d    & AzureWave Technology Inc.     & hap-29D71D              \\
					LiFX Bulb                 & d0:73:d5:01:83:08    & LIFI LABS MANAGEMENT PTY LTD  & LIFX Bulb               \\
					NEST Smoke Sensor         & 18:b4:30:25:be:e4    & Nest Labs Inc.                &                         \\
					Netatmo Camera            & 70:ee:50:18:34:43    & Netatmo                       & netatmo-welcome-183443  \\
					Netatmo Weather station   & 70:ee:50:03:b8:ac    & Netatmo                       &                         \\
					Phillip Hue Lightbulb     & 00:17:88:2b:9a:25    & Philips Lighting BV           & Philips-hue             \\
					Pixstart photo frame      & e0:76:d0:33:bb:85    & AMPAK Technology, Inc.        &                         \\
					Ring Door Bell            & 88:4a:ea:31:66:9d    & Texas Instruments             &                         \\
					Samsung Smart Cam         & 00:16:6c:ab:6b:88    & Samsung Electronics Co.,Ltd   &                         \\
					Smart Things              & d0:52:a8:00:67:5e    & Physical Graph Corporation    & SmartThings             \\
					TP-Link Camera            & f4:f2:6d:93:51:f1    & TP-LINK TECHNOLOGIES CO.,LTD. & Little Cam              \\
					TP-Link Plug              & 50:c7:bf:00:56:39    & TP-LINK TECHNOLOGIES CO.,LTD. & HS110(US)               \\
					Triby Speaker             & 18:b7:9e:02:20:44    & Invoxia                       &                         \\
					Withings Baby Monitor     & 00:24:e4:10:ee:4c    & Withings                      & WBP-EE4C                \\
					Withings Scale            & 00:24:e4:1b:6f:96    & Withings                      &                         \\
					Withings sleep sensor     & 00:24:e4:20:28:c6    & Withings                      & WSD-28C6                \\ \hline
				\end{tabular}
			\end{adjustbox}
%			\vspace{-1em}			
		\end{table}

		% lack of IoT profile and traces
		In this chapter, we address the above problem by developing a robust framework that classifies each IoT device separately in addition to one class of non-IoT devices with high accuracy using statistical attributes derived from network traffic characteristics. Qualitatively, most IoT devices are expected to send short bursts of data sporadically. Quantitatively, our preliminary work in \cite{infocom17} was one of the first attempts to study how much traffic IoT devices send in a burst and how long they idle between activities. We also evaluated how much signaling they perform (e.g. domain lookups using DNS or time synchronization using NTP) in comparison to the data traffic they generate. This chapter significantly expands on our prior work by employing a more comprehensive set of attributes on trace data captured over a much longer duration (of 6 months) from a test-bed comprising 28 different IoT devices.

		% need: performance, security, ...
		There is no doubt that it is becoming increasingly important to understand the nature of IoT traffic. Doing so helps contain unnecessary multicast/broadcast traffic, reducing the impact they have on other applications. It also enables operators of smart cities and enterprises to dimension their networks for appropriate performance levels in terms of reliability, loss, and latency needed by environmental, health, or safety applications. However, the most compelling reason for characterizing IoT traffic is to detect and mitigate cyber-security attacks. It is widely known that IoT devices are by their nature and design easy to infiltrate \cite{M2Msec14,LoiIoTSP,Andrea2015,Moskvitch2017,Dhanjani2015,SmartThings16}. New stories are emerging of how IoT devices have been compromised and used to launch large-scale attacks \cite{Guardian16}. The large heterogeneity in IoT devices has led researchers to propose network-level security mechanisms that analyze traffic patterns to identify attacks (see \cite{Vyas2015} and our recent work \cite{ANTS16}); success of these approaches relies on a good understanding of what ``normal'' IoT traffic profile looks like.

		% our contributions
		\vspace{-0.5em}
		Our primary focus in this chapter is to establish a machine learning framework based on various network traffic characteristics to identify and classify the default (i.e. baseline) behavior of IoT devices on a network. This chapter fills an important gap in the literature relating to classification of IoT devices based on their network traffic characteristics. Our contributions are
		\vspace{-0.5em}
		\begin{enumerate}
			\item We instrument a living lab with 28 IoT devices emulating a smart environment. The devices include cameras, lights, plugs, motion sensors, appliances and health-monitors. We collect and synthesize data from this environment for a period of 6 months. A subset of our data is made available for the research community to use.
			\vspace{-0.5em}
			\item We identify key statistical attributes such as activity cycles, port numbers, signaling patterns and cipher suites, and use them to give insights into the underlying network traffic characteristics.
			\vspace{-0.5em}
			\item We develop a multi-stage machine learning based classification algorithm and demonstrate its ability to identify specific IoT devices with over 99\% accuracy based on their network behavior. 
			\vspace{-0.5em}
			\item We evaluate the deployment of the classification framework in real-time, by examining the trade-offs between costs, response time, and accuracy of the classifier.
		\end{enumerate}

		% chapter organisation
		The rest of this chapter is organized as follows: We present our IoT setup and data traces in \S\ref{sec:c1_data}, and in \S\ref{sec:c1_IoTprofile} characterize traffic attributes of the various IoT devices. In \S\ref{sec:c1_class} we propose a machine learning based multi-stage device classification method and evaluate its performance, followed by a discussion on the real-time operation of the proposed system in \S\ref{sec:c1_perfAcc}. The chapter is concluded in \S\ref{sec:c1_con}.

	\section{IoT Traffic Collection and Synthesis}\label{sec:c1_data}
		%Data Collection here.
		In this section, we describe our smart environment infrastructure for collecting and synthesizing traffic from various IoT devices.
		{\rev
		\begin{figure}[b!]
			\centering
			\includegraphics[width=0.8\textwidth]{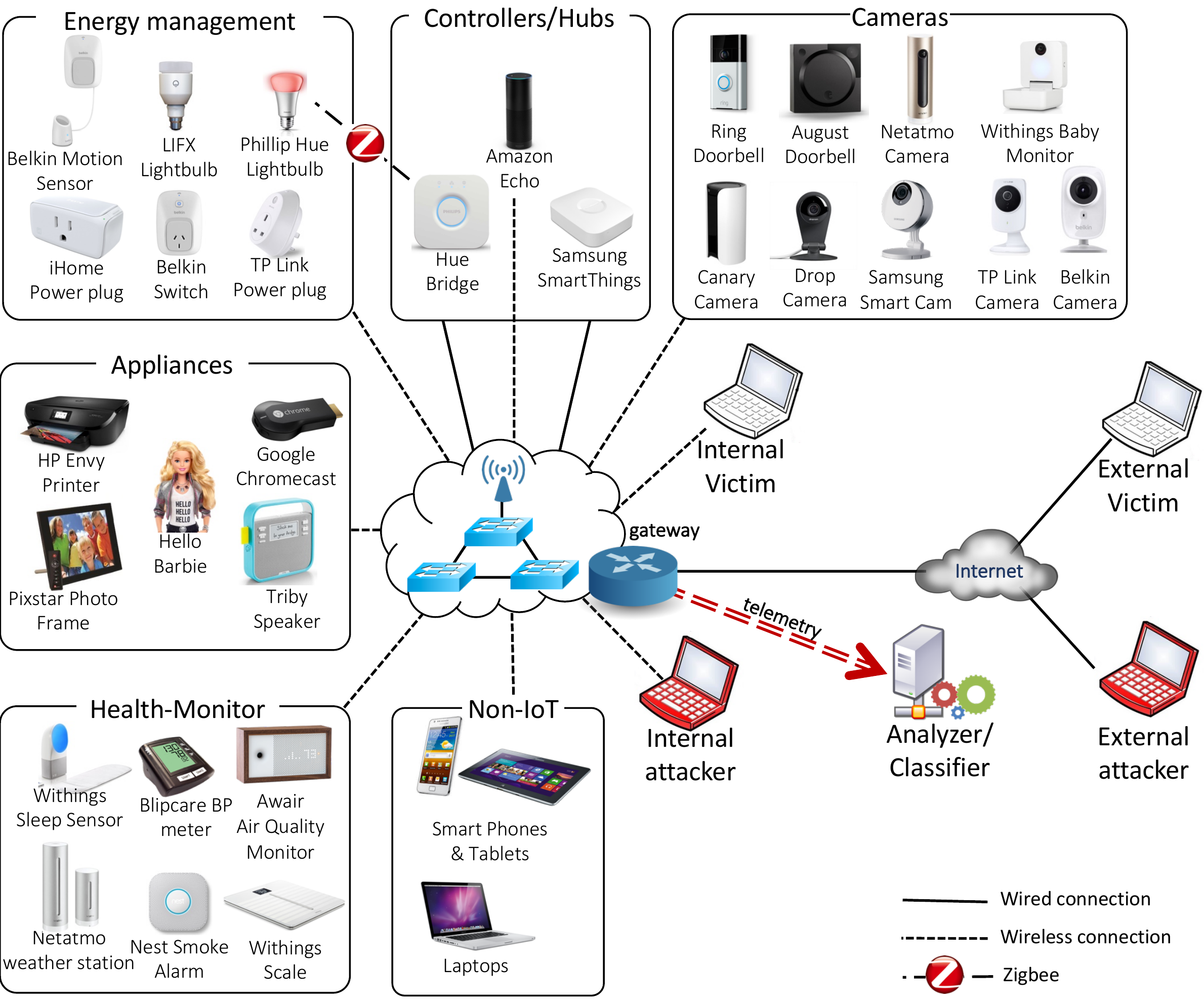}
			
			\caption{Testbed architecture showing connected 28 different IoT devices along with several non-IoT devices, and telemetry collected across the infrastructure is fed to our classification models.}
			\label{fig:labSetup}
		\end{figure}
		}
	
		\subsection{Experimental Test-bed}
			A real-life architecture of a ``smart environment'' is depicted in Fig.~\ref{fig:labSetup} that serves a wide range of IoT and non-IoT devices over its (wired/wireless) network infrastructure and allows them to communicate with the Internet servers via a gateway. Our lab setup is a specialized implementation of this architecture, housed at our campus facility, comprises one node of TP-Link Archer C7 v2 WiFi access point (representing internal switch) collocated with the Internet gateway. The TP-Link access point, flashed with the OpenWrt firmware release Chaos Calmer (15.05.1, r48532), serves as the gateway to the public Internet. We also installed additional OpenWrt packages on the gateway, namely \texttt{tcpdump (4.5.1-4)} for capturing traffic, \texttt{bash (4.3.39-1)} for scripting, \texttt{block-mount} package for mounting external USB storage on the gateway, \texttt{kmod-usb-core} and \texttt{kmod-usb-storage (3.18.23-1)} for storing the traffic trace data on the USB storage.

			In our lab setup, the WAN interface of the TP-Link access point is connected to the public Internet via the university network, while the IoT devices are connected to the LAN and WLAN interfaces respectively. Our smart environment has a total of 28 unique IoT devices representing different categories along with several non-IoT devices.
			Here, IoT refers to specific-purpose Internet connected devices (e.g. cameras and smoke sensors), while general-purpose devices (e.g. phones and laptops) fall into the non-IoT category.
			
			The IoT devices include cameras (Nest Dropcam, Samsung SmartCam, Netatmo Welcome, Belkin camera, TP-Link Day Night Cloud camera, Withings Smart Baby Monitor, Canary camera, August door bell, Ring door bell), switches and triggers (iHome, TP-Link Smart Plug, Belkin Wemo Motion Sensor, Belkin Wemo Switch), hubs (Smart Things, Amazon Echo), air quality sensors (NEST Protect smoke alarm, Netatmo Weather station, Awair air quality monitor), electronics (Triby speaker, PIXSTAR Photoframe, HP Printer, Hello barbie, Google Chromecast), healthcare devices (Withings Smart scale, Withings Aura smart sleep sensor, Blipcare blood pressure meter) and light bulbs (Phips Hue and LiFX Smart Bulb). Several non-IoT devices were also connected to the testbed, such as laptops, mobile phones and an Android tablet. The tablet was used to configure the IoT devices as recommended by the respective device manufacturers. 

		\subsection{Trace Data}
			All the traffic on the LAN side was collected using the \texttt{tcpdump} tool running on OpenWrt \cite{OpenWrt}. It is important to have a one-to-one mapping between a physical device and a known MAC address (by virtue of being in the same LAN) or IP address (i.e. without NAT) in the traffic trace. Capturing traffic on the LAN allowed us to use MAC address as the identifier for a device to isolate its traffic from the traffic mix comprising many other devices in the network. We developed a script to automate the process of data collection and storage. The resulting traces were stored as \texttt{pcap} files on an external USB hard drive of 1 TB storage attached to the gateway.  This setup permitted continuous logging of the traffic across several months. 
		
			We started logging the network traffic in our smart environment from 1-Oct-2016 to 13-Apr-2017, i.e. over a period of 26 weeks. The raw trace data contains packet headers and payload information. The process of data collection and storage begins at midnight local time each day using the \texttt{Cron} job on OpenWrt. We wrote a monitoring script on the OpenWrt to ensure that data collection/storage was proceeding smoothly. The script checks the processes running on the gateway at 5 second intervals. If the logging process is not running, then the script immediately restarts it, thereby limiting any data loss event to only 5 seconds. To make the trace data publicly available, we set up an Apache server on a virtual machine (VM) in our university data center and wrote a script to periodically transfer the trace data from the previous day, stored on the hard drive, onto the VM. The trace data from two weeks is openly available for download at: \url{http://iotanalytics.unsw.edu.au/}. The size of the daily logs varies between 61 MB and 2 GB, with an average of 365 MB.

	\section{IoT Traffic Characterization} \label{sec:c1_IoTprofile}
	
		We now present our observations using passive packet-level analysis of traffic from 28 IoT devices over the course of 26 weeks. We study a broad range of IoT traffic characteristics including activity patterns (e.g. distribution of volume/times during active/sleep periods), and signalling (e.g. domain names requested, server-side port numbers used and TLS handshake exchanges).
	
		IoT traffic constitutes (i) traffic generated by the devices autonomously -- e.g. DNS, NTP, etc. that are unaffected by human interaction, as well as (ii) traffic generated due to users interacting with the devices -- e.g. Belkin Wemo sensor responding to detection of movement, Amazon Echo responding to voice commands issued by a user, LiFX lightbulb changing colour and intensity upon user request, Netatmo Welcome camera detecting an occupant and instructing the LiFX light bulb to turn on with a specific colour, and so on. Our dataset well captures these two types of IoT traffic from a lab that represents a living smart environment (i.e. covering periods over which humans are present or absent in the environment). 
	
		To provide insights into the IoT traffic characteristics, we show in Fig.~\ref{fig:sankey} a Sankey plot of network traffic seen over a 24 hour period for Amazon Echo and LiFX lightbulb. These devices are chosen just for illustrative purposes. Each plot depicts the flow-level information generated by the respective device. Flows are: (a) either unicast or multicast/broadcast, (b) destined to either local hosts (LAN) or Internet servers (WAN), and (c) tied to protocols (TCP, UDP, ICMP or IGMP) and port numbers.
		
		\begin{figure}[t]
			\begin{center}
				\mbox{
					\subfloat[Amazon Echo.]{
						{\includegraphics[width=0.6\textwidth]{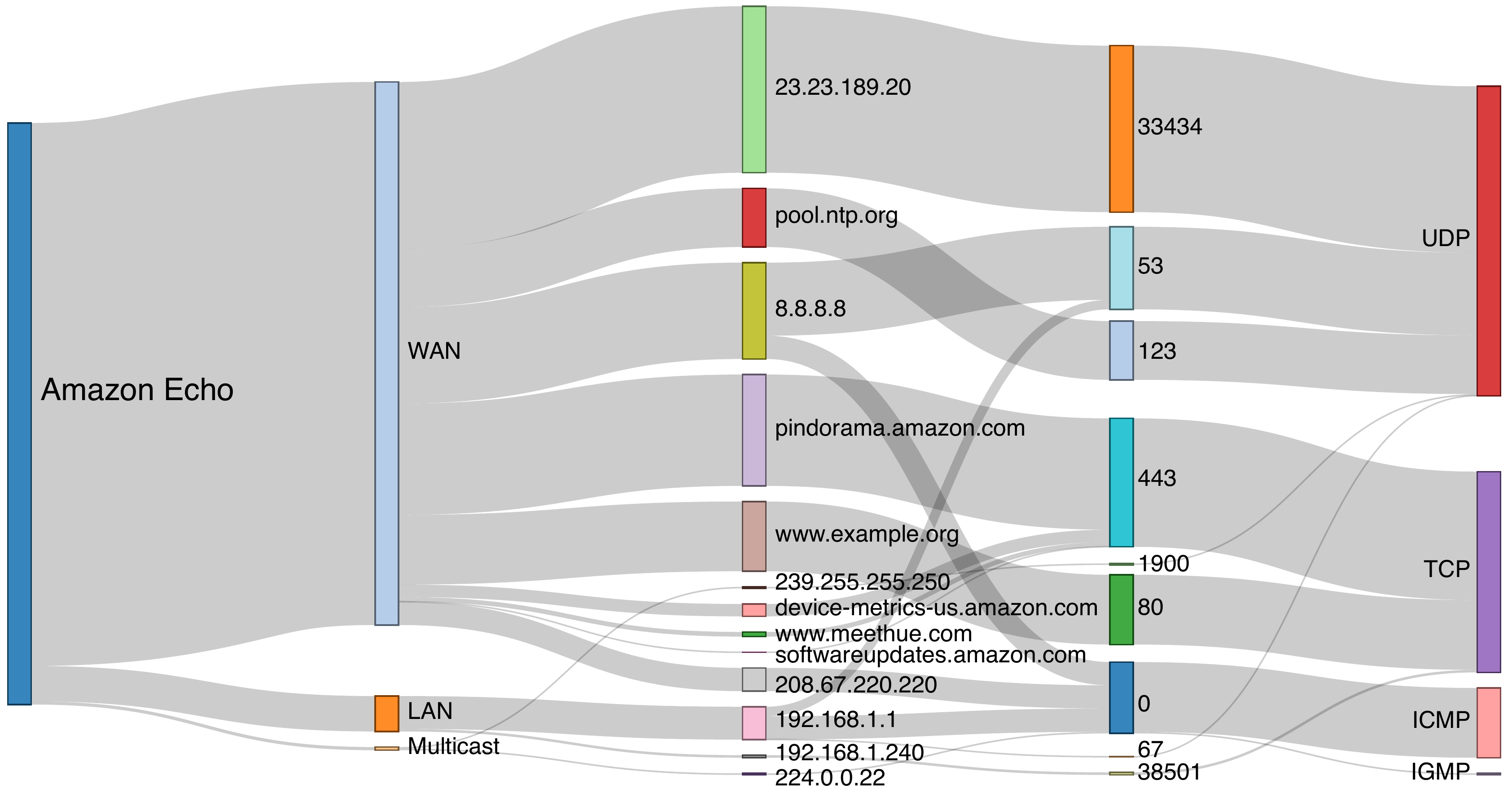}}\quad
						\label{fig:sankeyAmazonEcho}
					}
				}
				\mbox{
					\hspace{-2mm}
					\subfloat[LiFX lightbulb.]{
						{\includegraphics[width=0.6\textwidth]{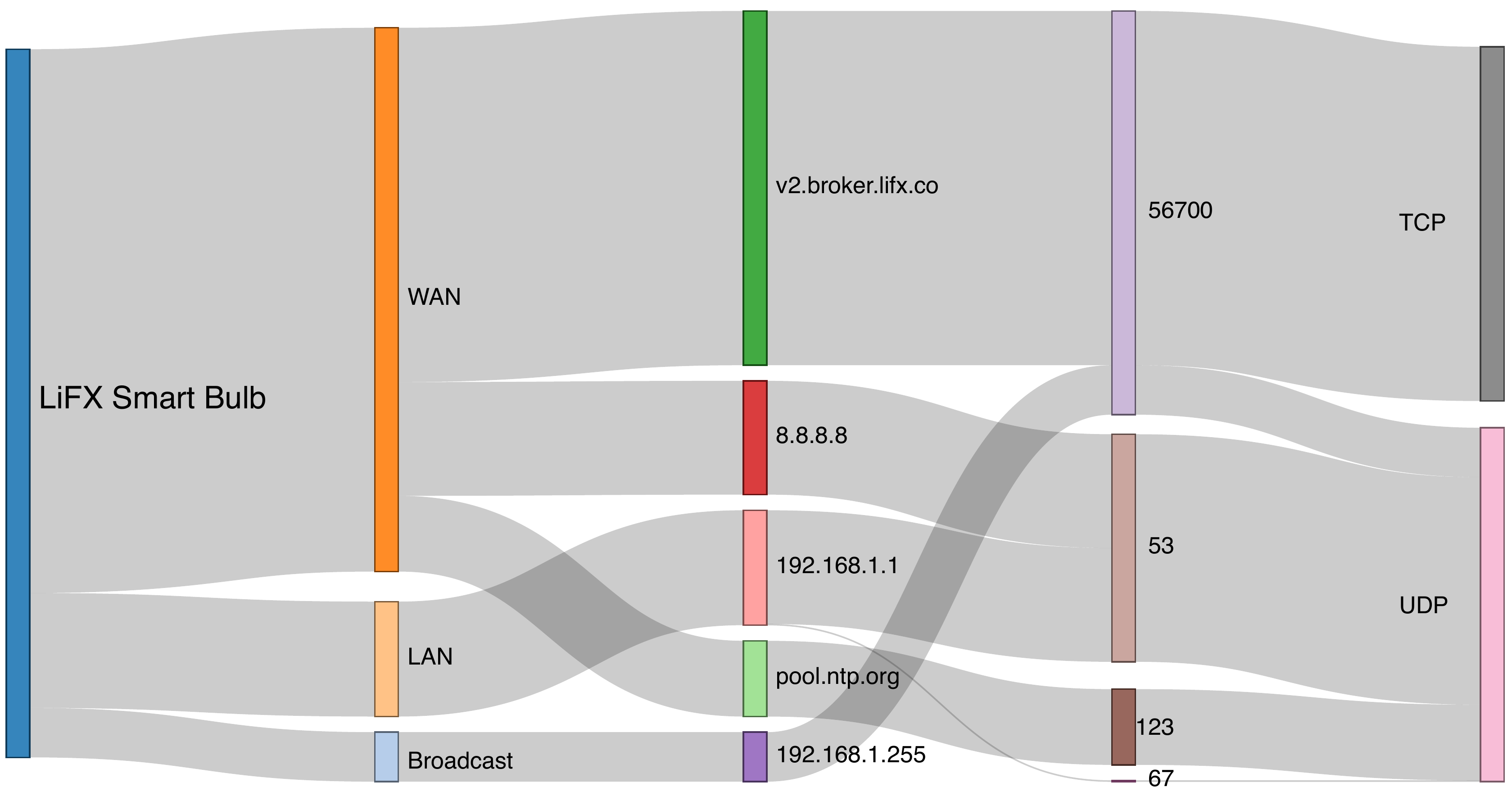}}\quad
						\label{fig:sankeyLiFX}
					}
					
				}
				\caption{Sankey diagram of daily network activity for two representative IoT devices, Amazon Echo and LiFX lightbulb. A clear distinction is observed in terms of their communication patterns, i.e. the servers they talk to, and the port numbers and protocols used for data exchange.}
				\label{fig:sankey}
			\end{center}
		\vspace{-1em}
		\end{figure}
		
		Fig.~\ref{fig:sankey} provides a visual aid depicting the underlying traffic signature exhibited by the two devices. For example, DNS (port number 53) and NTP (port number 123) are used by both Amazon Echo and LiFX lightbulb. While Amazon Echo uses HTTP (port number~80), HTTPS (port number~443) and ICMP (port number~0), LiFX lightbulb does not use any of these applications. Further, each device seems to communicate to a unique port number on a WAN server; TCP 33434 for Amazon Echo and UDP 56700 for LiFX lightbulb, as shown by the top flow in Figures~\ref{fig:sankeyAmazonEcho}~and~\ref{fig:sankeyLiFX}. Finally, we observe that Amazon Echo accesses a number of domain names including \texttt{softwareupdates.amazon.com}, 
		\texttt{device-metrics-su.amazon.com}, 
		\texttt{example.org}, 
		\texttt{pindorama.amazon.com} and \texttt{pool.ntp.org}. However, LiFX lightbulb communicates with only two domains, i.e. \texttt{v2.broker.lifx.co} and \texttt{pool.ntp.org}. 

		\subsection{IoT Activity and Volume Pattern}\label{sec:c1_IoTactive}
			
			We start with the \textit{activity} pattern of IoT devices that is defined by the properties of their traffic flows. % -- a flow is identified by a TCP connection or UDP 5-tuple with inactive timeout. 
			We define four key attributes at a per-flow level to characterize IoT devices based on their network activity: \textbf{flow volume} (i.e. sum total of download and upload bytes), \textbf{flow duration} (i.e. time between the first and the last packet in a flow), \textbf{average flow rate} (i.e. flow volume divided by the flow duration), and \textbf{device sleep time} (i.e. time interval over which the IoT device has no active flow).  
			
			We plot in Fig.~\ref{fig:IoTactivity} the probability distribution of the above four attributes for a chosen set of IoT devices using the trace data collected over 26 weeks. It can be observed from Fig.~\ref{fig:loTvol} that each IoT device tends to exchange a small amount of data per flow. For the case of the LiFX lightbulb (depicted by red bars), $26$\% of flows transfer between [$130$, $140$] bytes and $20$\% between [$120$, $130$] bytes. The flow volume for the Belkin motion sensor (depicted by green bars) is slightly higher; over $35$\% of flows transfer between [$2800$, $3800$] bytes. For the Amazon Echo (depicted by blue bars), over $95$\% of flows transfer less than $1000$ bytes. Though we present the flow volume histogram for only a few devices, most of our IoT devices exhibit a similar predictable pattern.
			
			\begin{figure}[t]
				\begin{center}
					\mbox{
						\subfloat[Flow volume.]{
							{\includegraphics[width=0.45\textwidth,height=0.29\textwidth]{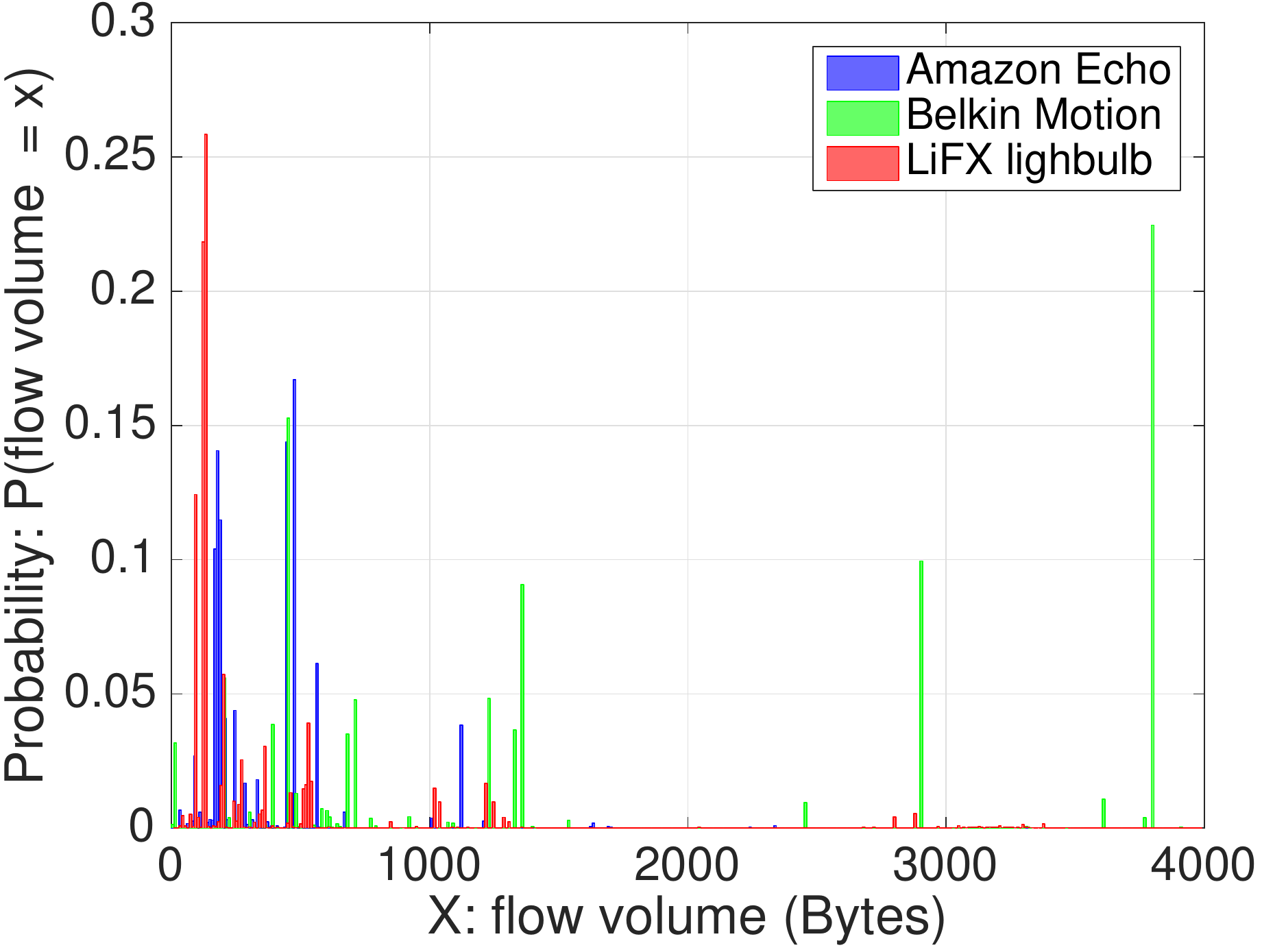}}\quad
							\label{fig:loTvol}
						}
						\subfloat[Flow duration.]{
							{\includegraphics[width=0.45\textwidth,height=0.29\textwidth]{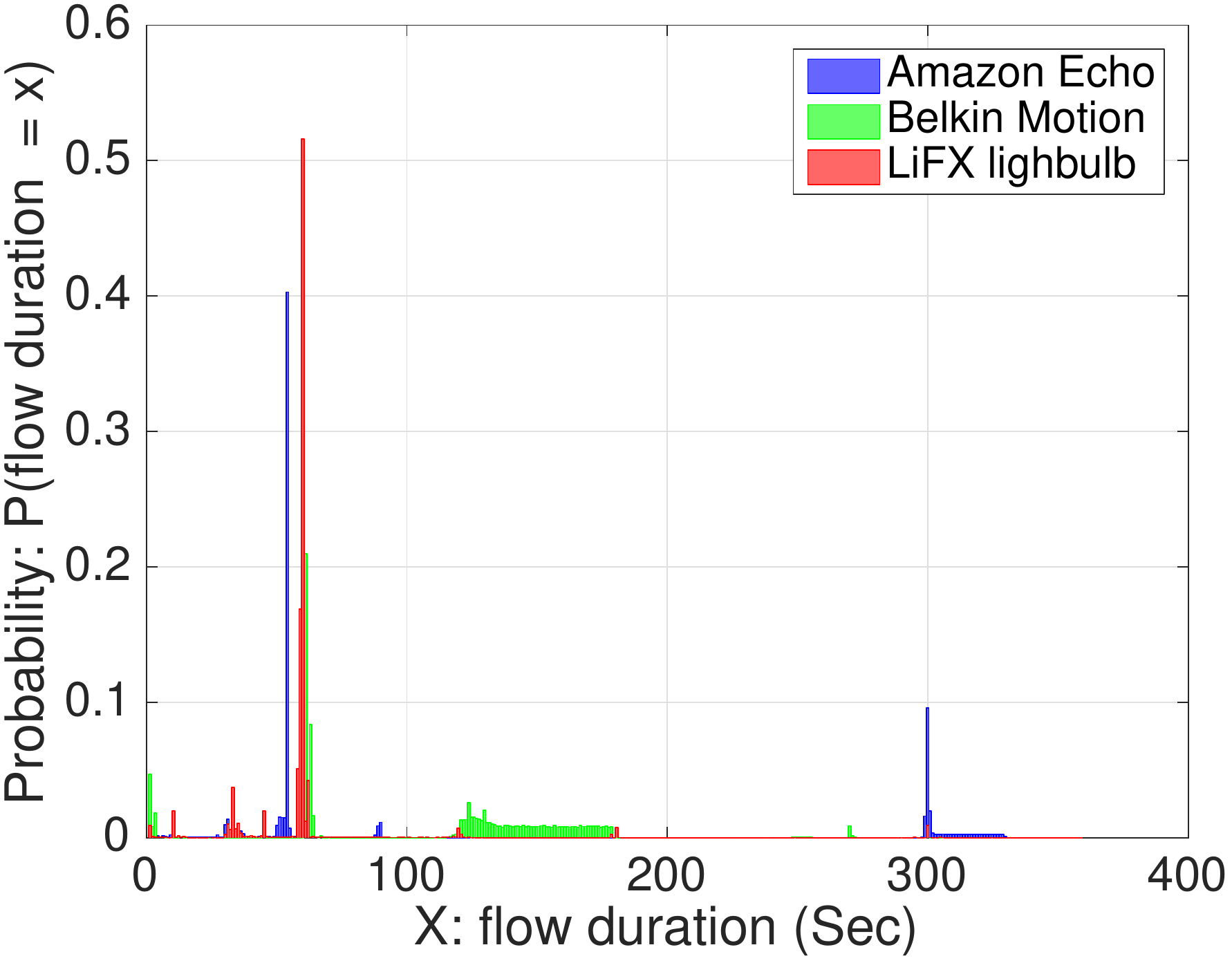}}\quad
							\label{fig:IoTdur}
						}
					}
					
					\mbox{
						\subfloat[Average flow rate.]{
							{\includegraphics[width=0.45\textwidth,height=0.29\textwidth]{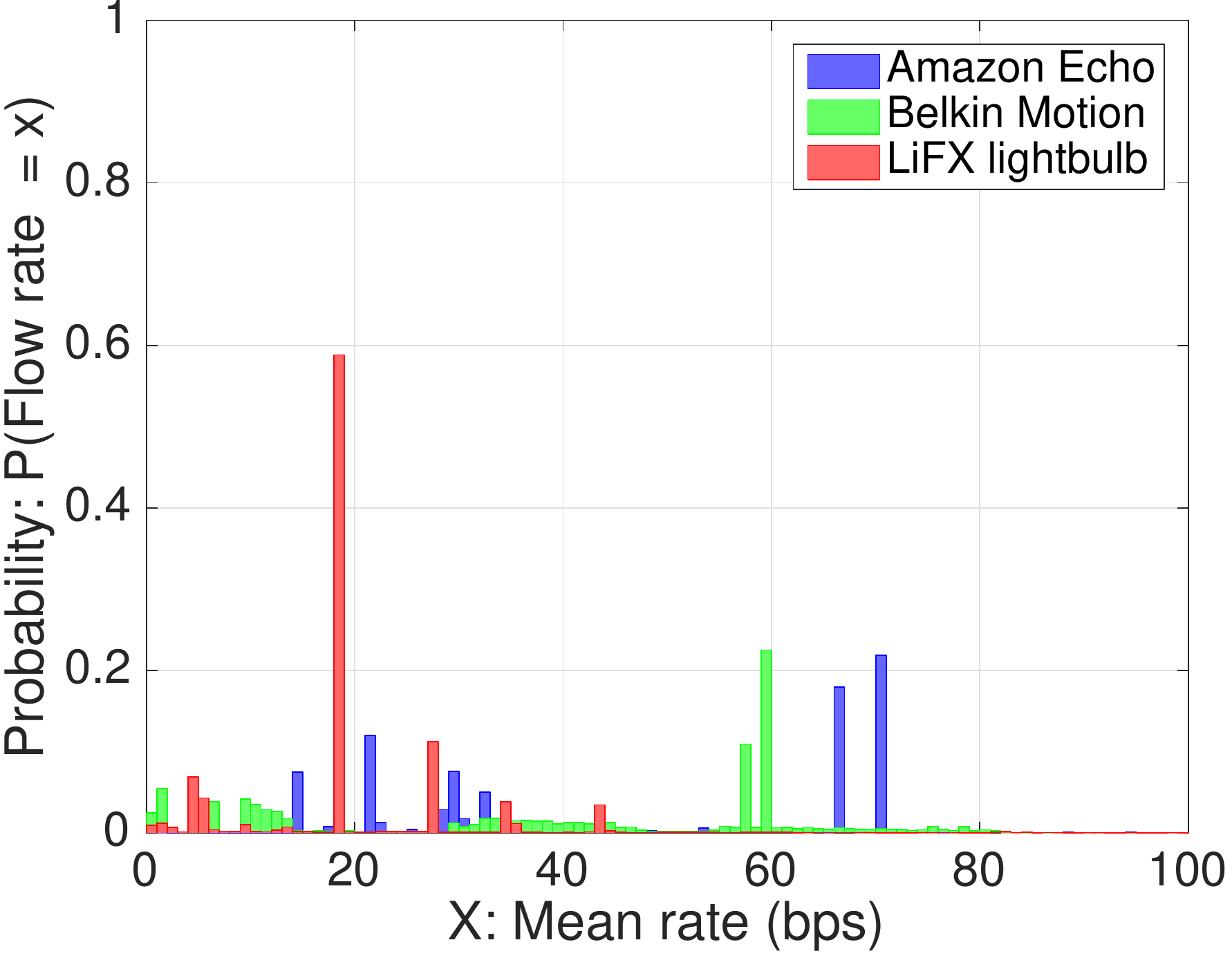}}\quad
							\label{fig:IoTmeanrate}
						}
						\subfloat[Device sleep time.]{
							{\includegraphics[width=0.45\textwidth,height=0.29\textwidth]{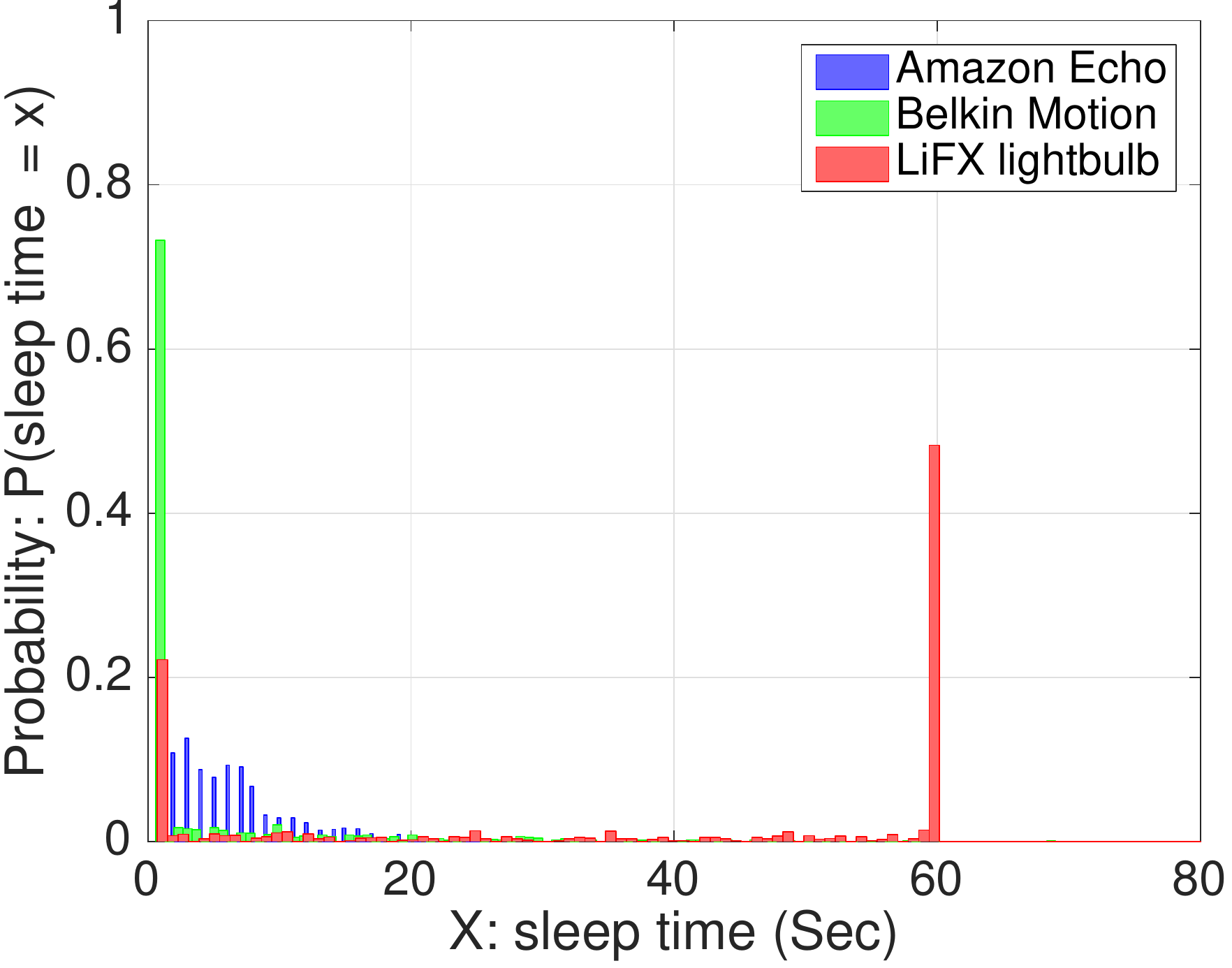}}\quad
							\label{fig:IoTsleep}
						}
					}
					
					\caption{Distribution of IoT activity pattern: (a) flow volume, (b) flow duration, (c) average flow rate and (d) device sleep time.}
					
					\label{fig:IoTactivity}
				\end{center}
			\end{figure}
			
			A similar pattern emerges for the flow duration as well. Referring to Fig.~\ref{fig:IoTdur}, we note that the flow duration of $53$ seconds is seen in more than $40$\% of flows for Amazon Echo, while a duration of $60$ seconds is seen for the LiFX lightbulb and Belkin motion sensor  with a probability of $50$\% and $21$\% respectively.     
			
			For the average flow rate attribute, Fig.~\ref{fig:IoTmeanrate} shows that the mean rate is rather small, in the bits-per-second range as one would qualitatively expect. Quantitatively, the figure shows that the LiFX lightbulb has an average flow rate of $18$ bits-per-second nearly $60$\% of the time. Nearly $30$\% of Belkin flows have a bit rate in the range $59$ to $60$ bits-per-second while nearly $40$\% Amazon Echo flows have a bit range in the range $70$ to $71$ bits-per-second.
			
			Lastly, in terms of the sleep time for the devices Fig.~\ref{fig:IoTsleep} shows that the Belkin motion sensor and the LiFX lightbulb exhibit a distinct sleep pattern. The duration is $1$ second and $60$ seconds with probability $73$\% and $48$\% respectively. However, multiple sleep times with small probabilities are observed for the Amazon Echo. This is because Amazon Echo keeps its TCP connections alive and goes to sleep only when it disconnects from the Internet. Other devices in our test-bed also perform like the Echo and do not seem to have a dominant sleep pattern.
			
		\subsection{IoT Signaling Pattern}\label{sec:c1_IoTprotocol}
			We now focus on the application layer protocols, inferred using the port numbers, that IoT devices mostly use to communicate locally in the LAN and/or externally with servers on the public Internet.
	
			\subsubsection{Server port numbers}\label{sec:c1_IoTprotocolPorts}
				\begin{figure}[t]
					\begin{center}
						\mbox{
							\subfloat[Amazon Echo \{10\}.]{
								{\includegraphics[width=0.3\textwidth]{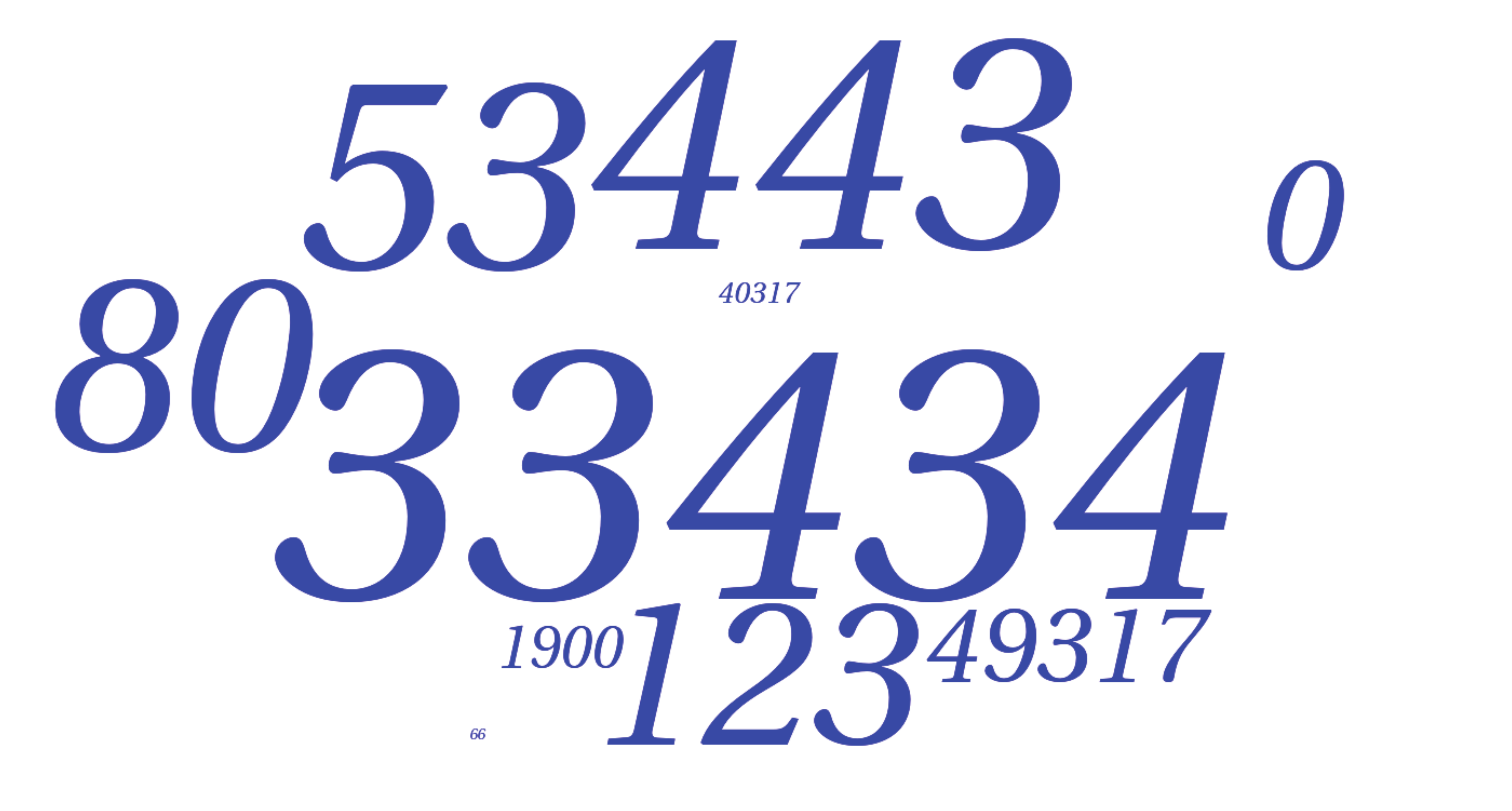}}\quad
								\label{fig:PortcloudAmazon}
							}
							\hspace{-2mm}
							\subfloat[LiFX lightbulb \{5\}.]{
								{\includegraphics[width=0.3\textwidth]{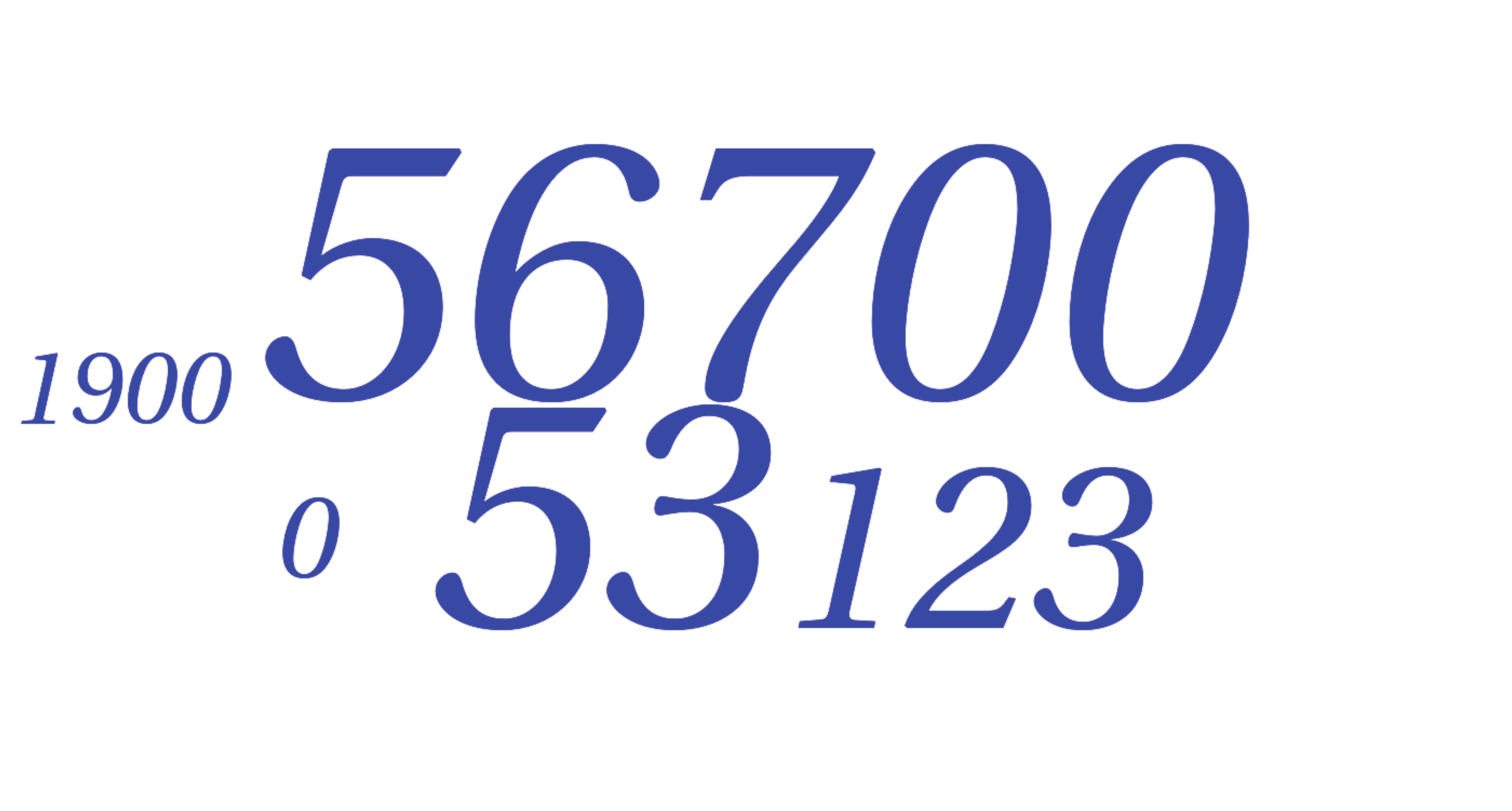}}\quad
								\label{fig:PortcloudLiFX}
							}
							\hspace{-2mm}
							\subfloat[Awair air monitor \{7\}.]{
								{\includegraphics[width=0.3\textwidth]{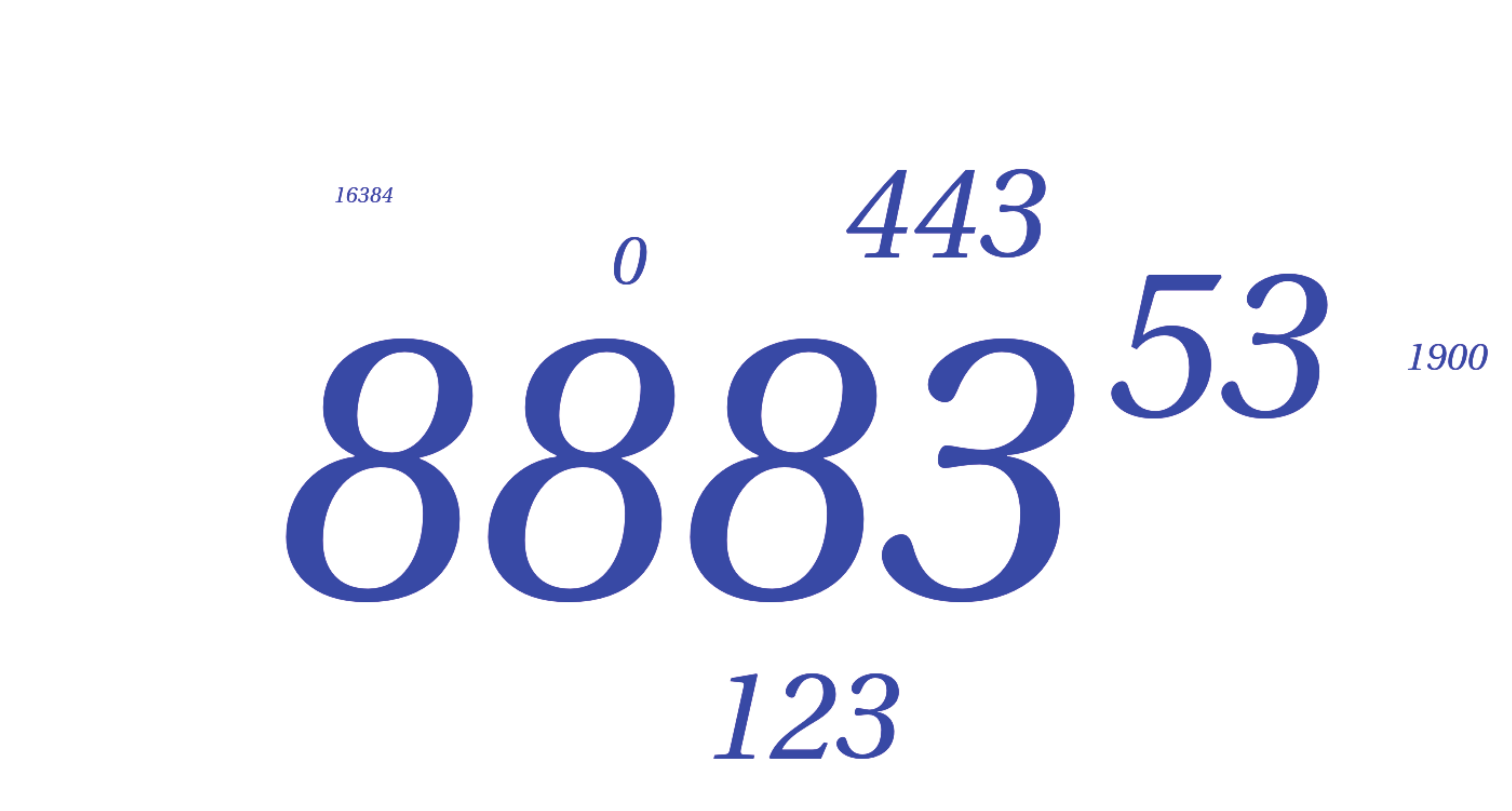}}\quad
								\label{fig:PortcloudAwair}
							}
						}
						\mbox{
							\subfloat[Belkin motion sensor \{7\}.]{
								{\includegraphics[width=0.3\textwidth]{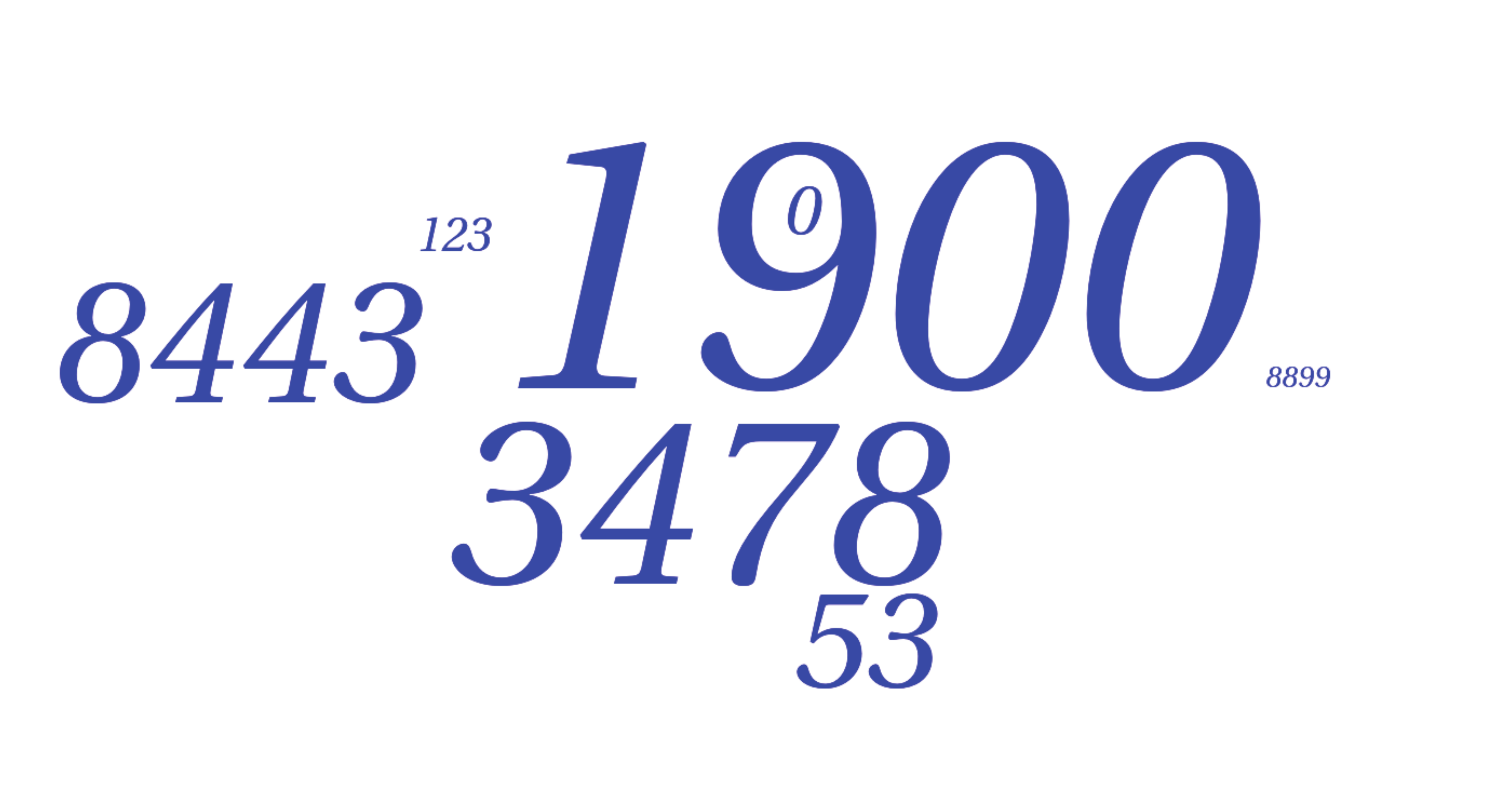}}\quad
								\label{fig:PortcloudBelkinMotion}
							}
							\hspace{-2mm}
							\subfloat[Belkin power switch \{7\}.]{
								{\includegraphics[width=0.3\textwidth]{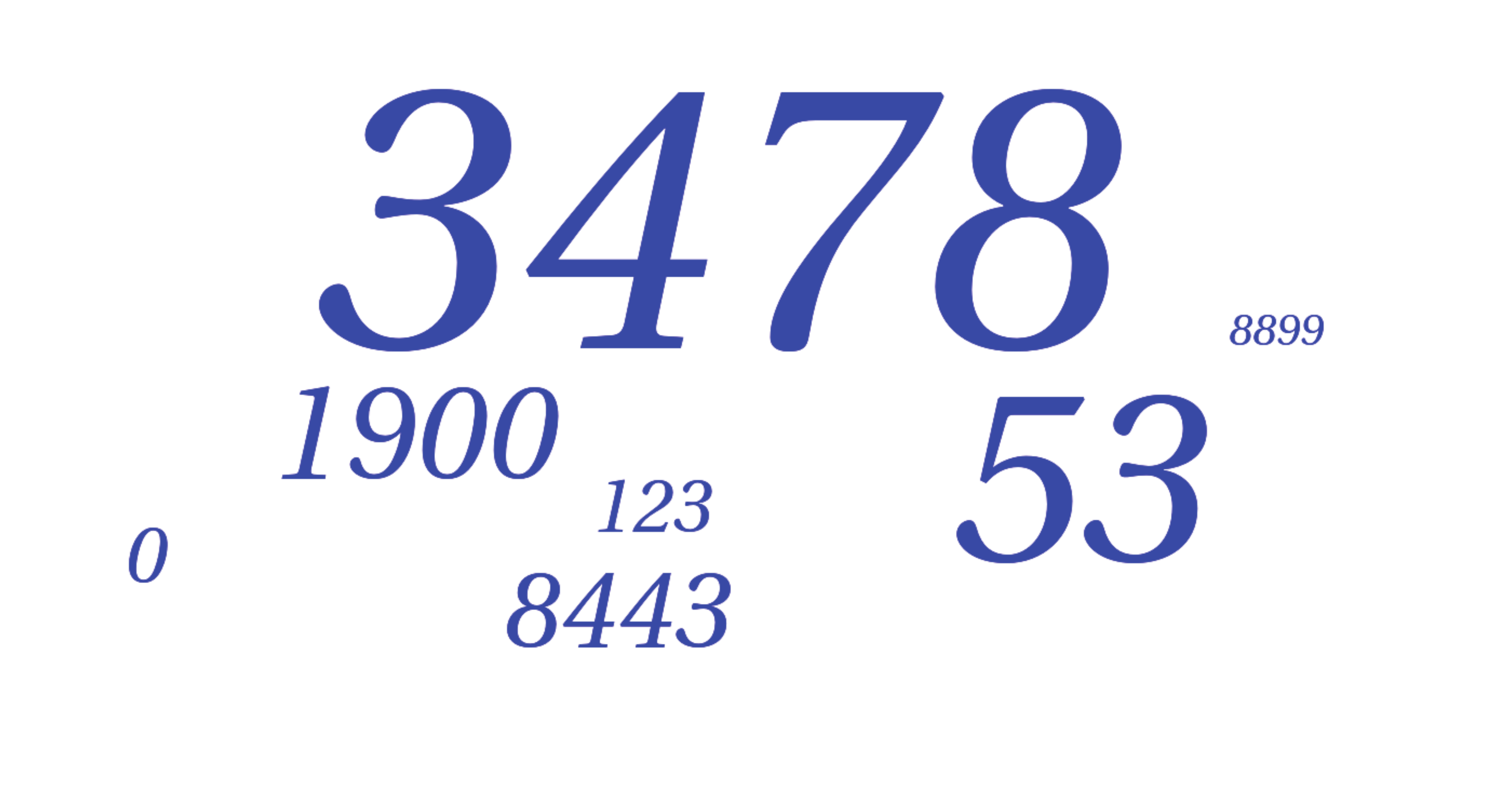}}\quad
								\label{fig:PortcloudBelkinSwitch}
							}
							\hspace{-2mm}
							\subfloat[Belkin camera \{9\}.]{
								{\includegraphics[width=0.3\textwidth]{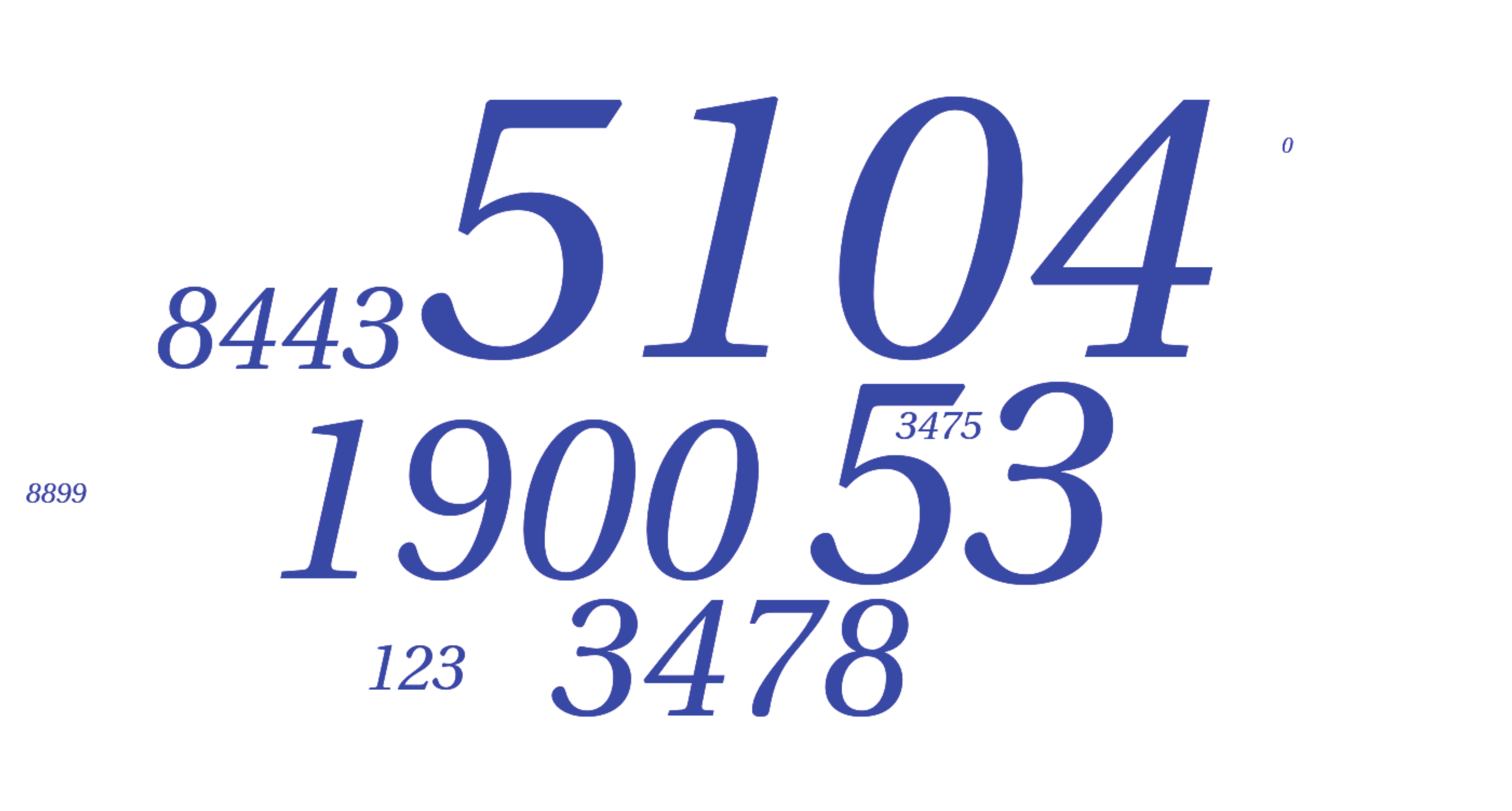}}\quad
								\label{fig:PortcloudBelkinCam}
							}
						}
						\mbox{
							\hspace{-2mm}
							\subfloat[Netatmo weather station \{4\}.]{
								{\includegraphics[width=0.3\textwidth]{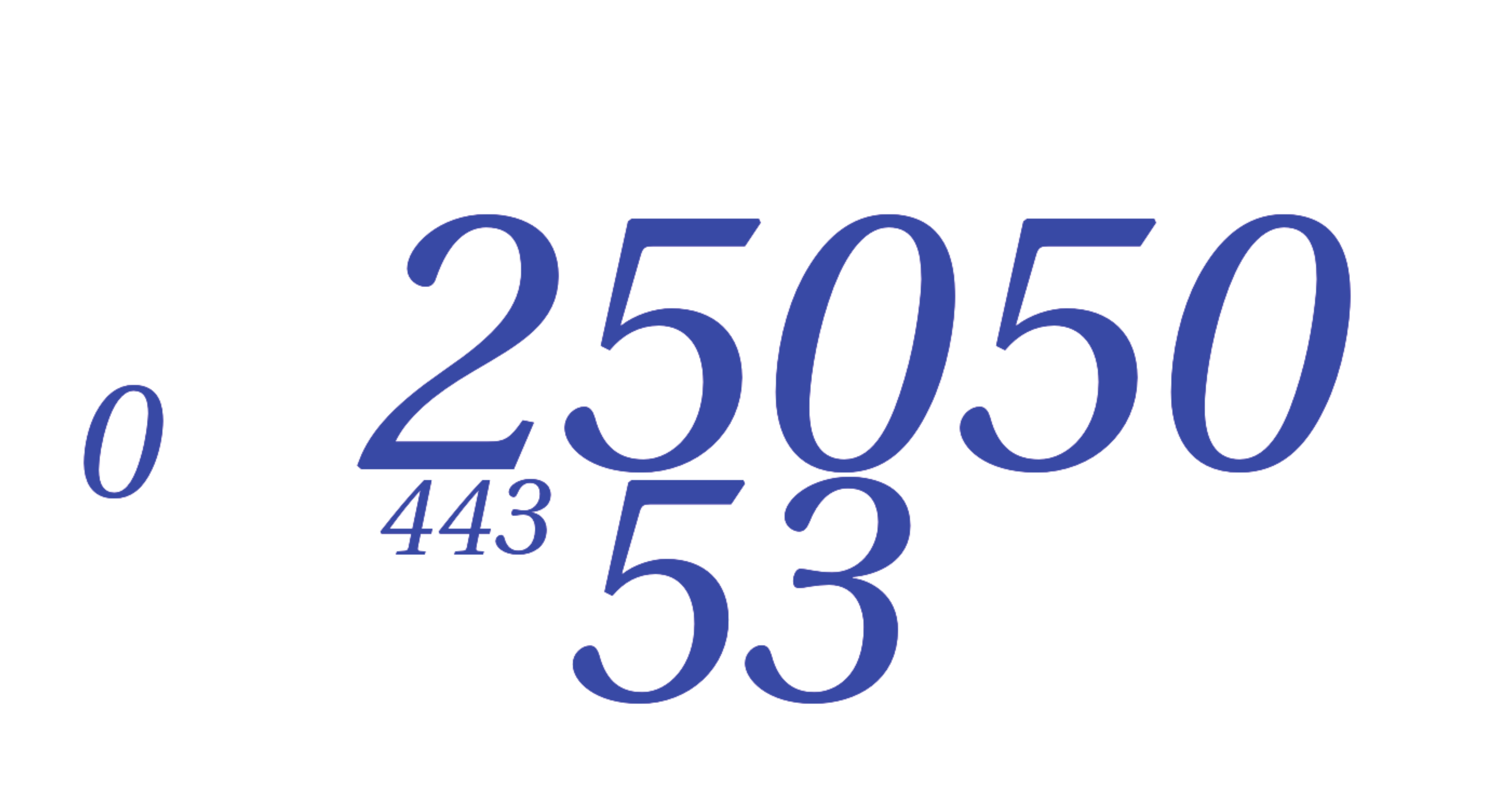}}\quad
								\label{fig:PortcloudNetAtmoWeather}
							}
							\hspace{-2mm}
							\subfloat[Non-IoT \{2382\}.]{
								{\includegraphics[width=0.3\textwidth]{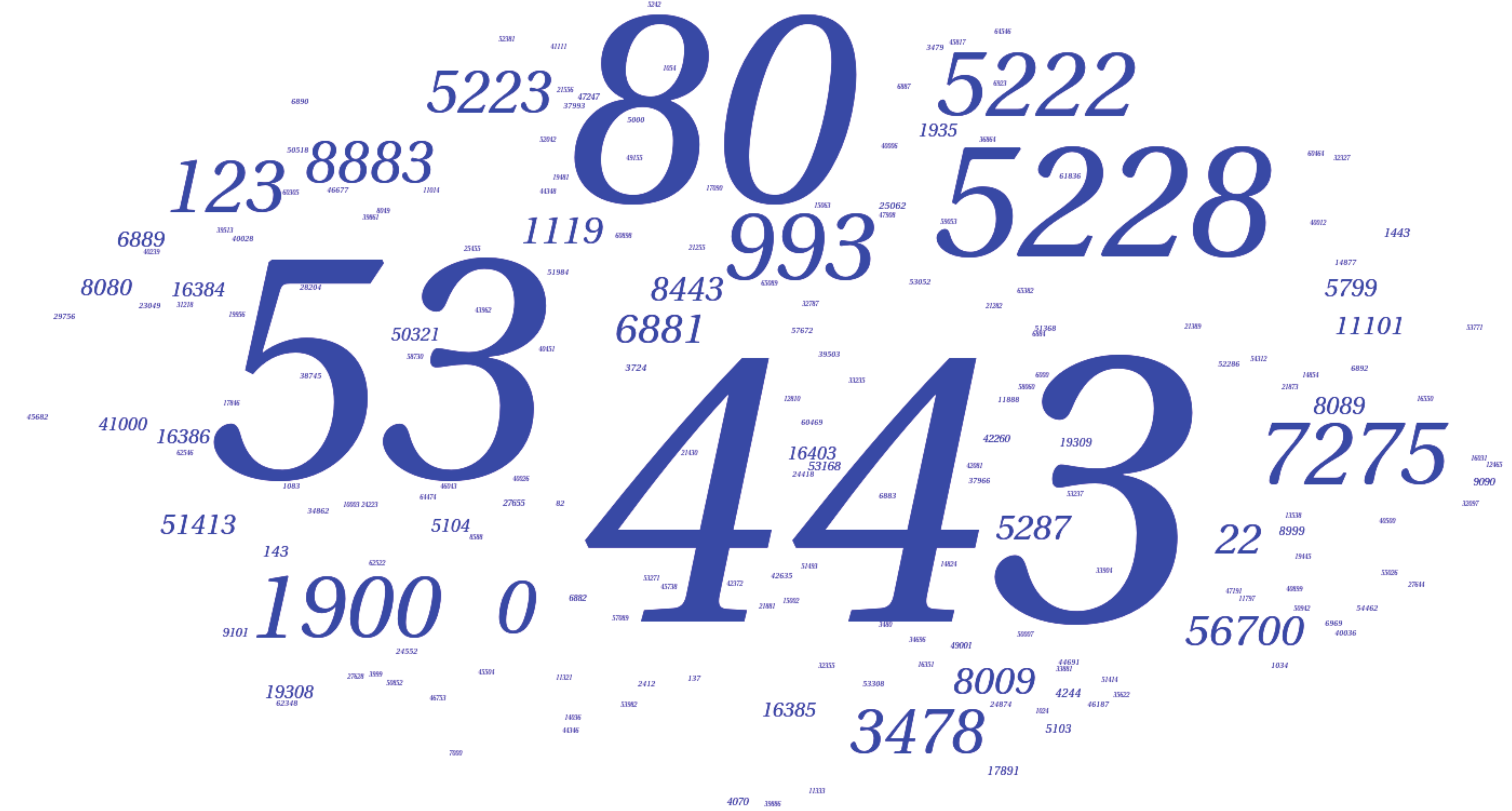}}\quad
								\label{fig:PortcloudNoneIoT}
							}
						}
						
						\caption{Word-cloud of server ports (total count of unique ports is shown in \{sub-captions\} next to the device name).}
						
						\label{fig:cloudPorts}
					\end{center}
				\end{figure}
			
				Fig.~\ref{fig:cloudPorts} shows the word cloud of server-side port numbers of all flows initiated from a variety of IoT devices. For each device, if a port is used more frequently then it  is shown by a larger font-size in the respective word cloud. Sub-captions (i.e. numbers within \{\}) report the number of unique server ports for each device. It can be seen that IoT devices each uniquely communicate with a handful of server ports whereas non-IoT devices use a much wider range of services (i.e. 2382 unique ports are shown in Fig.~\ref{fig:PortcloudNoneIoT} and many of them are very infrequent). We observe that non-standard ports 33434, 56700, 8883, and 25050 are prominently seen in traffic originating from Amazon Echo, LiFX lightbulb, Awair air quality monitor, and Netatmo weather station respectively, as shown in the top row of Fig.~\ref{fig:cloudPorts}. Further, we note devices from the same manufacturer share certain ports. For example, port numbers 8443 and 3478 are common between Belkin's motion sensor, power switch, and camera, as shown in Figures~\ref{fig:PortcloudBelkinMotion}-\ref{fig:PortcloudBelkinCam}. We also note that well-known standard port numbers such as 53 (DNS), 123 (NTP), 0 (ICMP) and 1900 (SSDP) are used by many of the IoT devices as well as the non-IoTs with various frequencies, as shown in Fig.~\ref{fig:cloudPorts}. Moreover, the server-side port number of 443 (TLS/SSL) is also used by many of the IoT devices.
				     
			\vspace{-1em}	
			\subsubsection{DNS queries}\label{sec:c1_IoTprotocolDNS}
			\vspace{-1em}
				
				\begin{figure}[t]
					\begin{center}
						\mbox{
							\subfloat[Amazon Echo \{30\}.]{
								{\includegraphics[width=0.28\textwidth]{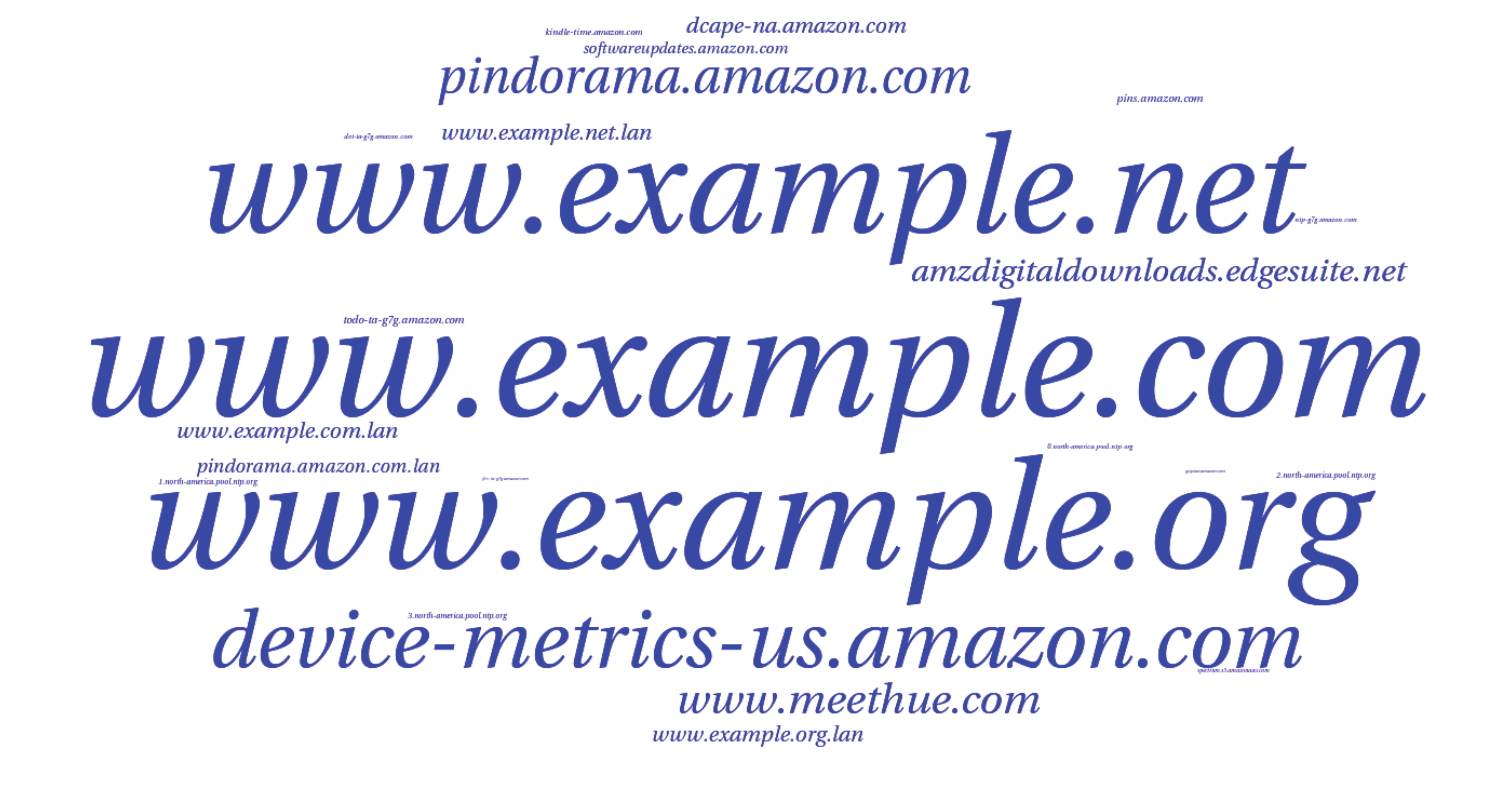}}\quad
								\label{fig:DNScloudAmazon}
							}
							\hspace{-2mm}
							%			\subfloat[Blipcare blood-pressure monitor.]{
							%				{\includegraphics[width=0.29\textwidth]{images/TMC/data/cloudwords/dns/DNS/BlipcareBP}}\quad
							%				\label{fig:DNScloudBlipcareBP}
							%			}
							\subfloat[Google Dropcam \{5\}.]{
								{\includegraphics[width=0.28\textwidth]{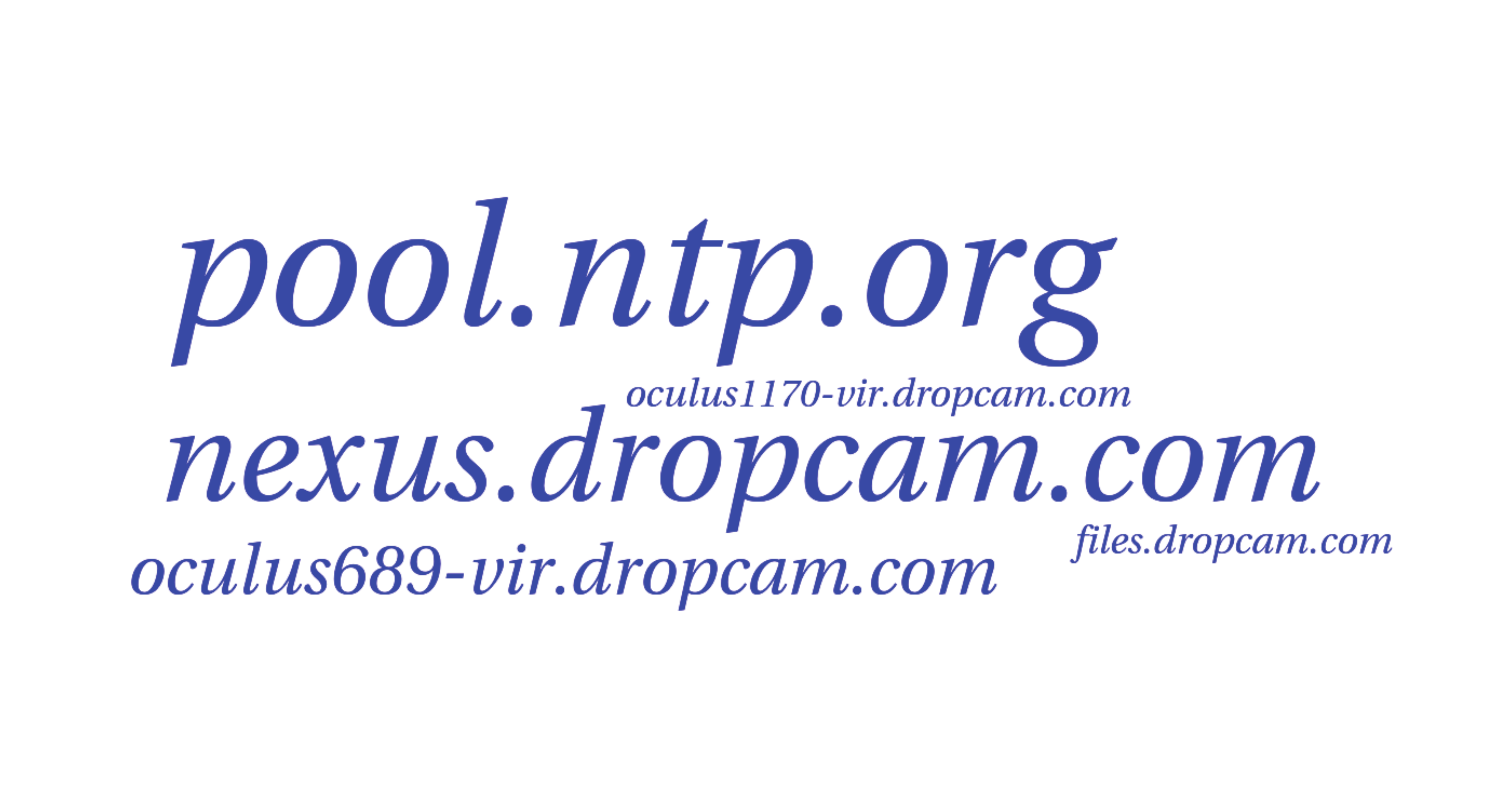}}\quad
								\label{fig:DNScloudDropcam}
							}
							\hspace{-2mm}
							\subfloat[HP printer \{6\}.]{
								{\includegraphics[width=0.28\textwidth]{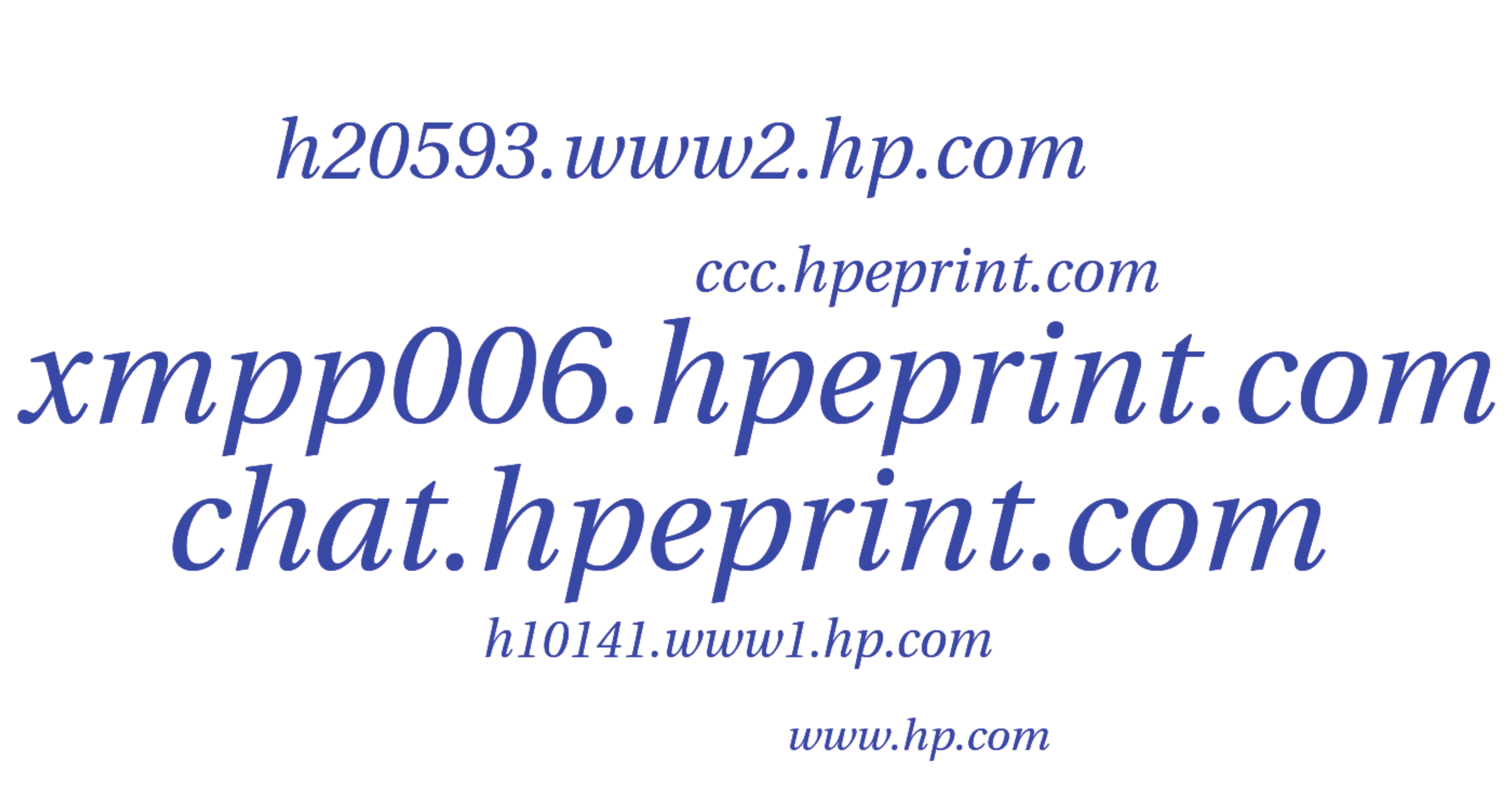}}\quad
								\label{fig:DNScloudHP}
							}
						}
						\mbox{
							\subfloat[Belkin camera \{11\}.]{
								{\includegraphics[width=0.28\textwidth]{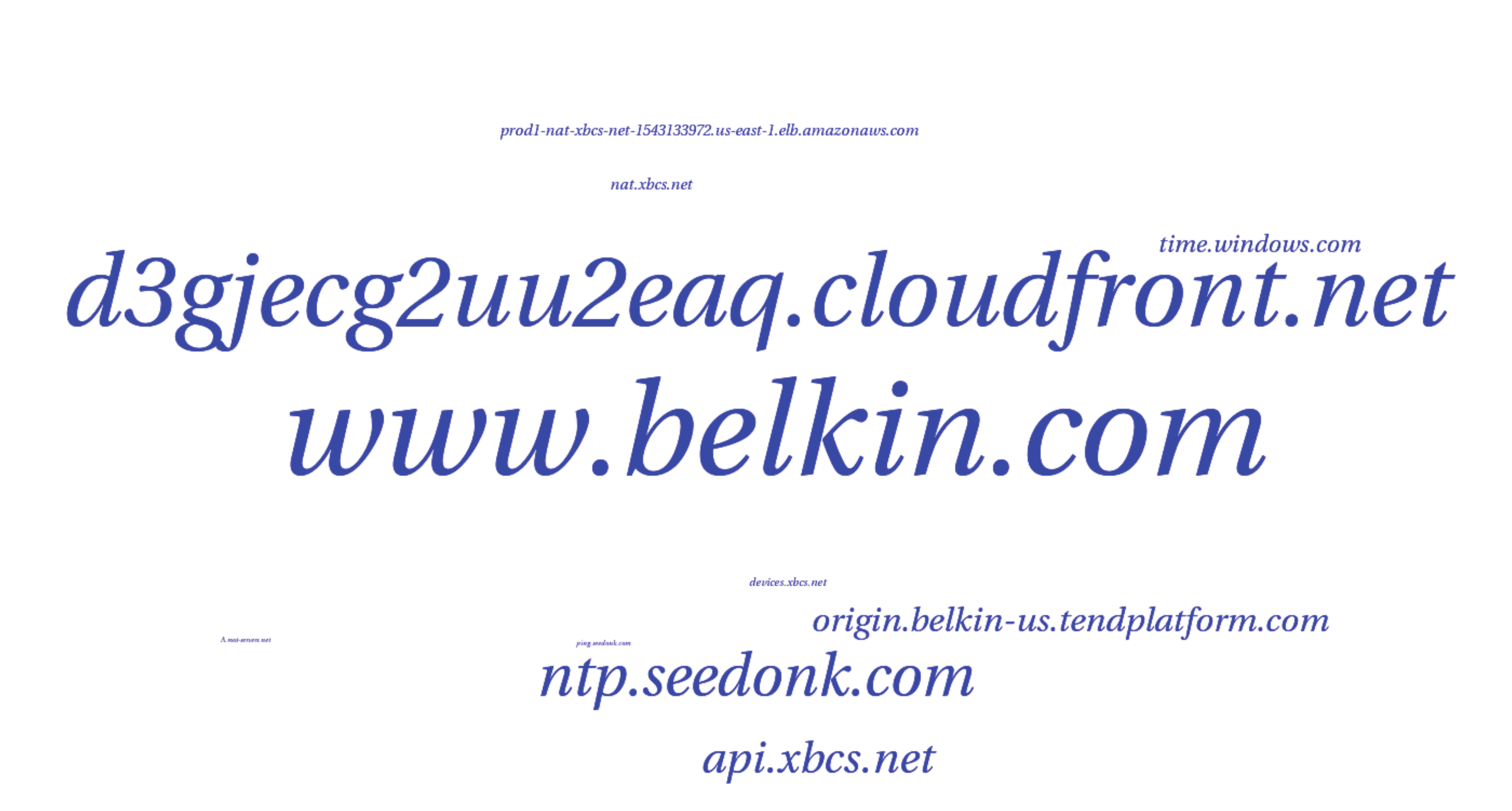}}\quad
								\label{fig:DNScloudBelkinCam}
							}
							\hspace{-2mm}
							\subfloat[Belkin motion sensor \{5\}.]{
								{\includegraphics[width=0.28\textwidth]{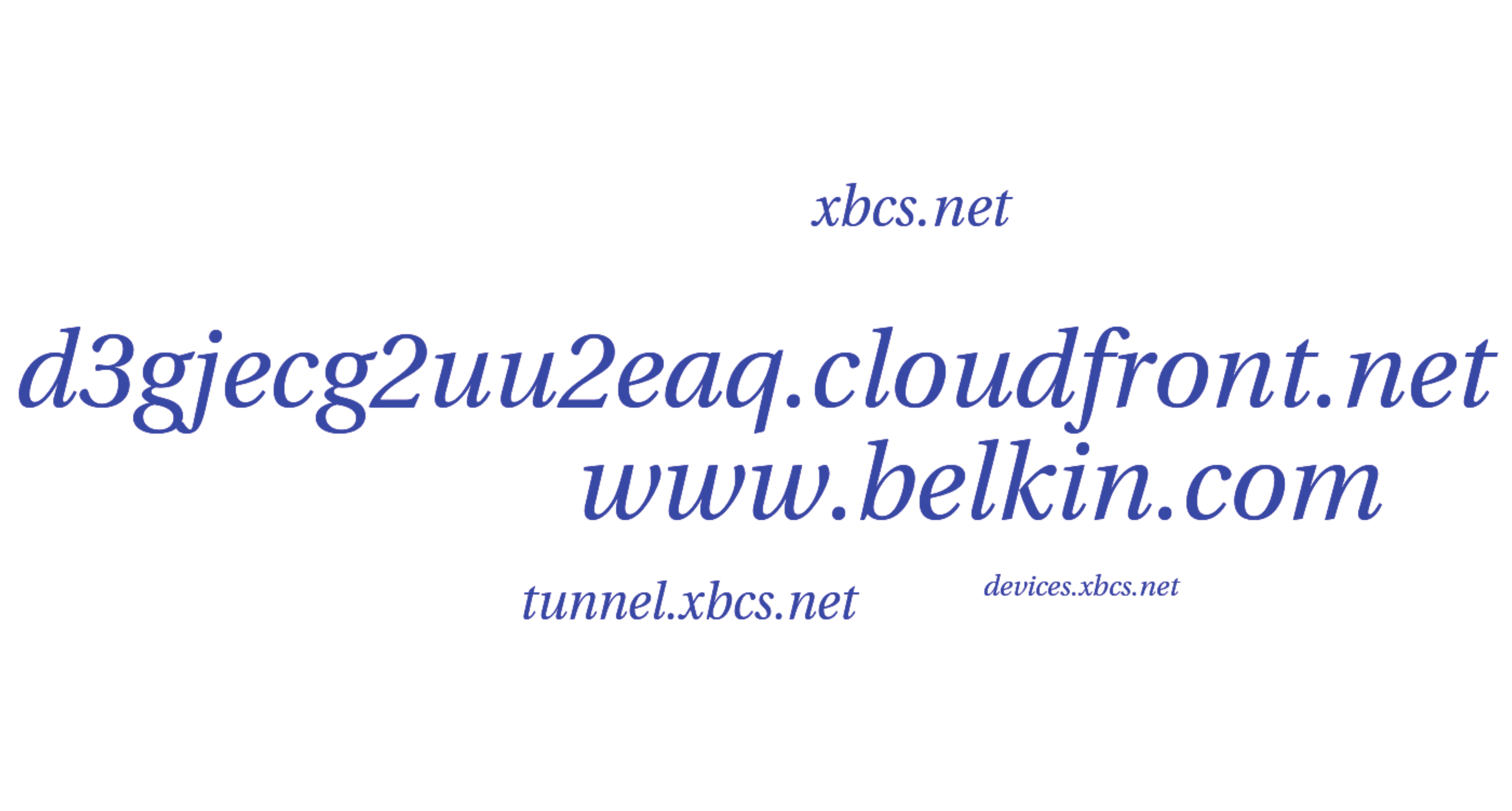}}\quad
								\label{fig:DNScloudBelkinMotion}
							}
							\hspace{-2mm}
							\subfloat[Belkin power switch \{8\}.]{
								{\includegraphics[width=0.28\textwidth]{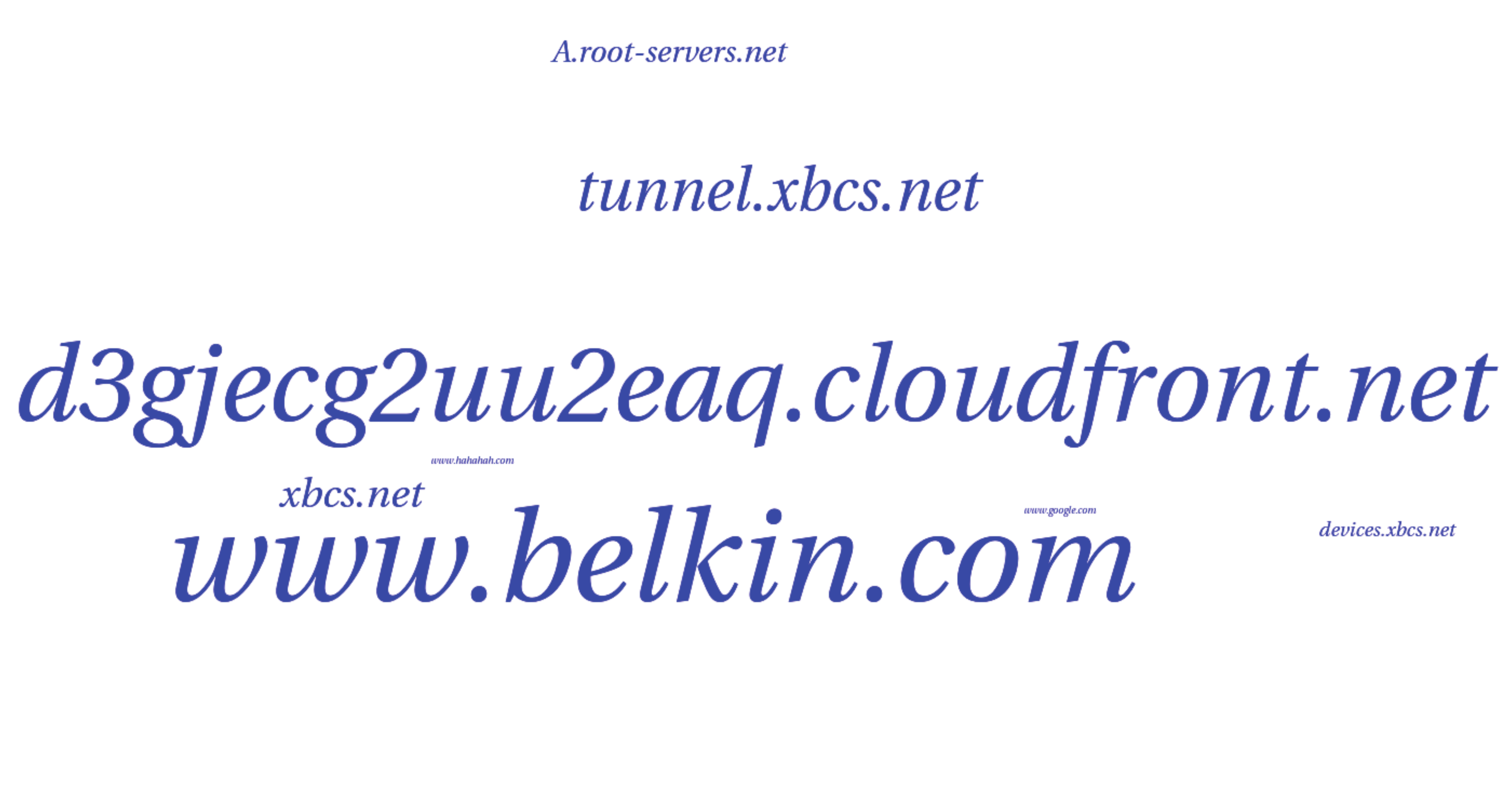}}\quad
								\label{fig:DNScloudBelkinSwitch}
							}
						}
						\mbox{
							
							\hspace{-2mm}
							\subfloat[Awair air quality~\{5\}.]{
								{\includegraphics[width=0.28\textwidth]{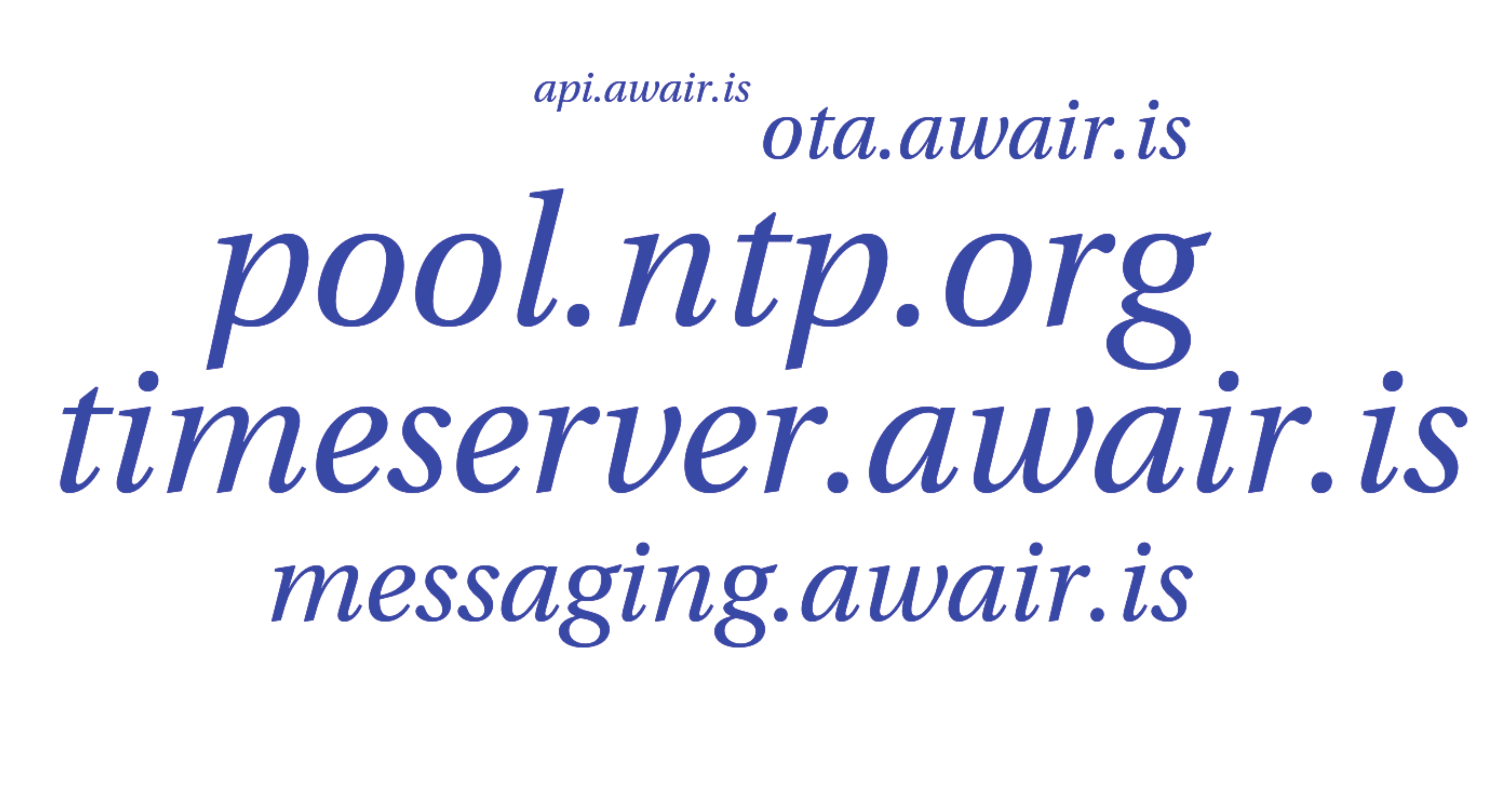}}\quad
								\label{fig:DNScloudAwair}
							}
							\hspace{-2mm}
							\subfloat[LiFX lightbulb \{2\}.]{
								{\includegraphics[width=0.28\textwidth]{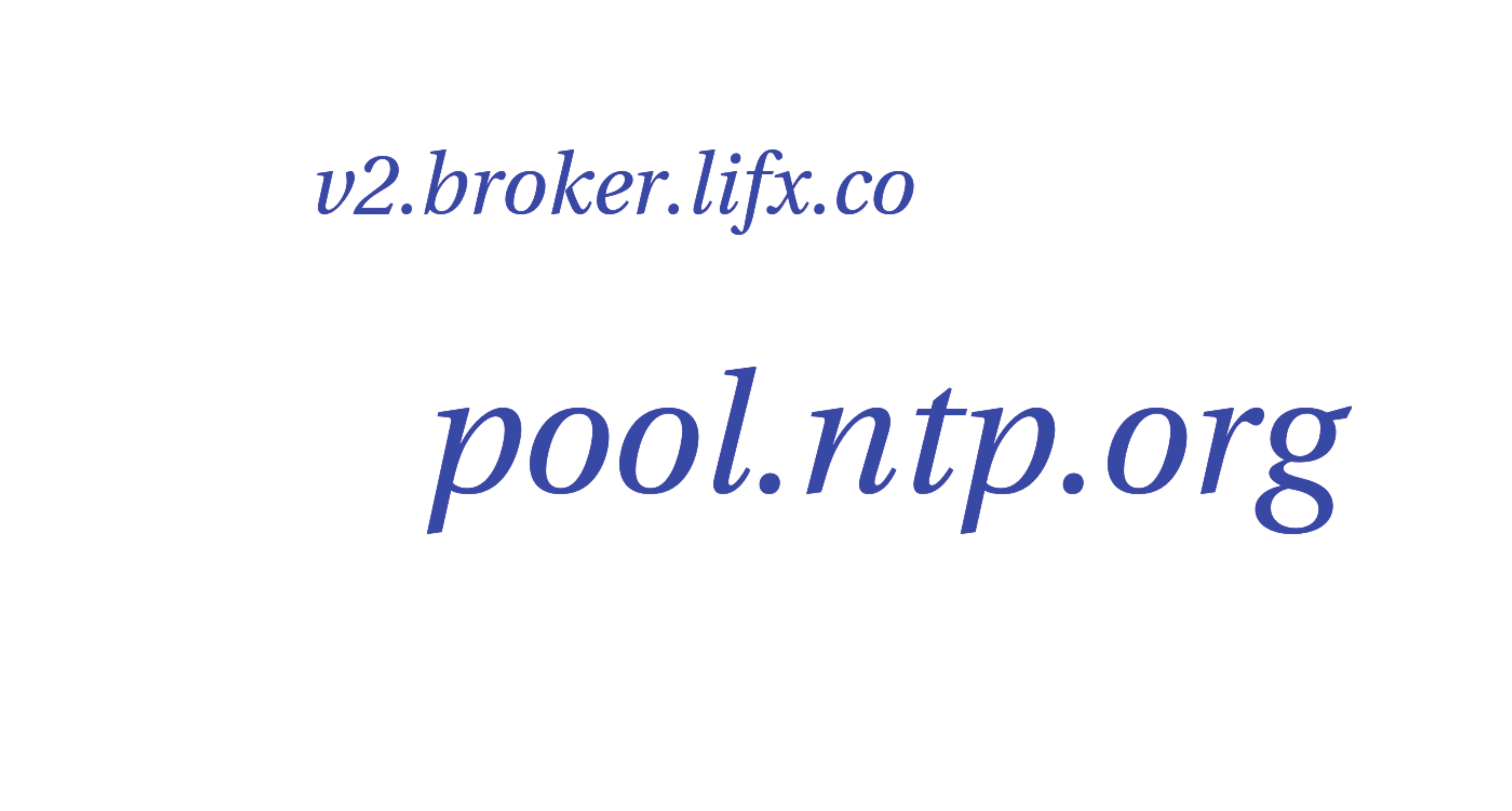}}\quad
								\label{fig:DNScloudLifx}
							}
							\hspace{-2mm}
							\subfloat[Non-IoT \{11927\}.]{
								{\includegraphics[width=0.28\textwidth]{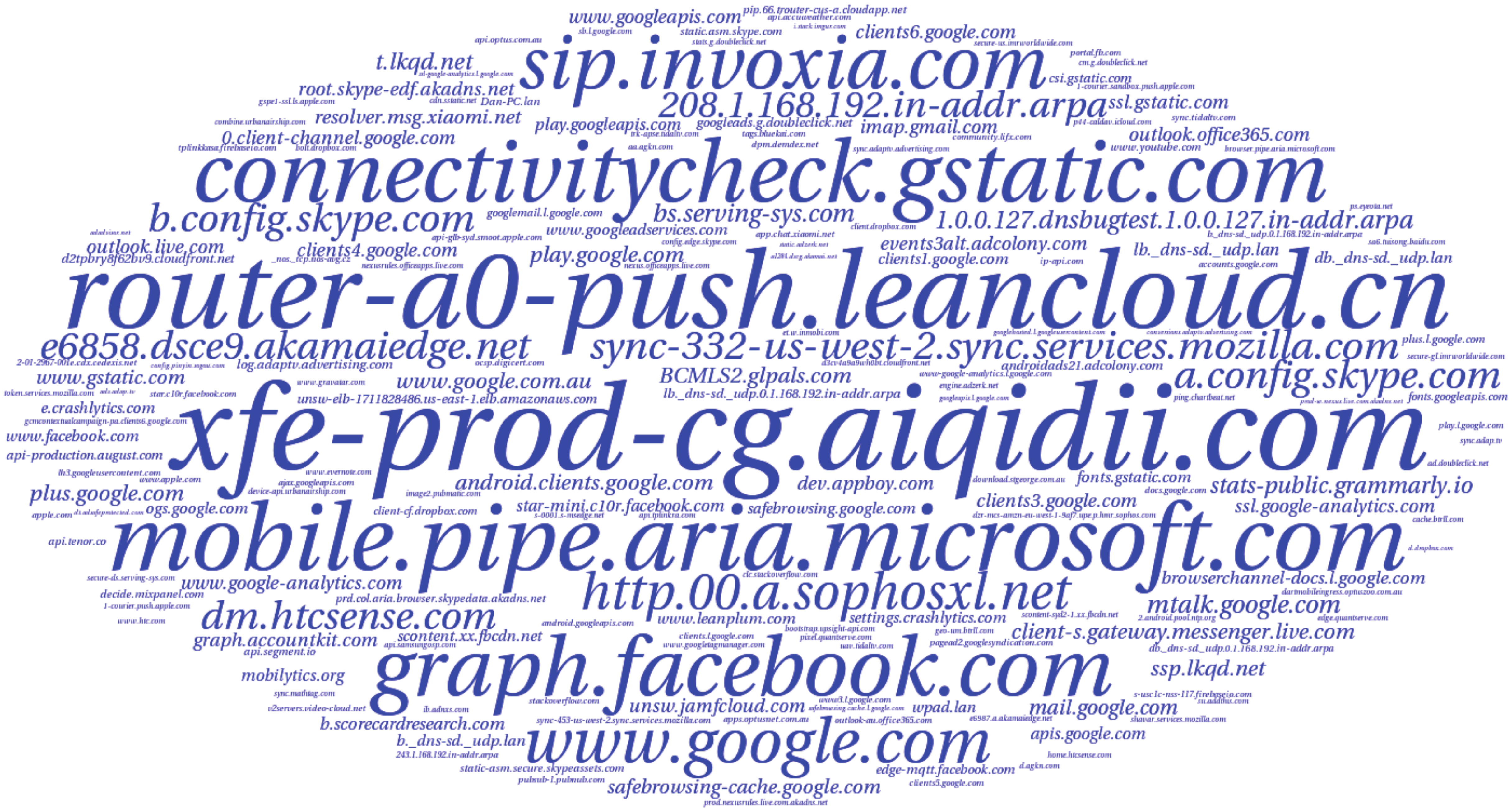}}\quad
								\label{fig:DNSnoneIoT}
							}
						}
						\caption{Word-cloud of domain names (total count of unique domains is shown in \{sub-captions\} next to the device name).}
						\label{fig:cloudDNS}
					\end{center}
				\end{figure}

				DNS is a common application used by almost all networked devices. Since IoT devices are custom-designed for specific purposes, they access a limited number of domains corresponding to their vendor-specific end-point servers. We plot in Fig.~\ref{fig:cloudDNS} the word cloud of domain names accessed by several IoT devices as well as non-IoTs. It is seen that IoT devices are fairly distinguishable by the domain names they communicate with. For example, as depicted in Figures~\ref{fig:DNScloudAmazon}-\ref{fig:DNScloudHP}, domains such as \verb|example.com|, \verb|example.net|, and \verb|example.org| are frequently requested by Amazon Echo; sub-domains of \verb|hp.com| and \verb|hpeprint.com| are seen in DNS queries from the HP printer. However, we also see that some prominent domain names are shared between the different devices. For example, \verb|belkin.com| and \verb|d3gjecg2uu2faq.cloudfront.net| are commonly used by Belkin devices (i.e. camera, motion sensor and power switch) as shown in Figures~\ref{fig:DNScloudBelkinCam}-\ref{fig:DNScloudBelkinSwitch}; or \verb|pool.ntp.org| is prominent in traffic flows generated from Google Dropcam, Awair air quality monitor and LiFX lightbulb, as shown in Figures~\ref{fig:DNScloudDropcam}-\ref{fig:DNScloudLifx}. Again considering non-IoTs in Fig.~\ref{fig:DNSnoneIoT}, we see about 12000 unique domains visited which is far diverse compared to IoT devices with only a handful of domains accessed repeatedly.
				We also found that IoT devices differ from one other in how often the DNS protocol is used. We have observed from our traffic traces that IoT devices generate DNS queries during different stages of its operation; for example only during the boot-up phase (e.g. Google Dropcamp) or when interacting with a user (e.g.  Hello Barbie) or periodically (e.g. Amazon Echo). As shown in Fig.~\ref{fig:HistDNSinterval},  certain IoT devices exhibit a characteristic signature in the frequency of their DNS queries. The LiFX lightbulb and Amazon Echo send DNS queries very frequently (i.e. every 5 minutes) but a device like the Belkin motion sensor requests domain names only once every 30 minutes. 
				
				\vspace{-1em}
			\subsubsection{NTP queries}\label{sec:c1_IoTprotocolNTP}
				\vspace{-1em}
				As mentioned earlier, NTP is another popular protocol used by IoT devices because precise and verifiable timing is crucial for IoT operations \cite{NIST}. Many IoT devices tend to use NTP protocol (UDP port 123) in a periodic manner in order to synchronize their time with publicly available NTP servers. For example, Awair air quality monitor, LiFX lightbulb and Google Dropcam obtain the IP address of time servers from \verb|pool.ntp.org|. We also find that time synchronization occurs repeatedly in our test-bed and many IoT devices exhibit a recognizable pattern in the use of NTP. For example, the Belkin power switch, LiFX lightbulb and SmartThings hub send NTP requests every 60, 300 and 600 seconds respectively, as shown in histogram plot of Fig.~\ref{fig:HistNTPinterval}.
				
				\begin{figure}[t]
					\begin{center}
						\mbox{
							
							\subfloat[Histogram of DNS interval.]{
								{\includegraphics[width=0.48\textwidth,height=0.3\textwidth]{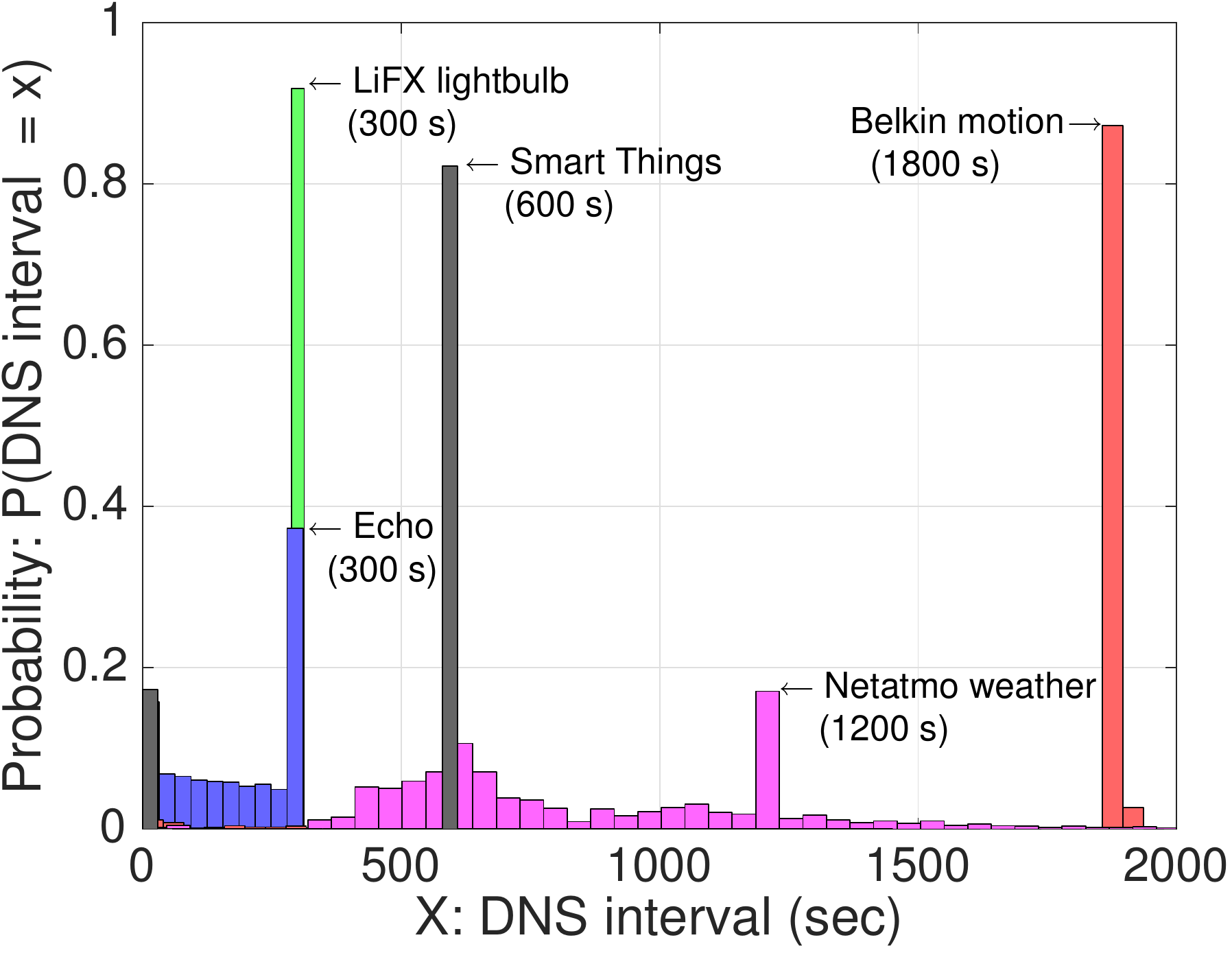}}\quad
								\label{fig:HistDNSinterval}
							}
							
							\subfloat[Histogram of NTP interval.]{
								{\includegraphics[width=0.48\textwidth,height=0.3\textwidth]{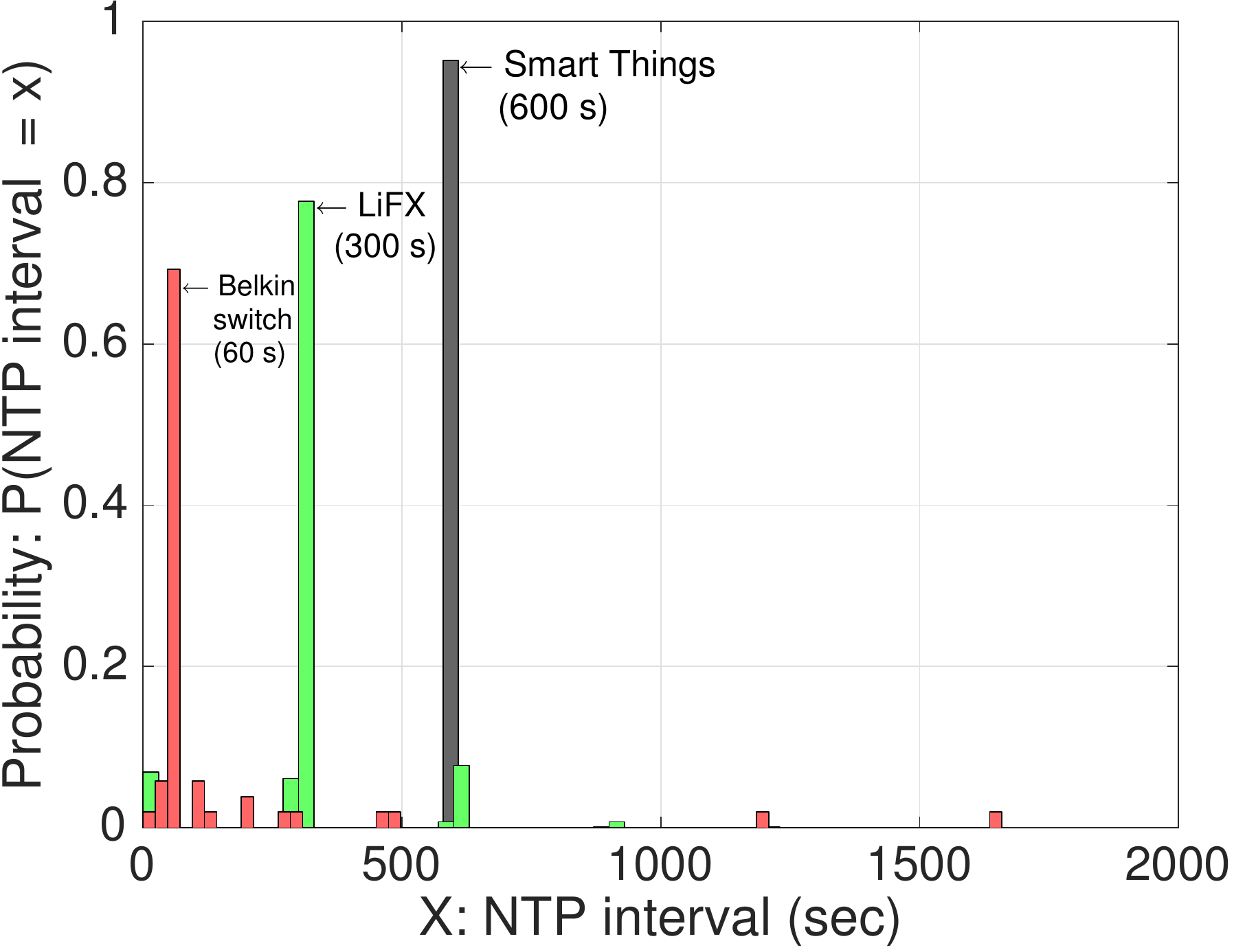}}\quad
								\label{fig:HistNTPinterval}
							}
						}
						\caption{Distribution of IoT signaling pattern: (a) DNS interval, (b) NTP interval.}
						
						\label{fig:IoTsignaling}
					\end{center}
				\end{figure}
	
			\subsubsection{Cipher suite}\label{sec:c1_IoTprotocolCS}
				A number of IoT devices use TLS/SSL protocol (port number 443) to communicate with their respective servers on the Internet \cite{EncryptedMalware16}. In order to initiate the TLS connection and negotiate the security algorithms with servers, devices start handshaking by sending a ``Client Hello'' packet with a list of  ``cipher suites'' that they can support, in the order of their preference. For example, Figures~\ref{fig:cs1}~and~\ref{fig:cs2} depict cipher suites that Amazon Echo offers to two different Amazon servers. Each cipher suite (i.e. 4-digit code) can take one of 380 possible values and represents algorithms for key exchange, bulk encryption and message authentication code (MAC). For example, the cipher \verb|002f| negotiated by an Amazon server uses RSA, AES\_128\_CBC, and SHA protocols for key exchange, bulk encryption and message authentication, respectively.
				
				\begin{figure}[t]
					\begin{center}
						\mbox{
							\subfloat[cs1 of Amazon Echo.]{
								{\includegraphics[width=0.80\textwidth]{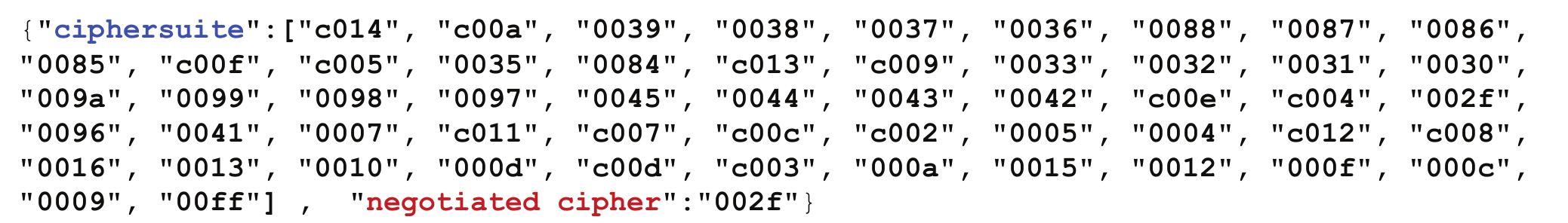}}\quad
								\label{fig:cs1}
							}
						}
						\mbox{
							\subfloat[cs2 of Amazon Echo.]{
								{\includegraphics[width=0.80\textwidth]{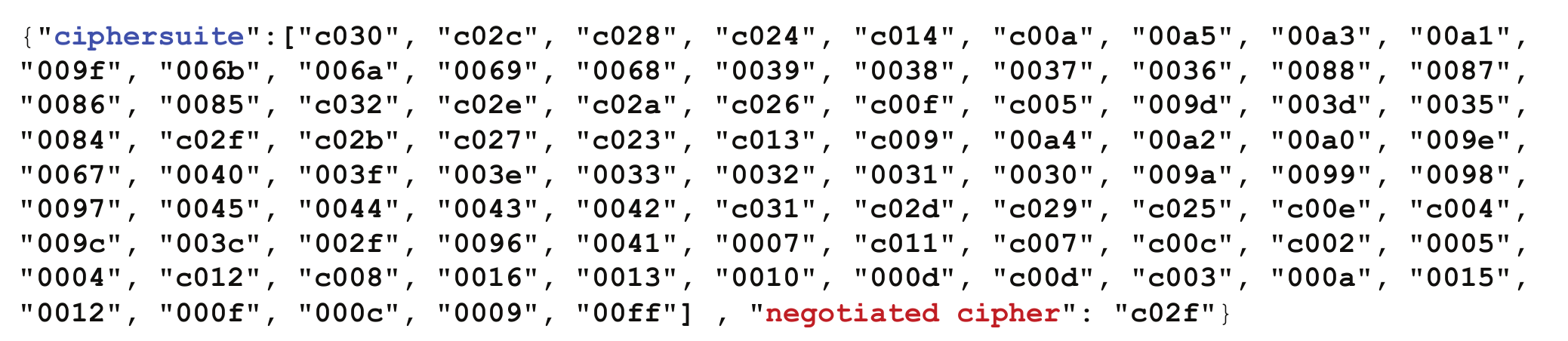}}\quad
								\label{fig:cs2}
							}
						}
						\caption{Signature of cipher suite.}
						\label{fig:CS}
					\end{center}
					\vspace{-1em}
				\end{figure}

				We find that 17 out of the 28 IoT devices in our setup, including the Amazon Echo, August Doorbell Cam, Awair air quality monitor, Belkin Camera, Canary Camera, Dropcam, Google Chromecast, Hello Barbie, HP ENVY Printer, iHome, Netatmo Welcome camera, Philips Hue lightbulb, Pixtar photoframe, Ring Door Bell, Triby, Withings Aura smart sleep sensor and Withings Scale, use TLS/SSL for communication. We find that Amazon Echo uses total of five different cipher suite strings when communicating SSL to different servers, Triby speaker uses two strings, while the Pixtar photoframe uses only one string for all of its SSL communications. We plot unique cipher suite strings from these three devices in Fig.~\ref{fig:CSsignature} as discrete signals: x-axis is the order of 4-digit cipher codes that appear in the offered suite, and y-axis is the index of the individual cipher codes  (i.e. a value from \{1, 2, ..., 380\}). It is seen that the collection of cipher suite signals enunciates a unique signature for each IoT device. Exceptionally, we found that Pixtar photoframe shares its single cipher suite with one of 18 suites that are used by August door-bell -- we will see in \S\ref{sec:c1_perfEval} that relying only on cipher suite attribute would not be effective in classifying Pixtar photo-frame traffic.
				
				\begin{figure}[t]
					\begin{center}
						\mbox{
							\subfloat[Amazon Echo.]{
								{\includegraphics[width=0.4\textwidth]{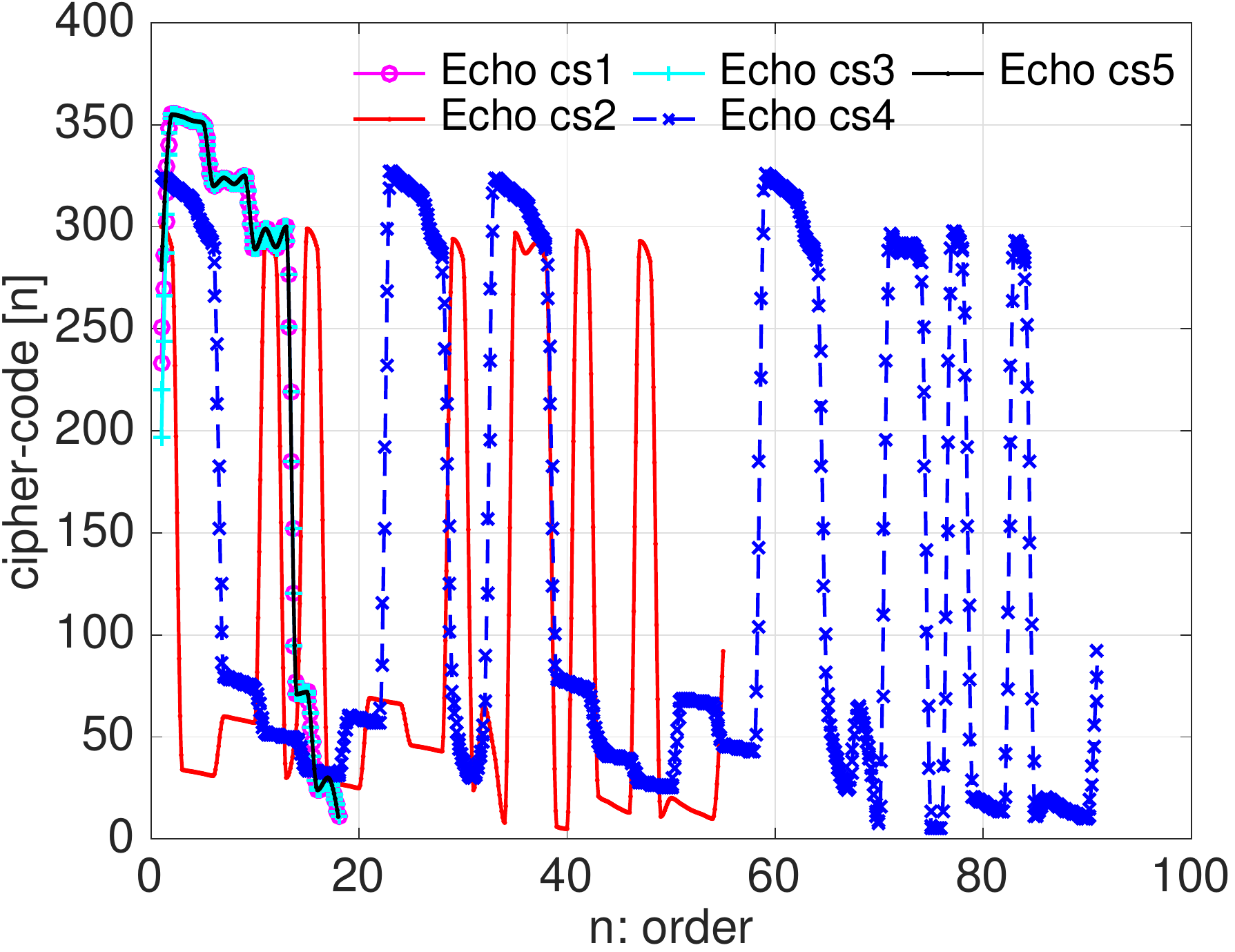}}\quad
								\label{fig:csSigAmazon}
							}
							\subfloat[Triby.]{
								{\includegraphics[width=0.4\textwidth]{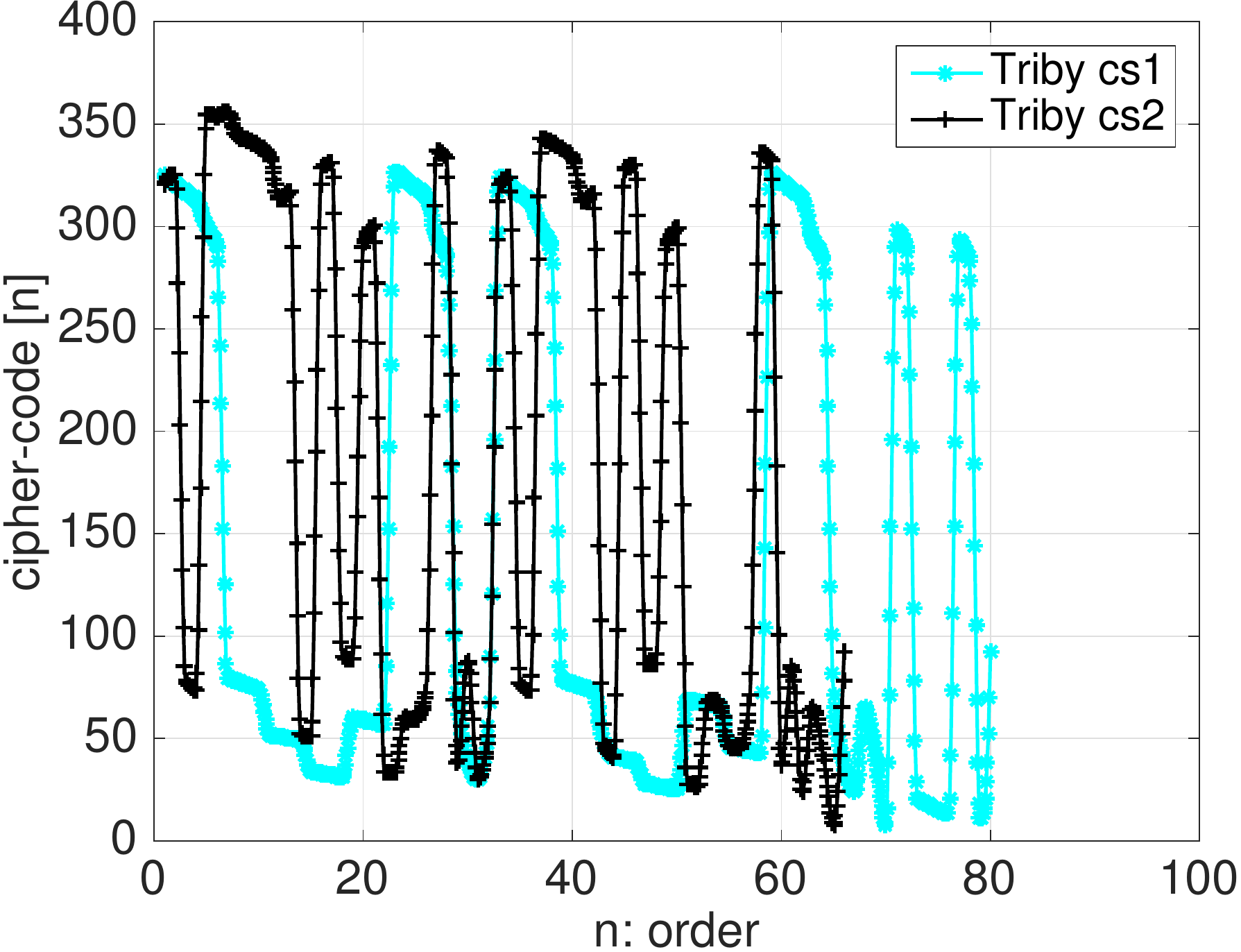}}\quad
								\label{figcsSigTriby}
							}
						}
						\mbox{
							\subfloat[Pixtar photoframe.]{
								{\includegraphics[width=0.4\textwidth]{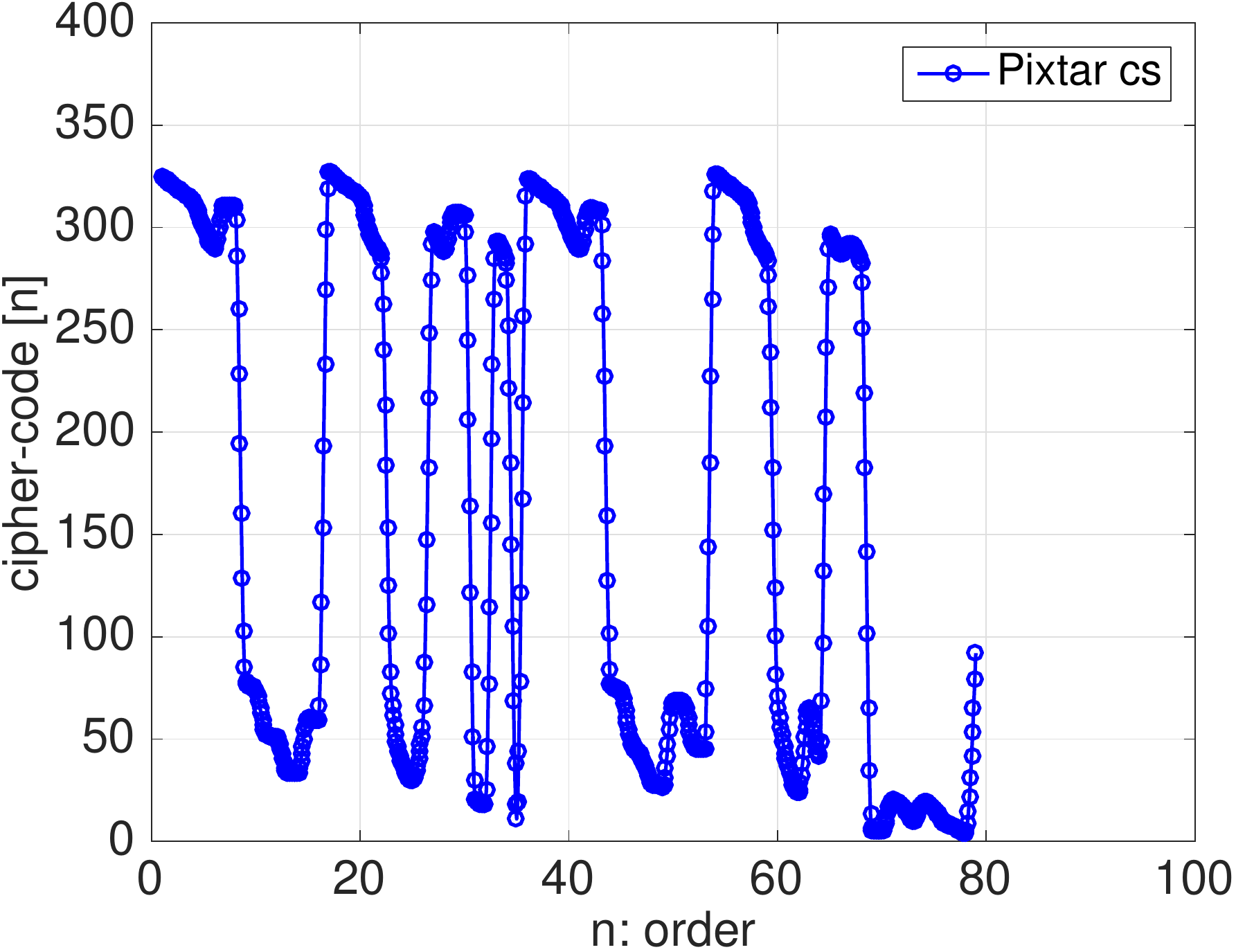}}\quad
								\label{fig:csSigPixtar}
							}
						}

						\caption{Signature of cipher suite.}
						
						\label{fig:CSsignature}
					\end{center}
				\end{figure}

				There are however many devices that rarely exchange cipher suites but instead prefer to keep their TLS connections alive for a long period. For example, Google Dropcam establishes a TLS connection to its own server whenever it boots up and maintains this connection as long as it has network connectivity, while Amazon Echo and Pixstar photoframe initiate on average 1 and 2 TLS connections respectively every hour. 
	
				{\bf Summary:} In this section, we have identified 8 key attributes based on the underlying network traffic characteristics of IoT devices. They are flow volume, flow duration, average flow rate, device sleep time, server port numbers, DNS queries, NTP queries and cipher suites. Although, some devices (e.g. Amazon Echo, or LiFX lightbulb) can be uniquely identified by considering just one or two traffic attributes such as the list of domain-names, port-numbers, or cipher suites, these come with challenges. For example, a strong attribute like the list of cipher-suites is observed very infrequently in the traffic (e.g. only once a day). As another example, different types of devices from the same vendor visit similar domains and use the same port numbers to access cloud servers. Capturing aspects such as the number of occurrences for these attributes (e.g. number of times a domain is accessed	or number of streams that use the port), in combination	with other attributes, vastly improves the prediction capability to distinguish between devices from the same manufacturer. In the next section, we develop a multi-stage machine learning based algorithm using combinations of these attributes to help classify IoT devices with high accuracy.
				
	\section{Machine Learning Based Classification}\label{sec:c1_class}
		In order to synthesize the attributes from our trace data, we first convert the raw pcap files into flows on an hourly basis using the Joy tool \cite{joy}. Then, for a given IoT device, we compute the traffic activity and signalling attributes defined in the previous section over the hourly instances. The number of instances for each device obtained from the trace spanning 26 weeks varies depending on factors such as the duration for which a device is online, or how a device generates traffic (autonomously or interactively). For example, there were only 13 hourly instances for the Blipcare BP monitor since it generates traffic only when the device is used by a user. On the other hand, we collected 4177 instances for Google Dropcam.

		\subsection{Multi-Stage Device Classification Architecture}\label{sec:c1_classArch}
			We note that three of our attributes namely ``set of domain names'', ``set of remote port numbers'' and ``set of cipher suites'' are nominal (i.e. are not treated as numeric values) and multi-valued (for example, \{``53'':3, ``123'':1, ``443'':2\} represents a set of remote port numbers with three occurrences of port number 53, two occurrences of port 123, and one occurrence of port number 443). Our remaining attributes including flow volume/duration, flow rate, sleep time, and DNS/NTP intervals contain single quantitative and continuous values.  We therefore employ a two-stage hierarchical architecture for our IoT classifier as shown in Fig.~\ref{fig:archClass}.
			
			\begin{figure}[t]
				\centering
				\includegraphics[width=1\textwidth]{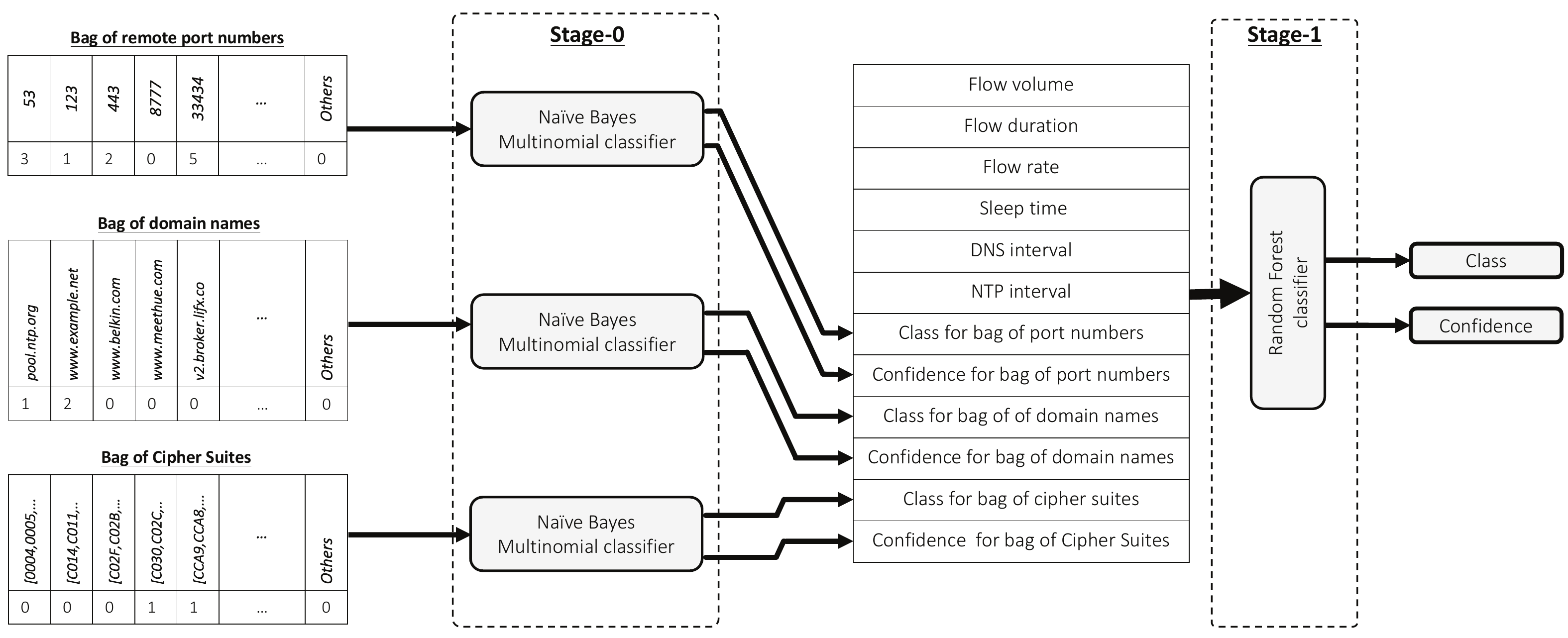}
				
				\caption{System architecture of the multi-stage classifier.}
				
				\label{fig:archClass}
			\end{figure}

			In this architecture, we first feed each multi-valued attribute to its corresponding stage-0 classifier in the form of a ``bag of words''. A bag of words is a matrix whose rows represent labeled instances, and columns represent unique words. This matrix has $M$ rows (i.e. total number of instances) and $N$ columns (i.e. number of unique words). We observed 356, 421 and 54 unique words for domain-names, remote port numbers and cipher suite strings, as shown in Fig.~\ref{fig:archClass}. In addition to these unique words, we aggregated all corresponding words for non-IoT devices as ``others'' - a column called ``others'' in each Stage-0 matrix represents words not seen in IoT traffic. Each cell of this matrix is the number of occurrences of such unique words in a given instance. 

			As shown in Fig.~\ref{fig:archClass}, each classifier of Stage-0 generates two outputs, namely a tentative class and a confidence level, which together with other single-valued quantitative attributes (i.e. flow volume, duration, rate, sleep time, DNS, NTP intervals) are fed into a Stage-1 classifier that produces the final output (i.e. the device identification with a confidence level).

			\subsubsection{Stage-0: Bag-of-words Classifiers} 
				We employ a Naive Bayes Multinomial classifier to analyze each bag of words in the stage-0 of our machine. It has been shown \cite{McCallum1998} that this classifier performs well in text classification when dealing with a large number of unique words. During the training phase, the classifier takes the distribution of words, e.g. individual unique domain names, and computes the probability of each word given a class using:
				\begin{equation}\label{eq:training}
					\Pr(w_{j}^{train}|c_i)=\frac{1+\sum\limits_{l=1}^D {n_{l,c_i,w_j}}^{train}}{N+\sum\limits_{k=1}^N\sum\limits_{l=1}^D {n_{l,c_i,w_k}}^{train}}
				\end{equation}
				where $w_j$ is a unique word in the training dataset (e.g. port number 56700); $c_i$ is a class label (e.g. LiFX lightbulb); $D$ is the total number of instances; ${n_{l,c_i,w_j}}^{train}$ is the number of $w_j$ occurrences in each of instances with class label of $c_i$; $N$ is the total number of unique words (e.g. we have $N=421$ unique port numbers in our dataset).

				During the testing phase, the classifier needs to compute the following probability for all possible classes:
				\begin{equation}\label{eq:testing}
					\Pr( c_i | W^{test} ) = \Pr(c_{i}^{train}) \prod\limits_{j=1}^N \Pr{({w_{j}}^{train}|c_i)}^{n_{j}^{test}}
				\end{equation}
				where $W^{test}$ is a set represented by $\{w_1:n_1^{test},w_2:n_2^{test},...,w_N:n_N^{test}\}$; $n_j^{test}$ is the occurrence number of individual unique words $w_j$ in a given test instance; $\Pr(c_{i}^{train})$ is the presence probability of a class  $c_{i}$ in the whole training dataset (i.e. number of $c_{i}$ training instances divided by total number of all training instances). The classifier finally chooses the class that gives the maximum probability in \eqref{eq:testing} for a given set of words along with their occurrences. Note that a Naive Bayes Multinomial classifier performs well if training instances are fairly distributed among various classes \cite{McCallum1998}.

			\vspace{-1em}
			\subsubsection{Stage-1 Classifier } 
			\vspace{-1em}
				We have a stage-1 classifier that takes all quantitative attributes along with the pair of outputs from each stage-0 classifier. Since the stage-1 attributes are not linearly separable and the outputs of stage-0 classifiers are nominal values, we use a Random Forest based stage-1 classifier. Another reason for selecting the Random Forest is its high tolerance to over-fitting compared to other decision tree classifiers.

		\subsection{Performance Evaluation}\label{sec:c1_perfEval} 
			We use the Weka \cite{Weka} tool for our IoT device classification. We have collected a total of 50,378 labeled instances from our traffic traces. As mentioned earlier, we have a number of instances from different devices -- those that generate traffic when triggered by user interaction have small number of instances (e.g. 13 for Blipcare BP monitor, 21 for Google Chromecast) and those that autonomously generate traffic have a fairly large number of instances (e.g. 2,868 for Samsung Smart Things or 2,247 for Amazon Echo). We have randomly split instances into two groups, one containing 70\% of the instances for ``training'' and another containing 30\% of the instances for ``testing''. 
			
			Table~\ref{tab:performance} shows the performance of our classifier under various scenarios, each captured by a pair of columns. For a given scenario, we measure the true positive rate (i.e. fraction of test instances that are correctly classified) and false positive rate (i.e. fraction of test instances that are incorrectly classified) for every device corresponding to the rows in Table~\ref{tab:performance}. We also obtain the average confidence level (i.e. a number between 0 and 1 depicted within square brackets in each cell) of our classifier for correctly classified and incorrectly classified instances. In addition, we aggregate the performance of individual classes and compute the overall accuracy (i.e. total true positive rate) along with the overall root relative squared error (RRSE) as measures of performance for our classifier. These measures are reported in the top row of each scenario in Table~\ref{tab:performance}. Note that our objective is to achieve a high accuracy (close to 100\%) with a fairly low error (close to zero).   
			
						\begin{table}[t!]
	\centering
	\caption{Performance of the proposed IoT device classifier under different sets of attributes.}
	\label{tab:performance}
	
	\begin{adjustbox}{max width=1\textwidth}
		\def\arraystretch{1.4}
		\begin{tabular}{|l|C{1.2cm}|L{5cm}|C{1.2cm}|L{5cm}|C{1.2cm}|L{5cm}|C{1.2cm}|L{5cm}|C{1.2cm}|L{5cm}| }  \hline
			
			\multirow{ 3}{*}{Devices}&\multicolumn{2}{c|}{Port Numbers}&\multicolumn{2}{c|}{Domain Names}&\multicolumn{2}{c|}{Cipher Suite}&\multicolumn{2}{c|}{Combined stage-0}&\multicolumn{2}{c|}{Final}\\\cline{2-11}
			
			{}&\multicolumn{2}{l|}{\makecell{Accuracy: 92.13\%\\RRSE: 39.93\%} }&\multicolumn{2}{l|}{\makecell{Accuracy: 79.48\%\\RRSE: 57.56\%} }&\multicolumn{2}{l|}{\makecell{Accuracy: 36.15\%\\RRSE: 86.73\%} }&\multicolumn{2}{l|}{\makecell{Accuracy: 97.39\%\\RRSE: 18.24\%} }&\multicolumn{2}{l|}{\makecell{Accuracy: 99.88\%\\RRSE: 5.06\%}}\\ \cline{2-11}

			{}& \rotatebox{90}{\parbox{2cm}{Correctly\\Classified}}
			& Incorrectly Classified& \rotatebox{90}{\parbox{2cm}{Correctly\\Classified}} & Incorrectly Classified& \rotatebox{90}{\parbox{2cm}{Correctly\\Classified}} & Incorrectly Classified& \rotatebox{90}{\parbox{2cm}{Correctly\\Classified}} & Incorrectly Classified& \rotatebox{90}{\parbox{2cm}{Correctly\\Classified}} & Incorrectly Classified\\ \hline
			
			\devEcho & \cellcolor[HTML]{15783E}\color{white} \makecell{\textsf{100.0\%}\\ \textsf{[1.00]}} & \begin{tabular}[l]{@{}l@{}}\ \\\ \end{tabular} & \cellcolor[HTML]{15783E}\color{white} \makecell{\textsf{99.6\%}\\ \textsf{[1.00]}} & \begin{tabular}[l]{@{}l@{}}\devDropcam: \textsf{0.4\%} \textsf{[0.08]}\\\ \end{tabular} & \cellcolor[HTML]{15783E}\color{white} \makecell{\textsf{100.0\%}\\ \textsf{[1.00]}} & \begin{tabular}[l]{@{}l@{}}\ \\\ \end{tabular} & \cellcolor[HTML]{15783E}\color{white} \makecell{\textsf{99.9\%}\\ \textsf{[1.00]}} & \begin{tabular}[l]{@{}l@{}}\devHPprinter: \textsf{0.1\%} \textsf{[0.52]}\\\ \end{tabular} & \cellcolor[HTML]{15783E}\color{white} \makecell{\textsf{99.7\%}\\ \textsf{[1.00]}} & \begin{tabular}[l]{@{}l@{}}\devNonIoT: \textsf{0.1\%} \textsf{[0.37]}\\\devDropcam: \textsf{0.1\%} \textsf{[0.43]}\end{tabular}\\ \hline
			\devAugust & \cellcolor[HTML]{15783E}\color{white} \makecell{\textsf{99.0\%}\\ \textsf{[1.00]}} & \begin{tabular}[l]{@{}l@{}}\deviHome: \textsf{0.6\%} \textsf{[1.00]}\\Others: \textsf{0.4\%} \textsf{[0.65]}\end{tabular} & \cellcolor[HTML]{15783E}\color{white} \makecell{\textsf{100.0\%}\\ \textsf{[1.00]}} & \begin{tabular}[l]{@{}l@{}}\ \\\ \end{tabular} & \cellcolor[HTML]{a2d889} \makecell{\textsf{78.8\%}\\ \textsf{[1.00]}} & \begin{tabular}[l]{@{}l@{}}\devPixstar: \textsf{21.2\%} \textsf{[1.00]}\\\ \end{tabular} & \cellcolor[HTML]{15783E}\color{white} \makecell{\textsf{100.0\%}\\ \textsf{[1.00]}} & \begin{tabular}[l]{@{}l@{}}\ \\\ \end{tabular} & \cellcolor[HTML]{15783E}\color{white} \makecell{\textsf{100.0\%}\\ \textsf{[1.00]}} & \begin{tabular}[l]{@{}l@{}}\ \\\ \end{tabular}\\ \hline
			\devAwair & \cellcolor[HTML]{15783E}\color{white} \makecell{\textsf{97.6\%}\\ \textsf{[1.00]}} & \begin{tabular}[l]{@{}l@{}}\devNonIoT: \textsf{2.0\%} \textsf{[0.32]}\\\devEcho: \textsf{0.4\%} \textsf{[0.53]}\end{tabular} & \cellcolor[HTML]{15783E}\color{white} \makecell{\textsf{99.2\%}\\ \textsf{[1.00]}} & \begin{tabular}[l]{@{}l@{}}\devSmartThings: \textsf{0.4\%} \textsf{[0.49]}\\\devDropcam: \textsf{0.4\%} \textsf{[0.08]}\end{tabular} & \cellcolor[HTML]{15783E}\color{white} \makecell{\textsf{99.2\%}\\ \textsf{[0.63]}} & \begin{tabular}[l]{@{}l@{}}\devDropcam: \textsf{0.8\%} \textsf{[0.08]}\\\ \end{tabular} & \cellcolor[HTML]{15783E}\color{white} \makecell{\textsf{100.0\%}\\ \textsf{[1.00]}} & \begin{tabular}[l]{@{}l@{}}\ \\\ \end{tabular} & \cellcolor[HTML]{15783E}\color{white} \makecell{\textsf{100.0\%}\\ \textsf{[1.00]}} & \begin{tabular}[l]{@{}l@{}}\ \\\ \end{tabular}\\ \hline
			\devBelkinCam & \cellcolor[HTML]{15783E}\color{white} \makecell{\textsf{95.5\%}\\ \textsf{[1.00]}} & \begin{tabular}[l]{@{}l@{}}\devBelkinMotion: \textsf{3.0\%} \textsf{[0.94]}\\Others: \textsf{1.5\%} \textsf{[0.67]}\end{tabular} & \cellcolor[HTML]{ffffe5} \makecell{\textsf{39.4\%}\\ \textsf{[0.99]}} & \begin{tabular}[l]{@{}l@{}}\devBelkinMotion: \textsf{59.8\%} \textsf{[0.62]}\\\devNonIoT: \textsf{0.8\%} \textsf{[1.00]}\end{tabular} & \cellcolor[HTML]{ffffe5} \makecell{\textsf{0.0\%}\\ \textsf{[-]}} & \begin{tabular}[l]{@{}l@{}}\devDropcam: \textsf{100.0\%} \textsf{[0.08]}\\\ \end{tabular} & \cellcolor[HTML]{15783E}\color{white} \makecell{\textsf{97.7\%}\\ \textsf{[0.99]}} & \begin{tabular}[l]{@{}l@{}}\devNonIoT: \textsf{1.5\%} \textsf{[0.74]}\\\devDropcam: \textsf{0.8\%} \textsf{[1.00]}\end{tabular} & \cellcolor[HTML]{15783E}\color{white} \makecell{\textsf{97.7\%}\\ \textsf{[0.99]}} & \begin{tabular}[l]{@{}l@{}}\devNonIoT: \textsf{1.5\%} \textsf{[0.60]}\\\devNetatmoCam: \textsf{0.8\%} \textsf{[0.57]}\end{tabular}\\ \hline
			\devBelkinMotion & \cellcolor[HTML]{15783E}\color{white} \makecell{\textsf{99.8\%}\\ \textsf{[1.00]}} & \begin{tabular}[l]{@{}l@{}}\devNonIoT: \textsf{0.2\%} \textsf{[1.00]}\\\ \end{tabular} & \cellcolor[HTML]{ffffe5} \makecell{\textsf{0.0\%}\\ \textsf{[-]}} & \begin{tabular}[l]{@{}l@{}}\devBelkinSwitch: \textsf{100.0\%} \textsf{[0.57]}\\\ \end{tabular} & \cellcolor[HTML]{ffffe5} \makecell{\textsf{0.0\%}\\ \textsf{[-]}} & \begin{tabular}[l]{@{}l@{}}\devDropcam: \textsf{100.0\%} \textsf{[0.08]}\\\ \end{tabular} & \cellcolor[HTML]{15783E}\color{white} \makecell{\textsf{99.5\%}\\ \textsf{[1.00]}} & \begin{tabular}[l]{@{}l@{}}\devSamsungcam: \textsf{0.3\%} \textsf{[0.79]}\\\devNonIoT: \textsf{0.2\%} \textsf{[1.00]}\end{tabular} & \cellcolor[HTML]{15783E}\color{white} \makecell{\textsf{99.8\%}\\ \textsf{[1.00]}} & \begin{tabular}[l]{@{}l@{}}\devNonIoT: \textsf{0.2\%} \textsf{[0.97]}\\\ \end{tabular}\\ \hline
			\devBelkinSwitch & \cellcolor[HTML]{15783E}\color{white} \makecell{\textsf{99.5\%}\\ \textsf{[1.00]}} & \begin{tabular}[l]{@{}l@{}}\devBelkinMotion: \textsf{0.2\%} \textsf{[0.77]}\\Others: \textsf{0.3\%} \textsf{[0.75]}\end{tabular} & \cellcolor[HTML]{15783E}\color{white} \makecell{\textsf{99.7\%}\\ \textsf{[0.57]}} & \begin{tabular}[l]{@{}l@{}}\devDropcam: \textsf{0.2\%} \textsf{[0.08]}\\\devBlipcareBP: \textsf{0.1\%} \textsf{[0.79]}\end{tabular} & \cellcolor[HTML]{ffffe5} \makecell{\textsf{0.0\%}\\ \textsf{[-]}} & \begin{tabular}[l]{@{}l@{}}\devDropcam: \textsf{100.0\%} \textsf{[0.08]}\\\ \end{tabular} & \cellcolor[HTML]{15783E}\color{white} \makecell{\textsf{99.8\%}\\ \textsf{[1.00]}} & \begin{tabular}[l]{@{}l@{}}\devBelkinMotion: \textsf{0.2\%} \textsf{[1.00]}\\\ \end{tabular} & \cellcolor[HTML]{15783E}\color{white} \makecell{\textsf{99.8\%}\\ \textsf{[1.00]}} & \begin{tabular}[l]{@{}l@{}}\devBelkinMotion: \textsf{0.2\%} \textsf{[0.93]}\\\ \end{tabular}\\ \hline
			\devCanaryCam & \cellcolor[HTML]{15783E}\color{white} \makecell{\textsf{100.0\%}\\ \textsf{[1.00]}} & \begin{tabular}[l]{@{}l@{}}\ \\\ \end{tabular} & \cellcolor[HTML]{15783E}\color{white} \makecell{\textsf{100.0\%}\\ \textsf{[1.00]}} & \begin{tabular}[l]{@{}l@{}}\ \\\ \end{tabular} & \cellcolor[HTML]{15783E}\color{white} \makecell{\textsf{100.0\%}\\ \textsf{[1.00]}} & \begin{tabular}[l]{@{}l@{}}\ \\\ \end{tabular} & \cellcolor[HTML]{15783E}\color{white} \makecell{\textsf{100.0\%}\\ \textsf{[1.00]}} & \begin{tabular}[l]{@{}l@{}}\ \\\ \end{tabular} & \cellcolor[HTML]{15783E}\color{white} \makecell{\textsf{100.0\%}\\ \textsf{[1.00]}} & \begin{tabular}[l]{@{}l@{}}\ \\\ \end{tabular}\\ \hline
			\devDropcam & \cellcolor[HTML]{15783E}\color{white} \makecell{\textsf{98.1\%}\\ \textsf{[0.33]}} & \begin{tabular}[l]{@{}l@{}}\devSmartThings: \textsf{0.6\%} \textsf{[0.99]}\\Others: \textsf{1.3\%} \textsf{[0.59]}\end{tabular} & \cellcolor[HTML]{15783E}\color{white} \makecell{\textsf{100.0\%}\\ \textsf{[0.09]}} & \begin{tabular}[l]{@{}l@{}}\ \\\ \end{tabular} & \cellcolor[HTML]{15783E}\color{white} \makecell{\textsf{100.0\%}\\ \textsf{[0.09]}} & \begin{tabular}[l]{@{}l@{}}\ \\\ \end{tabular} & \cellcolor[HTML]{a2d889} \makecell{\textsf{74.0\%}\\ \textsf{[0.96]}} & \begin{tabular}[l]{@{}l@{}}\devHPprinter: \textsf{25.7\%} \textsf{[0.52]}\\Others: \textsf{0.3\%} \textsf{[0.41]}\end{tabular} & \cellcolor[HTML]{15783E}\color{white} \makecell{\textsf{100.0\%}\\ \textsf{[1.00]}} & \begin{tabular}[l]{@{}l@{}}\ \\\ \end{tabular}\\ \hline
			\devLifx & \cellcolor[HTML]{15783E}\color{white} \makecell{\textsf{100.0\%}\\ \textsf{[1.00]}} & \begin{tabular}[l]{@{}l@{}}\ \\\ \end{tabular} & \cellcolor[HTML]{15783E}\color{white} \makecell{\textsf{99.7\%}\\ \textsf{[0.70]}} & \begin{tabular}[l]{@{}l@{}}\devSmartThings: \textsf{0.1\%} \textsf{[0.42]}\\\devDropcam: \textsf{0.1\%} \textsf{[0.08]}\end{tabular} & \cellcolor[HTML]{ffffe5} \makecell{\textsf{0.0\%}\\ \textsf{[-]}} & \begin{tabular}[l]{@{}l@{}}\devDropcam: \textsf{100.0\%} \textsf{[0.08]}\\\ \end{tabular} & \cellcolor[HTML]{15783E}\color{white} \makecell{\textsf{100.0\%}\\ \textsf{[1.00]}} & \begin{tabular}[l]{@{}l@{}}\ \\\ \end{tabular} & \cellcolor[HTML]{15783E}\color{white} \makecell{\textsf{100.0\%}\\ \textsf{[1.00]}} & \begin{tabular}[l]{@{}l@{}}\ \\\ \end{tabular}\\ \hline
			\devNestSmoke & \cellcolor[HTML]{15783E}\color{white} \makecell{\textsf{100.0\%}\\ \textsf{[1.00]}} & \begin{tabular}[l]{@{}l@{}}\ \\\ \end{tabular} & \cellcolor[HTML]{15783E}\color{white} \makecell{\textsf{100.0\%}\\ \textsf{[1.00]}} & \begin{tabular}[l]{@{}l@{}}\ \\\ \end{tabular} & \cellcolor[HTML]{ffffe5} \makecell{\textsf{0.0\%}\\ \textsf{[-]}} & \begin{tabular}[l]{@{}l@{}}\devDropcam: \textsf{100.0\%} \textsf{[0.08]}\\\ \end{tabular} & \cellcolor[HTML]{15783E}\color{white} \makecell{\textsf{100.0\%}\\ \textsf{[1.00]}} & \begin{tabular}[l]{@{}l@{}}\ \\\ \end{tabular} & \cellcolor[HTML]{15783E}\color{white} \makecell{\textsf{100.0\%}\\ \textsf{[1.00]}} & \begin{tabular}[l]{@{}l@{}}\ \\\ \end{tabular}\\ \hline
			\devNetatmoWeather & \cellcolor[HTML]{15783E}\color{white} \makecell{\textsf{99.8\%}\\ \textsf{[1.00]}} & \begin{tabular}[l]{@{}l@{}}\devDropcam: \textsf{0.2\%} \textsf{[0.08]}\\\ \end{tabular} & \cellcolor[HTML]{15783E}\color{white} \makecell{\textsf{99.9\%}\\ \textsf{[1.00]}} & \begin{tabular}[l]{@{}l@{}}\devDropcam: \textsf{0.1\%} \textsf{[0.08]}\\\ \end{tabular} & \cellcolor[HTML]{ffffe5} \makecell{\textsf{0.0\%}\\ \textsf{[-]}} & \begin{tabular}[l]{@{}l@{}}\devDropcam: \textsf{100.0\%} \textsf{[0.08]}\\\ \end{tabular} & \cellcolor[HTML]{15783E}\color{white} \makecell{\textsf{100.0\%}\\ \textsf{[1.00]}} & \begin{tabular}[l]{@{}l@{}}\ \\\ \end{tabular} & \cellcolor[HTML]{15783E}\color{white} \makecell{\textsf{100.0\%}\\ \textsf{[1.00]}} & \begin{tabular}[l]{@{}l@{}}\ \\\ \end{tabular}\\ \hline
			\devNetatmoCam & \cellcolor[HTML]{15783E}\color{white} \makecell{\textsf{95.4\%}\\ \textsf{[1.00]}} & \begin{tabular}[l]{@{}l@{}}\devDropcam: \textsf{2.0\%} \textsf{[0.97]}\\Others: \textsf{2.6\%} \textsf{[0.70]}\end{tabular} & \cellcolor[HTML]{15783E}\color{white} \makecell{\textsf{97.8\%}\\ \textsf{[1.00]}} & \begin{tabular}[l]{@{}l@{}}\devDropcam: \textsf{2.2\%} \textsf{[0.08]}\\\ \end{tabular} & \cellcolor[HTML]{15783E}\color{white} \makecell{\textsf{99.7\%}\\ \textsf{[0.92]}} & \begin{tabular}[l]{@{}l@{}}\devDropcam: \textsf{0.3\%} \textsf{[0.08]}\\\ \end{tabular} & \cellcolor[HTML]{15783E}\color{white} \makecell{\textsf{99.8\%}\\ \textsf{[1.00]}} & \begin{tabular}[l]{@{}l@{}}\devPixstar: \textsf{0.1\%} \textsf{[0.54]}\\\devDropcam: \textsf{0.1\%} \textsf{[0.60]}\end{tabular} & \cellcolor[HTML]{15783E}\color{white} \makecell{\textsf{99.9\%}\\ \textsf{[1.00]}} & \begin{tabular}[l]{@{}l@{}}\devHuebulb: \textsf{0.1\%} \textsf{[0.37]}\\\ \end{tabular}\\ \hline
			\devPixstar & \cellcolor[HTML]{15783E}\color{white} \makecell{\textsf{99.7\%}\\ \textsf{[1.00]}} & \begin{tabular}[l]{@{}l@{}}\devDropcam: \textsf{0.3\%} \textsf{[0.08]}\\\ \end{tabular} & \cellcolor[HTML]{15783E}\color{white} \makecell{\textsf{99.3\%}\\ \textsf{[1.00]}} & \begin{tabular}[l]{@{}l@{}}\devDropcam: \textsf{0.7\%} \textsf{[0.08]}\\\ \end{tabular} & \cellcolor[HTML]{ffffe5} \makecell{\textsf{0.0\%}\\ \textsf{[-]}} & \begin{tabular}[l]{@{}l@{}}\devAugust: \textsf{99.7\%} \textsf{[0.71]}\\\devDropcam: \textsf{0.3\%} \textsf{[0.08]}\end{tabular} & \cellcolor[HTML]{15783E}\color{white} \makecell{\textsf{100.0\%}\\ \textsf{[1.00]}} & \begin{tabular}[l]{@{}l@{}}\ \\\ \end{tabular} & \cellcolor[HTML]{15783E}\color{white} \makecell{\textsf{100.0\%}\\ \textsf{[1.00]}} & \begin{tabular}[l]{@{}l@{}}\ \\\ \end{tabular}\\ \hline
			\devSamsungcam & \cellcolor[HTML]{15783E}\color{white} \makecell{\textsf{99.4\%}\\ \textsf{[1.00]}} & \begin{tabular}[l]{@{}l@{}}\devBelkinMotion: \textsf{0.6\%} \textsf{[1.00]}\\\ \end{tabular} & \cellcolor[HTML]{ffffe5} \makecell{\textsf{14.5\%}\\ \textsf{[1.00]}} & \begin{tabular}[l]{@{}l@{}}\devDropcam: \textsf{73.4\%} \textsf{[0.10]}\\\devSmartThings: \textsf{12.0\%} \textsf{[0.43]}\end{tabular} & \cellcolor[HTML]{ffffe5} \makecell{\textsf{0.0\%}\\ \textsf{[-]}} & \begin{tabular}[l]{@{}l@{}}\devDropcam: \textsf{100.0\%} \textsf{[0.08]}\\\ \end{tabular} & \cellcolor[HTML]{15783E}\color{white} \makecell{\textsf{100.0\%}\\ \textsf{[1.00]}} & \begin{tabular}[l]{@{}l@{}}\ \\\ \end{tabular} & \cellcolor[HTML]{15783E}\color{white} \makecell{\textsf{100.0\%}\\ \textsf{[1.00]}} & \begin{tabular}[l]{@{}l@{}}\ \\\ \end{tabular}\\ \hline
			\devSmartThings & \cellcolor[HTML]{15783E}\color{white} \makecell{\textsf{97.5\%}\\ \textsf{[1.00]}} & \begin{tabular}[l]{@{}l@{}}\devLifx: \textsf{1.9\%} \textsf{[0.99]}\\Others: \textsf{0.5\%} \textsf{[0.68]}\end{tabular} & \cellcolor[HTML]{a2d889} \makecell{\textsf{79.9\%}\\ \textsf{[0.50]}} & \begin{tabular}[l]{@{}l@{}}\devLifx: \textsf{20.1\%} \textsf{[0.50]}\\\ \end{tabular} & \cellcolor[HTML]{ffffe5} \makecell{\textsf{0.0\%}\\ \textsf{[-]}} & \begin{tabular}[l]{@{}l@{}}\devDropcam: \textsf{100.0\%} \textsf{[0.08]}\\\ \end{tabular} & \cellcolor[HTML]{15783E}\color{white} \makecell{\textsf{99.8\%}\\ \textsf{[1.00]}} & \begin{tabular}[l]{@{}l@{}}\devLifx: \textsf{0.1\%} \textsf{[0.88]}\\\devDropcam: \textsf{0.1\%} \textsf{[0.97]}\end{tabular} & \cellcolor[HTML]{15783E}\color{white} \makecell{\textsf{99.8\%}\\ \textsf{[1.00]}} & \begin{tabular}[l]{@{}l@{}}\devLifx: \textsf{0.1\%} \textsf{[0.71]}\\\devDropcam: \textsf{0.1\%} \textsf{[0.67]}\end{tabular}\\ \hline
			\devTPcam & \cellcolor[HTML]{15783E}\color{white} \makecell{\textsf{100.0\%}\\ \textsf{[1.00]}} & \begin{tabular}[l]{@{}l@{}}\ \\\ \end{tabular} & \cellcolor[HTML]{15783E}\color{white} \makecell{\textsf{99.7\%}\\ \textsf{[1.00]}} & \begin{tabular}[l]{@{}l@{}}\devDropcam: \textsf{0.3\%} \textsf{[0.08]}\\\ \end{tabular} & \cellcolor[HTML]{ffffe5} \makecell{\textsf{0.0\%}\\ \textsf{[-]}} & \begin{tabular}[l]{@{}l@{}}\devDropcam: \textsf{100.0\%} \textsf{[0.08]}\\\ \end{tabular} & \cellcolor[HTML]{15783E}\color{white} \makecell{\textsf{100.0\%}\\ \textsf{[1.00]}} & \begin{tabular}[l]{@{}l@{}}\ \\\ \end{tabular} & \cellcolor[HTML]{15783E}\color{white} \makecell{\textsf{100.0\%}\\ \textsf{[1.00]}} & \begin{tabular}[l]{@{}l@{}}\ \\\ \end{tabular}\\ \hline
			\devTPswitch & \cellcolor[HTML]{15783E}\color{white} \makecell{\textsf{99.7\%}\\ \textsf{[1.00]}} & \begin{tabular}[l]{@{}l@{}}\devDropcam: \textsf{0.3\%} \textsf{[0.08]}\\\ \end{tabular} & \cellcolor[HTML]{15783E}\color{white} \makecell{\textsf{99.7\%}\\ \textsf{[0.99]}} & \begin{tabular}[l]{@{}l@{}}\devDropcam: \textsf{0.3\%} \textsf{[0.08]}\\\ \end{tabular} & \cellcolor[HTML]{ffffe5} \makecell{\textsf{0.0\%}\\ \textsf{[-]}} & \begin{tabular}[l]{@{}l@{}}\devDropcam: \textsf{100.0\%} \textsf{[0.08]}\\\ \end{tabular} & \cellcolor[HTML]{15783E}\color{white} \makecell{\textsf{100.0\%}\\ \textsf{[1.00]}} & \begin{tabular}[l]{@{}l@{}}\ \\\ \end{tabular} & \cellcolor[HTML]{15783E}\color{white} \makecell{\textsf{100.0\%}\\ \textsf{[1.00]}} & \begin{tabular}[l]{@{}l@{}}\ \\\ \end{tabular}\\ \hline
			\devTriby & \cellcolor[HTML]{15783E}\color{white} \makecell{\textsf{98.0\%}\\ \textsf{[1.00]}} & \begin{tabular}[l]{@{}l@{}}\devNetatmoWeather: \textsf{1.2\%} \textsf{[0.37]}\\Others: \textsf{0.8\%} \textsf{[0.49]}\end{tabular} & \cellcolor[HTML]{15783E}\color{white} \makecell{\textsf{100.0\%}\\ \textsf{[1.00]}} & \begin{tabular}[l]{@{}l@{}}\ \\\ \end{tabular} & \cellcolor[HTML]{ffffe5} \makecell{\textsf{41.2\%}\\ \textsf{[0.99]}} & \begin{tabular}[l]{@{}l@{}}\devDropcam: \textsf{54.8\%} \textsf{[0.08]}\\\devNetatmoWeather: \textsf{4.0\%} \textsf{[0.16]}\end{tabular} & \cellcolor[HTML]{15783E}\color{white} \makecell{\textsf{99.9\%}\\ \textsf{[1.00]}} & \begin{tabular}[l]{@{}l@{}}\devNonIoT: \textsf{0.1\%} \textsf{[1.00]}\\\ \end{tabular} & \cellcolor[HTML]{15783E}\color{white} \makecell{\textsf{99.9\%}\\ \textsf{[1.00]}} & \begin{tabular}[l]{@{}l@{}}\devNonIoT: \textsf{0.1\%} \textsf{[0.84]}\\\ \end{tabular}\\ \hline
			\devWithingsSleep & \cellcolor[HTML]{15783E}\color{white} \makecell{\textsf{96.8\%}\\ \textsf{[1.00]}} & \begin{tabular}[l]{@{}l@{}}\devNonIoT: \textsf{1.9\%} \textsf{[0.99]}\\Others: \textsf{1.2\%} \textsf{[0.55]}\end{tabular} & \cellcolor[HTML]{15783E}\color{white} \makecell{\textsf{99.6\%}\\ \textsf{[1.00]}} & \begin{tabular}[l]{@{}l@{}}\devDropcam: \textsf{0.4\%} \textsf{[0.08]}\\\ \end{tabular} & \cellcolor[HTML]{ffffe5} \makecell{\textsf{23.5\%}\\ \textsf{[1.00]}} & \begin{tabular}[l]{@{}l@{}}\devDropcam: \textsf{76.5\%} \textsf{[0.08]}\\\ \end{tabular} & \cellcolor[HTML]{15783E}\color{white} \makecell{\textsf{100.0\%}\\ \textsf{[1.00]}} & \begin{tabular}[l]{@{}l@{}}\ \\\ \end{tabular} & \cellcolor[HTML]{15783E}\color{white} \makecell{\textsf{100.0\%}\\ \textsf{[1.00]}} & \begin{tabular}[l]{@{}l@{}}\ \\\ \end{tabular}\\ \hline
			\devHuebulb & \cellcolor[HTML]{a2d889} \makecell{\textsf{88.8\%}\\ \textsf{[1.00]}} & \begin{tabular}[l]{@{}l@{}}\devSamsungcam: \textsf{11.1\%} \textsf{[0.45]}\\\devBelkinMotion: \textsf{0.1\%} \textsf{[1.00]}\end{tabular} & \cellcolor[HTML]{a2d889} \makecell{\textsf{89.0\%}\\ \textsf{[1.00]}} & \begin{tabular}[l]{@{}l@{}}\devDropcam: \textsf{11.0\%} \textsf{[0.08]}\\\ \end{tabular} & \cellcolor[HTML]{ffffe5} \makecell{\textsf{0.8\%}\\ \textsf{[0.71]}} & \begin{tabular}[l]{@{}l@{}}\devDropcam: \textsf{99.2\%} \textsf{[0.08]}\\\ \end{tabular} & \cellcolor[HTML]{15783E}\color{white} \makecell{\textsf{99.9\%}\\ \textsf{[1.00]}} & \begin{tabular}[l]{@{}l@{}}\devNonIoT: \textsf{0.1\%} \textsf{[0.57]}\\\ \end{tabular} & \cellcolor[HTML]{15783E}\color{white} \makecell{\textsf{99.9\%}\\ \textsf{[1.00]}} & \begin{tabular}[l]{@{}l@{}}\devNonIoT: \textsf{0.1\%} \textsf{[0.47]}\\\ \end{tabular}\\ \hline
			\devChromeCast & \cellcolor[HTML]{a2d889} \makecell{\textsf{62.5\%}\\ \textsf{[1.00]}} & \begin{tabular}[l]{@{}l@{}}\devEcho: \textsf{25.0\%} \textsf{[0.52]}\\\devNonIoT: \textsf{12.5\%} \textsf{[0.60]}\end{tabular} & \cellcolor[HTML]{15783E}\color{white} \makecell{\textsf{100.0\%}\\ \textsf{[1.00]}} & \begin{tabular}[l]{@{}l@{}}\ \\\ \end{tabular} & \cellcolor[HTML]{15783E}\color{white} \makecell{\textsf{100.0\%}\\ \textsf{[0.98]}} & \begin{tabular}[l]{@{}l@{}}\ \\\ \end{tabular} & \cellcolor[HTML]{a2d889} \makecell{\textsf{87.5\%}\\ \textsf{[1.00]}} & \begin{tabular}[l]{@{}l@{}}\devDropcam: \textsf{12.5\%} \textsf{[0.69]}\\\ \end{tabular} & \cellcolor[HTML]{a2d889} \makecell{\textsf{87.5\%}\\ \textsf{[0.98]}} & \begin{tabular}[l]{@{}l@{}}\devDropcam: \textsf{12.5\%} \textsf{[0.57]}\\\ \end{tabular}\\ \hline
			\devHPprinter & \cellcolor[HTML]{a2d889} \makecell{\textsf{61.5\%}\\ \textsf{[0.99]}} & \begin{tabular}[l]{@{}l@{}}\devDropcam: \textsf{38.0\%} \textsf{[0.16]}\\Others: \textsf{0.6\%} \textsf{[0.86]}\end{tabular} & \cellcolor[HTML]{ffffe5} \makecell{\textsf{3.8\%}\\ \textsf{[1.00]}} & \begin{tabular}[l]{@{}l@{}}\devDropcam: \textsf{96.2\%} \textsf{[0.08]}\\\ \end{tabular} & \cellcolor[HTML]{ffffe5} \makecell{\textsf{2.5\%}\\ \textsf{[0.45]}} & \begin{tabular}[l]{@{}l@{}}\devDropcam: \textsf{97.1\%} \textsf{[0.08]}\\Others: \textsf{0.4\%} \textsf{[0.75]}\end{tabular} & \cellcolor[HTML]{15783E}\color{white} \makecell{\textsf{99.3\%}\\ \textsf{[0.82]}} & \begin{tabular}[l]{@{}l@{}}\devDropcam: \textsf{0.4\%} \textsf{[0.85]}\\Others: \textsf{0.2\%} \textsf{[0.39]}\end{tabular} & \cellcolor[HTML]{15783E}\color{white} \makecell{\textsf{99.8\%}\\ \textsf{[0.99]}} & \begin{tabular}[l]{@{}l@{}}\devNonIoT: \textsf{0.1\%} \textsf{[0.67]}\\\devDropcam: \textsf{0.1\%} \textsf{[0.28]}\end{tabular}\\ \hline
			\deviHome & \cellcolor[HTML]{a2d889} \makecell{\textsf{79.2\%}\\ \textsf{[0.90]}} & \begin{tabular}[l]{@{}l@{}}\devDropcam: \textsf{10.2\%} \textsf{[0.34]}\\Others: \textsf{10.6\%} \textsf{[0.42]}\end{tabular} & \cellcolor[HTML]{a2d889} \makecell{\textsf{87.5\%}\\ \textsf{[0.97]}} & \begin{tabular}[l]{@{}l@{}}\devDropcam: \textsf{12.5\%} \textsf{[0.08]}\\\ \end{tabular} & \cellcolor[HTML]{ffffe5} \makecell{\textsf{18.0\%}\\ \textsf{[0.57]}} & \begin{tabular}[l]{@{}l@{}}\devDropcam: \textsf{82.0\%} \textsf{[0.08]}\\\ \end{tabular} & \cellcolor[HTML]{a2d889} \makecell{\textsf{89.8\%}\\ \textsf{[1.00]}} & \begin{tabular}[l]{@{}l@{}}\devDropcam: \textsf{9.8\%} \textsf{[0.96]}\\\devHPprinter: \textsf{0.4\%} \textsf{[0.52]}\end{tabular} & \cellcolor[HTML]{15783E}\color{white} \makecell{\textsf{100.0\%}\\ \textsf{[0.99]}} & \begin{tabular}[l]{@{}l@{}}\ \\\ \end{tabular}\\ \hline
			\devWithingsBaby & \cellcolor[HTML]{a2d889} \makecell{\textsf{58.2\%}\\ \textsf{[1.00]}} & \begin{tabular}[l]{@{}l@{}}\devNonIoT: \textsf{41.8\%} \textsf{[1.00]}\\\ \end{tabular} & \cellcolor[HTML]{15783E}\color{white} \makecell{\textsf{100.0\%}\\ \textsf{[1.00]}} & \begin{tabular}[l]{@{}l@{}}\ \\\ \end{tabular} & \cellcolor[HTML]{ffffe5} \makecell{\textsf{0.0\%}\\ \textsf{[-]}} & \begin{tabular}[l]{@{}l@{}}\devDropcam: \textsf{100.0\%} \textsf{[0.08]}\\\ \end{tabular} & \cellcolor[HTML]{15783E}\color{white} \makecell{\textsf{100.0\%}\\ \textsf{[1.00]}} & \begin{tabular}[l]{@{}l@{}}\ \\\ \end{tabular} & \cellcolor[HTML]{15783E}\color{white} \makecell{\textsf{100.0\%}\\ \textsf{[1.00]}} & \begin{tabular}[l]{@{}l@{}}\ \\\ \end{tabular}\\ \hline
			\devWithingsScale & \cellcolor[HTML]{a2d889} \makecell{\textsf{74.8\%}\\ \textsf{[0.98]}} & \begin{tabular}[l]{@{}l@{}}\devNonIoT: \textsf{15.3\%} \textsf{[0.56]}\\Others: \textsf{9.9\%} \textsf{[0.19]}\end{tabular} & \cellcolor[HTML]{ffffe5} \makecell{\textsf{41.4\%}\\ \textsf{[0.79]}} & \begin{tabular}[l]{@{}l@{}}\devWithingsSleep: \textsf{56.8\%} \textsf{[0.96]}\\\devDropcam: \textsf{1.8\%} \textsf{[0.08]}\end{tabular} & \cellcolor[HTML]{ffffe5} \makecell{\textsf{42.3\%}\\ \textsf{[0.33]}} & \begin{tabular}[l]{@{}l@{}}\devDropcam: \textsf{57.7\%} \textsf{[0.08]}\\\ \end{tabular} & \cellcolor[HTML]{15783E}\color{white} \makecell{\textsf{99.1\%}\\ \textsf{[1.00]}} & \begin{tabular}[l]{@{}l@{}}\devDropcam: \textsf{0.9\%} \textsf{[0.54]}\\\ \end{tabular} & \cellcolor[HTML]{15783E}\color{white} \makecell{\textsf{100.0\%}\\ \textsf{[1.00]}} & \begin{tabular}[l]{@{}l@{}}\ \\\ \end{tabular}\\ \hline
			\devRingdoor & \cellcolor[HTML]{ffffe5} \makecell{\textsf{0.6\%}\\ \textsf{[0.98]}} & \begin{tabular}[l]{@{}l@{}}\devNetatmoWeather: \textsf{95.8\%} \textsf{[0.18]}\\Others: \textsf{3.6\%} \textsf{[0.60]}\end{tabular} & \cellcolor[HTML]{15783E}\color{white} \makecell{\textsf{100.0\%}\\ \textsf{[0.98]}} & \begin{tabular}[l]{@{}l@{}}\ \\\ \end{tabular} & \cellcolor[HTML]{ffffe5} \makecell{\textsf{7.8\%}\\ \textsf{[1.00]}} & \begin{tabular}[l]{@{}l@{}}\devDropcam: \textsf{92.2\%} \textsf{[0.08]}\\\ \end{tabular} & \cellcolor[HTML]{15783E}\color{white} \makecell{\textsf{100.0\%}\\ \textsf{[1.00]}} & \begin{tabular}[l]{@{}l@{}}\ \\\ \end{tabular} & \cellcolor[HTML]{15783E}\color{white} \makecell{\textsf{100.0\%}\\ \textsf{[1.00]}} & \begin{tabular}[l]{@{}l@{}}\ \\\ \end{tabular}\\ \hline
			\devBlipcareBP & \cellcolor[HTML]{ffffe5} \makecell{\textsf{20.0\%}\\ \textsf{[0.54]}} & \begin{tabular}[l]{@{}l@{}}\devRingdoor: \textsf{80.0\%} \textsf{[0.41]}\\\ \end{tabular} & \cellcolor[HTML]{ffffe5} \makecell{\textsf{40.0\%}\\ \textsf{[0.79]}} & \begin{tabular}[l]{@{}l@{}}\devHPprinter: \textsf{60.0\%} \textsf{[0.44]}\\\ \end{tabular} & \cellcolor[HTML]{ffffe5} \makecell{\textsf{0.0\%}\\ \textsf{[-]}} & \begin{tabular}[l]{@{}l@{}}\devDropcam: \textsf{100.0\%} \textsf{[0.08]}\\\ \end{tabular} & \cellcolor[HTML]{15783E}\color{white} \makecell{\textsf{100.0\%}\\ \textsf{[0.90]}} & \begin{tabular}[l]{@{}l@{}}\ \\\ \end{tabular} & \cellcolor[HTML]{15783E}\color{white} \makecell{\textsf{100.0\%}\\ \textsf{[0.85]}} & \begin{tabular}[l]{@{}l@{}}\ \\\ \end{tabular}\\ \hline
			\devHelloBarbie & \cellcolor[HTML]{ffffe5} \makecell{\textsf{0.0\%}\\ \textsf{[-]}} & \begin{tabular}[l]{@{}l@{}}\devDropcam: \textsf{71.4\%} \textsf{[0.08]}\\Others: \textsf{28.6\%} \textsf{[0.50]}\end{tabular} & \cellcolor[HTML]{ffffe5} \makecell{\textsf{21.4\%}\\ \textsf{[1.00]}} & \begin{tabular}[l]{@{}l@{}}\devDropcam: \textsf{71.4\%} \textsf{[0.08]}\\\devHPprinter: \textsf{7.1\%} \textsf{[0.45]}\end{tabular} & \cellcolor[HTML]{ffffe5} \makecell{\textsf{21.4\%}\\ \textsf{[0.99]}} & \begin{tabular}[l]{@{}l@{}}\devDropcam: \textsf{78.6\%} \textsf{[0.08]}\\\ \end{tabular} & \cellcolor[HTML]{ffffe5} \makecell{\textsf{14.3\%}\\ \textsf{[0.97]}} & \begin{tabular}[l]{@{}l@{}}\devHPprinter: \textsf{78.6\%} \textsf{[0.52]}\\\devDropcam: \textsf{7.1\%} \textsf{[0.61]}\end{tabular} & \cellcolor[HTML]{15783E}\color{white} \makecell{\textsf{92.9\%}\\ \textsf{[0.99]}} & \begin{tabular}[l]{@{}l@{}}\devHuebulb: \textsf{7.1\%} \textsf{[0.35]}\\\ \end{tabular}\\ \hline
			\devNonIoT & \cellcolor[HTML]{a2d889} \makecell{\textsf{74.2\%}\\ \textsf{[0.98]}} & \begin{tabular}[l]{@{}l@{}}\devTriby: \textsf{16.6\%} \textsf{[0.90]}\\Others: \textsf{9.2\%} \textsf{[0.69]}\end{tabular} & \cellcolor[HTML]{a2d889} \makecell{\textsf{66.9\%}\\ \textsf{[0.97]}} & \begin{tabular}[l]{@{}l@{}}\devDropcam: \textsf{29.7\%} \textsf{[0.08]}\\Others: \textsf{3.4\%} \textsf{[0.73]}\end{tabular} & \cellcolor[HTML]{a2d889} \makecell{\textsf{59.5\%}\\ \textsf{[0.79]}} & \begin{tabular}[l]{@{}l@{}}\devDropcam: \textsf{36.3\%} \textsf{[0.08]}\\Others: \textsf{4.2\%} \textsf{[0.73]}\end{tabular} & \cellcolor[HTML]{15783E}\color{white} \makecell{\textsf{98.8\%}\\ \textsf{[1.00]}} & \begin{tabular}[l]{@{}l@{}}\devHPprinter: \textsf{1.1\%} \textsf{[0.56]}\\\devDropcam: \textsf{0.2\%} \textsf{[0.75]}\end{tabular} & \cellcolor[HTML]{15783E}\color{white} \makecell{\textsf{99.7\%}\\ \textsf{[0.99]}} & \begin{tabular}[l]{@{}l@{}}\devHPprinter: \textsf{0.3\%} \textsf{[0.55]}\\\ \end{tabular}\\ \hline
			
		\end{tabular}
		
	\end{adjustbox}
	\
\end{table}

			\subsubsection{Performance of Stage-0: Port Numbers Attribute}\label{sec:c1_perfS0}
				The first three columns correspond to those cases in which we consider only nominal attributes of stage-0 (i.e. bag of words corresponding to port numbers, domain names and cipher suites). The first column shows that when we only use a list of server-side port numbers for device classification, a reasonable accuracy of 92.13\% is achieved, but RRSE is poor (at 39.93\%). Inspecting the individual classes, we observe that certain classes highlighted by yellow or light-green (e.g. Ring door bell, Blipcare BP monitor, Hello Barbie, and Google chromecast) are poorly classified. We explain the reason behind this misclassification next.

				{\bf Ring door bell:} Out of 486 instances, 465 contain a single occurrence of the DNS query (i.e. remote port number 53). We see that $95.8$\% of test instances are incorrectly classified as Netatmo weather station. This is because of two reasons: (i) there are 2451 training instances of Netatmo compared to 323 of Ring door bell, which makes $\Pr(c_{i}^{train})$ of Netatmo larger than that of Ring door bell, and (ii) many Netatmo instances contain several (on average 4 times) occurrences of port 53 as opposed to only one for Ring Door bell, which also contributes to $\Pr( w_j | c_i)$ of Netatmo being greater than that for Ring door bell in \eqref{eq:training}. Thus, Ring door bell instances get classified as Netatmo weather station, warranting a second stage of classification with additional attributes for improved accuracy.

				{\bf Blipcare BP monitor:} It uses only two remote port numbers, namely 8777 and 53, in a total of 13 instances - the port numbers appear only once or twice in each instance. Surprisingly, we see that 80\% of Blipcare test instances are incorrectly classified as Ring Door Bell though the remote port number of 8777 is unique to the Blipcare BP monitor. This is because there are only a very small number of Blipcare instances in our dataset, which results in a fairly small value of $\Pr(``53" | Blipcare)=0.0203$ and $\Pr( ``8777" | Blipcare)=0.0294$ in \eqref{eq:training}, and a negligible value of $\Pr(Blipcare^{train})=0.0003$ in \eqref{eq:testing}. On the other hand, $\Pr( ``8777" | Ring)$ becomes very small as the remote port number 8777 is never used by the Ring Door Bell in our dataset. However, the probability of $\Pr( ``8777" | Ring)=0.0011$ in \eqref{eq:training} is sufficient enough to maximize the classifier probability $\Pr( Ring | \{``53":1, ``8777":1\})$ in \eqref{eq:testing}, given $\Pr(Ring^{train})=0.0097$.

				{\bf Other devices:}  Server-side port numbers are empty in $72$\% of instances for Hello Barbie, since it communicates with local devices instead of Internet-based end-points. Similarly for HP printer (38\%) and iHome power plug (10\%). The lack of server-side port number information explains why these devices are classified as Dropcam, which has the highest value of $\Pr(Dropcam^{train})=0.0828$ in \eqref{eq:testing}. We note that the confidence level of our stage-0 classifier is fairly low (i.e. less than $0.4$) in these cases, suggesting that the classifier chooses the most probable class given empty attribute (i.e. all $n_j^{test}$ are zero).

			\subsubsection{Performance of Stage-0: Domain Names Attribute}
				We now focus on the stage-0 machine that uses only a bag of domain-names, which yields an accuracy of $79.48$\% with a fairly high RRSE value of $57.56$\%, as shown in the second column in Table~\ref{tab:performance}. In this scenario, more classes suffer from misclassification (i.e. those with yellow coloured cells) compared to the previous scenario where only remote port numbers were considered. The  reasons behind the misclassification are threefold: (i) since devices from the same manufacturer share a collection of domain names, as discussed in \S\ref{sec:c1_IoTprotocolDNS}, $59.8$\% of Belkin camera test instances are misclassified as Belkin Motion sensor and $100$\% Belkin Motion sensor instances are misclassified as Belkin switch. Similarly, $56.8$\% of Withings scale instances are incorrectly classified as Withings sleep sensor, and $12$\% of Samsung smart cam are misclassified as Samsung Smartthings. (ii) a significant number of instances from select devices contain no DNS query entries (e.g. $96.2$\% of HP printer, $73.4$\%  of Samsung Smart Cam, $71.4$\% of Hello Barbie, $12.5$\% of iHome power plug, $11$\% of Hue bulb) and are thus incorrectly classified as a Dropcam, which also rarely generates DNS packets. (iii) the low number of training instances with domain names leads to poor performance (e.g. Blipcare BP meter and Hello Barbie).

			\subsubsection{Performance of Stage-0: Cipher Suite Attribute}
				Considering only the cipher suite attribute, this stage-0 classifier results in a fairly low accuracy of $36.15$\% with a high RRSE of $86.73$\%, as shown in the third column in Table~\ref{tab:performance}. Again, the main reason for such poor performance is the scarcity of cipher suite attribute in the training instances, though this attribute carries a very strong signature to uniquely identify an IoT device. Note that many of the IoT devices do not use secure communication at all and are thus devoid of this attribute (i.e. have an empty field for it). Unsurprisingly, instances of devices that exchange cipher suite fairly frequently including Amazon Echo, Awiar air quality monitor, Canary camera, Google Chromecast and Netatmo camera are correctly classified, as shown by the dark-green color cells in the corresponding column in Table~\ref{tab:performance}. In addition, we find that August doorbell cam is sharing one of its cipher suite strings (out of total 18) with Pixstar photoframe, which has a single cipher suite string. Thus, $21.2$\% of August door bell instances are misclassified as Pixstar photoframe and almost all instances of Pixstar photoframe are classified as August doorbell.

			\subsubsection{Performance of Stage-0: Combination of Attributes}
				We expect the combination of the three bags of words (port numbers, domain names, and cipher suites) to significantly enhances the accuracy of our classifier, as indeed shown by the fourth column titled ``Combined stage-0'' in Table~\ref{tab:performance}. The overall accuracy reaches to $97.39\%$ with RRSE of $18.24\%$. It can be seen that the majority of test instances are correctly classified, except for Hello Barbie. This is because most of the Hello Barbie attributes are empty in stage-0 and thus it is classified as Dropcam, as mentioned earlier. 

				Interestingly, we see that all test instances of Blipcare BP monitor are classified correctly though the accuracy of individual stage-0 was fairly poor. This is because our decision-tree-based classifier in stage-1 sees a strong correlation between the outputs of stage-0 classifiers and the actual class of training instance, even though those outputs (tentative class) are incorrect -- e.g. having the tentative output from remote port number classifier as Ring door bell, having the tentative output from cipher suite classifier as Dropcam, and having the confidence level from domain name classifier less than $0.66$ collectively is a strong indication of Blipcare instance.

			\subsubsection{Overall Performance}
				As the last step, we incorporate the outputs from the stage-0 classifiers into stage-1 (without the latter having any notion of the quantitative attributes from the former), and additionally include quantitative attributes (flow volume, duration, rate, sleep time, DNS and NTP intervals). The last column of Table~\ref{tab:performance} shows the overall performance of the classification framework. In this case, the accuracy reaches a remarkably high value of $99.88$\%, with almost all classes labeled correctly with a very small value of RRSE at $5.06$\%. Fig.~\ref{fig:confMap} shows the full confusion matrix of our classification when all the attributes are used in conjunction, and corroborates that the diagonal entries (corresponding to correct classification) are all at or very close to 100\%, with just two exceptions -- the Google Chromecast and the Hello Barbie. As explained earlier, the Chromecast gets classified as the Dropcam in some instances, while the Hello Barbie gets classified as a Hue bulb.

				\begin{figure}[t!]
					\centering
					\includegraphics[width=1\textwidth]{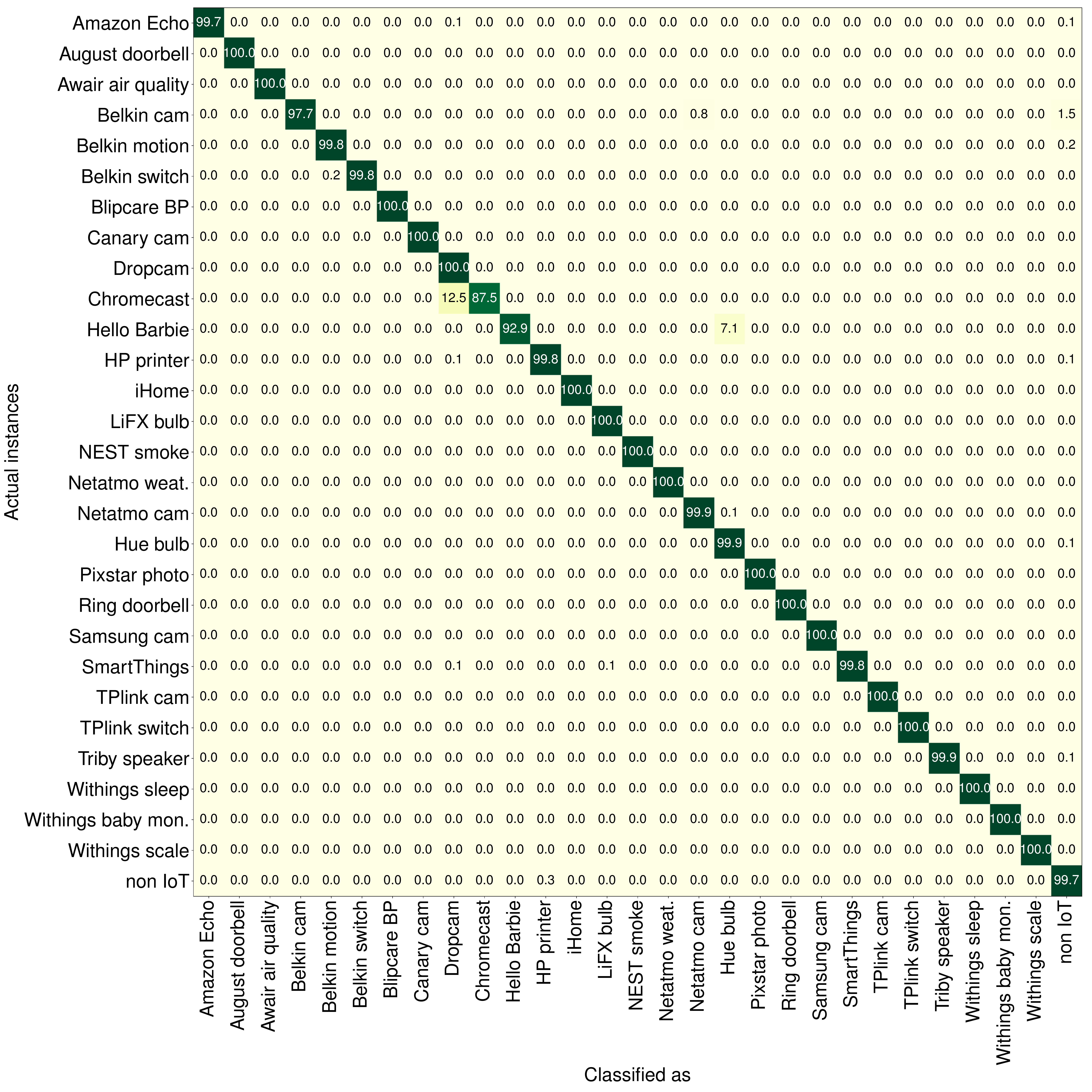}
					
					\caption{Confusion matrix of our IoT device classification using all attributes (accuracy: 99.88\%, RRSE: 5.06\%).}
					
					\label{fig:confMap}
				\end{figure}

	\section{Real-Time Operation in a Network}\label{sec:c1_perfAcc}
		Thus far, we have examined the performance of our multi-stage classifier using off-line analysis on captured traffic traces (i.e. pcap files). In this section, we discuss how one can realize a real-time implementation of our system taking into account the various stages involved in the analysis, namely attribute collection, machine training, and interpreting the classifier's output.

		\subsection{Computing Attributes}\label{sec:c1_attrbtMerit}
			Extracting the attributes on-the-fly requires infrastructure that has sufficient visibility into the traffic flowing on the network. Flow related attributes such as flow volume, flow duration and flow rate can be extracted relatively easily using network switches that are instrumented with special hardware-accelerated flow-level analyzers, e.g. NetFlow capable devices \cite{Vyncke2008}. We therefore deem the extraction cost of flow related attributes to be fairly low, and show them via blue color bars in Fig.~\ref{fig:merit} that depicts the relative costs and merits of the various attributes.

			\begin{figure}[t!]
				\vspace{-1em}
				\begin{center}
					\includegraphics[width=0.45\textwidth]{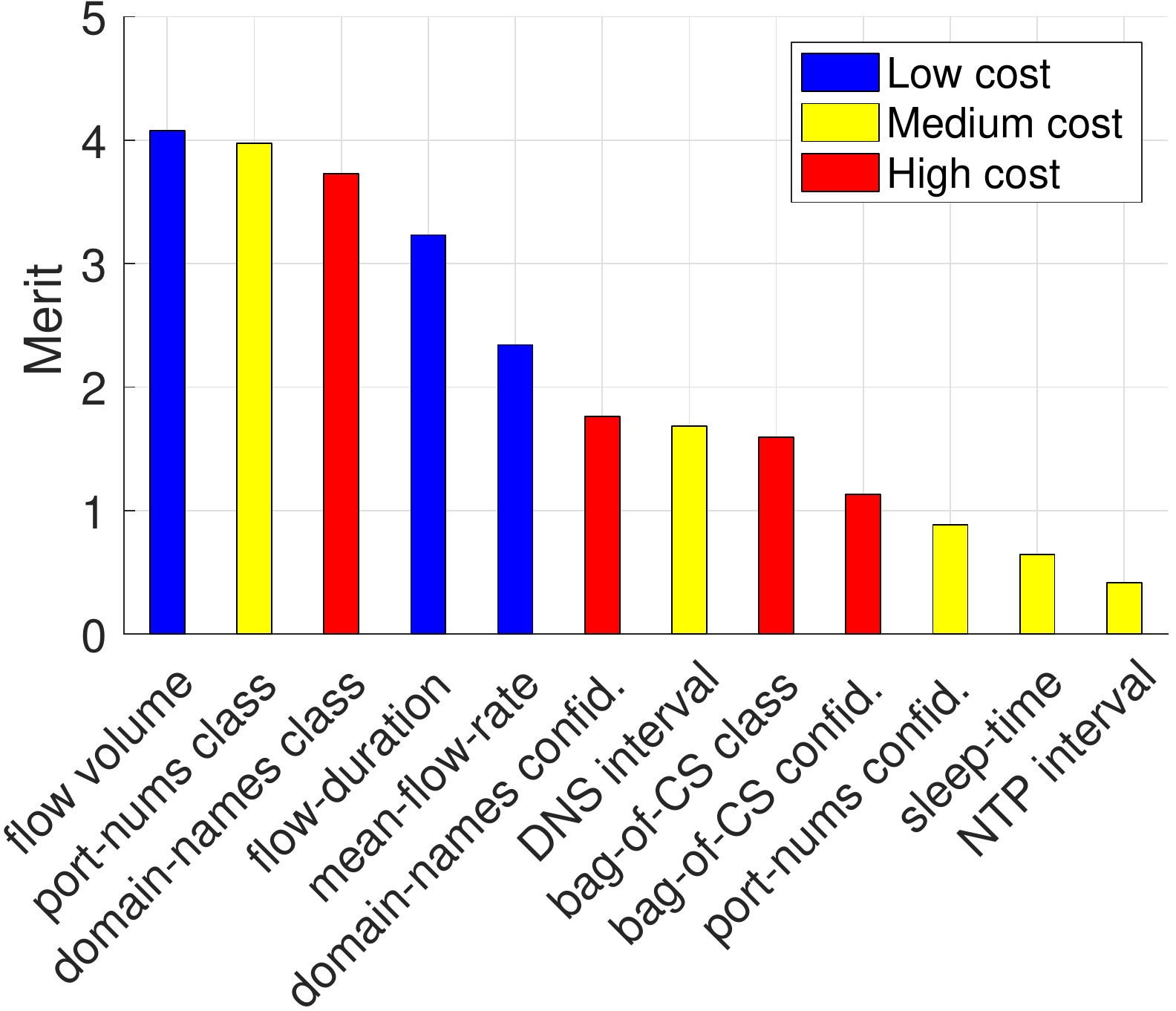}
					\caption{Merit of attributes.}
					\label{fig:merit}
				\end{center}
				\vspace{-2em}
			\end{figure}

			Attributes including bag of port numbers, sleep-time, and frequency of DNS/NTP requests can be extracted using flow-aware network switches with extra computation and state management. For example, remote port numbers of all flows associated with a given IoT device need to be recorded for the bag of port numbers. However, this specific state is not captured by default in commodity switches. Similarly, time intervals between successive UDP packets of NTP/DNS should be recorded, which requires additional computation. We therefore associate these attributes with medium cost, and shown as yellow color bars in Fig.~\ref{fig:merit}. 

			\vspace{-0.5em}
			Lastly, two of our attributes, namely bag of domain names and bag of cipher suite strings, can only be extracted by looking inside the payload of the appropriate packets, which imposes considerable cost on processing. Thus, we associate these attributes with high collection cost, and shown them via red color bars in Fig.~\ref{fig:merit}.
			
			\vspace{-0.5em}
			Having understood the extraction cost of various attributes, let us now examine the relative importance of the attributes in classifying the IoT devices. We quantify the importance of each attribute by employing the \textit{select attributes} tool in Weka with \textit{InfoGain} attribute evaluator and \textit{Ranker} search method. Fig.~\ref{fig:merit} shows the attributes in decreasing order of merit score. A high merit score translates to superior strength in identifying the class of an instance.  We can see that the ``flow-volume'' is the most important attribute, followed by ``bag of remote port numbers'', ``bag of domain names'' and ``flow duration'' respectively. The sleep-time and NTP interval are the attributes with the lowest merit.

			Knowing the relative cost and merit of each attribute allows us to evaluate the performance of our classifier using: (a) only low cost attributes, (b) combination of low and medium cost attributes, and (c) all attributes. The classifier accuracy and RRSE are shown in Table ~\ref{tab:perf}. It is seen that using only low-cost attributes results in $97.85$\% accuracy with an RRSE value of $18.63$\%; the additional use of medium-cost attributes increases accuracy to $99.68$\% and significantly reduces the RRSE error to $7.7$\%; while including all attributes yields an overall accuracy of $99.88$\% and RRSE of $5.06$\%. The method can therefore be tuned to achieve appropriate balance between attribute collection cost and accuracy/error of classification.

			\begin{table}[bt]
				\centering
				\caption{Impact of attributes combination on performance of classifier.}
				\label{tab:perf}
				
				\begin{adjustbox}{max width=\textwidth}
					\def\arraystretch{1.5}
					\begin{tabular}{l|l|l|}
						\cline{2-3}
						& \textbf{Accuracy} & \textbf{RRSE}  \\ \hline
						\multicolumn{1}{|l|}{all attributes}             & 99.88\%    & 5.06\%  \\ \hline
						\multicolumn{1}{|l|}{low- and medium-cost attributes} & 99.68\%    & 7.70\%   \\ \hline
						\multicolumn{1}{|l|}{only low-cost attributes}        & 97.85\%    & 18.63\% \\ \hline
					\end{tabular}
				\end{adjustbox}
				\vspace{-1em}
			\end{table}

			\vspace{-2em}
		\subsection{Training the Machine}\label{sec:c1_trainImpact}
			\vspace{-1em}
			The duration of the training data set is another source of cost incurred by our classification. In Fig.~\ref{fig:trainImpact}, we plot the accuracy of the classifier on the left y-axis and the RRSE on the right y-axis as a function of the number of days involved in collecting the training data set. Note that the x-axis is in log-scale and each day represents 24 instances. 

			\vspace{-0.5em}
			It can be seen that the classifier achieves an overall accuracy is $99.28$\% with only one day of training and saturates at $99.76$\% when trained over 16 days. On the other hand, RRSE drops from $14.43$\% to $7.5$\% when the training duration is increased from 1 day to 16 days. It further falls to $5.82$\% when we train using $70$\% of all instances from 128 days. As mentioned in \S\ref{sec:c1_class}, the RRSE value is sensitive to the accuracy of individual classes. We therefore believe that if there is a balanced number of instances from various classes, our classifier would perform better in terms of RRSE. 
			
			\begin{figure}[t]
				\begin{center}
					\mbox{
						\subfloat[Impact of training data.]{
							\includegraphics[width=0.58\textwidth]{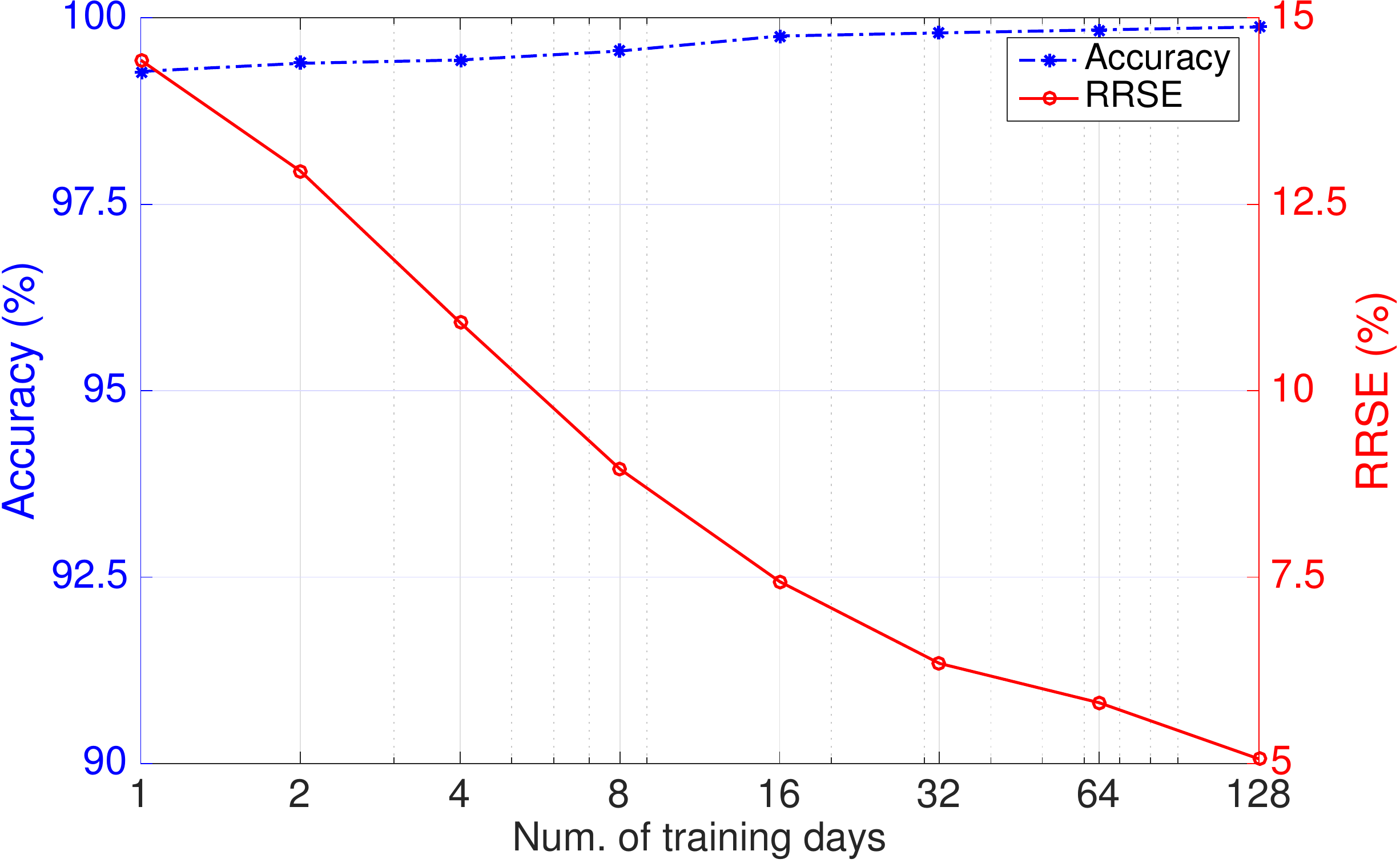}\quad
							\label{fig:trainImpact}
						}
					}
					\mbox{
						\hspace{-2mm}
						\subfloat[Confidence of classifier.]{
							\includegraphics[width=0.51\textwidth]{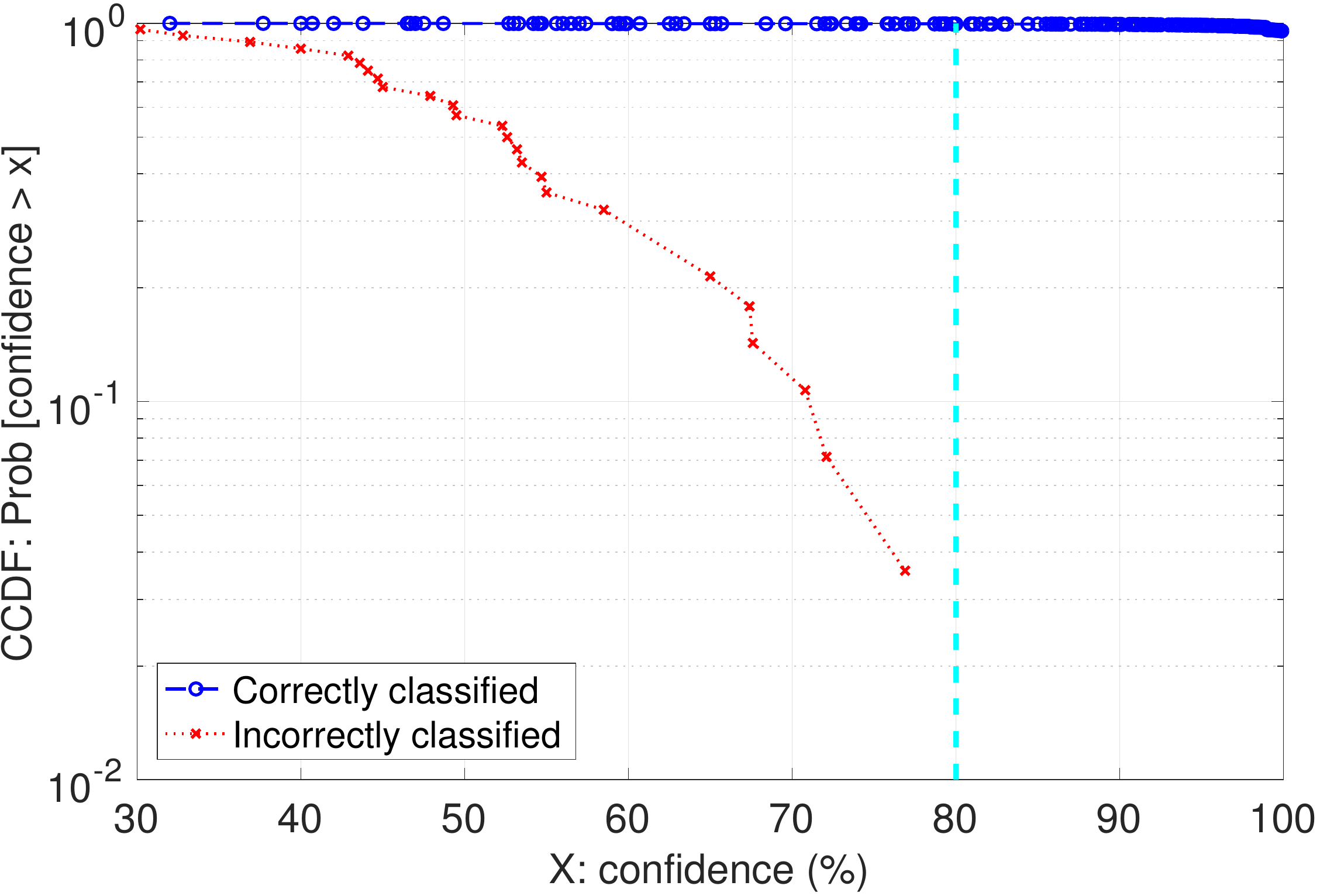}\quad
							\label{fig:confidence}
						}
					}	
					
					\caption{Operational insights for real-time implementation of our device classifier: (a) impact of training and (b) confidence-level for correct/incorrect classification.}
					
					\label{fig:perf}
				\end{center}
			\end{figure}
		
		\subsection{Interpreting the Output of Classifier}\label{sec:c1_perfConf}

			As discussed in \S\ref{sec:c1_classArch}, our classifier generates a confidence level during the testing phase. This can be used as a measure of reliability for our classifier. If adequate information is not provided by a test instance then the classifier will choose a random class (as discussed in \S\ref{sec:c1_perfS0}) with a low confidence level - this can be interpreted as an ``unknown'' class. For example, given instances with an empty value for the cipher suite attribute, the corresponding stage-0 classifier will output Dropcam class with a confidence value of less than $10$\% - even for Dropcam instances that are classified correctly the confidence level is low within the same range.
			\vspace{-0.7em}

			We plot the CCDF of confidence level of our stage-1 classifier in Fig.~\ref{fig:confidence} for instances classified as correct and incorrect. It is clearly seen that the confidence level is always below $80$\% when an instance is incorrectly classified, as shown by the red dotted line - the average confidence level for incorrectly classified instances is $54.22$\%.  On the other hand, our classifier has an average $99.74$\% confidence level for instances that are correctly classified. We note that for only a negligible fraction of correctly classified instances (i.e. $0.37$\%) the confidence level is less than 80\% as shown by the blue dashed line. This suggests that we can comfortably rely on our classifier's output for a device if it results in a confidence level of greater than $80$\%, otherwise we need to collect more traffic (and richer instances) from that device in order to increase the confidence level. 
			
			{\rev
				To demonstrate the ability of our classifier in detecting changes of normal behavior, we have launched UDP reflection and TCP SYN attacks of varying rates on the Samsung camera. When our classifier is fed these attributes during the attack, it incorrectly identifies the device, but its confidence-level drops to less than 50\%. We note that the confidence level is 100\% for normal traffic from Samsung camera, as shown in the last column of Table~\ref{tab:performance}. This is taken as a sign of anomalous behavior that warrants further investigation by the network operator. Note that the anomaly detection is not the primary focus of this chapter, and hence our evaluation is limited to only one attack type. Later in Chapter 5 we will study in detail several attacks when we evaluate our anomaly detection scheme.
			}

	\section{Conclusion}\label{sec:c1_con} 

		Despite the proliferation of IoT devices in smart homes, enterprises, campuses, and cities around the world, operators of such environments lack visibility into what IoT devices are connected to their networks, what their traffic characteristics are, and whether the devices are functioning appropriately free from security compromises. This work is the first to systematically characterize and classify IoT devices at run-time. We instrumented a smart environment with 28 unique IoT devices and collected traffic traces continuously over 26 weeks. We then statistically characterized the traffic in terms of activity cycles, signalling patterns, communication protocols and cipher suites. We developed a multi-stage machine learning based classification framework that uniquely identifies IoT devices with over 99\% accuracy. Finally, we evaluated the real-time operational cost, {\rev response time}, and accuracy trade-offs of our classification method. This chapter shows that IoT devices can be identified with high accuracy based on their network behavior, and sets the stage for detecting misbehaviors resulting from security breaches in the smart environment. However, obtaining all of these characteristics using specialized hardware accelerators in real-time becomes more expensive, and unscalable due to the need of deep packet inspection. In the following chapter, we will discuss an approach to monitor the devices with low-cost attribute extraction.
\chapter{Behavioral Monitoring using Low-Cost Attributes}
\label{chap:telemetry}
\minitoc

	%Cyber-security risks for Internet of Things (IoT) devices sourced from a diversity of vendors and deployed in large numbers, are growing rapidly. Therefore, management of these devices is becoming increasingly important to network operators.
	%Our study in the previous chapter showed the technical feasibility of dynamic fast- lane and slow-lane creation driven by content providers (CPs), using software defined networking (SDN)platforms.
	In the previous chapter, we distinguish the types of IoT devices based on their network characteristics. However to secure the IoT devices, we need to detect their anomalous activities and behavioral drifts. This turns out to be non-trivial as it (a) requires an inference engine that monitors the fine-grained activities of the devices (b) should incur a low-cost computational cost for telemetry. The existing IoT traffic monitoring techniques use specialized acceleration on network switches, or full inspection of packets in software, which can be complex, expensive, inflexible, and unscalable. In this chapter, we use an SDN paradigm combined with machine learning to leverage the benefits of programmable flow-based telemetry with flexible data-driven models to manage IoT devices based on their network activity. Our contributions are three-fold: (1) We analyze traffic traces of 17 real consumer IoT devices collected in our lab over a six months period and identify a set of traffic flows (per-device) whose time-series attributes computed at multiple timescales (from a minute to an hour) characterize the network behavior of various IoT device types, and their operating states (\ie booting, actively interacted with user, or being idle); (2) We develop a multi-stage architecture of inference models that use flow-level attributes to automatically distinguish IoT devices from non-IoTs, classify individual types of IoT devices, and identify their states during normal operations. We train our models and validate their efficacy using real traffic traces; and (3) We quantify the trade-off between performance and cost of our solution, and demonstrate how our monitoring scheme can be used in operation for detecting behavioral changes (firmware upgrade or cyber attacks). Parts of this chapter have been published in ~\cite{ANTS16} and ~\cite{TNSM19}.

	\section{Introduction}\label{sec:c2_introduction}
		The lack of effective security on IoTs \cite{Suo2012, Jing2014, Andrea2015} presents a number of challenges for network operators of large organizations who are looking to bring these devices online at scale. Implementing device-level security would definitely help protect against automated attacks \cite{ZDNetBotnets2018}, but its efficacy can vary across manufacturers and device types depending upon devices capabilities and their mode of operation \cite{Sivaraman2015WiMob}. In a parallel effort, IETF has approved an Internet standard called ``Manufacturer Usage Description" (MUD) \cite{ietfMUD18} to protect IoT devices. This framework allows manufacturers to formally specify the intended behavior of their devices that can be used to generate and enforce access control lists (ACLs) \cite{IoTSnP18-mudids} for IoT devices, limiting their network behavior to only a tight set of services. Although MUD policies can reduce the surface of attacks on IoTs they are still insufficient, since ACL rules do not restrict temporal variation of traffic flows (\eg traffic with unwanted volume or pattern cannot be prevented if endpoints and protocols conform to MUD rules).

		Therefore, it is crucial for organizations to maximize visibility into their IoT infrastructure \cite{TMC18}, and thus better manage security risks of these vulnerable devices \cite{Cisco2017}. Network administrators need to know all connected devices and their expected operations on the network, and continuously monitor their activities ensuring IoTs behave ``normally'' \cite{Cisco2017}. Existing traffic monitoring solutions are either purely software-based (hence unscalable to high traffic rates) or customized hardware-based (hence inflexible and expensive) \cite{FlowRadar2016}. Network operators, today, widely use NetFlow \cite{rfcNetFlow} (an embedded switch instrumentation) to obtain aggregate measurement of traffic flows. However, it comes at cost of CPU resources on the switch \cite{NetflowCPU2010} for generating, collating, and exporting flow records. To reduce this overhead, operators statistically mirror packet samples (\eg sFlow \cite{sFlow}) to a remote collector for extracting flow information that inevitably leads to reduced accuracy. On the other hand, special-purpose hardware appliances (\ie deep packet inspection engines) offer both accuracy and performance in traffic monitoring but they are prohibitively expensive for many network operators.

		In this chapter, we aim to monitor behavior of IoT devices on the network using a combination of Software Defined Networking (SDN) telemetry and machine learning methods. We believe that the SDN paradigm by its nature provides flow-level isolation and visibility in a low-cost and scalable manner. For accurate detection of devices and tracking of their dynamic behaviors, we employ machine learning algorithms to learn key patterns of traffic flows. Our first contribution is to identify a set of TCP and UDP flows (for each IoT device) and highlight characteristics attributes, computed from time-series of flows at multiple time-scales, distinguishing various IoT device types and their behavioral states (booting, active, or idle) on the network. Our second contribution develops a multi-stage architecture consisting of a set of inferencing models that use flow-level attributes to automatically recognize traffic of IoT devices from non-IoTs, classify types of IoT devices, and identify operating states of each IoT during normal operation. We train our models and validate their performance to obtain high accuracy using real traffic traces. Finally, we demonstrate the efficacy of our scheme in detecting network behavioral changes due to firmware upgrade or cyber-attacks. Also, we quantify the trade-off between performance and cost of our monitoring solution for real-time deployment.  

		% chapter organization
		The rest of this chapter is organized as follows: In \S\ref{sec:c2_attbt} we present our dataset and traffic flows, and characterize attributes of various IoT devices and their operating states. We propose the architecture of IoT traffic inference and evaluate its performance in \S\ref{sec:c2_classification}, followed by a discussion on the operational trade-off and use of the proposed system in \S\ref{sec:c2_op}. The chapter is concluded in \S\ref{sec:c2_con}.

	\section{Traffic Flows and Attributes}\label{sec:c2_attbt}
		In this section, we begin by analyzing real traffic traces collected in our lab. We then identify traffic attributes to distinguish IoT devices from non-IoTs, classify individual IoTs, and determine their operating states. 

		\subsection{Traffic Trace Dataset}\label{sec:c2_data}
			We used two sets of full PCAP traffic traces collected from our testbed. The first dataset (\ie DATA1) was collected from a network consisting of more than thirty IoT and non-IoT devices for a duration of six months  (\ie 01-Oct-2016 to 31-Mar-2017)~\cite{TMC18}. We select 17 IoT devices, those whose trace was present for at least 60 days in packet traces. These devices include Amazon Echo, August doorbell, Awair air quality, Belkin motion sensor, Belkin switch, Dropcam, HP printer, LiFX bulb, NEST smoke sensor, Netatmo weather, Netatmo camera, Hue bulb, Samsung smart camera, Smart Things, Triby speaker, Withings sleep sensor, and Withings scale. Note that our dataset contains traffic traces of six non-IoT devices including Android phone, Android tablet, Windows laptop, MacBook, and two iPhones.

			The second dataset (\ie DATA2) consists of traces with state annotation for selected IoT devices including Amazon Echo, Belkin switch, Dropcam, and LiFX bulb. We developed a software tool to automatically interact with these four devices over two days and annotate their traffic traces. Annotations indicate three operating states of IoT devices, namely ``\textit{boot}'' (\ie getting connected to the network), ``\textit{active}''  (\ie interacting with users), and ``\textit{idle}''  (\ie not being booted or actively used).
			{\rev
				The main purpose of state classification is to monitor the activity of IoT devices at a fine-grained level, augmenting the device classification model. We believe that boot, active, and idle states are generic and across all IoT devices. Hence, the state classifier is able to capture minor variations in the activity patterns of a device regardless of various attack models.  
			}

			For the boot state, we used a TP-Link HS110 smart plug (whose traffic is not considered in our analysis) supplying power to these four devices. We wrote a script to automatically turn off/on this smart plug resulting a boot state for the subjected IoT device. For the active state, we used an app called ``{\myverb{RepetiTouch Pro}}'' and a text to-speech engine called {\myverb{espeak}}\cite{espeak}. The former records and replays interactions of a real user with three IoT devices including Belkin switch, Google Dropcam camera, and LiFX lightbulb via their manufacturer app -- the user interactions (\ie  turning on/off the switch, streaming video from the camera, and turning on/off the bulb) were recorded on an Android tablet which was connected to the local network of our testbed. The latter periodically asks scripted questions (\eg ``How is the weather in Sydney, Australia'') from Amazon Echo. For the idle state, we used all traffic traces that were annotated as neither boot nor active, during the data collection period.
			{\rev
				The difficulty of capturing all possible user interactions in the active state, led us to limit the number of devices studied to only four.
			}

		\subsection{Traffic Flows and Attributes}\label{sec:c2_att}
			{\rev
				In Chapter~\ref{chap:characterization} we showed that individual IoT devices exhibit identifiable patterns in their traffic flows such as DNS/NTP/SSDP signaling profiles, activity cycles, and volume patterns. Although, these attributes contain a rich set of information to fingerprint IoTs and their activities, they incur high cost of extraction from the network in real-time.
			}
		
			{\rev
				In this chapter, our attributes are computed for individual 2-tuple and/or 3-tuple flows (coarse-grained). Note that this differs from our attributes in Chapter~\ref{chap:characterization} where IoT traffic attributes such as activity volume and average flow rates were computed for individual 5-tuple flows (fine-grained). The attributes employed in Chapter~\ref{chap:characterization} provide a richer set of information, and hence the trained models yield a very high accuracy. However, those attributes are expensive to extract and compute from the network traffic in real-time. Instead, the attributes we identify and use in this chapter are largely at aggregate level (more cost-effective) but give a slightly lower accuracy (still reasonably acceptable) when compared to results in Chapter~\ref{chap:characterization}. 
			}
		
			{\rev
				As we mention in the Chapter~\ref*{chap:survey}, flow-level telemetry provided by SDN APIs~\cite{McKeown2008} enables us to dynamically measure specific traffic flows at low-cost with reasonable resolutions. Also, we note that SDN-enabled switches that are currently available in the market typically support a large number of flow rules without experiencing performance degradation. For example, a NoviSwitch provides massive table with up to 1 million flow rules in TCAM for wildcard matches while offering up to 400 Gbps throughput. This means that with insertion of 8 OpenFlow rules, one switch can essentially manage monitoring of more than 100K IoT devices.
			}
			
			Inspired by recent proposals \cite{ANTS16,TeleScopeArXiv2018} on network telemetry using SDN, we consider a set of flow rules that collectively characterize traffic signature of IoT devices.
			{\rev
				We choose eight low-cost flow entries that can cover a subset of most commonly used signaling and activity patterns of the IoT device that we studied in Chapter~\ref{chap:characterization}.
			}
			For each device, these flow rules are pro-actively inserted into SDN-enabled switch(es) to which IoT devices are connected, as shown in Fig.~\ref{fig:testbed}. We use MAC address as the identifier of a device -- one may use IP address (without NAT), physical port number, or VLAN for a one-to-one mapping of a physical device to its traffic trace. For real-time monitoring, counters of these flow rules are periodically (\ie every minute) measured via the SDN controller that will form traffic attributes of each device.
			\begin{figure}[t]
				\vspace{-3mm}
				\centering
				\includegraphics[width=0.7\textwidth]{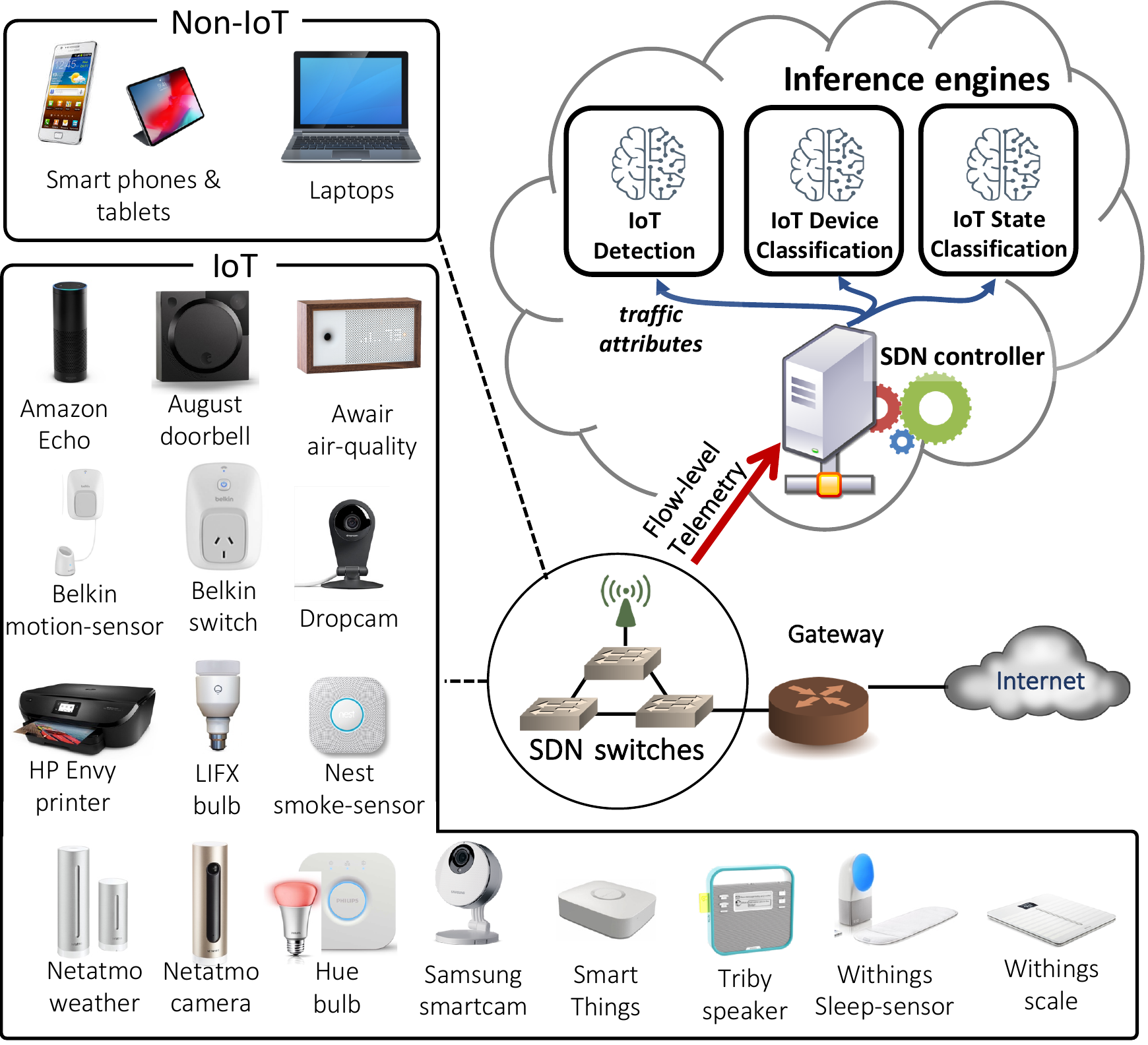}
				%\vspace{-3mm}
				\caption{System architecture of network telemetry and inference engines.}
				%\vspace{-4mm}
				\label{fig:testbed}
			\end{figure}

			Table~\ref{tab:flows} shows eight flow rules which we use to measure network traffic of each IoT device with the following order: \textbf{(1,2)} DNS outgoing queries and incoming responses on UDP 53; \textbf{(3,4)} NTP outgoing queries and incoming responses on UDP 123; \textbf{(5)} SSDP outgoing queries on UDP 1900; \textbf{(6,7)} other ``remote'' (\eg Internet) traffic outgoing from and incoming to the device that passes through the gateway; and \textbf{(8)} all ``local'' (\ie LAN) traffic incoming to the device. Note that we do not measure incoming SSDP traffic  to IoT devices in order to avoid capturing (and mixing with) discovery activities of other devices on the local network. Note that rules priority (the second last column in Table~\ref{tab:flows}) are used to split the traffic of each device into three levels: signaling packets (\ie priority 100), other remote packets (\ie priority 10), and local packets (\ie priority 1).

			\begin{table}[b]
				\centering
				\caption{Flow rules specific to each device, proactively inserted into SDN switch for real-time telemetry.}
				\label{tab:flows}
				\begin{adjustbox}{max width=1\textwidth}
					\begin{tabular}{@{}lcccccccrc@{}}
						\toprule
						\textbf{Flow description} & \multicolumn{1}{l}{\textbf{srcETH}} & \multicolumn{1}{l}{\textbf{dstETH}} & \multicolumn{1}{l}{\textbf{srcIP}} & \multicolumn{1}{l}{\textbf{dstIP}} & \textbf{Protocol} & \multicolumn{1}{l}{\textbf{srcPort}} & \multicolumn{1}{l}{\textbf{dstPort}} & \multicolumn{1}{l}{\textbf{Priority}} & \multicolumn{1}{l}{\textbf{Action}} \\ \midrule
						DNS query (DNS$\uparrow$) & {\fontsize{8}{20}\usefont{OT1}{lmtt}{b}{n}\noindent <devMAC>} & * & * & * & 17 & * & 53 & 100 & forward \\
						DNS response (DNS$\downarrow$) & * & {\fontsize{8}{20}\usefont{OT1}{lmtt}{b}{n}\noindent <devMAC>} & * & * & 17 & 53 & * & 100 & forward \\
						NTP query (NTP$\uparrow$) & {\fontsize{8}{20}\usefont{OT1}{lmtt}{b}{n}\noindent <devMAC>} & * & * & * & 17 & * & 123 & 100 & forward \\
						NTP response (NTP$\downarrow$) & * & {\fontsize{8}{20}\usefont{OT1}{lmtt}{b}{n}\noindent <devMAC>} & * & * & 17 & 123 & * & 100 & forward \\
						SSDP query (SSDP$\uparrow$) & {\fontsize{8}{20}\usefont{OT1}{lmtt}{b}{n}\noindent <devMAC>} & * & * & * & 17 & * & 1900 & 100 & forward \\
						outgoing remote (Rem.$\uparrow$) & {\fontsize{8}{20}\usefont{OT1}{lmtt}{b}{n}\noindent <devMAC>} & {\fontsize{8}{20}\usefont{OT1}{lmtt}{b}{n}\noindent <gwMAC>} & * & * & * & * & * & 10 & forward \\
						incoming remote (Rem.$\downarrow$) & {\fontsize{8}{20}\usefont{OT1}{lmtt}{b}{n}\noindent <gwMAC>} &  {\fontsize{8}{20}\usefont{OT1}{lmtt}{b}{n}\noindent <devMAC>} & * &  * & * & * & * & 10 & forward \\ 
						incoming local (Loc.$\downarrow$) & * & {\fontsize{8}{20}\usefont{OT1}{lmtt}{b}{n}\noindent <devMAC>} & * & * & * & * & * & 1 & forward \\ \bottomrule
					\end{tabular}
				\end{adjustbox}
			\end{table}
			
			For each of the eight flows (mentioned above), we use two key attributes \cite{infocom17} namely \textbf{\textit{average packet size}} and \textbf{\textit{average rate}}. Also, note that traffic attributes can better characterize network behavior of individual devices if they are computed at multiple time-scales \cite{multiTime2005}. We, therefore, collect packet counts and byte counts per each flow every minute, and compute attributes at time-granularities of 1-, 2-, 4-, 8-, 16-, 32-, 64-minutes. This way, we generate fourteen attributes for each flow which means a total of 112 attributes per device.

			In order to synthesize flow rules, we wrote a native SDN simulator \cite{SDNsim} that takes an input PCAP trace, and performs packet-by-packet service (matching packet headers against flow table entries, updating statistics, applying required actions) inside a software SDN switch. The simulator records counters of flow bytes and packets periodically (\eg one minute). Using the output of the simulator, we use another script to generate instances of traffic attributes for each device every minute. An instance is a vector of 112 attributes with a label (\eg Amazon Echo:boot).

			\subsection{Traffic Characteristics of IoT Devices}
				We now highlight traffic characteristics of individual IoT devices that can be learned so as to distinguish them from non-IoTs, classify their device type, and identify their operating states.
			
			\begin{figure}[t]
				\begin{center}
					\mbox{
						\subfloat[Volume of remote traffic at 32-mins resolution.]{
							{\includegraphics[width=0.45\textwidth]{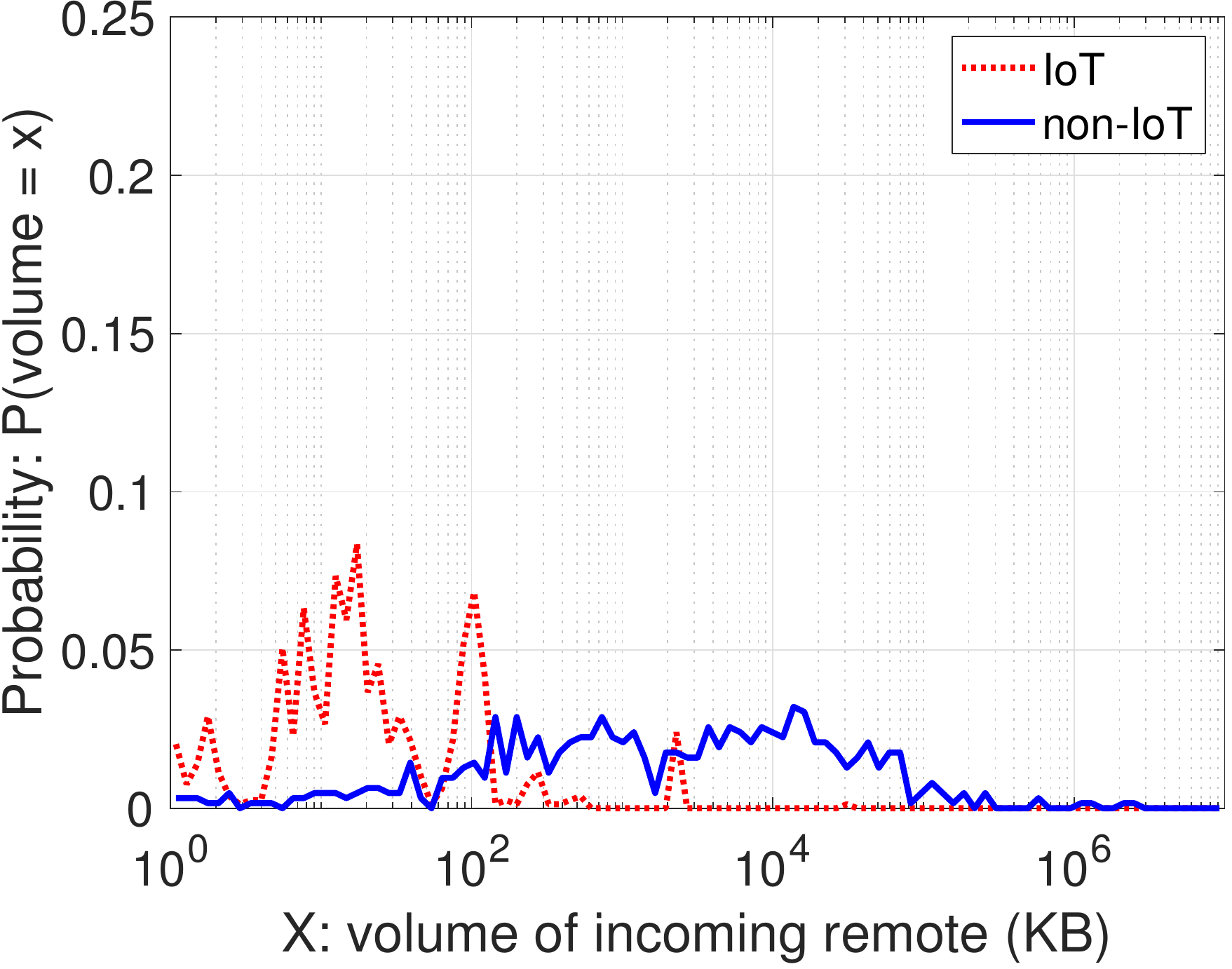}}\quad
							\label{fig:profileIoT?Internet}
						}
						\quad
						\subfloat[Count of DNS query packets at 64-mins resolution.]{
							{\includegraphics[width=0.45\textwidth]{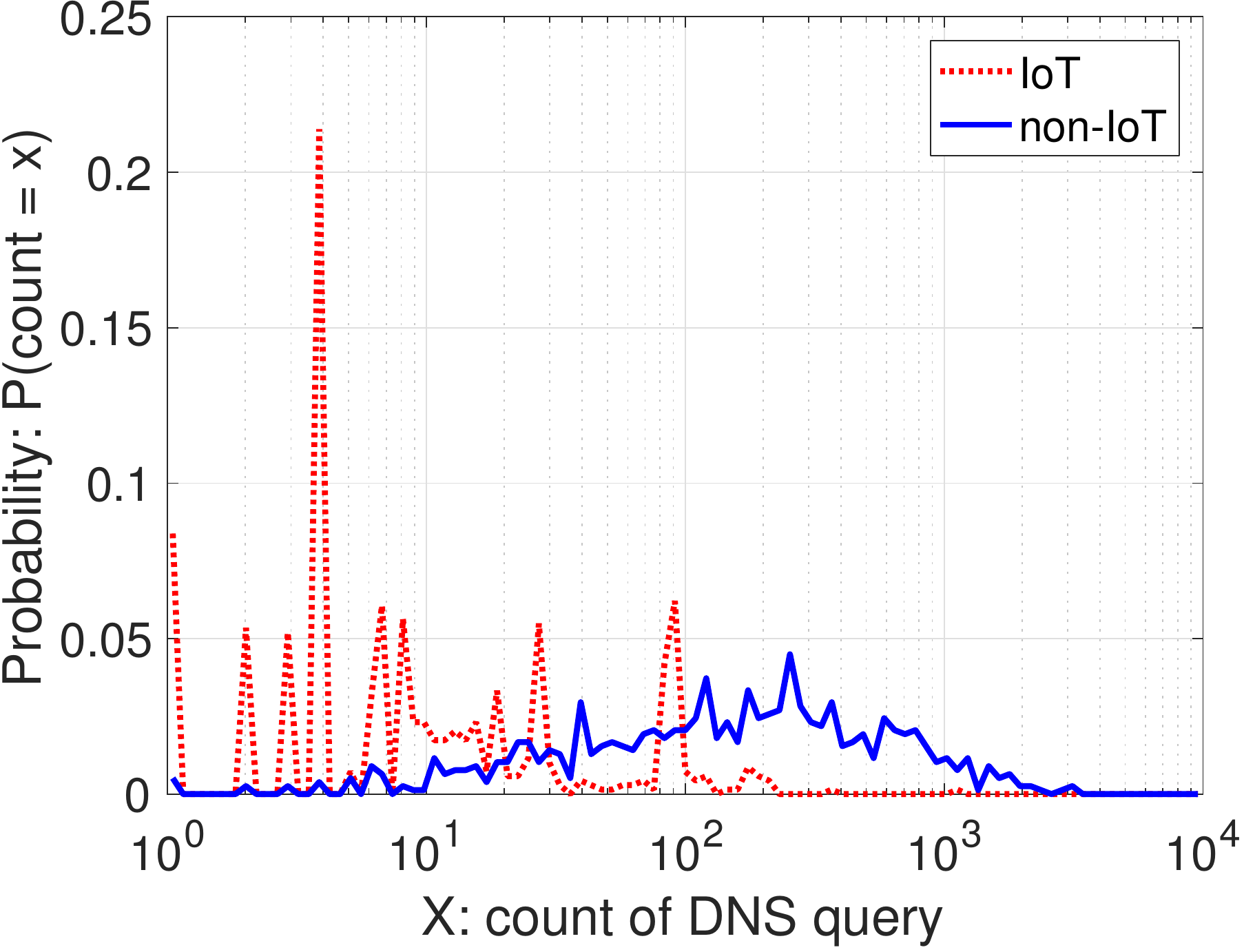}}\quad
							\label{fig:profileIoT?DNS}
						}
					}
					
					\vspace{-3mm}
					\caption{Histogram of traffic profile to compare IoT and non-IoT devices: (a) remote traffic volume over 32-minute; and (b) DNS query count over 64-minute.}
					\vspace{-3mm}
					\label{fig:profileIoT?} 
				\end{center}
			\end{figure}
			\textbf{IoT versus non-IoT:}
				We begin with traffic attributes that differentiate IoT devices from non-IoTs. Fig.~\ref{fig:profileIoT?} shows the probability density of two representative attributes, namely remote traffic volume at 32-minute resolution, and DNS query count at 64-minute resolution. We can see in Fig.~\ref{fig:profileIoT?Internet}  that IoT devices tend to transfer a small volume of traffic from remote (\ie Internet) network and $90$\% of instances they download less than $500$ KB every half-an-hour. However, for non-IoTs this value is widely spread between $10$ KB to $100$ MB and mostly they transfer more than $500$ KB. In terms of DNS activity in Fig.~\ref{fig:profileIoT?DNS}, IoTs display identifiable patterns of query count mostly less than $100$ (\eg $22$\% of instances with four queries per hour), while non-IoTs have a wider range of DNS query count (\ie $10$ to $3000$ DNS queries over an hour) with almost equal probabilities.
				\begin{figure}[t!]
					\begin{center}
						\mbox{
							\subfloat[Count of NTP response packets (16-min).]{
								{\includegraphics[width=0.45\textwidth]{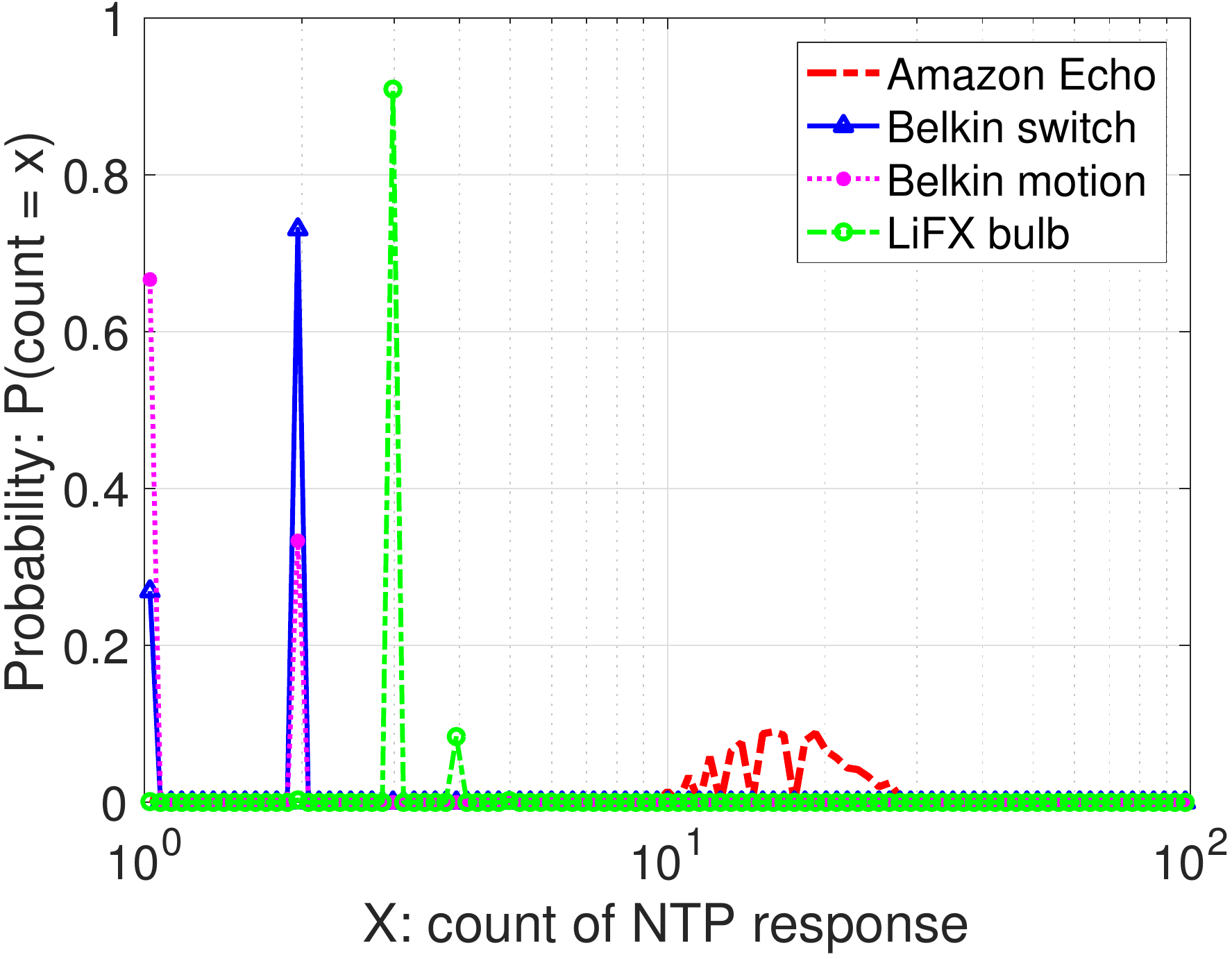}}\quad
								\label{fig:profileIoTdevices?NTP}
							}
							
							\subfloat[Volume of remote traffic (8-min).]{
								{\includegraphics[width=0.45\textwidth]{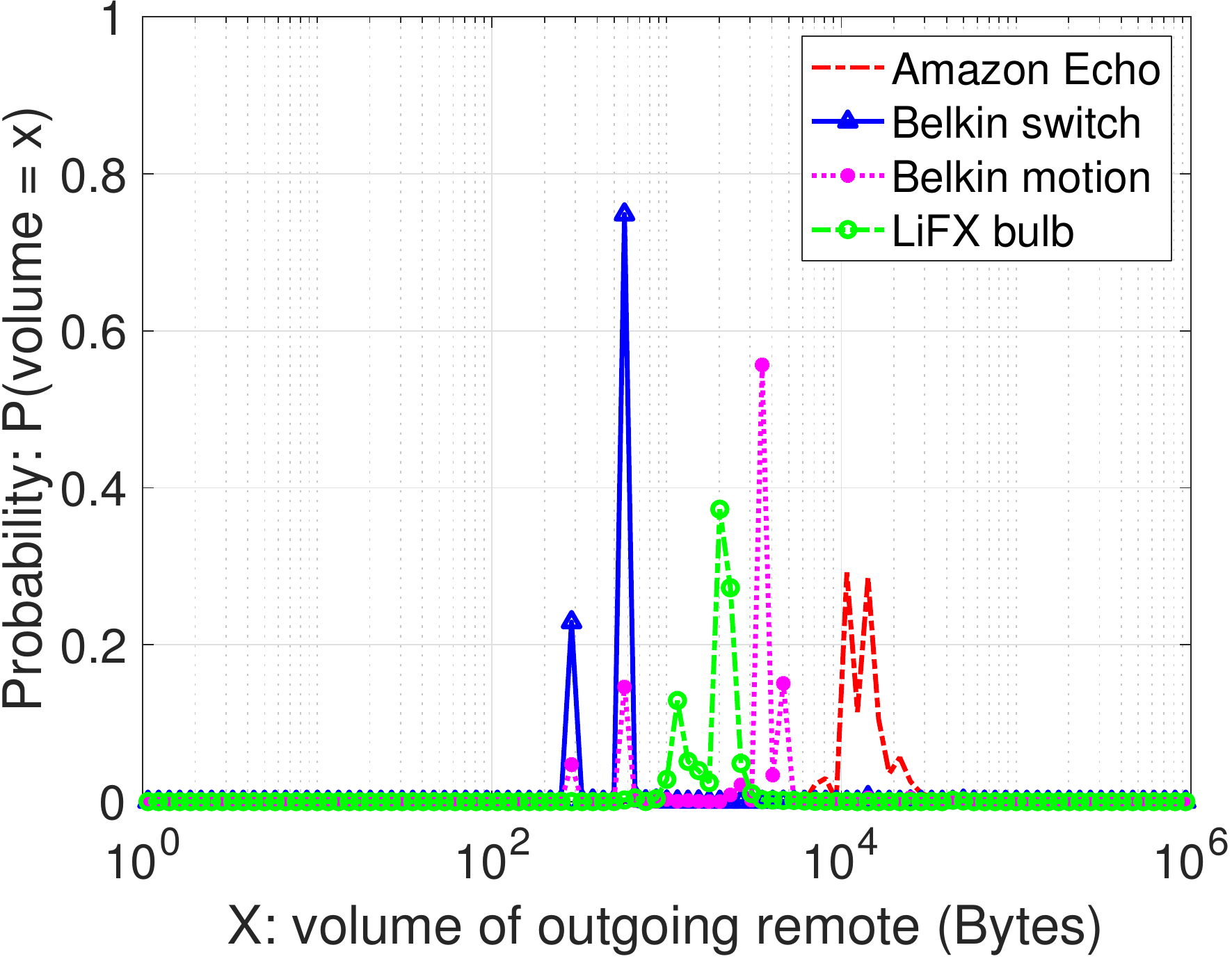}}\quad
								\label{fig:profileIoTdevices?Internet}
							}
						}
						\mbox{
							\subfloat[Volume of SSDP traffic (8-min).]{
								{\includegraphics[width=0.45\textwidth]{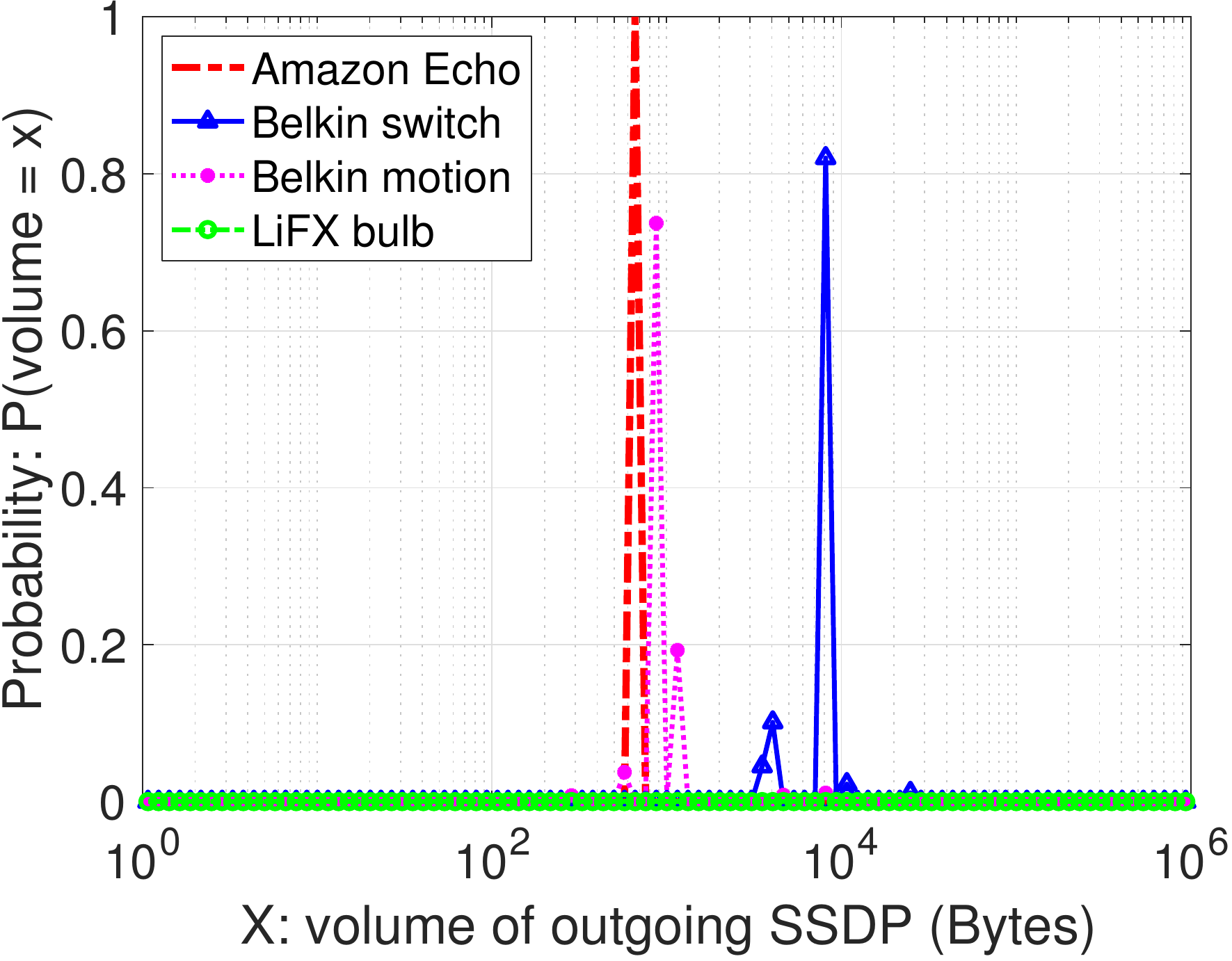}}\quad
								\label{fig:profileIoTdevices?SSDP}
							}
						}
						%\vspace{-3mm}
						\caption{Histogram of traffic profile for representative IoT devices: (a) NTP response count, (b) download volume of remote traffic, and (c) upload volume of SSDP traffic. }
						%\vspace{-3mm}
						\label{fig:profileIoTdevices?}
					\end{center}
				\end{figure}

			\textbf{IoT device types:}
				Focusing on IoT devices, we now consider three traffic attributes, namely NTP responses count at 16-min resolution, upload volume of remote traffic at 8-min resolution, and volume of SSDP responses at 8-min resolution, as shown in Fig.~\ref{fig:profileIoTdevices?}. We quantitatively compare traffic characteristics of four representative IoT devices from three different manufacturers (\ie Amazon, Belkin, and LiFX). It is observed from Fig.~\ref{fig:profileIoTdevices?NTP} that the LiFX bulb (depicted by solid green lines) sends three NTP responses every 16-minute interval for more than $90$\% of instances. This measure varies between 7 to 42 responses for Amazon Eco (depicted by dashed red lines). For Belkin power switch and motion sensor (depicted by solid blue and dotted pink lines), we see two significant peaks at a count of one and two NTP responses, each with a different probability -- Belkin switch seems more active (compared to Belkin motion), with $70$\% probability of generating 2 NTP responses at 16-minute resolution.   
			
				For download volume of remote traffic attribute at 8-minute resolution, shown in Fig.~\ref{fig:profileIoTdevices?Internet}, we see a relatively unique pattern in the probability density function for each of these four devices: for Belkin switch and Belkin motion it peaks at $573$ bytes and $3$KB, respectively, while the LiFX bulb and Amazon Echo each exhibits a range of values, [$0.5$, $3$] KB and [$7$, $33$] KB, respectively. 
			
				Considering the upload volume of SSDP traffic in Fig.~\ref{fig:profileIoTdevices?SSDP}, the Belkin switch seems distinctive from Belkin motion (\ie probability of $82$\% for $8$ KB volume in Belkin switch compared to $73$\% chance for $800$ bytes volume in Belkin motion). Amazon Echo displays a strong pattern with a peak of $100$\% at volume of $650$ bytes. Lastly, we observe that LiFX does not use SSDP protocol at all, and thus lacks this attribute in its traffic profile.   
			
			\textbf{Operating states of IoT:}
				We now look at selected traffic attributes of Amazon Echo, Belkin switch, and Dropcam at the three operating states, shown in Fig.~\ref{fig:profileStates?}.  
				We focus on download volume of remote traffic for Amazon Echo since it frequently communicates with its cloud servers; download volume of local traffic for Belkin switch since it receives command from user mobile app connected to the local network; and upload volume of remote traffic for Dropcam since it tends to send videos to its cloud servers. We can see that the three operating states are fairly distinct in chosen attributes shown in Fig.~\ref{fig:profileStates?}. It is observed that all three devices exchange smaller volume of traffic during their idle state (shown by dashed green lines) compared to active and boot states. As an example,for Amazon Echo, shown in Fig.~\ref{fig:profileStates?amazon}, $75$\% of idle instances receive between [$0.5$, $1$] KB from remote servers at 2-minute resolution, while the probability density function for boot and active instances peaks at $30$ KB and $70$ KB, respectively. Additionally, we observe for Belkin switch that the volume of local traffic during boot state is larger than active state. This is because this device sends SSDP discovery when it boots up and that results in the arrival of responses from all SSDP-capable devices on the network. Therefore, a peak at $110$ KB is seen for boot state (dotted red line) in Fig.~\ref{fig:profileStates?belkin}. 
				\begin{figure}[t]
					\begin{center}
						\mbox{
							\subfloat[Amazon Echo: remote traffic (2-min).]{
								{\includegraphics[width=0.45\textwidth]{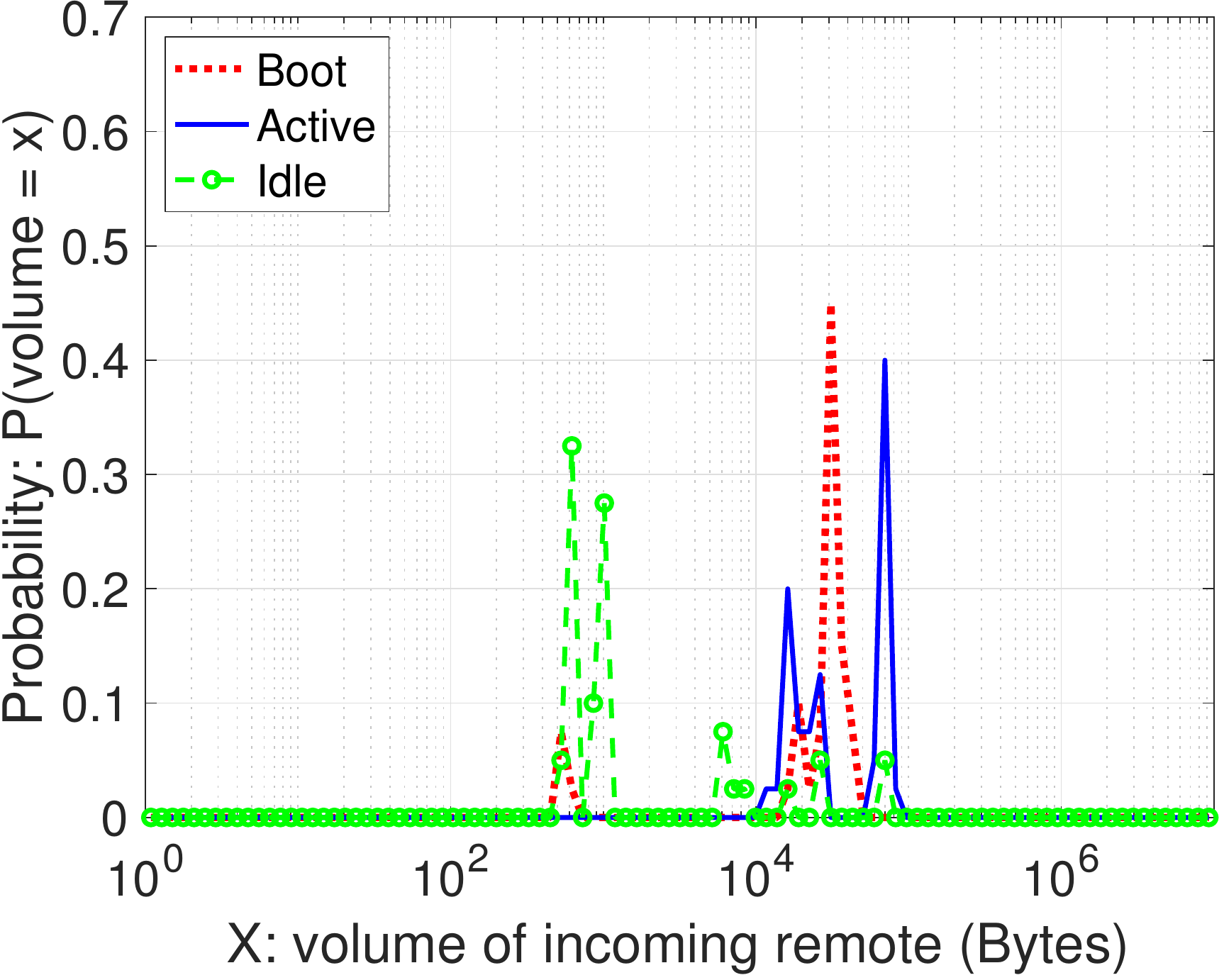}}\quad
								\label{fig:profileStates?amazon}
							}
							
							\subfloat[Belkin switch: local traffic (1-min).]{
								{\includegraphics[width=0.45\textwidth]{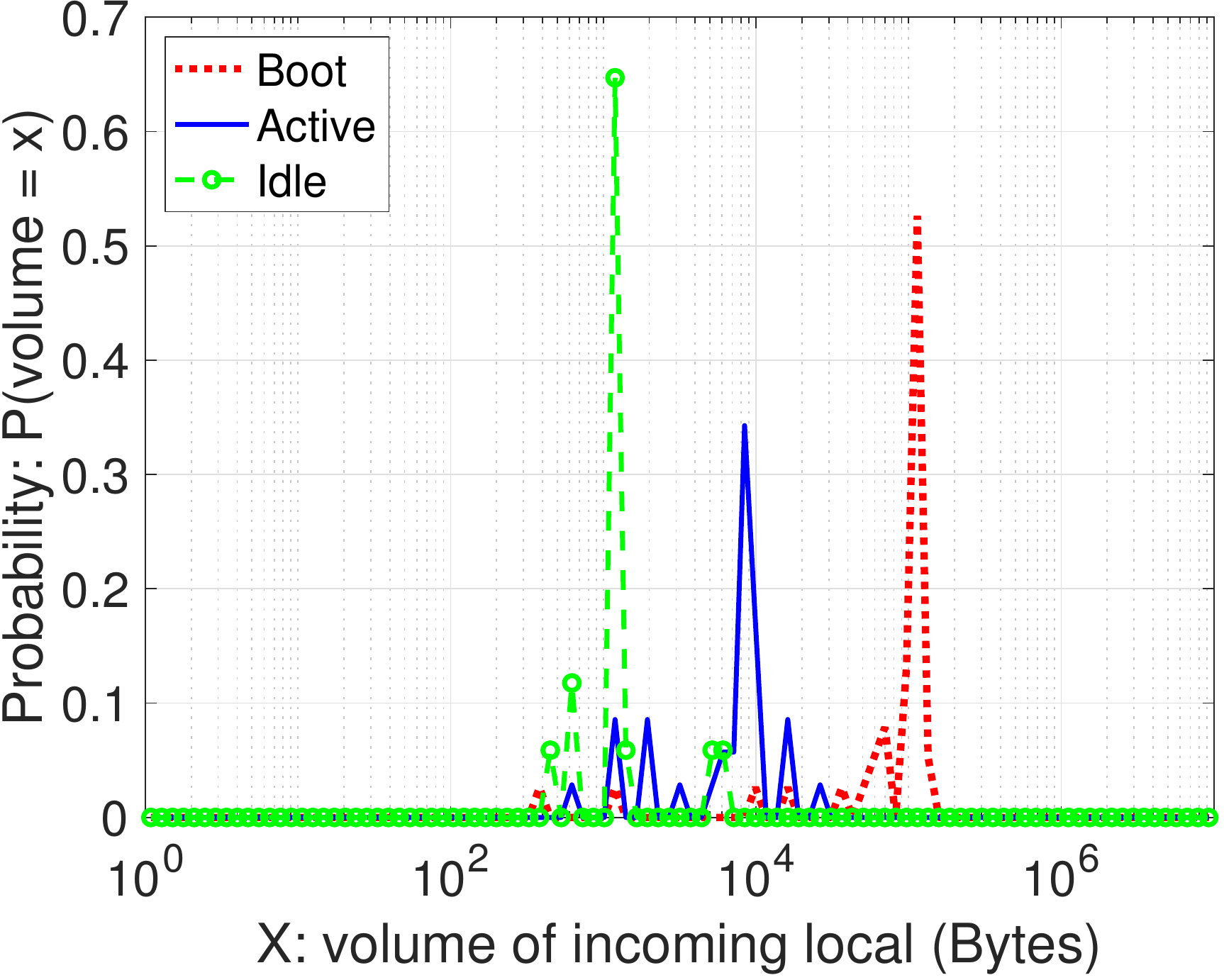}}\quad
								\label{fig:profileStates?belkin}
							}
						}
						\mbox{
							\subfloat[Dropcam: remote traffic (2-min)]{
								{\includegraphics[width=0.45\textwidth]{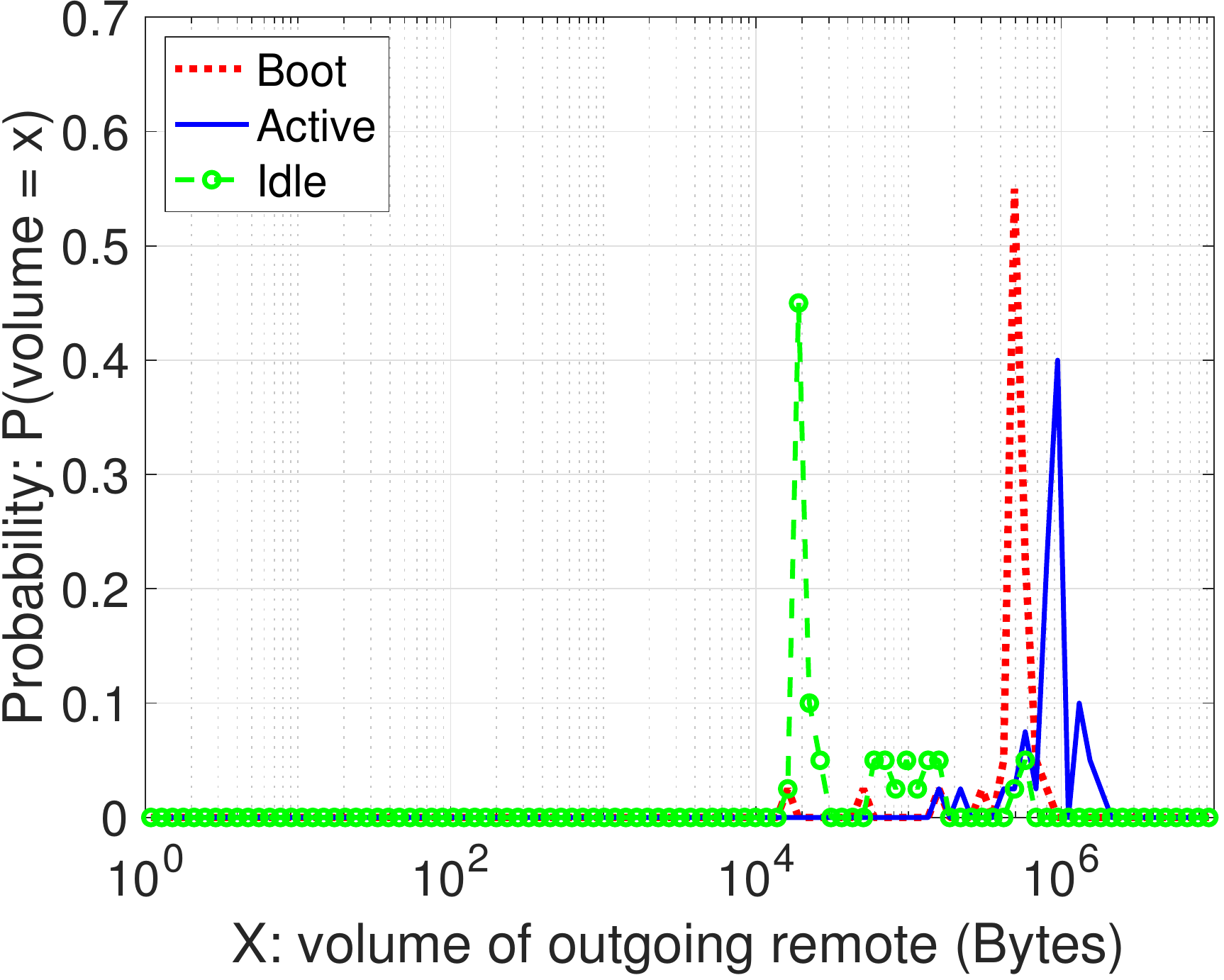}}\quad
								\label{fig:profileStates?dropcam}
							}
						}
						%\vspace{-3mm}
						\caption{Histogram of traffic profile for IoT devices at three operating states: (a) Amazon Echo, (b) Belkin switch, and (c) Dropcam.}
						%\vspace{-3mm}
						\label{fig:profileStates?}
					\end{center}
				\end{figure}
				
	\section{IoT Traffic Inference Engines}\label{sec:c2_classification}
		In this section, we develop a multi-stage architecture to automatically infer IoT traffic, train a set of models, and evaluate their performance. 

		\vspace{-1em}
		\subsection{Inference Architecture}
			For a given device on the network, we have three objectives: (a) to determine if the device is IoT or non-IoT, and if it is detected as IoT; (b) to classify its device type (\eg Amazon Echo, Dropcam); and (c) to identify the operating state of IoT (\ie boot, active, idle). To meet these objectives we need a set of trained models: a bi-class classifier to distinguish IoT devices from non-IoTs (\ie IoT detector); a multi-class classifier to determine the type of a given IoT device (\ie IoT classifier); and a set of multi-class classifiers to identify IoT operating states (\ie a state classifier for each device type). Note that state classifiers are specialized models and each learns traffic patterns of one device in the three states of operation. State classifiers tend to have narrower views, and hence become more sensitive to change of behavior for their respective devices (compared to the device classifier with a broader view). Therefore, these specialized models are able to enhance the visibility of network operators into subtle changes \cite{EY2017} in their IoT infrastructure. 

			There exist a number of techniques \cite{SupLearning06} such as Neural Networks, Support Vector Machines (SVMs), and Decision Trees that can be used to train models to infer predefined classes. Neural networks have proven to be very effective in classifying input data with high dimensions, but they demand a large amount of training data. Also, neural networks are seen as black box models since it becomes difficult to interpret their reasoning process. Performance of SVMs is very sensitive to the selection of hyper-parameters, and hence it becomes difficult to train an accurate model. On the other hand, decision tree-based techniques are widely used since it is easier to generate (reasonably) accurate models with a relatively small amount of data. Importantly, they generate trees which can be readily interpreted. Note that decision tree algorithms are prone to over-fitting which can be avoided by the use of ensemble decision trees. In this work, we employ Random Forest \cite{RandomForest2001} which builds an ensemble of decision trees, each uses a random subset of attributes. It is best known for its performance in various classification tasks \cite{trafficClass2015,TMC18,TeleScopeArXiv2018}. 

			\begin{figure}[t!]
				\centering
				\includegraphics[width=1\textwidth]{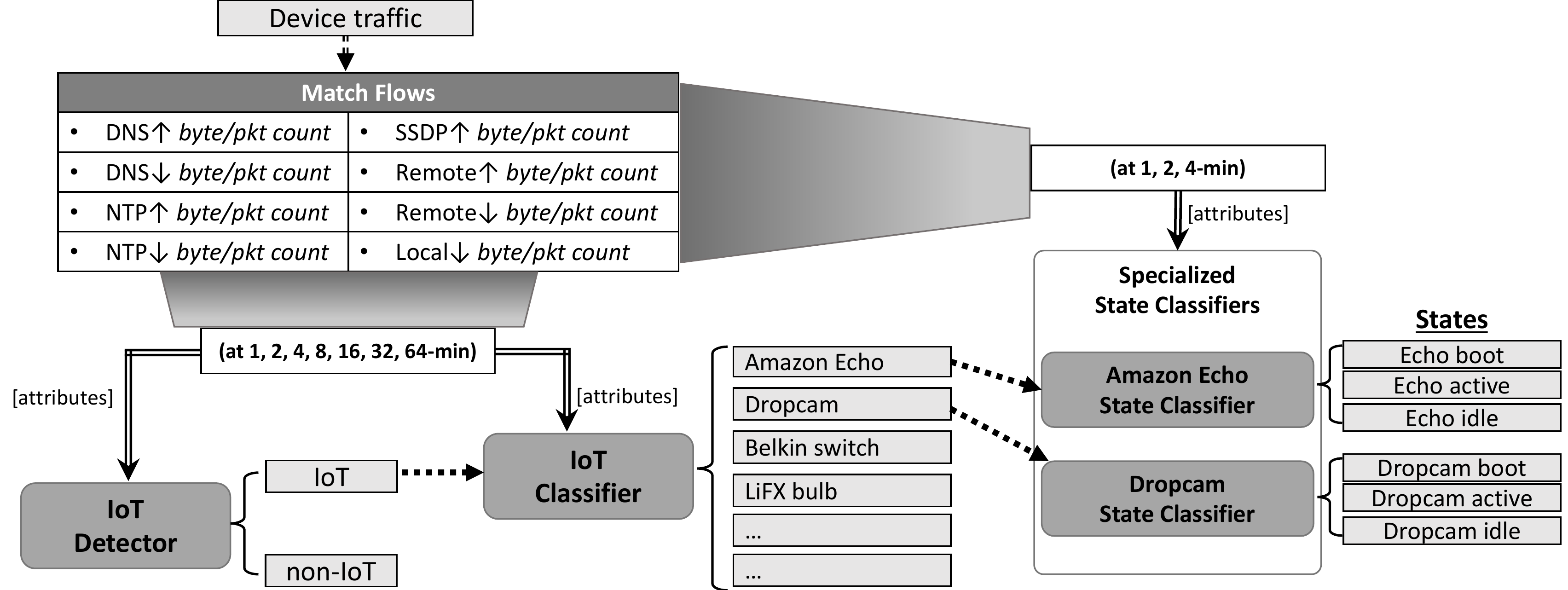}
				\caption{Hierarchical architecture of IoT traffic inference engines.}
				\label{fig:arch}
				
			\end{figure}

			%Hierarchical solution
			Fig.~\ref{fig:arch} illustrates our hierarchical architecture for IoT traffic inference. It consists of three layers of random-forest classifiers (\ie an IoT detector, an IoT classifier, and a set of IoT state classifiers). Once a new device connects to the network, the programmable switch is pushed by additional flow rules (Table~\ref{tab:flows}) pertinent to the device. We first feed the IoT detector model by full set of periodic flow-level attributes (\ie at time-scales of powers of two between 1-min to 64-min). Upon detection of an IoT device with sufficiently high confidence (say, more than 80\%), the second model (\ie IoT classifier) is called by the full set of attributes -- a device will not be checked by the second layer of inference, if it is detected as non-IoT at the first layer. The output of the IoT classifier triggers a pertinent state classifier at the third layer of our architecture. Our state classifier models consume a subset of attributes, however, it is only up to 4-minute resolution. This is because change of states (\eg boot) result in short-term effects on device traffic pattern -- considering long-term attributes may reduce the ability of the model to accurately detect the operating state in real-time.

		\subsection{Models Training and Performance Evaluation}\label{sec:c2_perf}
			We now label instances to train our classifiers, and generate models needed for the three layers of the inference architecture, as shown in Fig.~\ref{fig:arch}. We next evaluate their performance using test instances. 
			For both training and testing the traffic classifiers we use Weka \cite{Weka} tool.

			\textbf{Instances:} Recall from \S\ref{sec:c2_att} that our instances are computed every minute. Since two of our models consume full-set attributes (\ie 1-min to 64-min) we down sample our instances by the factor of 15 to avoid over-fitting for the IoT detector and the IoT classifier -- it is likely to have heavily-correlated instances generated within 15 minutes. Note that the risk of over-fitting is less for the state classifiers given that they only use short timescale attributes.    

			%Instance count
			We have collected a total of 115,237 instances of IoT and non-IoT devices from our DATA1 (in \S\ref{sec:c2_data}) and 10,423 instances of four  IoT devices (\ie  Amazon Echo, Belkin switch, Dropcam, and LiFX bulb) with state annotation from our DATA2 (in \S\ref{sec:c2_data}). We have a different number of instances across various devices in our dataset, depending upon their presence and activity on the testbed, and their interactions with the lab users. Among all devices, NEST smoke-sensor has the lowest number (\ie 865) of instances since it communicates once a day for a short period of time. The highest count belongs to Dropccam with 11,873 instances as it was online more than 90\% of days during the 6-month period of packet capture and it frequently communicates with its cloud-servers whenever it is on the network. 

			\textbf{Metrics:} Since classes are not evenly distributed in our datasets, we use three metrics including weighted ``\textit{precision}'', ``\textit{recall}'', and ``\textit{F\textsubscript{1} score}'' along with confusion matrix to evaluate the performance of each model. These metrics are defined as follows:
			\begin{equation}\label{eq:precision}
				precision = \frac{\TP} {\TP + \FP}
			\end{equation} 
			\begin{equation}\label{eq:recall}
				recall = \frac{\TP} {\TP + \FN}
			\end{equation} 
			\begin{equation}\label{eq:f1score}
				F_1 =\frac{2 \times precision \times recall} { precision + recall}
			\end{equation} 
			where \textit{TP} is the rate of true positive, \textit{FP} is the rate of false positive, and \textit{FN} is the rate of false negative. Note that \textit{F\textsubscript{1}} conveys the balance between precision and recall values and is computed by the harmonic mean of these two values in Eq.~\ref{eq:f1score}. All metrics take a value between 0 and 1.

			%Confident level
			In addition to the correctness of classification, we record confidence-level of our random-forest models for all instances, correctly classified and incorrectly classified ones. Ideally, we expect our models to display high confidence (\ie close to 1) when they predict a correct class for an input instance, and low confidence (\ie close to 0) when they predict an incorrect class. Lack of confidence indicates that the tested instance contains attributes different from those that were learned before (\ie new or unseen pattern). We next look at individual models at various layers of the inference architecture

			\textbf{IoT Detector:}
				%training set preparation
				In our DATA1, there exist 1212 instances labeled as non-IoT and 114,025 instances labeled as 17 types of IoT. For the training set, we randomly choose 800 instances (\ie 66\%) from non-IoT and 50 instances from each class of IoT device (\ie a total of 850). Remaining instances in DATA1 are used to test this bi-class classifier.

				%Performance evaluation
				Fig.~\ref{fig:confMapIoT} shows the confusion matrix of the IoT detector model. The rows show actual labels (\ie IoT or non-IoT) and columns show predicted labels -- cell numbers are in percentage. Table~\ref{tab:eval_iot} shows all performance metrics of this model. It is seen that $98.7$\% of IoT test instances and $97.8$\% of non-IoT test instances are correctly classified, as shown by diagonal values of the confusion matrix in Fig.~\ref{fig:confMapIoT}. Looking at the last two columns of Table~\ref{tab:eval_iot}, the confidence-level of this model is fairly high on average (\ie $0.968$ and $0.947$) for correct classification and is relatively low on average (\ie $0.635$ and $0.701$) for incorrect classification. Also, the three performance metrics; namely precision, recall, and F\textsubscript{1}; all indicate reasonable high values on average as $0.983$, $0.982$, and $0.983$ respectively.  
				\begin{figure}[t]
					\centering
					\includegraphics[width=0.3\textwidth]{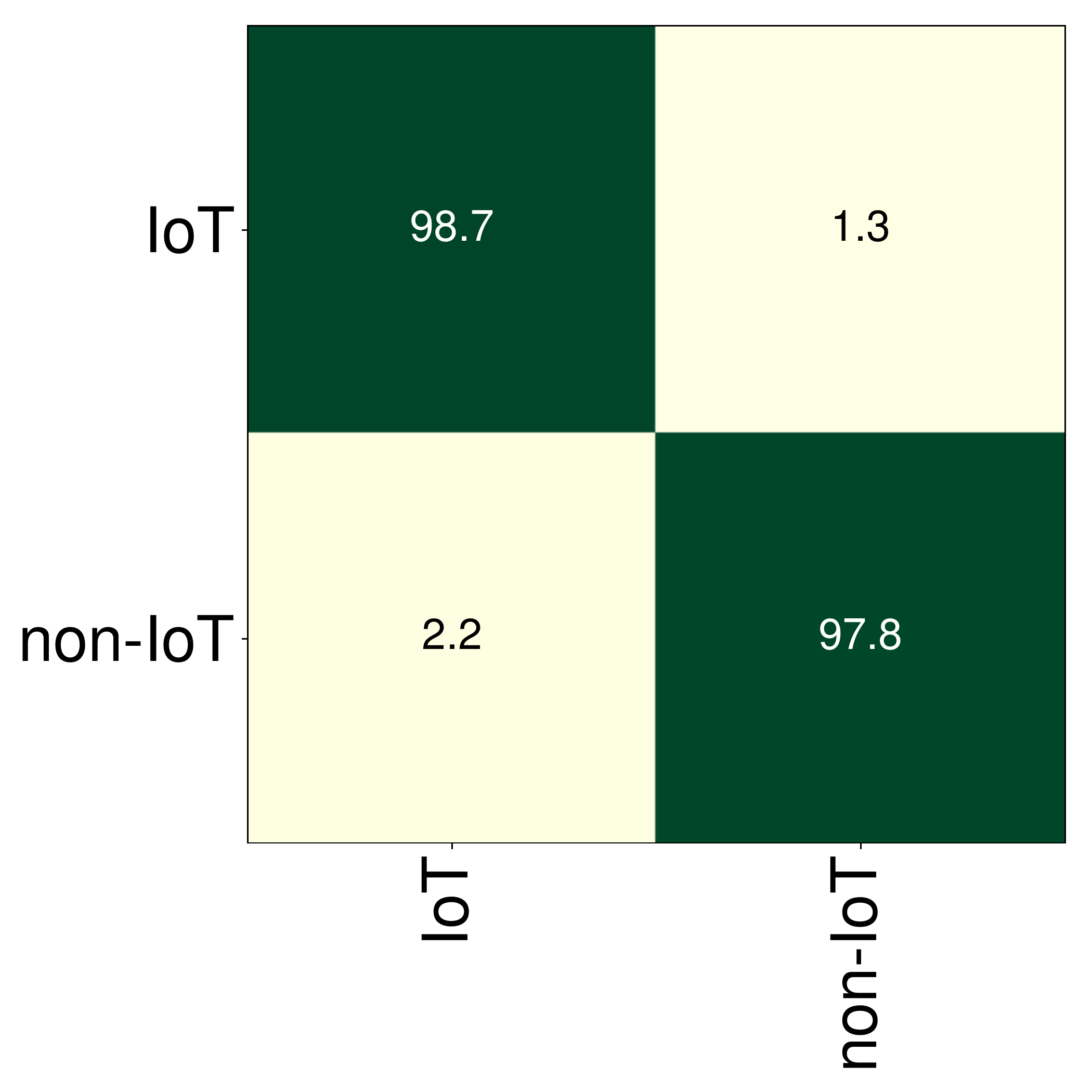}
					\caption{Confusion matrix of the  IoT detector.}
					\label{fig:confMapIoT}
				\end{figure}
				\begin{table}[b]
					\caption{Performance metrics of the IoT detector model.}
					\label{tab:eval_iot}
					\centering
					\begin{adjustbox}{max width=1\textwidth}
						\begin{tabular}{@{}lcccccccc@{}}
							\toprule
							\textbf{IoT/non-IoT} & \textbf{TP} & \textbf{FN} & \textbf{FP} & \textbf{Precision}&\textbf{Recall}&\textbf{$F_1$}& \textbf{\shortstack{TP avg.\\confidence}} & \textbf{\shortstack{FN avg.\\confidence}}\\ \midrule
							IoT&0.987 &0.013 &    0.022  &   0.979  &   0.987  &   0.983   &    0.968   &      0.635\\
							non-IoT & 0.978 &    0.022  &   0.013  &   0.987   &  0.978 &    0.983     &  0.947   &      0.701 \\ \bottomrule
						\end{tabular}
					\end{adjustbox}
				\end{table}

		\textbf{IoT Classifier:}
		
			%Training set preparation for device classifier
			For this model, we split IoT instances of the DATA1 into chronological sets of training and testing given a sufficient number of instances available in our dataset over six-month period. This way the performance of the model is evaluated over time. Therefore, we use instances collected during the first three months (\ie 01-Oct-2016 to 31-Dec-2016) for training and the remaining issuances (\ie collected between 01-Jan-2017 and 31-Mar-2017) for testing.
			
			%Misclassifications
			Fig.~\ref{fig:confMapDevice3month} depicts the confusion matrix of the IoT classifier. We observe that the model performs well in predicting most of classes. For example, the correct prediction rate for Amazon Echo, August doorbell, Belkin switch, or Dropcamp is more than $97$\%. However, the model performance does not seem acceptable for certain classes. For example, it is seen that $12.0$\% of NEST smoke-sensor instances are misclassified as Withings scale, and $39.8$\% of HP printer instances are misclassified as Belkin switch. Additionally, for the Awair air-quality sensor  only $77.1$\% of test instances are correctly classified while $8.6$\% and $13.0$\% are misclassified as August doorbell and Withings scale, respectively. We note that the model displays a low confidence on average for incorrect prediction of these three classes, \ie $0.485$ for Awair air-quality, $0.535$ for HP printer, and $0.389$ for NEST sensor -- due to space constraints we omit detailed table of performance metrics per individual classes. Additionally, we find that even though $94.0$\% of Hue bulb instances are correctly classified the average confidence of our model is $0.572$ (\ie undesirably low).
			\begin{figure}[t]
				\centering
				\includegraphics[width=0.97\textwidth]{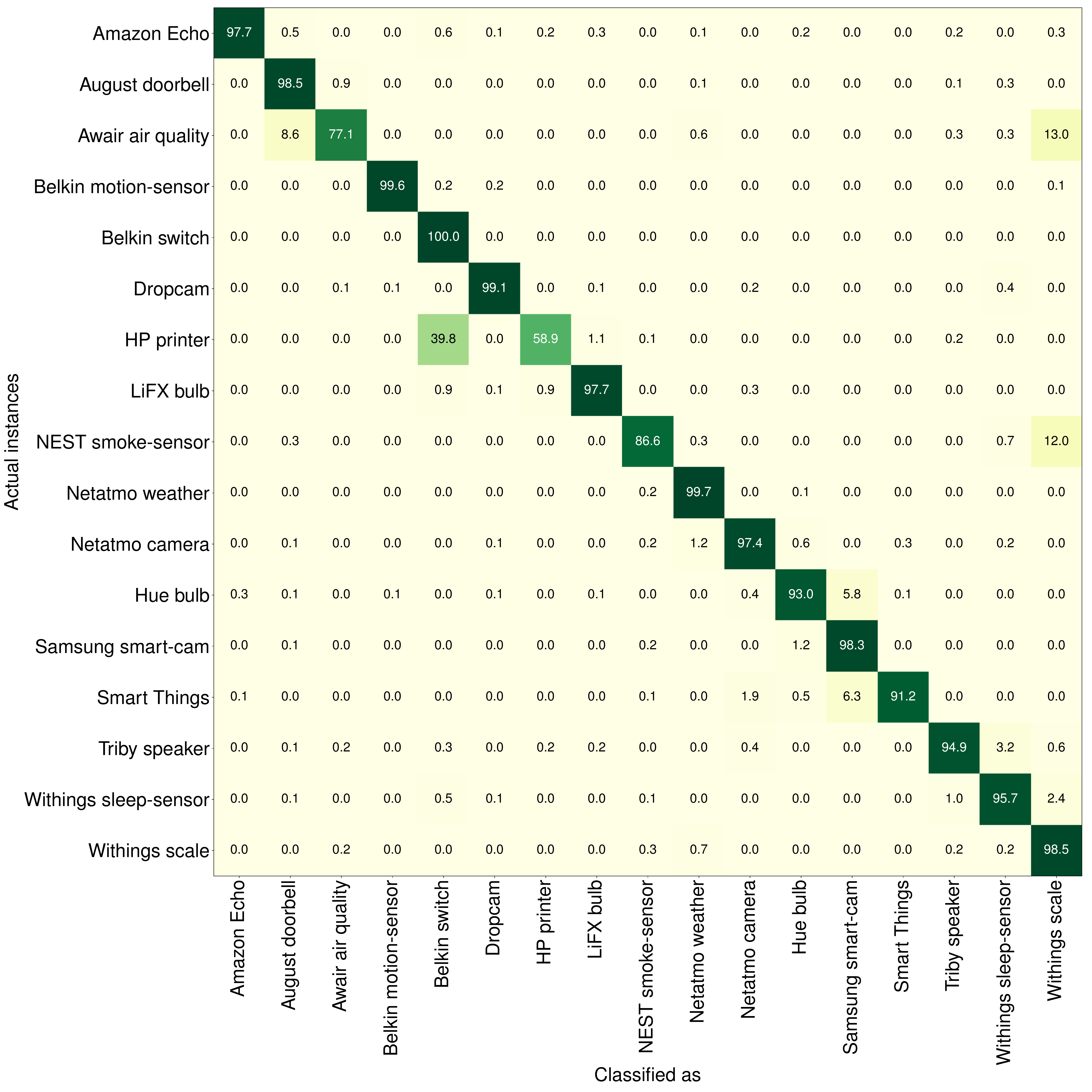}
				\vspace{-1mm}
				\caption{Confusion matrix of IoT classifier trained by the first three months' worth of data.}
				%\vspace{-4mm}
				\label{fig:confMapDevice3month}
			\end{figure}

			\begin{figure}[t]
				\begin{center}
					\mbox{
						\subfloat[HP printer.]{
							{\includegraphics[width=0.8\textwidth]{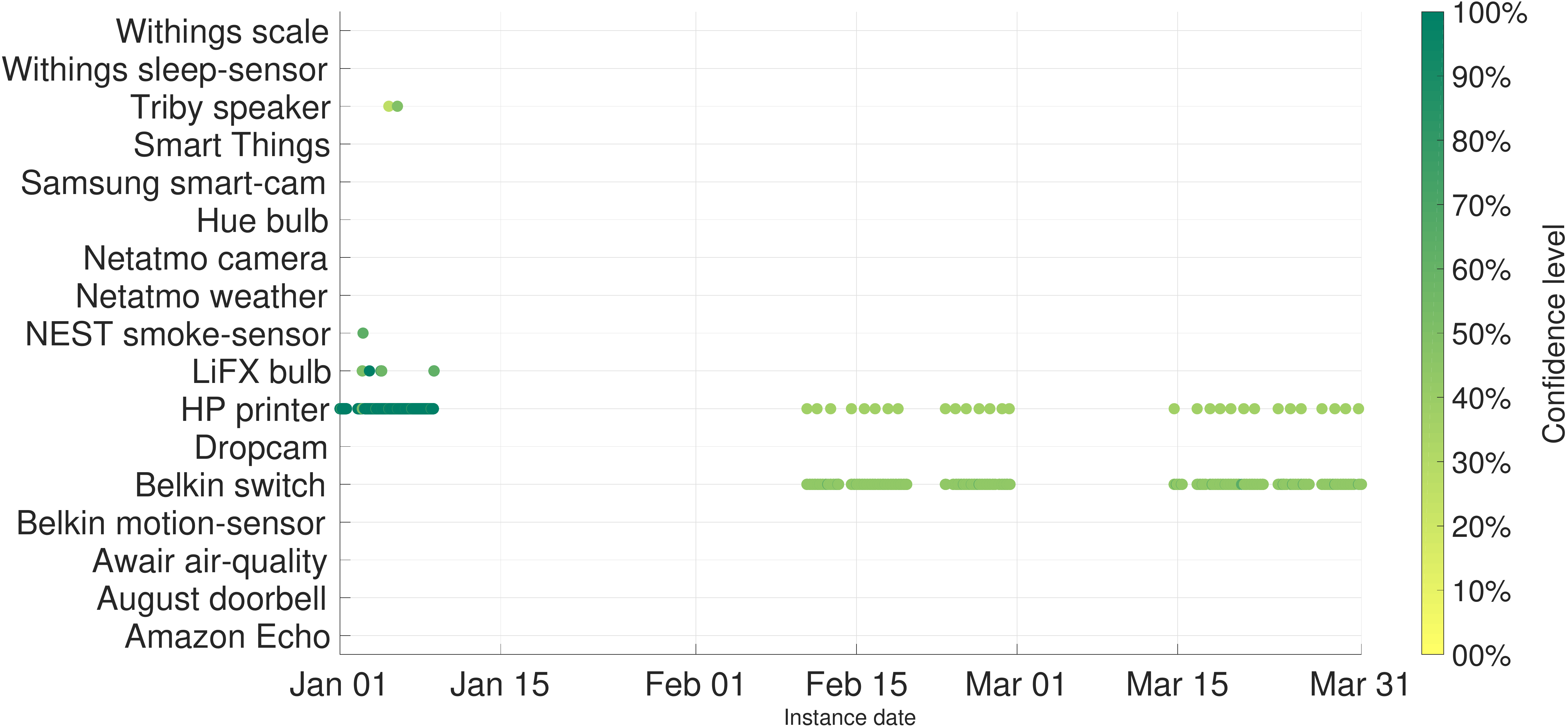}}\quad
							\label{fig:PerfHP}
						}
					}
					\mbox{
						\subfloat[Hue light-bulb.]{
							{\includegraphics[width=0.8\textwidth]{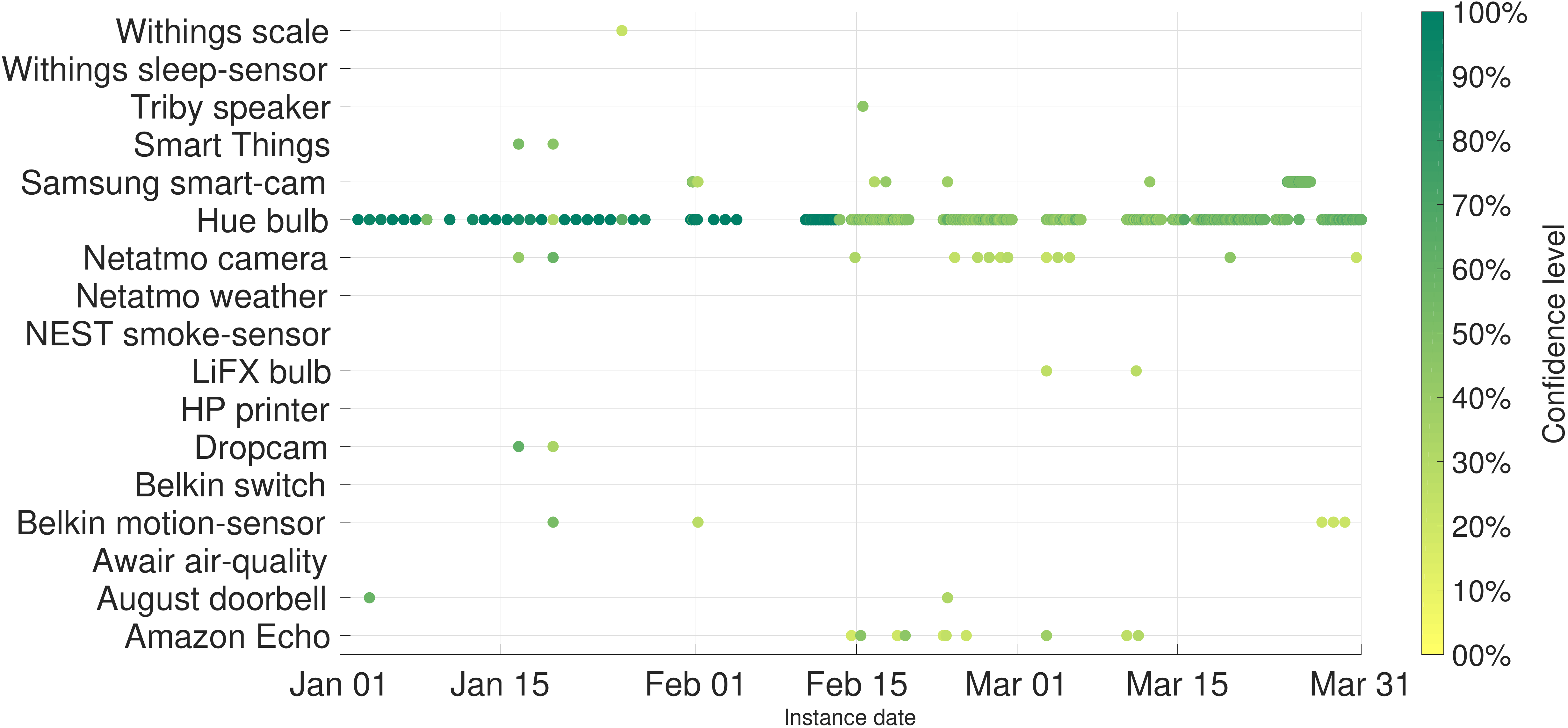}}\quad
							\label{fig:PerfHue}
						}
					}
					\caption{Time trace of IoT classifier outputs with test traffic instances from: (a) HP printer, and (b) Hue light-bulb.}
					\label{fig:PerfTimeTrace}
				\end{center}
			\end{figure}
			
			%Time domain analysis figures
			To better analyze the poor performance of the model in certain classes, we plot in Fig.~\ref{fig:PerfTimeTrace} the time trace of model outputs with test traffic instances from HP printer and Hue light-bulb. Each circle represents an instance and its color shows the model confidence. A color bar on the right side of plots shows the mapping of confidence values to colors -- dark green indicates high confidence and yellow indicates low confidence. Starting from Fig.~\ref{fig:PerfHP}, we observe that instances of HP printer from the first week of January are mostly classified correctly and are supported by high confidence levels (\ie dark green circles). The printer goes offline for about a month and comes back online on 11-Feb-2017 and this is when its traffic is mostly misclassified as Belkin switch with consistently low confidence levels from the model (\ie light green circles). This clearly shows that the behavior of HP printer changed when it restarted in mid-February -- we manually inspected traffic traces and verified that it was due to a legitimate firmware upgrade (\ie benign changes). 
			Moving to Fig.~\ref{fig:PerfHue}, we see classifier outputs for Hue bulb traffic instances during the whole testing period. Though instances are mostly predicted correctly, the confidence level starts falling, from an average of $0.93$ to average $0.50$, on 15-Feb-2017. Again this behavioral change led to a manual inspection by which we verified that it was legitimate.
			\begin{figure}[t]
				\centering
				\includegraphics[width=0.97\textwidth]{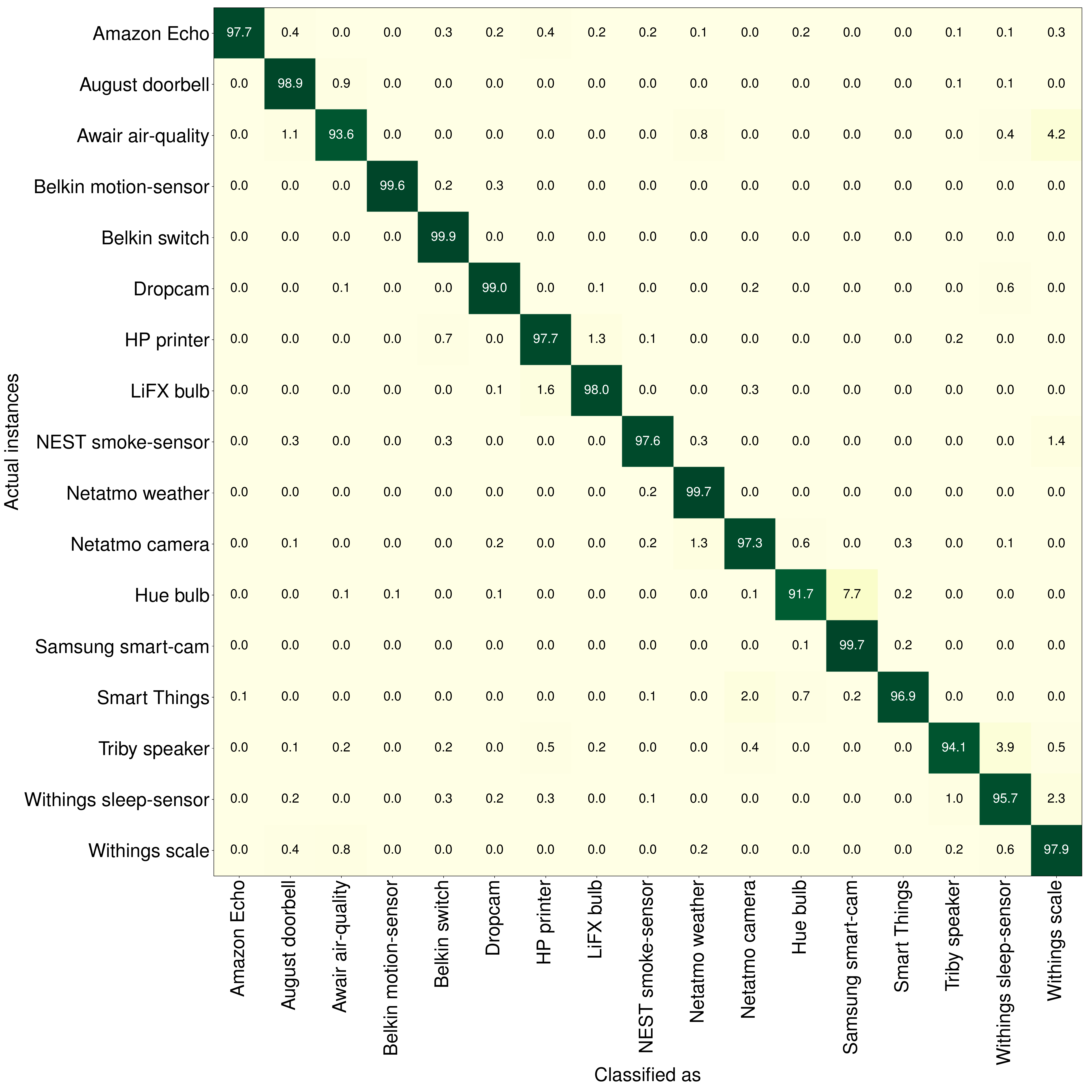}
				\vspace{-1mm}
				\caption{Confusion matrix of IoT device classification  re-trained by additional data from two weeks in February.}
				%\vspace{-4mm}
				\label{fig:confMapRetrain}
			\end{figure}

			Given these observations, we augment our training set with two weeks of data (\ie from 12-Feb-2017 to 25-Feb-2017) for duration over which new legitimate traffic patterns emerged. Fig.~\ref{fig:confMapRetrain} shows the performance of the IoT classifier after it is re-trained. It is seen that the confusion matrix is almost diagonal with the TP rate of more than $90$\% for all classes (\ie on average $97.4$\%). We list in Table~\ref{tab:iotdevice-retrain_tmp} all performance metrics of the IoT classifier after re-training. We observe that the model average confidence is boosted across all classes -- specifically it reaches to $0.975$ for Hue bulb instances. Also, three performance metrics consistently display an acceptable performance of classification with $0.998$, $0.973$, and $0.986$ for average precision, recall, and F1 score.

			\begin{table}[t]
				\centering
				%\vspace{-3mm}
				\caption{Performance metrics of the IoT classier model (after re-training).}
				\vspace{-2mm}
				\label{tab:iotdevice-retrain_tmp}
				\begin{adjustbox}{max width=0.95\textwidth}
					\begin{tabular}{@{}lcccccccc@{}}
						\toprule
						\textbf{IoT/Non-IoT} & \textbf{TP} & \textbf{FN} & \textbf{FP} & \textbf{Precision}&\textbf{Recall}&\textbf{$F_1$}& \textbf{TP avg. confidence} & \textbf{FN avg. confidence} \\ \midrule
						Amazon Echo&0.977&0.023&0.000&1.000&0.977&0.989&0.994&0.430\\
						August doorbell&0.989&0.011&0.001&0.999&0.989&0.994&0.974&0.509\\
						Awair air-quality&0.936&0.064&0.002&0.998&0.936&0.966&0.850&0.616\\
						Belkin motion-sensor&0.996&0.004&0.000&1.000&0.996&0.998&0.825&0.340\\
						Belkin switch&0.999&0.001&0.001&0.999&0.999&0.999&0.990&0.604\\
						Dropcam&0.990&0.010&0.001&0.999&0.990&0.994&0.987&0.462\\
						HP printer&0.977&0.023&0.001&0.999&0.977&0.988&0.985&0.591\\
						LiFX bulb&0.980&0.020&0.001&0.999&0.980&0.990&0.892&0.617\\
						NEST smoke-sensor&0.976&0.024&0.001&0.999&0.976&0.987&0.818&0.472\\
						Netatmo weather&0.997&0.003&0.002&0.998&0.997&0.997&0.935&0.824\\
						Netatmo camera&0.973&0.027&0.001&0.999&0.973&0.986&0.980&0.371\\
						Hue bulb&0.917&0.083&0.001&0.999&0.917&0.956&0.975&0.505\\
						Samsung smart-cam&0.997&0.003&0.006&0.994&0.997&0.996&0.989&0.367\\
						Smart Things&0.969&0.031&0.001&0.999&0.969&0.984&0.980&0.437\\
						Triby speaker&0.941&0.059&0.001&0.999&0.941&0.969&0.785&0.452\\
						Withings sleep-sensor&0.957&0.043&0.004&0.996&0.957&0.976&0.968&0.504\\
						Withings scale&0.979&0.021&0.003&0.997&0.979&0.988&0.953&0.560\\ \bottomrule
					\end{tabular}
				\end{adjustbox}
			\vspace{1em}
			\end{table}
		
		\vspace{1em}
		\textbf{IoT State Classifiers:}
			From our DATA2, we generated a different number of instances of four IoT devices with state labels including: Amazon Echo (boot: 208, active: 74, idle: 1795); Belkin switch (boot: 110, active: 84, idle: 2688); Dropcam (boot: 145, active: 98, idle: 2639); and LiFX bulb (boot: 160, active: 84, idle: 2338).
			To train individual device-specific models, we randomly choose 40 instances from each of their respective states -- remaining instances are used to test the models.
			{\rev 
				Note that the state classifier are specialized models trained independently for each device type, and hence adding more devices or increasing instances of a given device will not affect the performance and robustness of existing state classifiers.
			}
			
			Fig.~\ref{fig:confStates} shows the confusion maps of the four IoT state classifiers. Our first observation is that all four models  predict very well the active state -- $100.0$\% TP rate in  three models (Amazon Echo, Dropcamp, and LiFX bulb), and $95.5$\% TP rate in Belkin switch. Next, we see that boot instances are prone to be misclassified as idle, and vice versa (\eg $8.6$\% and $7.5$\% of boot instances respectively in Belkin switch and LiFX bulb are misclassified as idle). This misclassification could be possibly because instances pertinent to state transitions  (\eg boot to idle) are not precisely annotated.

			\begin{figure*}[t]
				\begin{center}
					\mbox{
						\subfloat[Amazon Echo.]{
							{\includegraphics[width=0.3\textwidth]{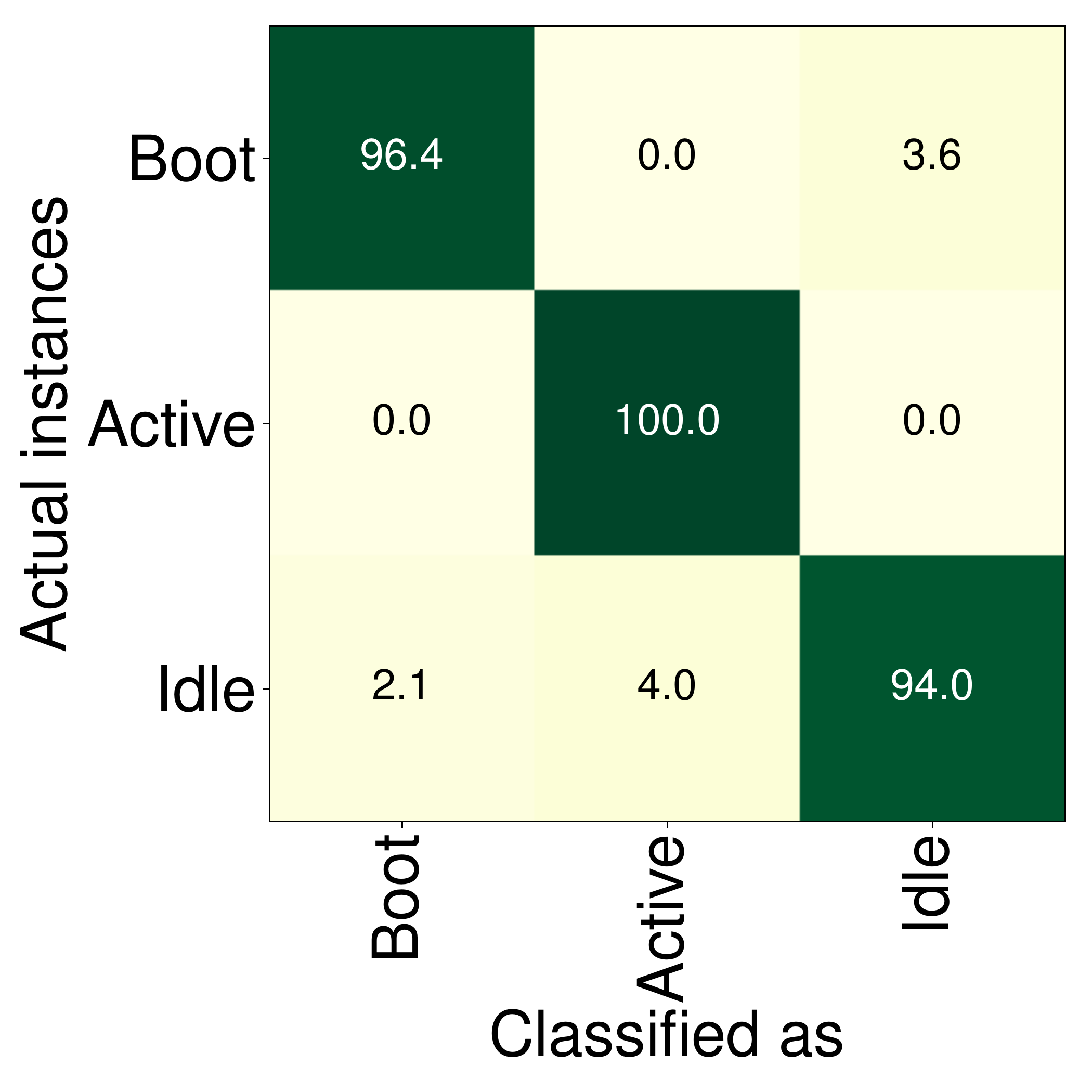}}
							\label{fig:state_echo}
						}
						\subfloat[Belkin switch.]{
							{\includegraphics[width=0.3\textwidth]{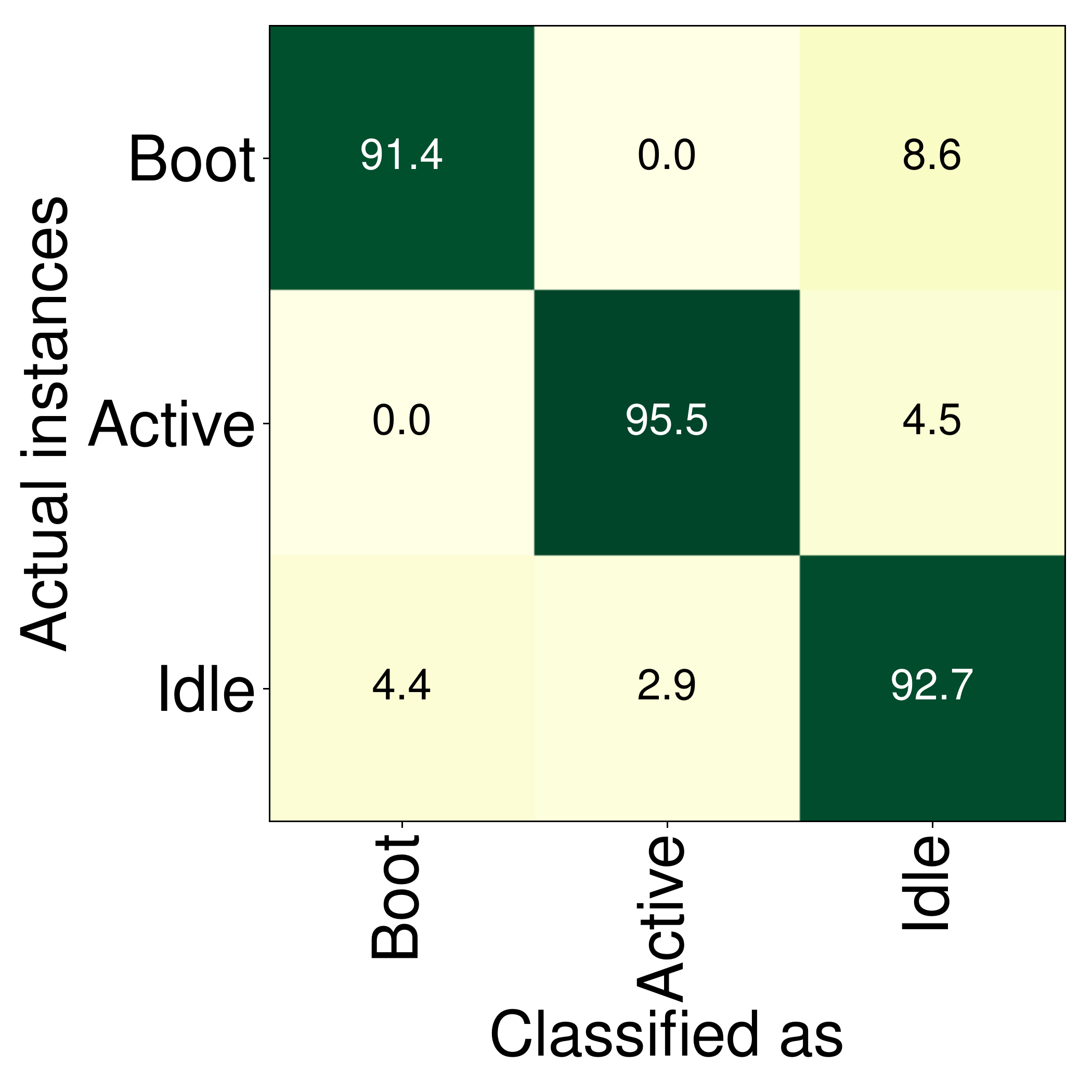}}
							\label{fig:state_belkin}
						}
					}
					\mbox{
						\subfloat[Dropcam.]{
							{\includegraphics[width=0.3\textwidth]{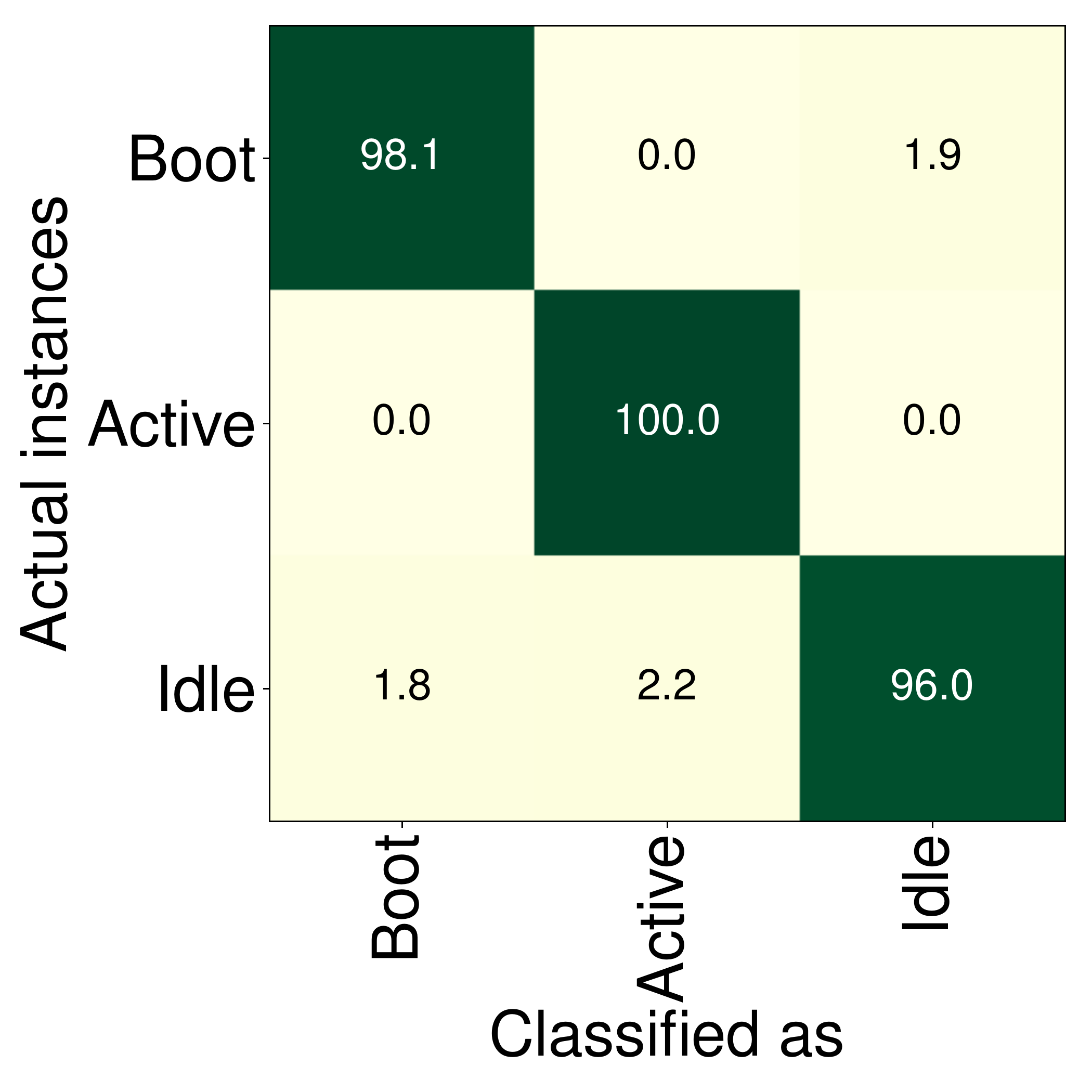}}
							\label{fig:state_dropcam}
						}
						\subfloat[LiFX.]{
							{\includegraphics[width=0.3\textwidth]{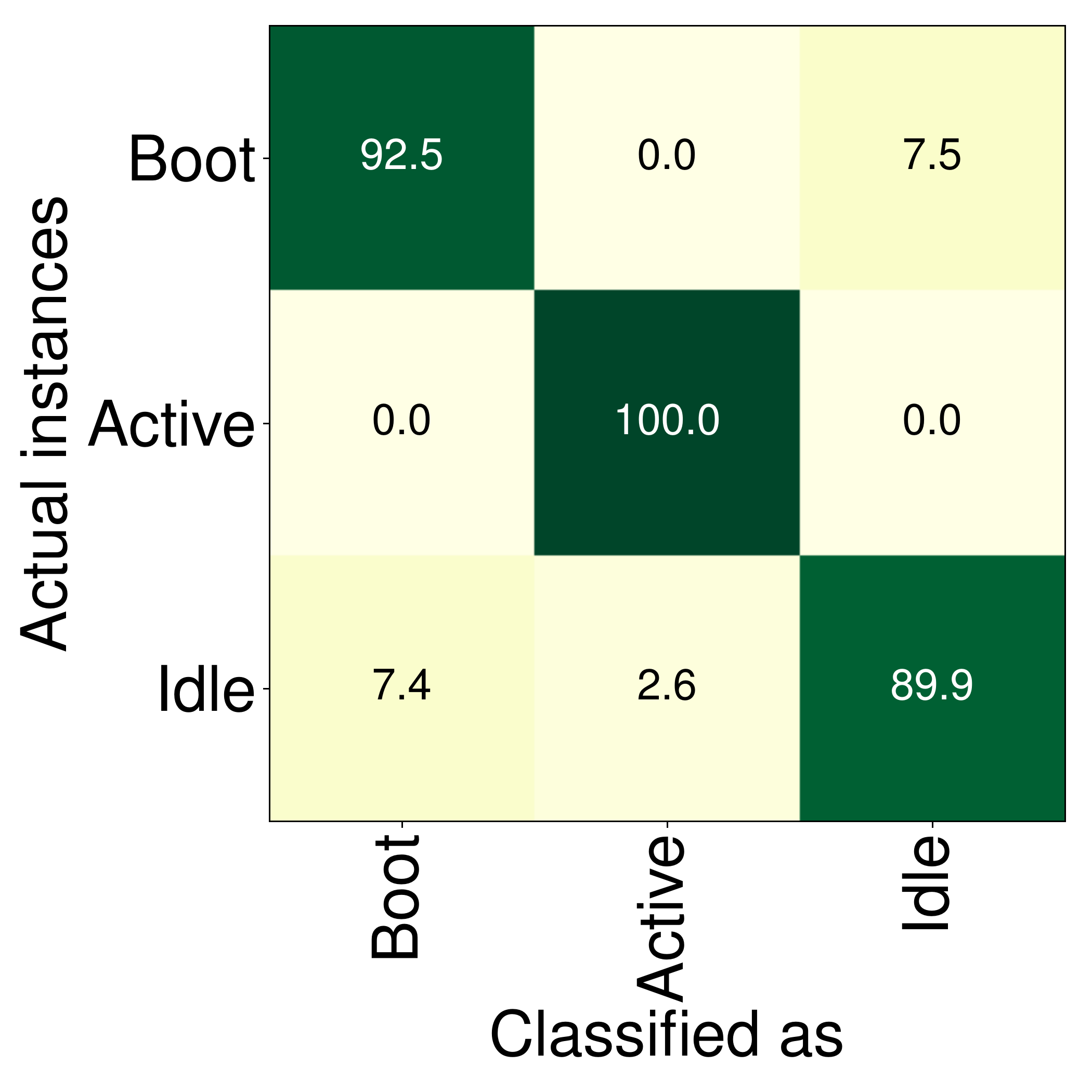}}
							\label{fig:state_lifx}
						}
					}
					\caption{Confusion matrix of IoT state classifiers: (a) Amazon Echo; (b) Belkin switch; (c) Dropcam; and (d) LiFX bulb.}
					\label{fig:confStates}
				\end{center}
			\end{figure*}
	
	\vspace{-1em}		
	\section{Practical and Operational Considerations}\label{sec:c2_op}
		\vspace{-1em}
		In the previous section, we evaluated the performance of our inference engines using all traffic attributes of devices during their normal operation. In this section, we first quantify the cost of our scheme in practice and show how we can optimize the trade-off between cost and performance.  Next, we demonstrate how network operators can interpret the outputs of inference engines, and therefore manage their cyber-security risk.
		
		\vspace{-1em}
		\subsection{Cost of Attributes }\label{sec:c2_attributeImpact}
			\vspace{-1em}
			In order to quantify the cost of our scheme, we begin by examining the impact of individual attributes on the performance of traffic inference. We have 112 attributes for the IoT detector and the IoT classifier, and 48 attributes for the IoT state classifiers. Note that some of these attributes could be highly correlated, and hence become redundant. Also, some attributes may not be very relevant to class prediction, and hence can be removed.  
		
			\textbf{Redundant Attributes:}\label{sec:c2_CFS}
				We use a selection algorithm called Correlation-based Feature Subset (CFS) \cite{Hall1998} with best-first searching method. CFS is a filter that uses a correlation-based heuristic to find a subset of attributes with the highest merit -- \ie attributes highly-correlated with the class, yet uncorrelated with each other.

			\textbf{Importance of Attributes:}\label{sec:c2_infoGain}
				In decision tree-based machine learning, the Information Gain (IG) method is used to measure the weight of various attributes in accurate prediction. Important attributes carry more information (\ie large IG value) to distinguish classes, and unrelated attributes have no information. We now compute the IG value of attributes used for each of the three classifier types.
				
				To better visualize the merit of various attributes, we illustrate in Fig.~\ref{fig:meritMatrixFull} the IG values computed for all 112 attributes used by the IoT classifier. Each cell represents an attribute (\ie rows are flow counters and columns are various timescales), and is labeled (and color coded) by its IG value -- the darker the cell, the higher the IG value. Fig.~\ref{fig:meritMatrixSub} shows a subset of 35 attributes selected by the CFS algorithm eliminating correlated (\ie redundant) attributes. Note that this subset still results in the same performance of prediction as presented in the previous section.
				
				\begin{figure}[t]
					\centering
					\mbox{
						\subfloat[All attributes.]{
							{\includegraphics[width=0.45\textwidth]{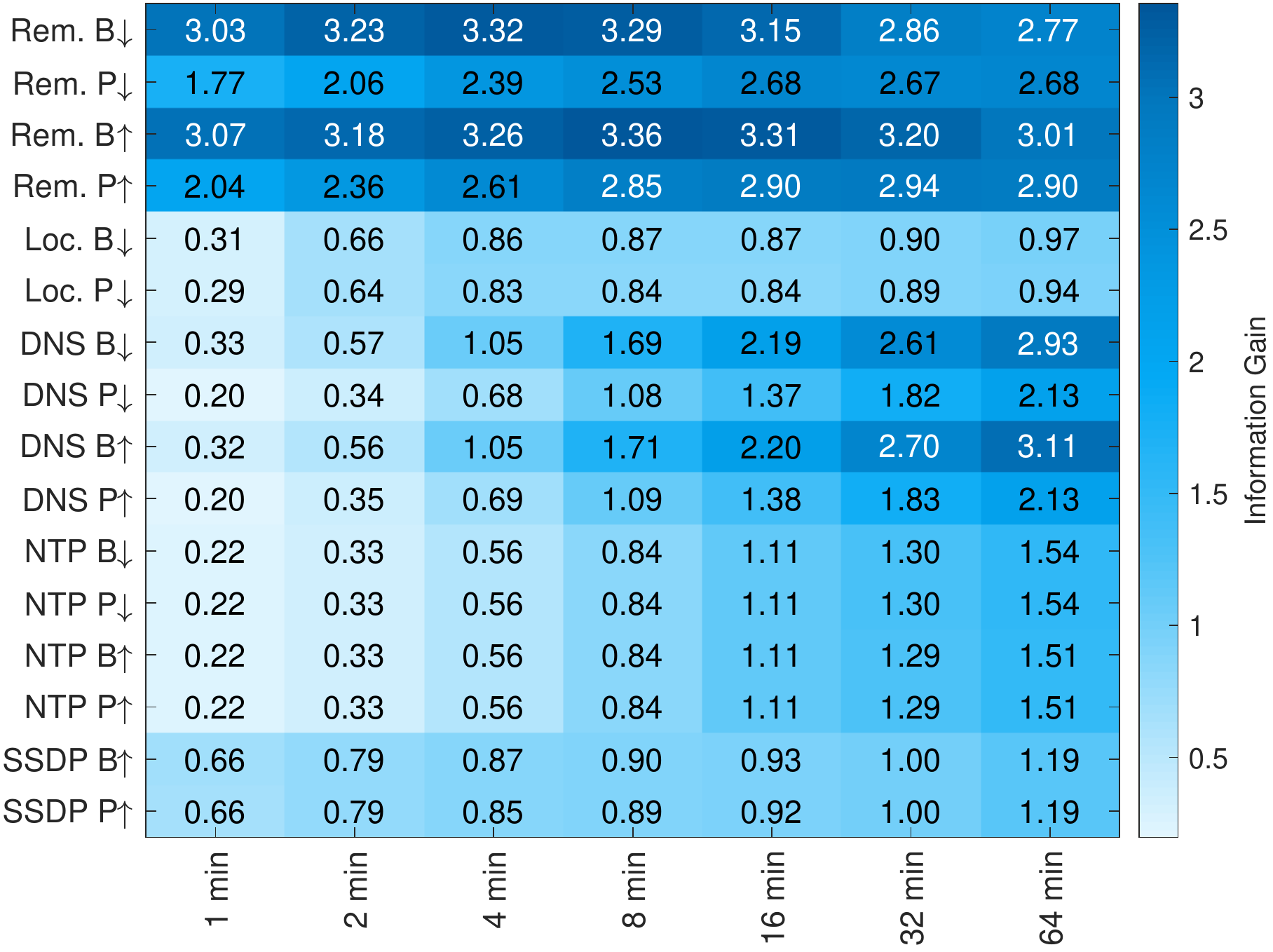}}
							\label{fig:meritMatrixFull}
						}
						
						\subfloat[Subset of nonredundant attributes.]{
							{\includegraphics[width=0.45\textwidth]{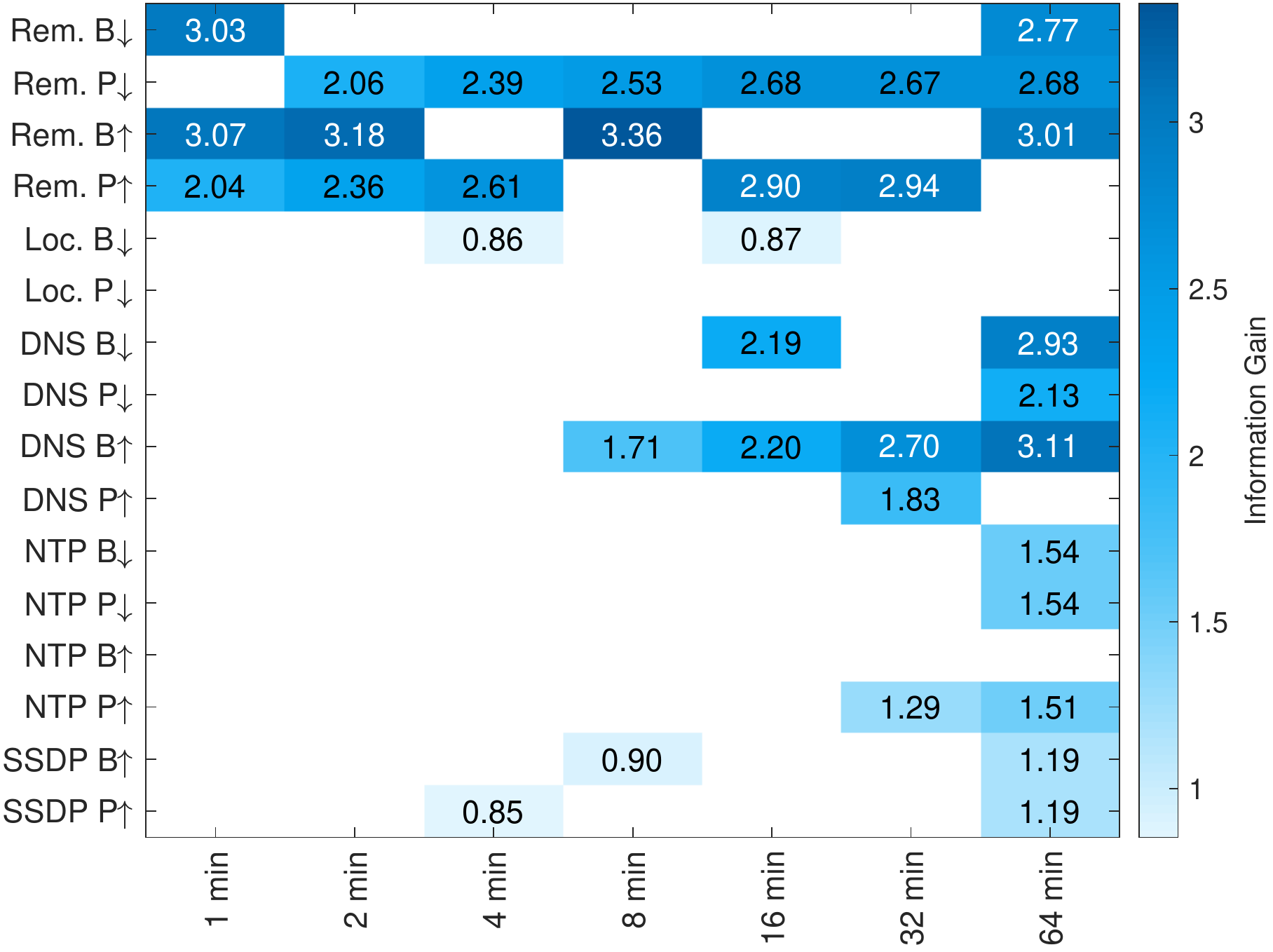}}	
							\label{fig:meritMatrixSub}	
						}
					}
					\vspace{-3mm}
					\caption{Information gain value of: (a) all attributes, and (b) CFS-selected attributes, for the IoT classifier.}	
					%\vspace{-4mm}
					\label{fig:meritMatrix}
				\end{figure}

				We observe that the highest IG value $3.36$ corresponds to  ``\textit{outgoing remote byte-count over 8-minute}'' followed by ``\textit{incoming remote byte-count over 4-minute}'' with IG $3.32$. Another observation is that byte-count of both incoming/outgoing remote over mid-term timescales (\ie 4-, 8-, 16-min) have higher information compared to other attributes, as shown by darker cells. Also, DNS counters over longer timescales (\ie 32- and 64-min) display a relatively high gain of information in predicting class of IoT devices.
				\vspace{-1em}

				Overall, attributes of two flow rules related to incoming local traffic and outgoing SSDP queries seem to have minimal impacts in IoT device classification. This is mainly because only a few of IoT devices in our lab (\eg Hue bulb, Bekin motion, Amazon Echo) communicate on the local network or send SSDP queries. Even though these flow rules (and associated attributes) may not seem important across all devices, they can precisely characterize and help identify devices which use them in their network traffic. 
				\vspace{-1em}

				\begin{figure}[t]
					\begin{center}
						\mbox{
							\subfloat[Amazon Echo.]{
								{\includegraphics[width=0.22\textwidth]{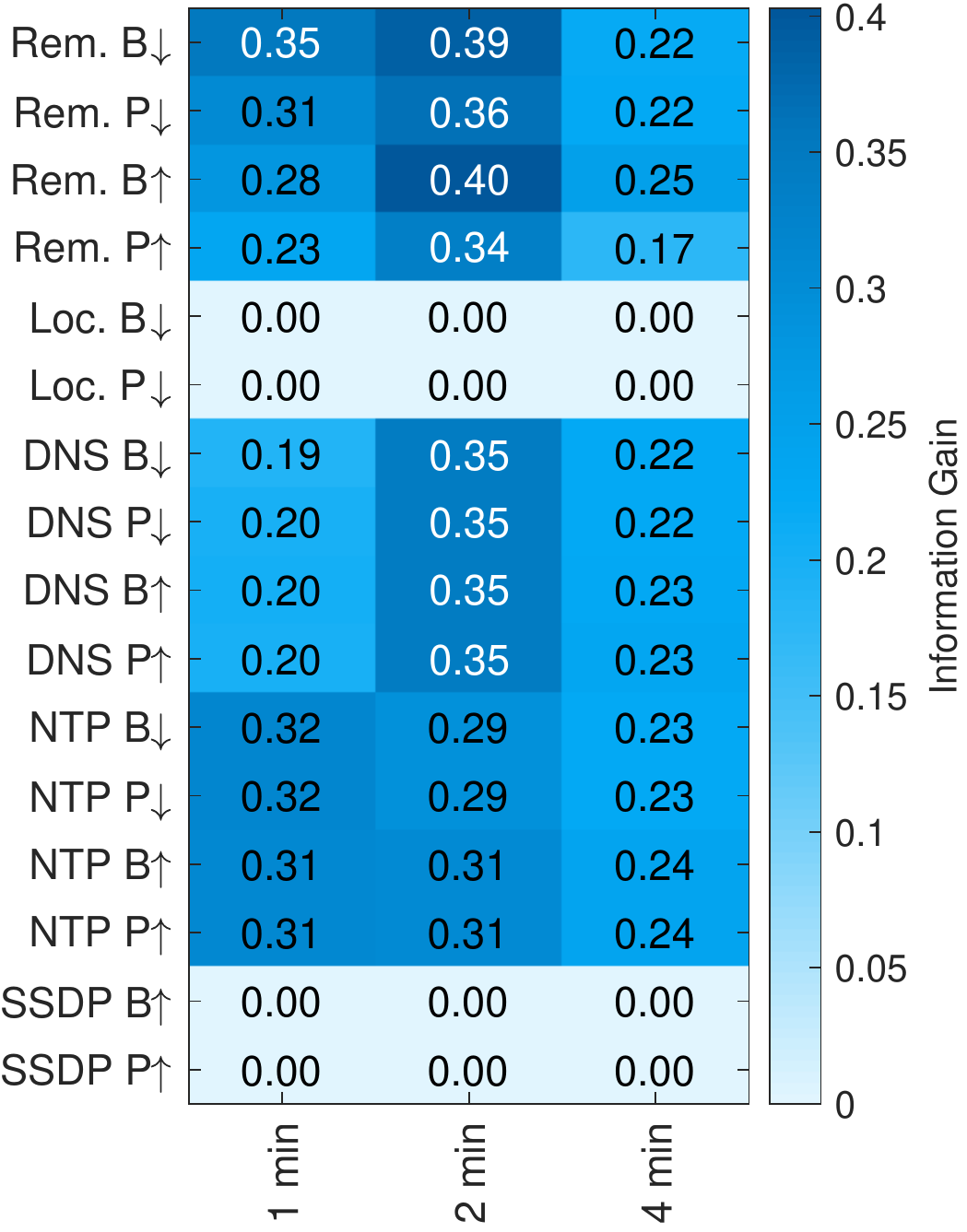}}\quad
								\label{fig:meritMatrixState_Echo_all}
							}
							
							\subfloat[Belkin switch.]{
								{\includegraphics[width=0.22\textwidth]{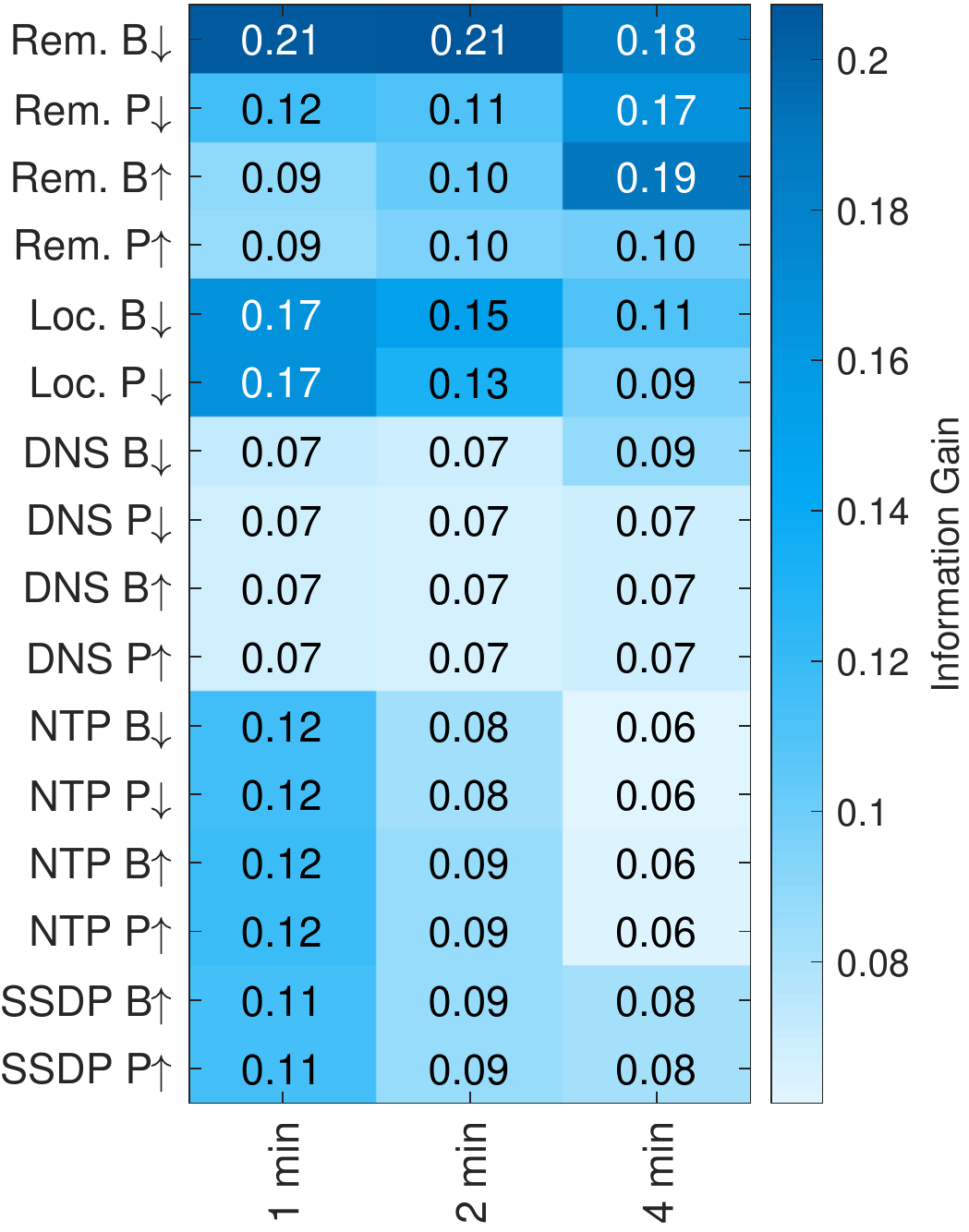}}\quad
								\label{fig:meritMatrixState_Belkin_all}
							}
							
							\subfloat[Dropcam.]{
								{\includegraphics[width=0.22\textwidth]{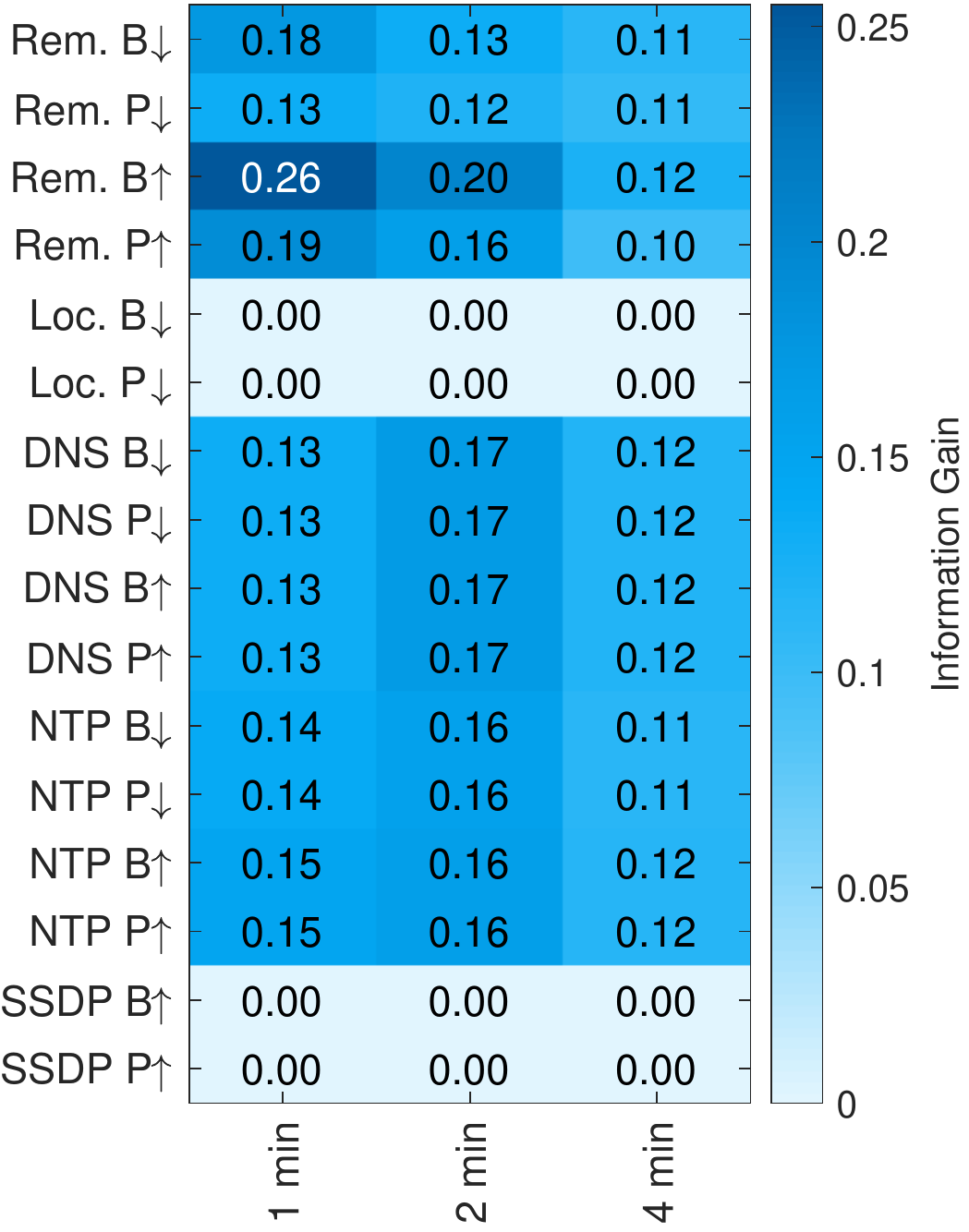}}\quad
								\label{fig:meritMatrixState_Dropcam_all}
							}
							
							\subfloat[LiFX bulb.]{
								{\includegraphics[width=0.22\textwidth]{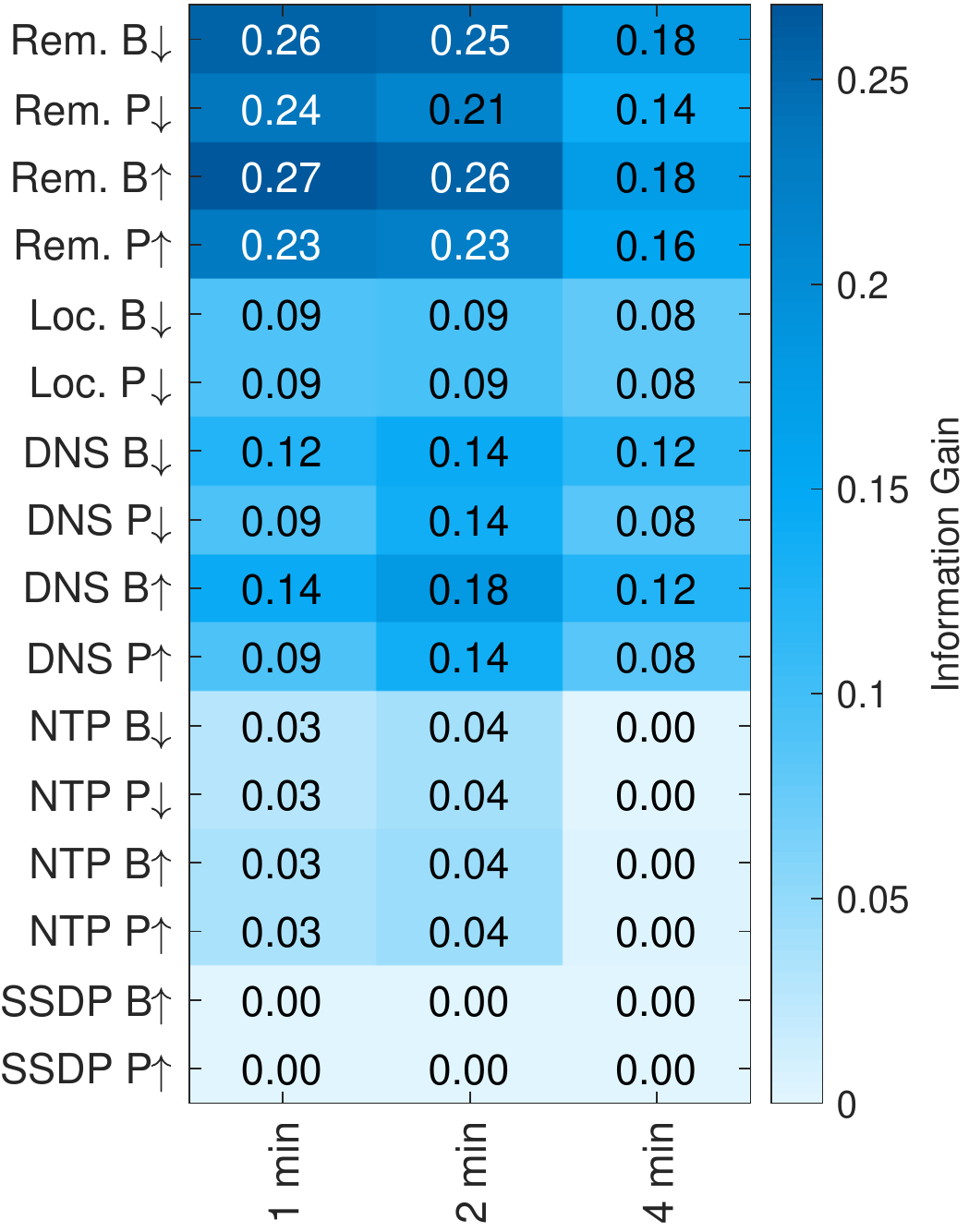}}\quad
								\label{fig:meritMatrixState_Lifx_all}
							}
							
						}
						\caption{Information gain of all attributes for state classier models: (a) Amazon Echo, (b) Belkin switch, (c) Dropcamp, and (d) LiFX bulb.}
						\label{fig:attStates_all}
					\end{center}
				\end{figure}
				\begin{figure}[b]
					\begin{center}
						\mbox{
							\subfloat[Amazon Echo.]{
								{\includegraphics[width=0.22\textwidth]{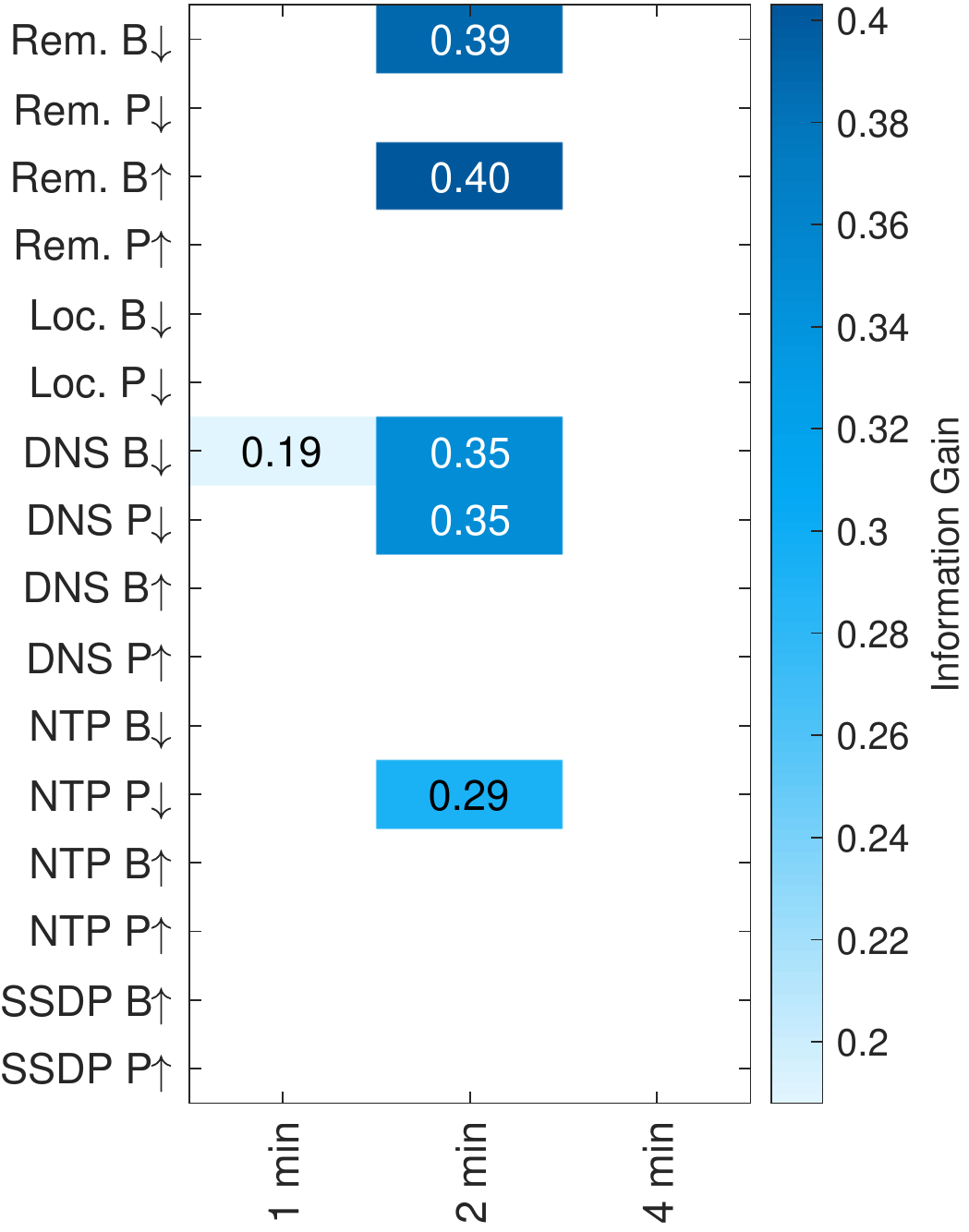}}\quad
								\label{fig:meritMatrixState_Echo}
							}
							
							\subfloat[Belkin switch.]{
								{\includegraphics[width=0.22\textwidth]{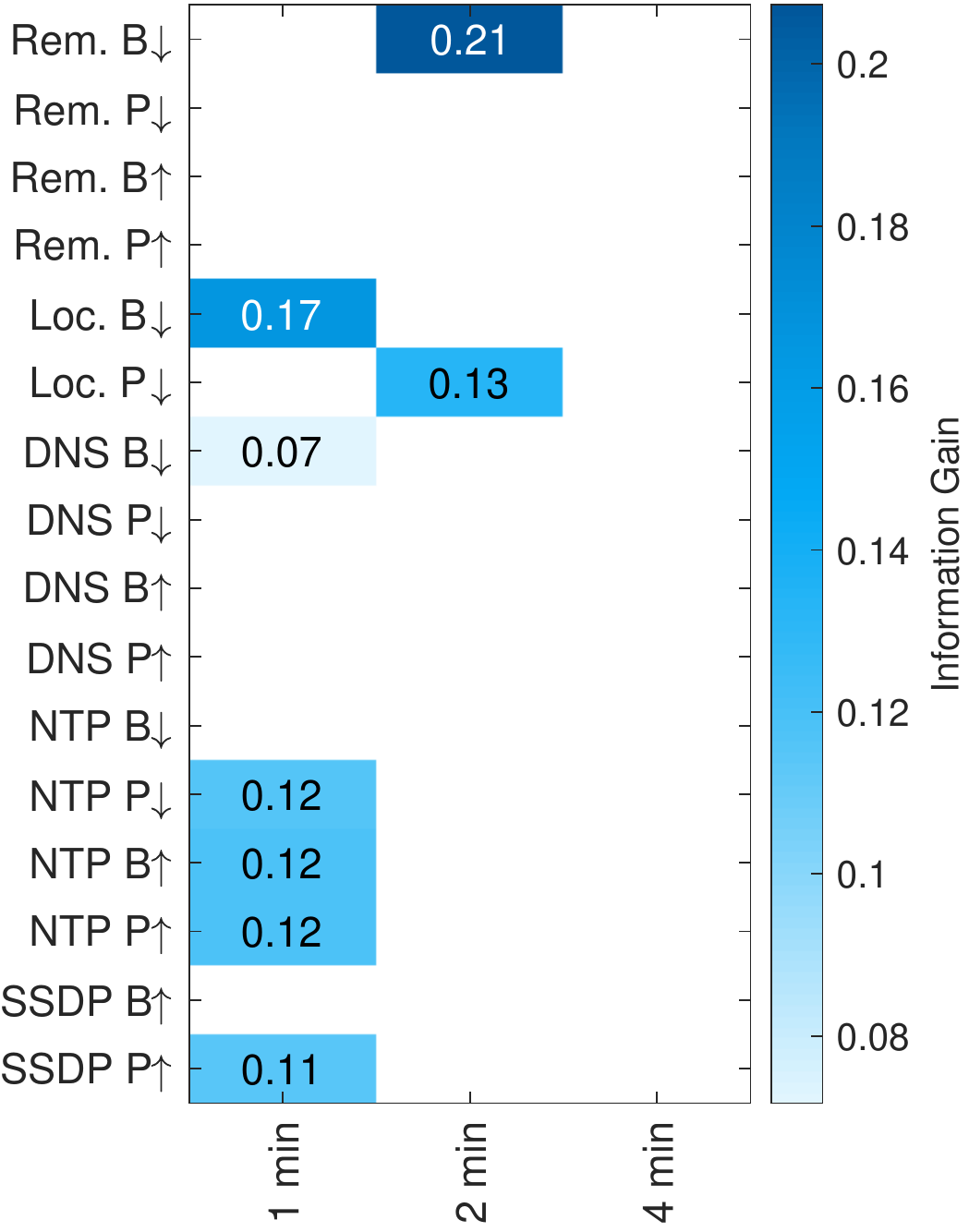}}\quad
								\label{fig:meritMatrixState_Belkin}
							}
							
							\subfloat[Dropcam.]{
								{\includegraphics[width=0.22\textwidth]{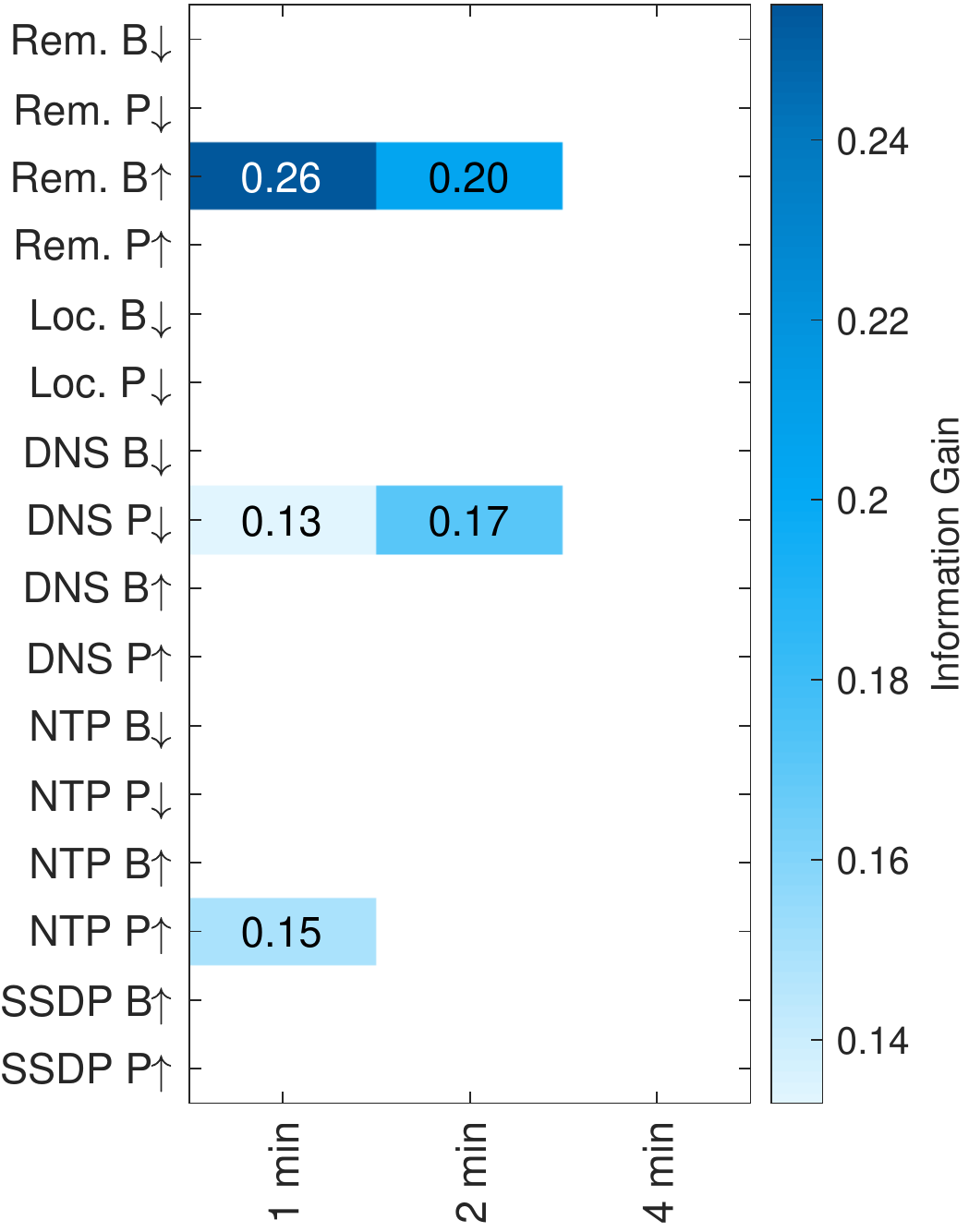}}\quad
								\label{fig:meritMatrixState_Dropcam}
							}
							
							\subfloat[LiFX bulb.]{
								{\includegraphics[width=0.22\textwidth]{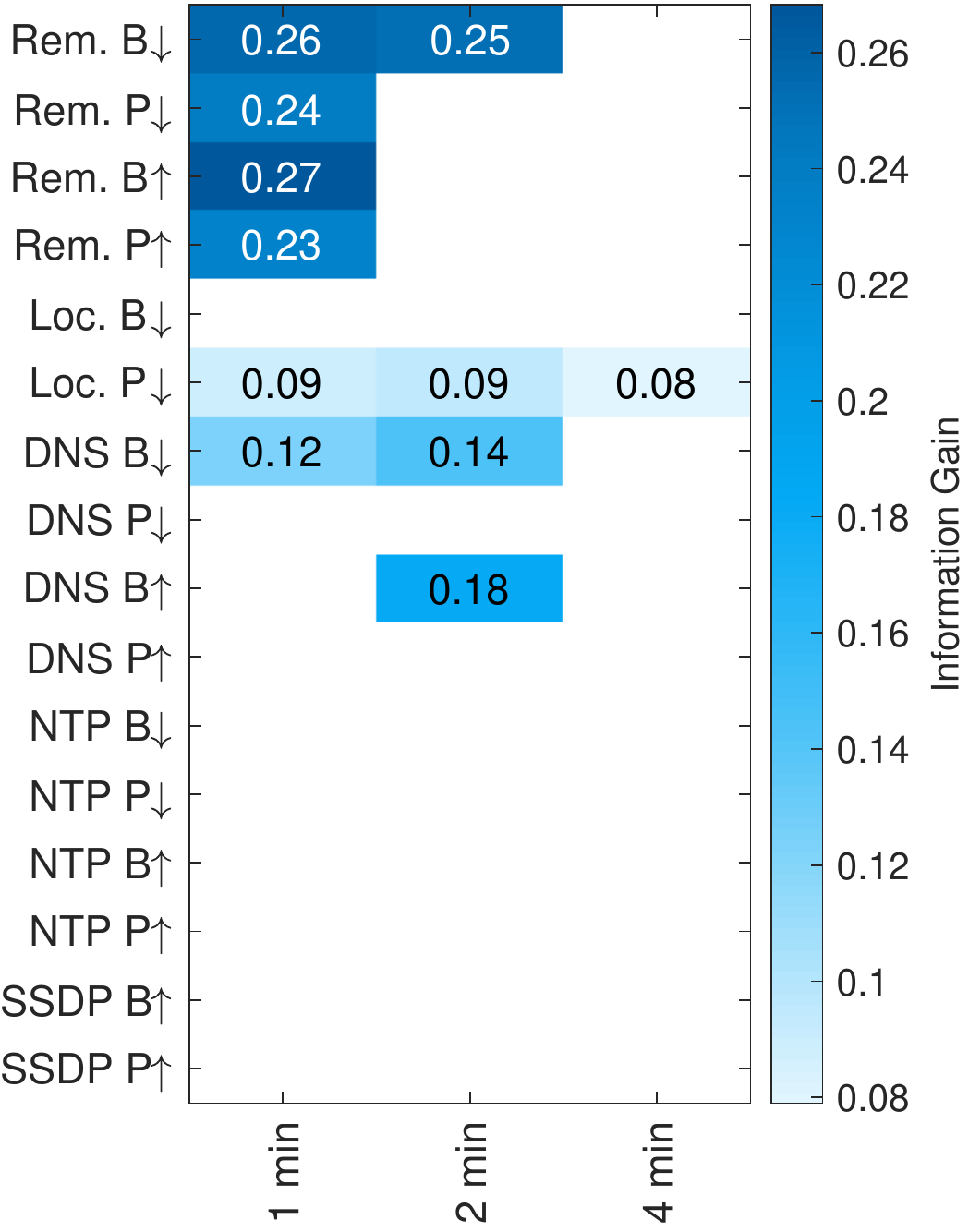}}\quad
								\label{fig:meritMatrixState_Lifx}
							}
							
						}
						\caption{Information gain of non-redundant attributes for state classier models: (a) Amazon Echo, (b) Belkin switch, (c) Dropcamp, and (d) LiFX bulb.}
						\label{fig:attStates}
					\end{center}
				\end{figure}
				%About other classifiers
				{\rev
					Moreover, we have analyzed the impact of attributes for two other types of classifiers. Fig.~\ref{fig:attStates_all} and Fig.~\ref{fig:attStates} show the information gain value of all and non-redundant attributes for state classification models. We observe, for example, in Fig.~\ref{fig:meritMatrixState_Echo} that attributes of only for flow rules (\ie incoming remote, outgoing remote, incoming DNS, incoming NTP) are needed for the state classification of Amazon Echo -- there is no attribute selected for the other four flows. Another observation is that attributes over very short timescales (\ie 1-min and 2-min) become important in classifying operating states of IoT devices. Also, we note that the variation of IG values for non-redundant attributes is less (\ie between $0.1$ and $0.3$) compared to the IoT classifier model (\ie between $0.86$ and $3.36$).  For the IoT detector model, we found that attributes over longer timescale (\ie outgoing/incoming byte-count over 32-min and 64-min) have higher impact. It is important to note that our observations and findings may change in different environments depending upon types of devices and their possible interactions.
				}
			
				\textbf{Cost versus Performance:}\label{sec:c2_attCost}
					There exist two sources of cost in our inference scheme: (1) number of flow entries; and (2) space complexity of computing attributes. Given the fixed size of TCAM on programmable (SDN) switches, efficient management of flow entries \cite{FlowMngTCAM2014} becomes crucial to scale of scheme for deployment in a network with a large number of IoT devices. Since our attributes are computed at multiple timescales up to 64-minutes, we need to maintain the time-series of  flow counters accordingly (\ie 64 data-points each corresponds to a minute).
				
					We, therefore, aim to reduce the cost by decreasing attributes, without significantly affecting performance. Table~\ref{tab:flowReducedAttributes} shows the number of flow entries needed by each inference model with non-redundant set of attributes -- check-marked cells indicate the flows needed for attributes of models in each row. It clearly shows room for optimizing our approach by dynamic management of flow entries on the programmable switch. For example, the IoT detector model only needs four flow rules per device. Once a device is detected as IoT, an additional four flows are needed by the IoT classifier (\ie a total of eight flows). Once the IoT device is successfully classified, it may need a reduced number of flows depending upon its specialized state classifier (some flows can be removed from the switch). The state classifier of Amazon Echo, Belkin switch, Dropcam, and LiFX respectively need 4, 6, 3, and 5 flow entries per each unit of device. 
					
					\begin{table}[b]
						\centering
						\caption{Flow entries (per-device) needed for non-redundant attributes set.}
						\label{tab:flowReducedAttributes}
						\begin{adjustbox}{max width=1\textwidth}
							\begin{tabular}{|l|c|c|c|c|c|c|c|c|c|}
								\hline
								Inference model & Rem.$\uparrow$ & Rem.$\downarrow$ & Loc.$\downarrow$ & DNS$\uparrow$ & DNS$\downarrow$ & NTP$\uparrow$ & NTP$\downarrow$ & SSDP$\uparrow$ & Num. of flow entries\\ \hline
								IoT detector & \checkmark & \checkmark & &\checkmark & \checkmark & & & \checkmark &5\\ \hline
								IoT classifier  & \checkmark & \checkmark & \checkmark &\checkmark & \checkmark & \checkmark& \checkmark & \checkmark &8\\ \hline
								State classifier - Amazon Echo & \checkmark & \checkmark & & & \checkmark& & \checkmark & &4\\ \hline
								State classifier - Belkin Switch & & \checkmark &\checkmark& & \checkmark & v & \checkmark & \checkmark &6\\ \hline
								State classifier - Dropcam & \checkmark & & & & \checkmark & \checkmark & & &3\\ \hline
								State classifier - LiFX & \checkmark & \checkmark & \checkmark & \checkmark & \checkmark & & & &5\\ \hline
							\end{tabular}
						\end{adjustbox}
					\end{table}
				
					%Futher cost reduction using trade-off
					We further optimize by a careful trade-off between cost of performance. We: (a) first sort non-redundant attributes in descending order; (b) then accumulate attributes one-by-one from the sorted list; and lastly (c) quantify the cost and performance at each step. Let us visualize this process for the IoT classifier in Fig.~\ref{fig:attImpact}. We plot performance metrics and cost signals, each as a function of cumulative set of high-merit attributes. With 35 non-redundant attributes, it is seen in Fig.~\ref{fig:perfAtt} that average weighted precision, recall, F\textsubscript{1} score reach to $97.5$\%, $97.3$\%, $97.4$\% respectively, and in Fig.~\ref{fig:costAtt} that the total cost per device would reach to 8KB of memory and 8 flow entries. We note that with the top 25  attributes we can achieve about  $97$\% in all performance metrics which can save four flow entries (\ie 50\% saving) and reduce the space complexity to 5KB (\ie 37\% reduction). 
					
					{\rev
						Devices experimented in our testbed, indeed, represent majority of IoT devices that are currently available in the market. We acknowledge that the importance of the attributes can vary in different environments. That’s why we begin by incorporating all attributes for our baseline evaluation without removing any attribute. Later, a network operator may use our method to perform cost-benefit analysis to identify and possibly remove low-impact (less important) attributes for their environment.
					}
					%Obviously, operators of IoT networks may choose different strategies to balance their cost and performance depending upon their environment and resources.  

					\begin{figure}[t]
						\begin{center}
							\mbox{
								\subfloat[Performance metrics of classification.]{
									{\includegraphics[width=0.449\textwidth]{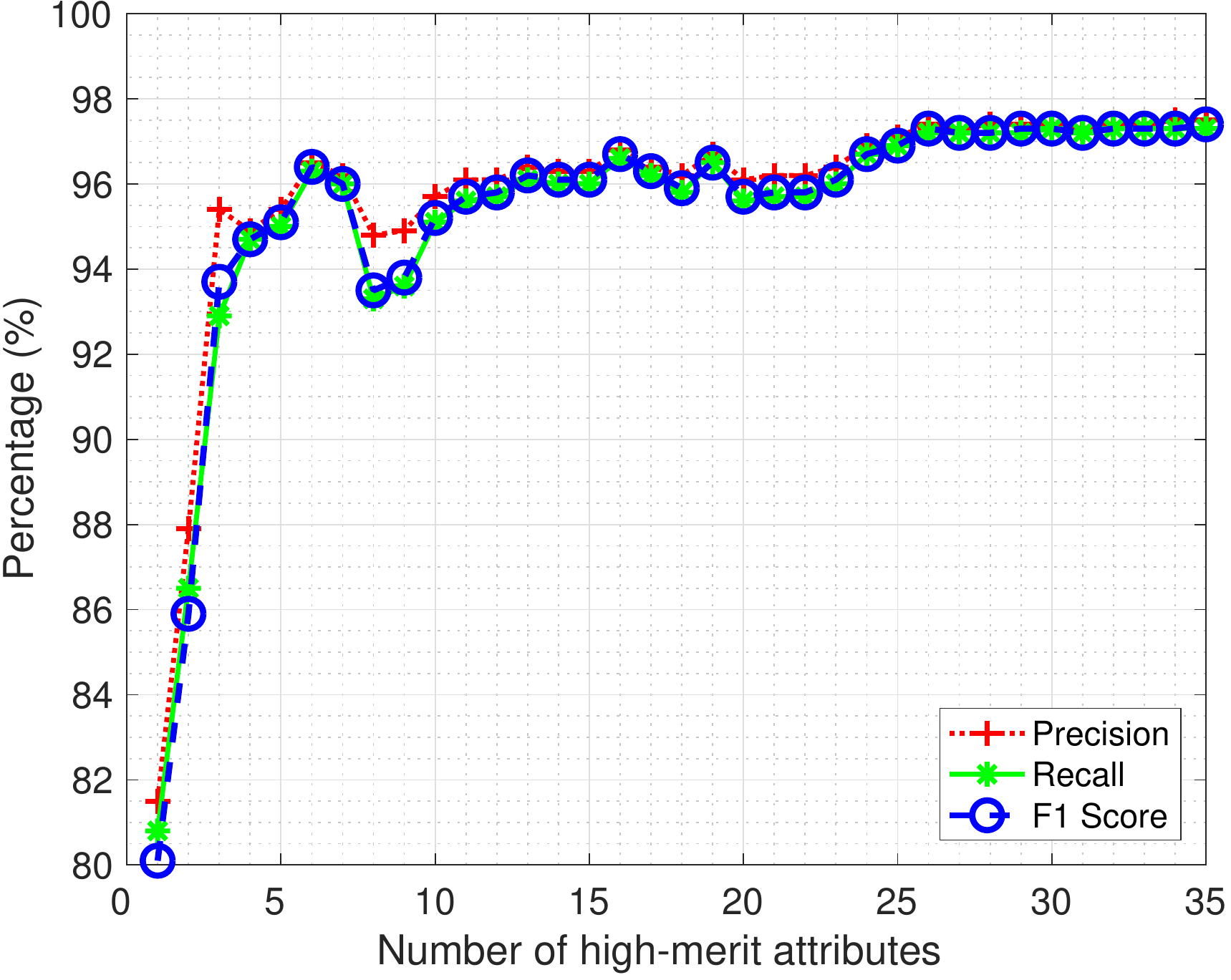}}\quad
									\label{fig:perfAtt}
								}
								
								\subfloat[Cost of attributes and flow rules.]{
									{\includegraphics[width=0.49\textwidth]{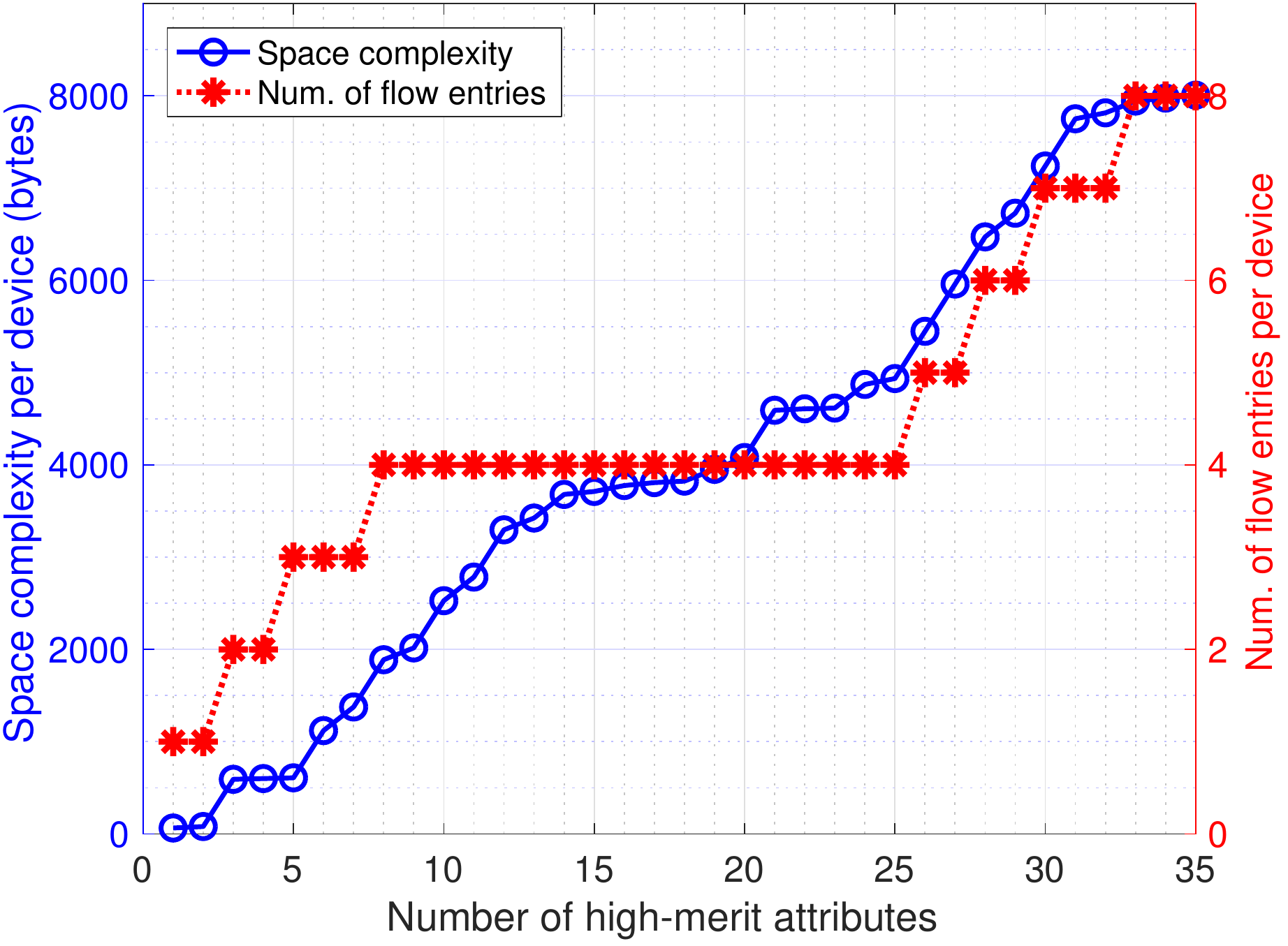}}\quad
									\label{fig:costAtt}
								}
							}
							
							\caption{Impact of attributes on: (a) performance; (b) cost, for the IoT classifier.}
							\label{fig:attImpact}
						\end{center}
					\vspace{-0.5cm}
					\end{figure}

		\vspace{-0.5cm}
		\subsection{Use of Inference Engines in Real-Time}
			We now demonstrate how our scheme can help network operators detect behavioral changes due to malicious network activities or cyber-attacks.

			Unlike traditional non-IoT devices, behavior profile of IoTs does not significantly change by interactions with users or environment. We discussed in \S\ref{sec:c2_perf} how legitimate firmware upgrades can be detected by our solution (\ie consistent misclassification and/or low confidence), as shown in Fig.~\ref{fig:PerfTimeTrace}. Note that sudden changes in outputs of inference models (if persist) for given device(s) can trigger an investigation by network administrators or inspection appliances.  

			We now test our classier models with attack traffic on IoT devices. We use a set of publicly available PCAP traces \cite{HamzaSOSR2019} that contain both benign and attack traffic (clearly annotated) corresponding to a few IoT devices we use in this work. 
			
			\begin{figure}[t]
				\begin{center}
					\mbox{
						\subfloat[IoT classifier with Belkin switch traffic.]{
							{\includegraphics[width=0.8\textwidth]{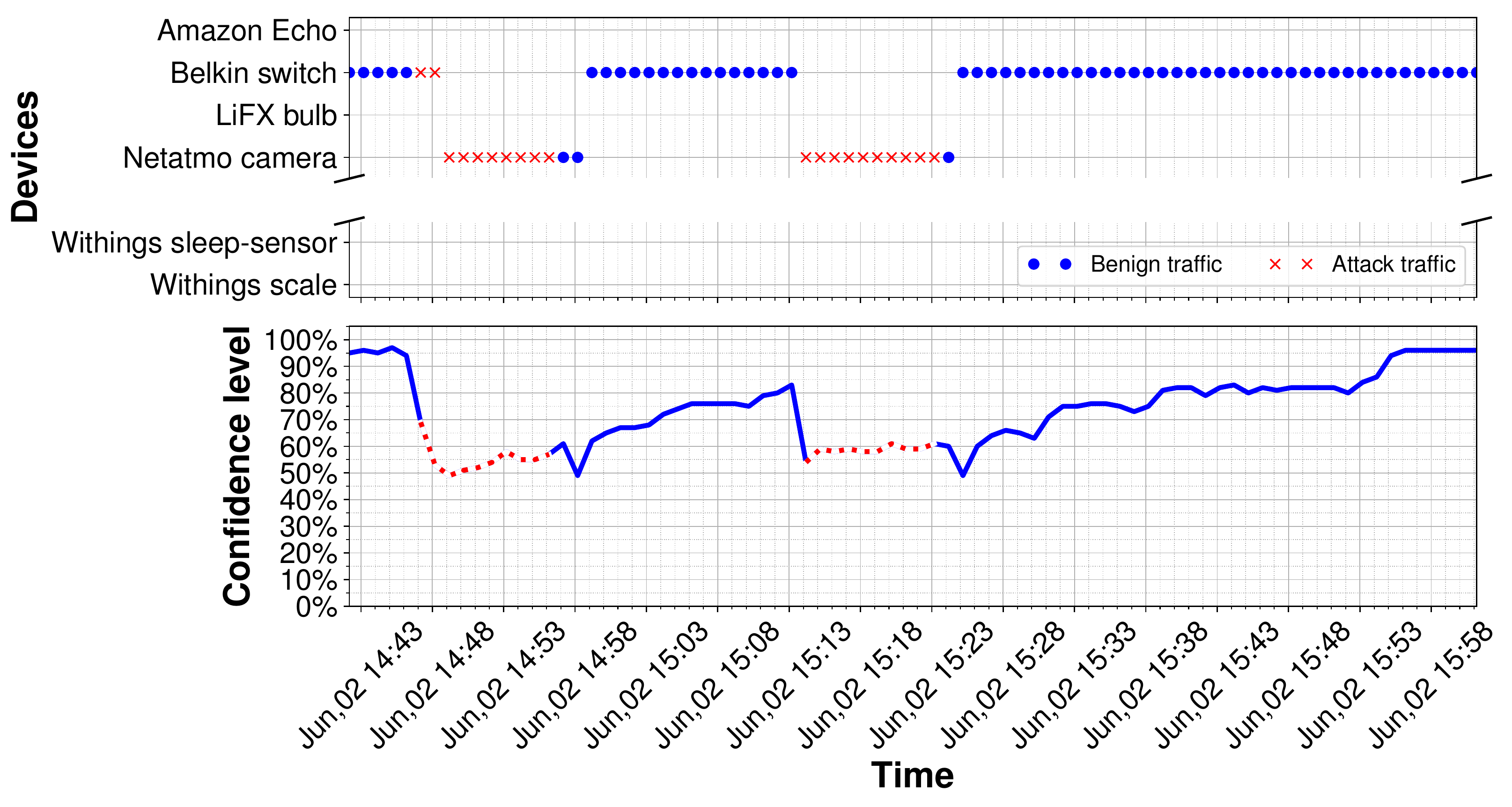}}\quad
							\label{fig:attackDeviceClassBelkin}
						}
					}
					\mbox{
						\subfloat[IoT classifier with Amazon Echo traffic.]{
							{\includegraphics[width=0.8\textwidth]{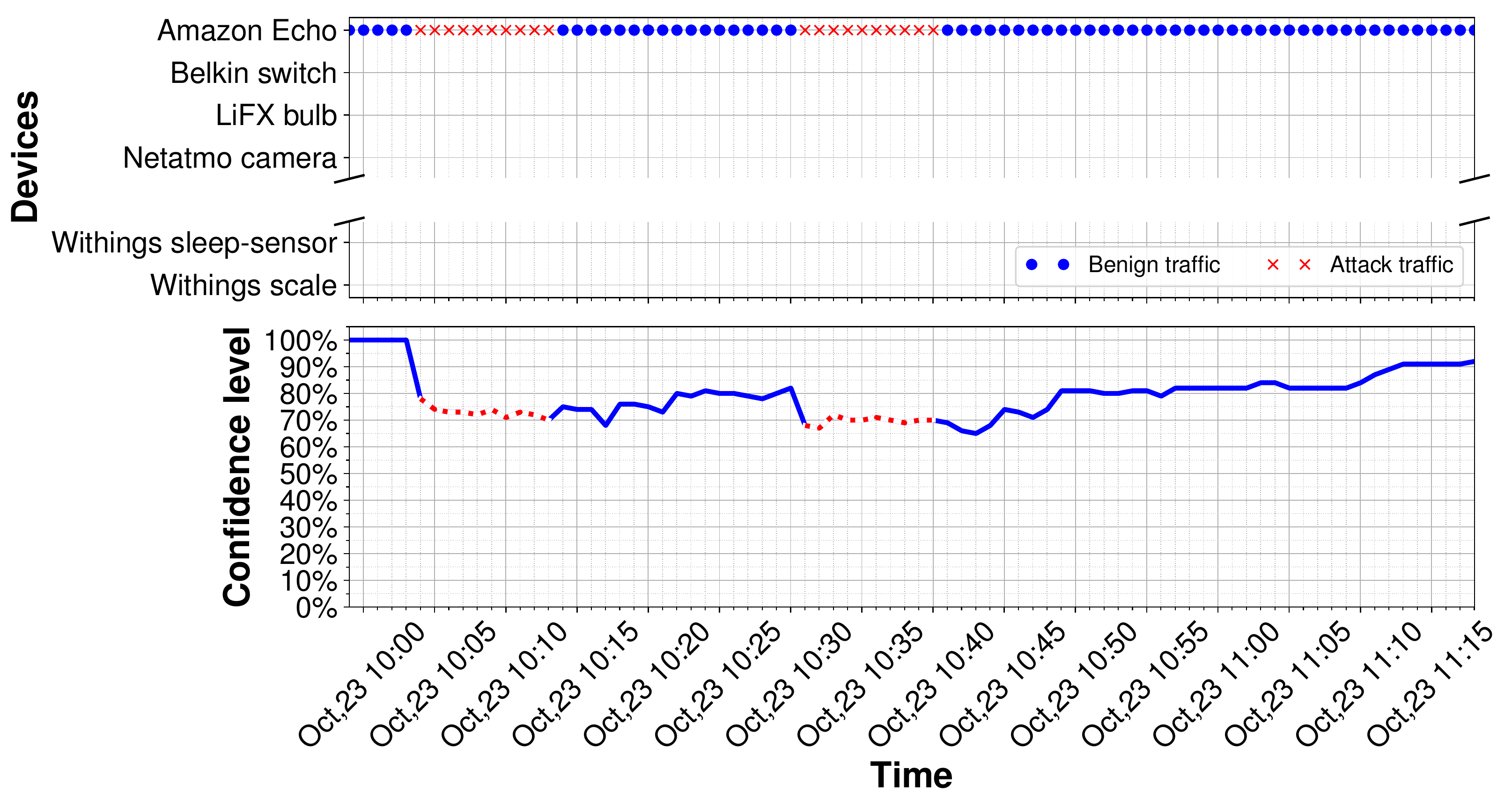}}\quad
							\label{fig:attackDeviceClassAmazon}
						}
						
					}
					
					%\vspace{-3mm}
					\caption{Time trace of device classifier outputs with benign and attack traffic of: (a) Belkin switch; and (b) Amazon Echo.}
					\label{fig:PerfAttackTraceAmazonBelkin}
				\end{center}
			\vspace{-2em}
			\end{figure}
			In Figures~\ref{fig:PerfAttackTraceAmazonBelkin} and~\ref{fig:PerfAttackTraceLiFx} we present results of three representative scenarios: (1) the output label of the IoT classifier changes persistently (\ie repeatedly misclassifying) accompanied by a sudden drop in confidence; (2) the output label of the IoT classifier does not change, but its confidence drops and persistently stays at low levels; and (3) the output of the IoT classifier remains normal (expected label with reasonable confidence), but the respective state classifier mis-behaves.   
			In these plots, red crosses indicate time periods over which attack traffic is launched to the respective IoT device, and blue circles show purely benign traffic instances.
			
			Fig.~\ref{fig:attackDeviceClassBelkin} illustrates the scenario 1 for Belkin switch. This time trace displays a situation where the IoT device experiences TCP SYN reflection attack twice, each for a duration of 10 minutes. 
			It is seen that during attack periods (shown by red cross markers) the predicted label changes from Belkin switch to Netatmo camera with confidence less than $70$\%. Right after the attack, the output comes back to its original label and the confidence starts rising gradually. 

			\begin{figure}[t]
				\begin{center}
					\mbox{
						\subfloat[IoT classifier with LiFX bulb traffic.]{
							{\includegraphics[width=0.8\textwidth]{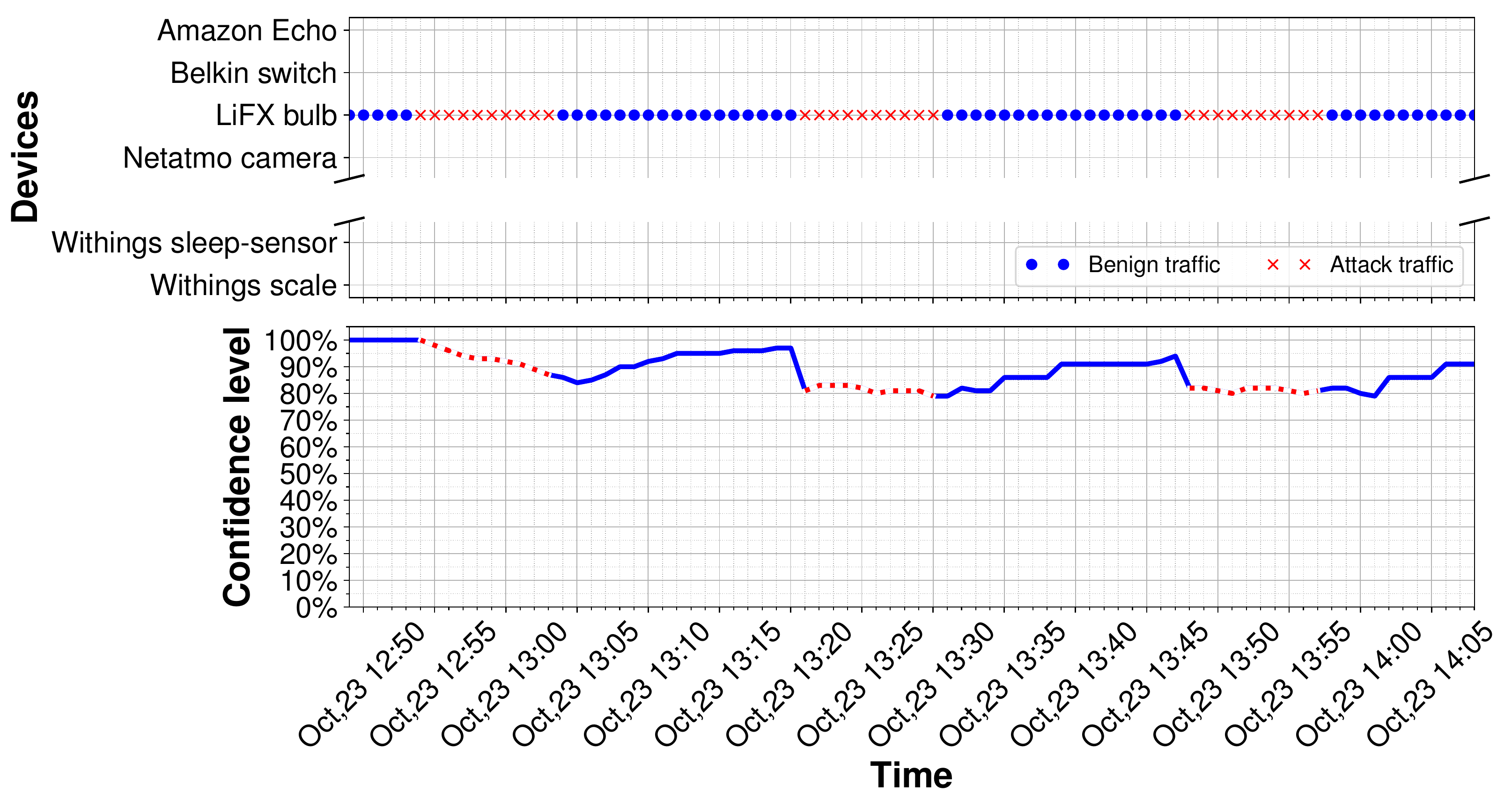}}\quad
							\label{fig:attackDeviceClassLiFx}
						}
					}
					\mbox{
						\subfloat[LiFX state classifier with LiFX bulb traffic.]{
							{\includegraphics[width=0.8\textwidth]{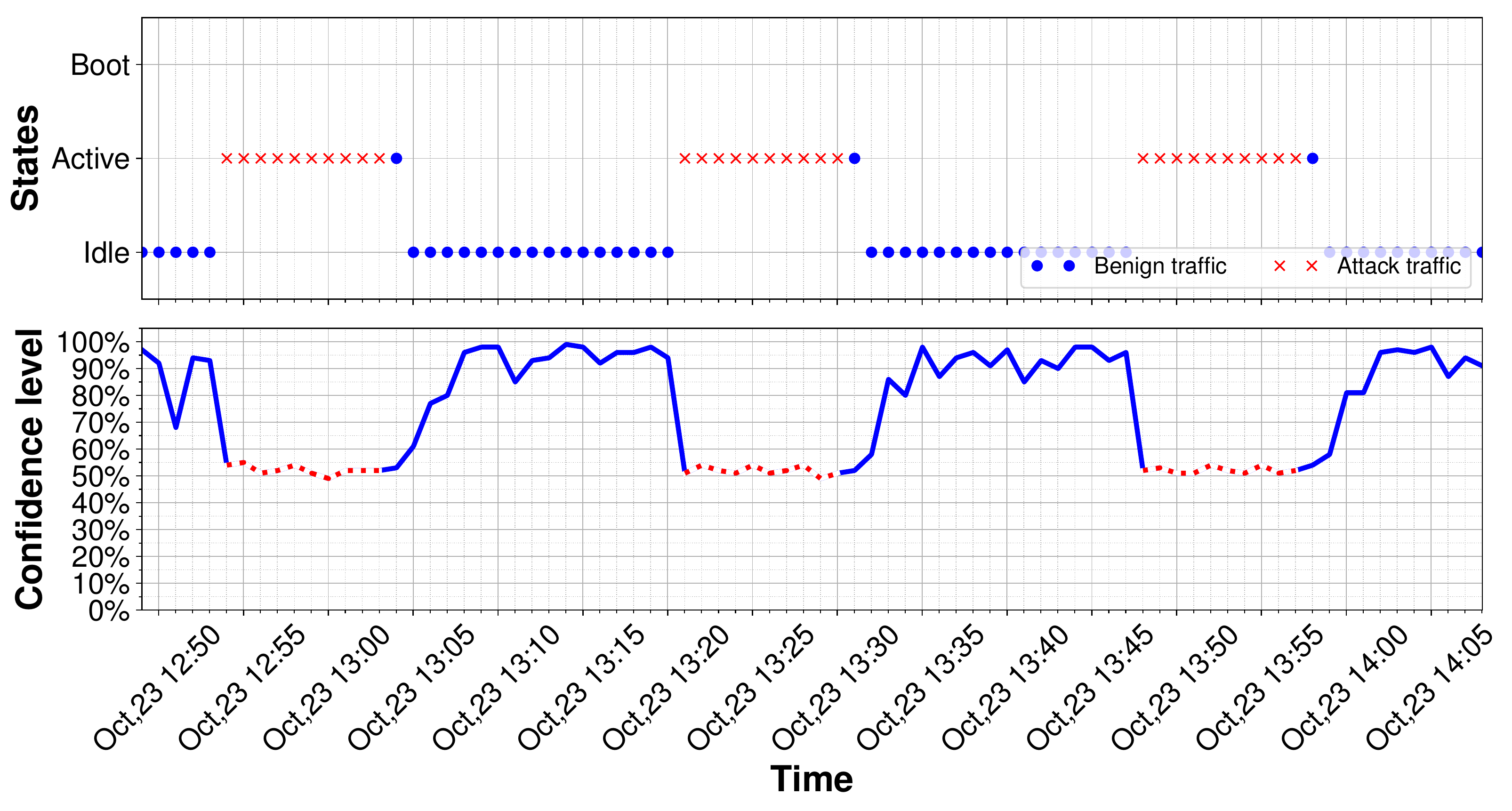}}\quad
							\label{fig:attackStateClassLiFx}
						}
						
					}
					
					\caption{Time trace of outputs for: (a) device classifier, and (b) state classifier, with benign and attack traffic of LiFx bulb.}
					\label{fig:PerfAttackTraceLiFx}
				\end{center}
				\vspace{-1em}
			\end{figure}
			
			Fig.~\ref{fig:attackDeviceClassAmazon} (representative of the Scenario 2) displays a time trace of our monitoring scheme for Amazon Echo under UDP-based DoS attack over two periods of 10-minutes each. We observe that these attacks do not change the predicted label of the IoT classifier, but cause the model confidence to decay rapidly (and remains below $80$\% persistently).
			
			Lastly, Fig.~\ref{fig:attackDeviceClassLiFx} (representative of the scenario 3) illustrates a situation where attack traffic is not intense (\ie ping of death attack on LiFX bulb), and hence does not affect the broad view of the IoT classifier model. However, the specialized model of the state classifier is significantly affected.  During the attack periods, the LiFX bulb is persistently seen in an active state which is not normal for a light-bulb since its activity (turning on/off or changing color) is expected to be relatively short. Additionally, the confidence of the state classifier quickly falls to a level of about $50$\% which is not normal again. 
			
			{\rev
				Although drops in the confidence level indicate traffic anomalies, it can be challenging for a network administrator to differentiate random traffic variations (e.g. noises) occurring over a very short time interval from a real attack which may persist for a considerable amount of time. Next chapter will develop a method to clearly differentiate anomalies from short term traffic variations.
			}
			
	\section{Conclusion}\label{sec:c2_con} 
		%Operators of smart environments face mounting pressure to enhance their visibility into their IoT infrastructure with many vulnerable devices.
		This chapter developed a real-time behavioral monitoring solution for IoT devices employing low-cost flow-level telemetry with the support of OpenFlow enabled SDN switches. We identified traffic flows that can collectively characterize the network behavior of IoT devices and their states such as booting, user interaction or idle. We then trained a set of classification models with supervised machine learning algorithms for a three-stage inference architecture using real traffic traces of 17 IoT devices collected over a period of six months. We validate their efficacy in detecting IoT devices from non-IoTs, classifying their type, and identifying their operating state. Lastly, we showed how we balance the trade-off between cost and performance of our scheme, and demonstrated how operators can use it to detect IoT behavioral changes (both legitimate and malicious).
		{\rev
			While this chapter mainly focused on low-cost attributes and optimizing the cost of network telemetry, we just scratch the surface of anomaly detection. The following chapter improves the sensitivity of inference engine for detecting behavioral changes in IoT network traffic, due to firmware upgrades or low-rate cyber-attacks, using clustering models.
		}
\chapter{Behavioral Change Detection using Clustering Algorithm}
	\label{chap:anomaly}
	\vspace{-5mm}
	\minitoc
	%The Internet-of-Things (IoT) is increasingly becoming a major challenge for network administrators to manage connected devices and sensors ranging from smart-lights to smoke-alarms and security-cameras, at scale. IoT devices use extensive variety of firmware, and provide little or no access to and management of their operating system and configuration. Operators, therefore, need to employ traffic classification models to automatically identify their IoT assets at network-level, and monitor their behavior in real-time.
	In the previous two chapters, we developed inference engines that help network operators automatically identify IoT assets via network-level traffic analysis, and monitor their behavior in real-time. However, IoT manufacturers often tend to release new firmware which improves device functionalities or even automatically perform upgrades from cloud servers to devices that are operational in the field. This becomes challenging for classification models to incorporate behavioral changes (or new classes) dynamically without retraining the entire model. 
	
	In this chapter, we develop a modular device classification architecture that allows us to dynamically accommodate legitimate changes in IoT assets, via either the addition of a new device profile or an upgrade of existing profiles, without replacing the entire set of models. Our contributions are threefold: (1) We develop an unsupervised one-class clustering method for each device to detect their normal network behavior. We use traffic attributes identified in the previous chapter that can be obtained from flow-level network telemetry to characterize behavior of individual IoT devices; (2) We tune individual device-specific clustering models and use them to classify IoT devices from network traffic in real-time. We enhance our classification by developing methods for automatic conflict resolution and model consistency monitoring mechanism; and (3) We evaluate the efficacy of our scheme by applying it to traffic traces (benign and attack) from 12 real IoT devices, and demonstrate its ability to detect behavioral changes with overall accuracy of more than 94\%. Parts of this chapter have been published in~\cite{LCN19} and~\cite{IOTJ19}.
	
\section{Introduction}\label{sec:c5_intro}

	% growth of IoT
	%The Internet-of-Things (IoT) continues to expand its reach into homes, offices, enterprise campuses, and even cities, as more devices are rapidly connected to networks for collecting and sharing data. 
	%Network operators today are unable to identify all IoT devices on their networks, lack real-time visibility into the network behavior of  known IoT assets \cite{Cisco2017}, and are unsure whether connected devices behave legitimately or not. As a result, unmonitored IoT devices have already caused data breaches or been hijacked to carry out large-scale attacks on the Internet \cite{HamzaSOSR2019}.  
	
	% problem
	IoT devices are typically purpose built with limited functionalities -- they communicate with a specific set of endpoints (\ie servers) using a small number of TCP/UDP flows. Therefore, a growing number of traffic classification proposals are emerging based on supervised machine-learning techniques (\eg multi-class decision-trees or neural-networks) that use packet-level \cite{Meidan2017}, flow-level \cite{Ortiz2019}, or a combination of packet-level and flow-level \cite{TMC18} traffic attributes for monitoring IoTs behavioral patterns on the network. In our prior work \cite{LCN19} we showed that generating the model for multi-class classifiers becomes practically challenging  when a new device type is added to the network or the behavior of existing device types legitimately changes (due to firmware upgrades by device manufacturers) -- it is needed to regenerate the entire model of all classes. 
	In order to avoid over-fitting the generated model to specific classes, we need to carefully balance (\ie representing classes equally) the training dataset comprising instances of all device types. However, certain devices need much more instances to capture their normal behavior. 
	
	% our contributions
 	In this chapter, we employ a set of one-class clustering models (one per IoT device), and each can be independently trained and updated. Our first contribution develops an inference engine using an unsupervised one-class clustering model for each device to detect their normal network behavior using low-cost traffic attributes that can be computed from real-time flow-level telemetry. Our second contribution tunes individual device-specific clustering models and uses them to classify IoT devices types from network traffic in real-time. We enhance our classification by developing methods for automatic conflict resolution and monitoring consistency of individual models. Finally, we evaluate the efficacy of our scheme by applying it to traffic traces (benign and attack) from 12 real IoT devices and demonstrate its ability to detect behavioral changes with an overall accuracy of more than 94\%. 

	% chapter organization
	The rest of this chapter is organized as follows: In \S\ref{sec:c5_ml}, we present our dataset and traffic attributes, and build unsupervised clusters to characterize network behavior of individual IoT devices. In \S\ref{sec:c5_class}, we design and implement an inference engine composed of classification models to identify the IoT devices and detect anomalies. We also devise a scoring technique to measure the consistency of those classification models. In \S\ref{sec:c5_operation}, we evaluate the efficacy of the inference engine in classifying device types and detecting the attack, followed by comparison on one-class classification with a multi-class classification method. The chapter is concluded in \S\ref{sec:c5_con}.

\section{Clustering Flow-Level Attributes}\label{sec:c5_ml}

	In this section, we first outline our IoT dataset, network telemetry, and traffic attributes. Next, we show how clusters of attributes will characterize network behavior of individual IoT devices.
	
	\subsection{Flow-Level Telemetry and Traffic Attributes}\label{sec:c5_telemtry}
		
		\textbf{Dataset:}
		{\rev
			We use two sets of packet traces for this work, namely DATA1 (benign traffic) and DATA2 (mix of benign and attack traffic).  The first dataset (\ie DATA1) was collected from a testbed consisting of more than 30 IoT devices for a duration of six months (\ie 01-Oct-2016 to 31-Mar-2017) \cite{TMC18}. We select 12 IoT devices namely the Amazon Echo, Belkin motion sensor, Belkin switch, Dropcam, HP printer, LiFX bulb, Netatmo weather station, Netatmo camera, Samsung camera, Smart Things, Triby speaker, and Withings sleep sensor as those showed significant activities during the early period of the dataset (\ie 1-Oct-2016 to 15-Nov-2016). 
			We assumed that DATA1 does not contain any attack data and the devices we used in our testbed are not compromised. However, we allowed the devices to update their firmware automatically. Also, we aware that these data may contain unintentional connection losses for some devices due to the downtimes of cloud service.
			We evaluate (in \S\ref{sec:c5_class}) on DATA1 the efficacy of our inference engine in classifying device profiles as well as detecting their behavioral changes.
		}
		
		{\rev
			Our second dataset (\ie DATA2) contains more than eight weeks' worth of PCAP traces \cite{HamzaSOSR2019} collected from ten IoT devices (in a different environment) over two months in 2018.
			DATA2 includes normal traffic (covering boot, active, and idle operating states) and also annotated attack traffic (direct and reflective) on these IoT devices. For this dataset, we assumed that no attacks occurred during the days other than the annotated ones. In order to reduce the unintentional behavioural changes, we minimize the automatic firmware updates by manually updating the possible devices before collecting the data. 
			We use DATA2 (in \S\ref{sec:c5_attack}) to evaluate the performance of our scheme in detecting cyber-attacks that cause behavioral changes in real-time.
		}
		
		\textbf{Flow-Level Telemetry and Attributes:} 
		We showed in \hyperref[chap:characterization]{Chapter~\ref*{chap:characterization}} that individual IoT devices exhibit identifiable patterns in their traffic flows such as activity cycles and volume patterns, and profiles of signaling protocols such as DNS, NTP, and SSDP. To monitor IoT behavior on the network in real-time, we identify a set of flows (specific to each device) that collectively capture its entire traffic. These flow rules can be programmed into an SDN-enabled switch \cite{ANTS16,TeleScopeArXiv2018} through which the traffic of IoT devices passes -- rules of different devices are distinguished by a match field corresponding to device identifier (\ie MAC or IP address). Counters of these flow rules are periodically (configurable, say, every minute) measured, and will form traffic attributes of individual devices.

		\begin{table}[b]
			\centering
			\caption{Flow rules (per-device) needed for network traffic telemetry.}
			\label{tab:c5_flows}
			
			\setlength\tabcolsep{3pt}
			\begin{adjustbox}{max width=0.8\columnwidth}
				\def\arraystretch{1.1}
				\begin{tabular}{@{}lccccccc@{}}
					\toprule
					\textbf{Flow description}               & \multicolumn{1}{l}{\textbf{srcETH}}                           & \multicolumn{1}{l}{\textbf{dstETH}}                           & \multicolumn{1}{l}{\textbf{srcIP}} & \multicolumn{1}{l}{\textbf{dstIP}} & \multicolumn{1}{l}{\textbf{srcPort}} & \multicolumn{1}{l}{\textbf{srcPort}}&\multicolumn{1}{l}{\textbf{proto}}\\ \midrule
					DNS$\uparrow$               & {\fontsize{8}{20}\usefont{OT1}{lmtt}{b}{n}\noindent <devMAC>} & *                                                             & *                                  & *                                  & *                                    & 53                                  &17\\
					DNS$\downarrow$          & *                                                             & {\fontsize{8}{20}\usefont{OT1}{lmtt}{b}{n}\noindent <devMAC>} & *                                  & *                                  & 53                                   & *                                   &17\\
					NTP$\uparrow$               & {\fontsize{8}{20}\usefont{OT1}{lmtt}{b}{n}\noindent <devMAC>} & *                                                             & *                                  & *                                  & *                                    & 123                                 &17\\
					NTP$\downarrow$         & *                                                             & {\fontsize{8}{20}\usefont{OT1}{lmtt}{b}{n}\noindent <devMAC>} & *                                  & *                                  & 123                                  & *                                   &17\\
					SSDP$\uparrow$          & {\fontsize{8}{20}\usefont{OT1}{lmtt}{b}{n}\noindent <devMAC>} & *                                                             & *                                  & *                                  & *                                    & 1900                                &17\\
					remote$\uparrow$   & {\fontsize{8}{20}\usefont{OT1}{lmtt}{b}{n}\noindent <devMAC>} & {\fontsize{8}{20}\usefont{OT1}{lmtt}{b}{n}\noindent <gwMAC>}  & *                                  & *                                  & *                                    & *                                   &*\\
					remote$\downarrow$& {\fontsize{8}{20}\usefont{OT1}{lmtt}{b}{n}\noindent <gwMAC>}  & {\fontsize{8}{20}\usefont{OT1}{lmtt}{b}{n}\noindent <devMAC>} & *                                  & *                                  & *                                    & *                                   &*\\ 
					local$\downarrow$      & * & {\fontsize{8}{20}\usefont{OT1}{lmtt}{b}{n}\noindent <devMAC>}                                                             & *                                  & *                                  & *                                    & *                                   &*\\ \bottomrule
				\end{tabular}
			\end{adjustbox}
			
		\end{table}

		Table~\ref{tab:c5_flows} shows eight flow rules that we use to measure network traffic of each IoT device with the following order:
		\textbf{(1,2)} DNS outgoing queries and incoming responses on UDP 53; \textbf{(3,4)} NTP outgoing queries and incoming responses on UDP 123; \textbf{(5)} SSDP outgoing queries on UDP 1900; \textbf{(6,7)} other ``remote'' traffic (\eg Internet) outgoing from and incoming to the device that passes through the gateway; and \textbf{(8)} all ``local'' traffic (\ie LAN) incoming to the device. Note that we do not monitor SSDP traffic incoming to IoT devices to avoid capturing (and mixing) the discovery activities of other devices on the local network. Also, we do not monitor local traffic coming to IoT device as this traffic is assumed to have originated from another IoT device locally -- this way, activity of local flows is counted only for one device (receiver). We have used MAC address as the identifier of a device -- one may use an IP address (\ie without NAT), physical port number, or VLAN for a one-to-one mapping of a physical device to its traffic trace.
		
		We use two key attributes \cite{infocom17} namely \textbf{\textit{average packet size}} and \textbf{\textit{average rate}}  for each of the eight flows mentioned above. We also note that traffic attributes can better characterize individual devices if they are computed at multiple time-scales \cite{multiTime2005} particularly in the characterization of long-range dependent traffic. We, therefore, collect per-flow packet and byte counts every minute, and compute attributes at time-granularities of 1, 2, 4, and 8 minutes. This way we generate eight attributes for each flow that means a total of 64 attributes per device.
		
		\textbf{Extracting Attributes:}
		In order to synthesize flow entries and thereby extract attributes from the traffic traces, we use our native packet-level parsing tool \cite{infocom17}. It takes raw PCAP files as input, develops a table of flows (like in an SDN switch) and exports byte/packet counters of each flow at a configurable resolution (\eg 60 sec). Lastly, we generate a stream of instances (a vector of attributes periodically generated every minute) corresponding to each of the individual devices. 
		
		We begin with DATA1, and use a month's worth of its data (\ie 01-Oct-2016 to 31-Oct-2016) for training and the following two weeks for testing our models -- the second column in Table~\ref{tab:c5_classifier_parm} summarizes the number of training/testing instances per each device type contained in this part of DATA1. Later in \S\ref{sec:c5_class}, we will use the rest of DATA1 (spanning a longer period of traffic traces) to show how our models detect changes in IoT behaviors. 
			
		\begin{table}[t]
			\centering
			\caption{Summary of partial DATA1 (benign traffic): Device instances and clustering parameters.}
			\label{tab:c5_classifier_parm}
			
			\begin{adjustbox}{width=0.6\columnwidth}
				\def\arraystretch{1.5}
				\begin{tabular}{|l|cc|cc|}\hline
					& \multicolumn{2}{c|}{\textbf{Instance count}} & \multicolumn{2}{c|}{\shortstack{\\\textbf{Unsupervised classifier}\\ \textbf{parameters}}}                                                                                                                                  \\\hline
					Device        &  \shortstack{Training \\(1-month)}            & \shortstack{\\Testing \\(2-week)}            & \shortstack{\# Principal\\components} &  \shortstack{\#\\clusters} \\\hline
					Amazon Echo           &  40843 &  18694 &                    19 &              256 \\
					Belkin motion         &  35153 &  18780 &                    17 &              256 \\
					Belkin switch         &  40991 &  18771 &                    18 &              256 \\
					Dropcam               &  41089 &  18787 &                     9 &              128 \\
					HP printer            &  40713 &  18693 &                    13 &              128 \\
					LiFX bulb                 &  36952 &  18707 &                    14 &              256 \\
					Netatmo cam           &  40788 &  18706 &                    15 &              512 \\
					Netatmo weather       &  24896 &  17473 &                     9 &              128 \\
					Samsung cam           &  40841 &  18696 &                    16 &              256 \\
					Smart Things          &  41073 &  18799 &                    13 &              256 \\
					Triby speaker         &  31898 &  18694 &                    15 &              256 \\
					Withings sleep sensor &  32033 &  10877 &                    12 &              128 \\ \hline   
				\end{tabular}
			\end{adjustbox}
			
		\end{table}
	
	\subsection{Attributes Clustering}
	
		Our primary objective is to train a number of models (one per IoT device) where each model recognizes traffic patterns of a particular device type (\ie class) and rejects data from all other classes -- \ie one-class classifier generates ``positive'' outputs for a known/normal instances, and ``negative'' otherwise. This approach enables us to re-train each model independently (in case of legitimate changes). Also, it has been shown that device-specialized models can better detect anomalous traffic patterns (outliers) \cite{HamzaSOSR2019}. There are a number of algorithms for one-class classification. One of the most common and efficient methods is K-means~\cite{Tegeler12} which finds groups of instances (\ie ``clusters'') for a given class that are similar to one another. Each cluster is identified by its centroid, and an instance is associated with a cluster if the instance is closer to the centroid of that cluster than any other centroids. 
		
		To provide insights into traffic characteristics of IoT devices, we show in Fig.~\ref{fig:c5_cluster} clusters of instances for three representative devices namely, the Amazon Echo, Belkin switch, and LiFX bulb from our dataset. Note that our instances are multi-dimensional (\ie 64 attributes), and thus can not be easily visualized. Therefore, we employ the Principal Component Analysis (PCA) to project data instances to a two-dimensional space just for illustration purposes -- data instances are shown as dots and cluster centroids are shown as crosses. Note that only 10\%  of instances are shown in each cluster for better visualization -- as an example, four dots in cluster A1 of Amazon Echo, shown in Fig.~\ref{fig:c5_cluster}.(a), approximately represent 40 instances. 
		
		\begin{figure}[t]
			\vspace{-1em}
			\begin{center}
				\centering
				\includegraphics[width=1\textwidth]{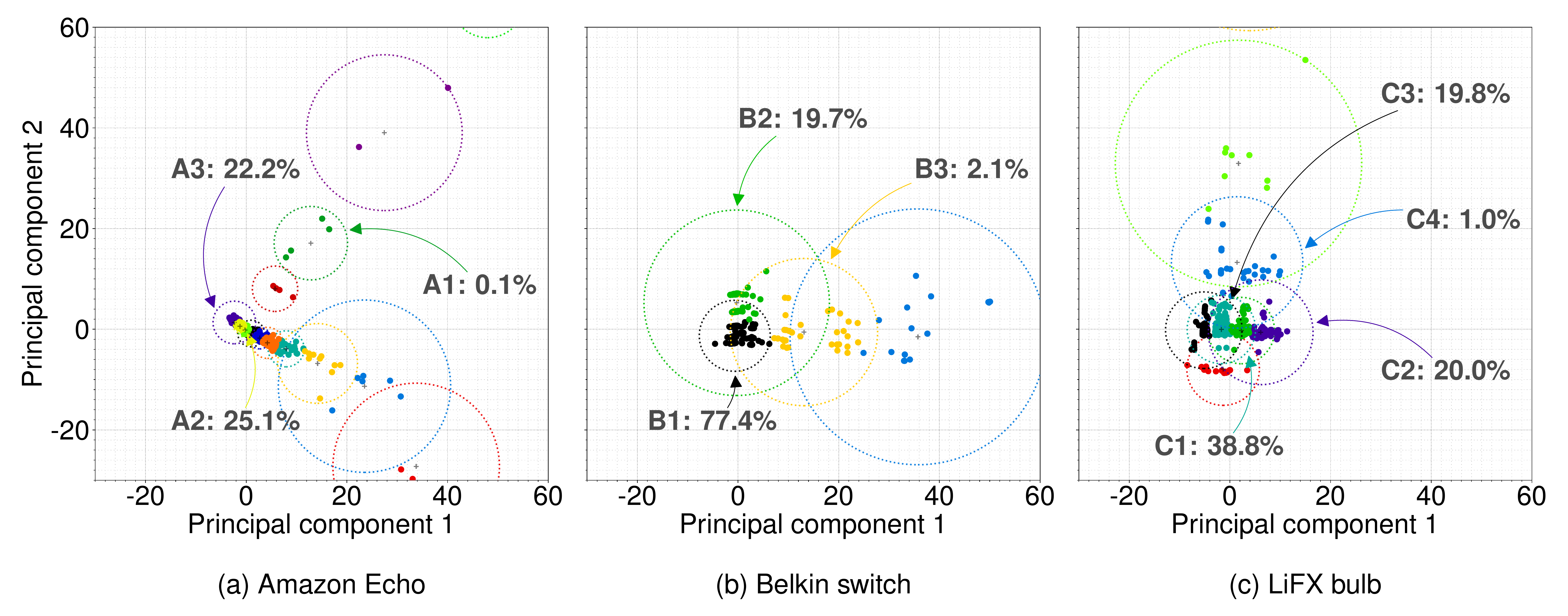}
				
				\captionof{figure}{Clusters of data instances in two-dimensional space for representative IoT devices: (a) Amazon Echo; (b) Belkin switch; and (c) LiFX bulb.}
				\label{fig:c5_cluster}
			\end{center}
			\vspace{-1em}
		\end{figure}		
		
		Dotted circles depict the boundary of clusters. These boundaries will be used to determine if a test instance belongs to clusters of a class or not. As per a rule of thumb for finding outliers \cite{miller1993tutorial}, a boundary for each cluster is chosen in a way to exclude data points whose distance from the centroid is relatively large (\ie values more than 1.5 times the interquartile range from the third quartile). In other words we define the boundary for each cluster that covers the first 97.5\%  \cite{Ruan2005} of data points closest to the cluster center and exclude farther instances to avoid impurities in our training dataset.
		
		It is important to note that an actual cluster forms a contour (enclosing associated data points) which could form a complex shape. Given that our individual models consist of tens of clusters we approximate the shape of their contours, and hence make our classification scheme computationally cost-effective and more efficient. Furthermore, K-Means algorithm attempts to partition the training dataset into spherical clusters when it is tuned optimally (\ie equal distance from centroids in all dimensions). This way, spherical boundaries will be easily used to determine if a test instance belongs to clusters of a class or not.
		
		From Fig.~\ref{fig:c5_cluster}, it is seen that instances of Amazon Echo, Belkin switch, and LiFX bulb are grouped into 16, 4, and 8 clusters, respectively. We observe that instance clusters of Amazon Echo are fairly spread across the 2D space. For Belkin switch, clusters are mainly spread across the principal-component-1 while their principal-component-2 is limited between $-20$ and $20$. Lastly, LiFX bulb instances are spread along the principal-component-2, while limited between $-20$ and $20$ in the principal-component-1. Note that each cluster of a class has a probability (``cluster likelihood'') of covering training instances from the corresponding device type, depending upon device traffic patterns seen in the training dataset. As annotated in Fig.~\ref{fig:c5_cluster},  highly  probable clusters for Amazon Echo are A2 ($25.1$\%) and A3 ($22.2$\%), for Belkin switch are B1 ($77.4$\%) and B2 ($19.7$\%), and for LiFX bulb are C1 ($38.8$\%) and C2 ($20.0$\%). These clusters highlight the dominant traffic characteristics of their respective device.
		\begin{figure}[t]
			\vspace{-1em}
			\begin{center}
				\centering
				\includegraphics[width=1\textwidth]{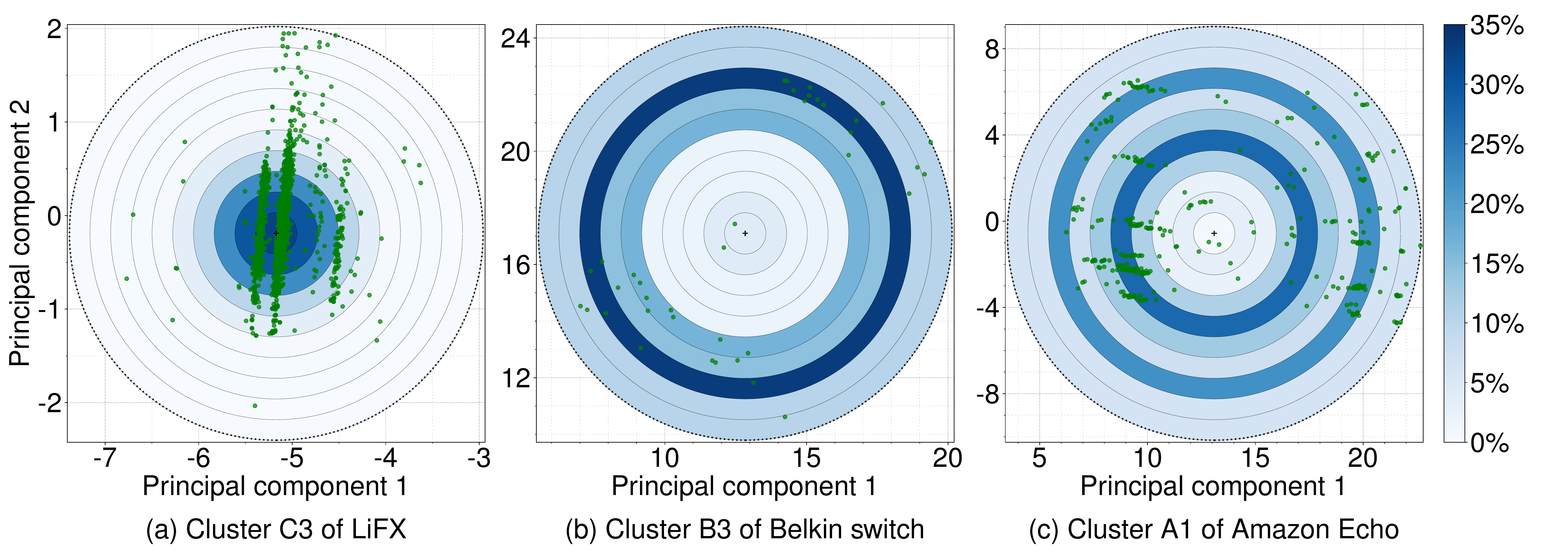}
				
				\captionof{figure}{Distance probability of clusters: (a) C3 of LiFX bulb;(b) B3 of Belkin switch; and (c) A1 of Amazon Echo.}
				\label{fig:c5_distp}
			\end{center}
			\vspace{-1em}
			
		\end{figure}	 
		
		We also note that distribution of instances within each cluster also varies across clusters. Fig.~\ref{fig:c5_distp} shows a zoomed version of one cluster from each of our three representative IoT devices -- instances are shown by green dots. Each cluster is divided into 10 equal bands starting from the centroid to the cluster boundary. For each band, we compute a probability that indicates the fraction of training instances it covers. The probability of bands is color coded on a linear scale (\eg dark blue indicates a higher probability).

		It can be seen in Fig.~\ref{fig:c5_distp}a that $95$\% of LiFX instances inside cluster C3 fall under four central bands of this cluster. Moving to B3 of the Belkin switch in Fig.~\ref{fig:c5_distp}b, we observe that $81$\% of instances fall in middle bands (from 4th to 8th). Lastly, looking at a less probable cluster of Amazon A1 in Fig.~\ref{fig:c5_distp}c, $85$\% of instances are covered by the last five bands far from the centroid. We would like to reiterate that the 2D space is used here for illustration purposes only. In our classification scheme, we employ a hyper-sphere in  64-dimensional space for clustering instances of IoT traffic attributes.

\section{Unsupervised Classification of IoT Devices}\label{sec:c5_class}

	In this section, we describe the architecture of our inference engine which consists of a set of one-class models for individual device types. Next, we develop methods to resolve conflicts between multiple models for device classification. Finally, we develop a scoring technique to measure the consistency of the models in classifying IoT devices, identify two monitoring phases namely initial and stable, and detect behavioral changes.
	
	\subsection{Clustering Models: Generation, Tuning, and Testing}
	
		Prior to generating clustering models, we need to pre-process our raw dataset.  
		First, we normalize each attribute independently to avoid outweighing large-value attributes (\eg average bytes rate of incoming remote traffic at 8-min timescale) over smaller attributes (\eg average packet size of outgoing NTP traffic at 1-min timescale) \cite{Mohamad2013} as the scale of value for different attributes varies significantly (\ie several orders of magnitude). We employ the Z-score method (\ie computing $\mu$ and $\sigma$ from the training dataset) to scale individual attributes. Second, we project data instances into a lower dimension space by using PCA \cite{Ding2004} which results in linearly uncorrelated principal components. This is because our data is 64-dimensional which can be computationally expensive for real-time prediction, and also affect the clustering performance (possibly getting biased towards less significant attributes). The orthogonal components enable K-means to detect clusters more clearly by removing redundant and noisy attributes of training dataset. We choose the number of PCA components to retain optimum ``cumulative variance'' \cite{achen_1990} for our dimension reduction engine.
		\begin{figure}[t]
			\begin{center}
				\centering
				\includegraphics[width=0.6\textwidth]{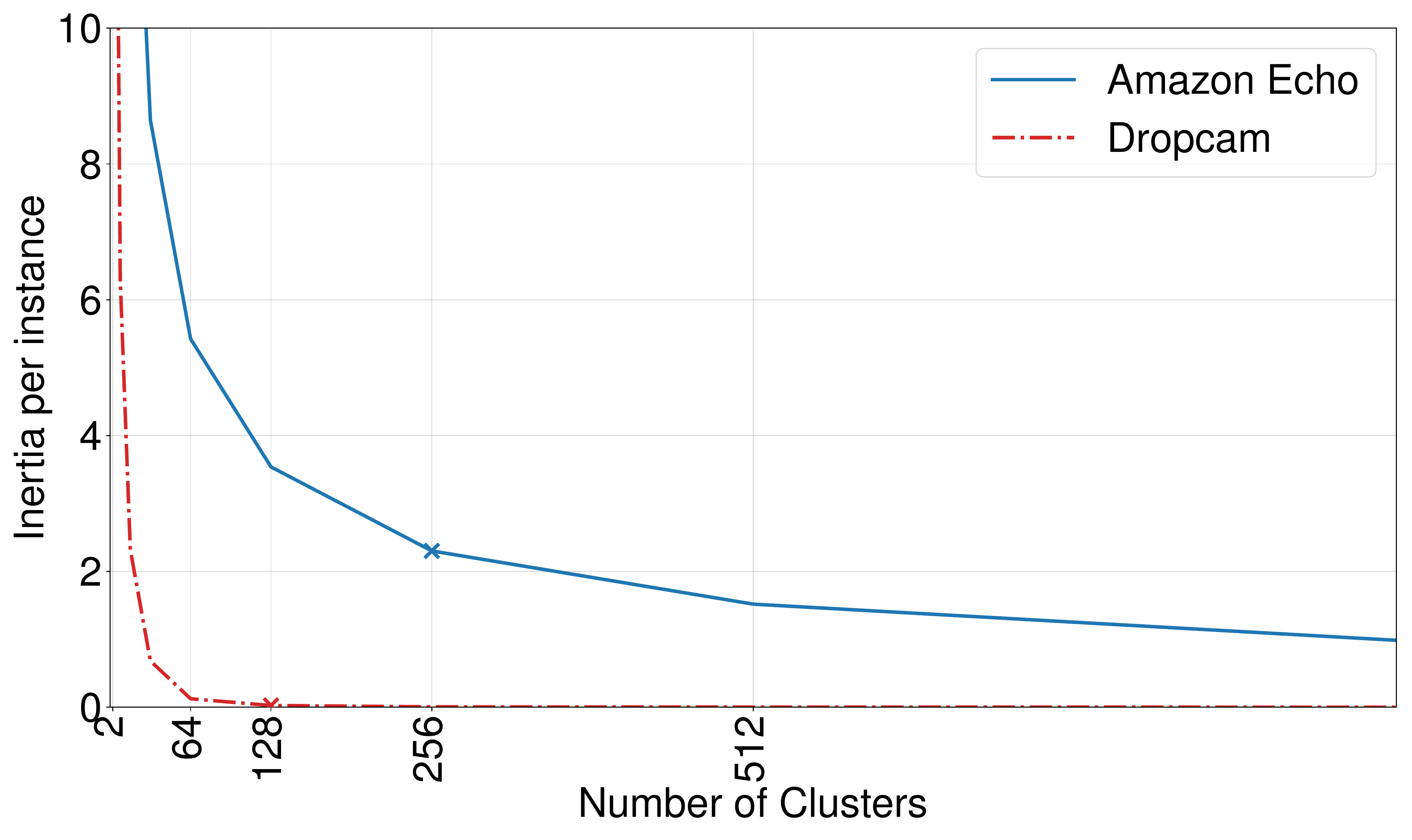}
				\captionof{figure}{Elbow method for selecting optimal number of clusters.}
				\label{fig:c5_elbow}
			\end{center}
			\vspace{-1em}
		\end{figure}
		
		Following dimension reduction, we apply K-means algorithm with varying $K$ values in the power-of-2 (\ie $2^i$ where $i=1,...,10$).  Note that setting $K$ to small values would not generate an accurate model of network behavior for IoT devices, and large values increase the computational cost in both training and testing phases. Also, a very large $K$ results in smaller size clusters, and hence a rigid classifier which cannot correctly detect normal (legitimate) instances with small deviations from the training data -- \ie over-fitting. We find the optimal number of clusters using the elbow method~\cite{David1996}. Fig.~\ref{fig:c5_elbow} shows the average square distance of instances from the cluster centers (\ie Inertia per instance) versus clusters count for two representative device types. The optimal cluster number (marked by `$\times$' on each curve) is chosen when the first derivative of inertia per instance exceeds a very small negative value  $-0.01$ (almost flat) when increasing clusters count. It can be seen that the model for Amazon Echo needs 256 clusters for optimal performance, and this measure is  128 clusters for Dropcam. We show in the rightmost  column of Table~\ref{tab:c5_classifier_parm}, the model parameters for individual device types that are obtained from the methods mentioned above.

		Having clustering models generated, we test an instance of IoT traffic attributes after scaling and dimension reduction, as shown by a sequence of steps in Fig.~\ref{fig:c5_oneclass}. The test instance is presented to all of the device-specific models to find the nearest centroid of each model -- the minimum of euclidean distances between the test instance and clusters centroid is chosen. Given a nearest centroid, the instance is checked against the corresponding cluster to determine if it falls inside or outside of that cluster boundary, and if inside, compute a confidence level. To better illustrate this process, let us consider the two-dimensional space of clusters we discussed earlier in Fig.~\ref{fig:c5_cluster}. Assume that a test instance has its principal component-1 and component-2 equal to 0 and 20, respectively.  The nearest centroids to this test instance are A1 of Amazon Echo, B2 of Belkin switch, and C4 of LiFX. Since the test instance falls outside of A1 boundary, the Amazon Echo model results in a negative output while the other two models give positive outputs. In what follows next, we show how to select the ``winner'' model in case of multiple positive outputs for a test instance.
		
		\begin{figure}[t]
			\begin{center}
				\includegraphics[width=0.40\textwidth]{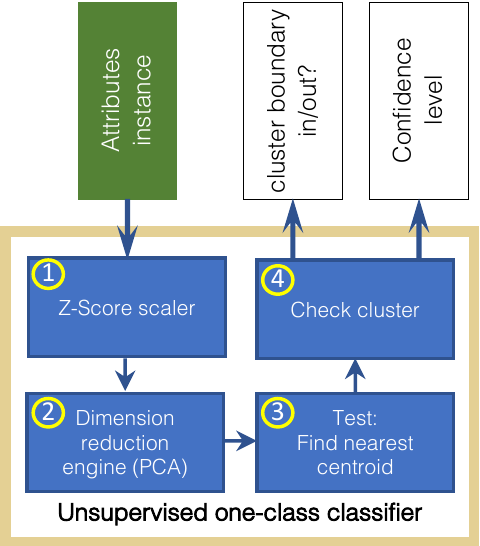}\quad
				\caption{Use of each clustering model for a test instance.}
				\label{fig:c5_oneclass}
			\end{center}
		\end{figure}
	
	\vspace{-1em}
	\subsection{Conflict Resolution}\label{sec:c5_confidence}
		Each model learns the normal behavior of one device type. Different devices may display a slightly similar traffic behavior (\eg DNS, NTP or SSDP) for a short period of time \cite{Ortiz2019}. This can result in multiple positive outputs generated by our clustering models for an instance. We address this issue by using the ``confidence'' of models which give positive output: the model with the highest confidence-level is selected as the winner.
	
	\textbf{Confidence-level:}
		We derive a probability value for test instances to be associated with a cluster of that model - we call it ``associate probability''. Given an instance $Ins$ receiving a positive output from a model $M_i$ and falling in a distance band $D_l$ of the nearest cluster $C_j$ (of the model $M_i$), the associate probability is estimated by:
		\begin{equation}
		\vspace{0.5em}
		P_{[Ins|M_i(C_j(D_l))]}^{test} = P_{[C_j|M_i]}^{train} \times P_{[D_l|C_j]}^{train}
		\vspace{0.5em}
		\end{equation}
		where $P_{[C_j|M_i]}^{train}$ is the likelihood of the nearest cluster $C_j$ within the model $M_i$ and $P_{[D_l|C_j]}^{train}$ is the probability of distance band $D_l$ inside the cluster $C_i$ -- both probability values are obtained from the training dataset. 
		We note that $P_{[C_j|M_i]}^{train}$ is always non-zero (by optimal tuning \cite{Pakhira2009}), but it is possible to have $P_{(D_l|C_j)}^{train}$ equal to zero when none of the training instances fall inside a band $D_l$ (\ie unexplored distance bands in the training data). To avoid a zero confidence for test instances, we slightly modify the band probability using the Laplacean prior\cite{McCallum1998}, priming each band instances count with a count of one, as given by:
		\begin{equation}
		\vspace{0.5em}
		P_{[D_l|C_j)]}^{train} = \dfrac{1+N_{D_l}} {L+N_{C_j}}
		\vspace{0.5em}
		\end{equation}
		
		where $N_{D_l}$ is the number of training instances inside the band $D_l$; $N_{C_j}$ is the total number of instances in the cluster $C_j$; $L$ is the total count of distance bands in the cluster  -- we use ten bands in every cluster ($L=10$).
		
		The associate probability, to some extent, indicates the model confidence. However, it becomes challenging to select the winner among multiple models giving positive output since the number of clusters and also the distribution of distance bands vary across models, and hence the associate probability is scaled differently. For example, models with large number of clusters may have relatively smaller values of $P_{(C_j|M_i)}^{train}$, or a cluster with highly sparse bands would result in smaller values of $P_{(D_l|C_j)}^{train}$. 
		\begin{figure}[t]
			\begin{center}
				\includegraphics[width=0.6\textwidth]{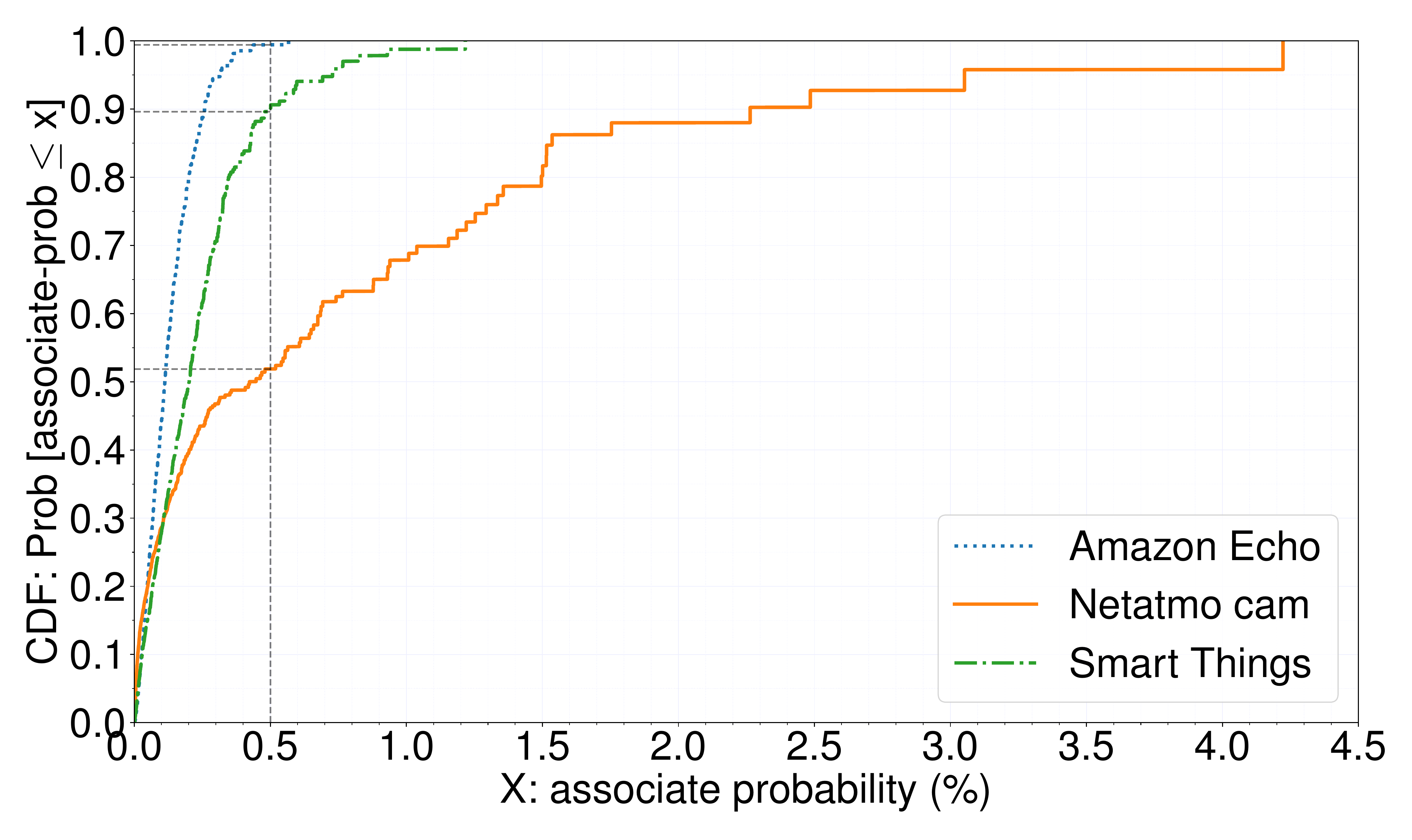}\quad
				\caption{Distribution of clustering probability for training instances of three device types.}
				\label{fig:c5_distp_score}
			\end{center}
		\end{figure}
	
		To obtain a metric of confidence for comparison across models, we scale the associate probability (computed above for a test instance) by using the distribution of this probability in a training dataset. To better illustrate this scaling process, we show in Fig.~\ref{fig:c5_distp_score} the cumulative distribution function (CDF) of the associate probability for training instances from three IoT models namely, Amazon Echo, Netatmo cam, and Smart Things. For example, we observe that the associate probability equals to 0.5\% is a high value for Amazon Echo model (shown by dotted blue lines) keeping it above more than $99$\% of training instances. However, this measures becomes $90$\% and $52$\% for Smart Things (dashed green lines) and Netatmo (solid orange) models, respectively. Therefore, given the associate probability (from a model that gives positive output for a test instance)  we derive the model confidence-level by computing the fraction of its training data that fall below the test instance (with respect to the model's empirical CDF of associate probability).

	\subsection{Consistency Score}\label{sec:c5_consistency}
	
		Ideally, for monitoring individual IoT devices we expect consistent outputs to be generated by the inference engine over time. It is important to note that a given device which is consistently and correctly classified by a model over a period of time (say, a week), may occasionally get missed (\ie negative output) by its intended model. To bootstrap the monitoring process for a newly connected (and possibly unknown) device, we initially need this device to consistently receive positive outputs from one of the existing models in order to accept the device and label it by a known class (``stable state''). Once a device becomes known (accepted) and is at its stable state, receiving negative outputs frequently from its indented model, which indicates a change (legitimate or illegitimate) in the device behavior and thus requires further investigations. 
		
		\begin{figure}[t]
			\begin{center}
				\includegraphics[width=0.7\textwidth]{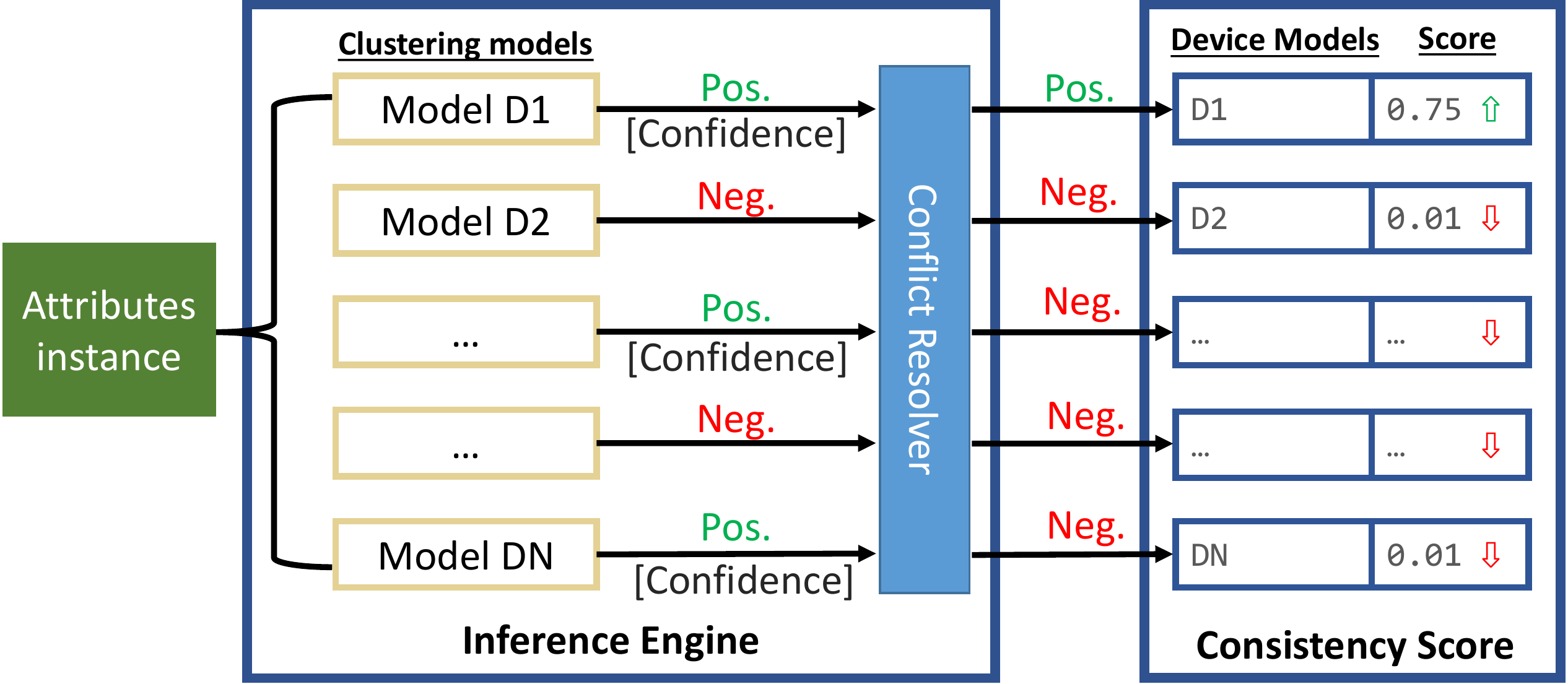}\quad
				\caption{Architecture of our inference engine.}
				\label{fig:c5_inferenceengine}
			\end{center}
		\end{figure}
	
		We develop a score (between 0 and 1) to track the consistency of our device classification -- we call it ``consistency score''. Fig~\ref{fig:c5_inferenceengine} shows the architecture of our inference engine  consisting of classification followed by consistency scoring. For instances of a given device, the consistency score is computed and tracked per each model and updated following classification of instance -- the consistency score of a model rises by its positive output and falls by its negative output over time. To better understand the dynamics of this score, let us begin with an example. We take three day's worth of  instances from Smart Things device, and present (real-time replay) this traffic to four trained models including Smart Things, Netatmo camera, Amazon Echo, and Withing Sleep sensor. We show in Fig.~\ref{fig:c5_consistency_demo} the consistency score of these four models in real-time. It can be seen that as we expect the score of intended model Smart Things (shown by solid lines) is dominant while the other three are negligible (and hence invisible). For Smart Things the score slowly rises and reaches to a high level of $0.8$ after about 30 hours. We also observe that sometimes the score of the intended model falls slightly and rises again -- this is because some instances may display patterns closer to other models. We zoom in to the gray band region (Nov 4, 11pm - Nov 5, 2am) to see the magnified score of other models. It is observed that once other models give positive output their score quickly spikes, but soon after drops back to zero  (shown by dotted red lines for Netatmo cam) as the intended model Smart Things wins again.  
		
		\begin{figure}[t!]
			\begin{centering}
				\includegraphics[width=1\textwidth]{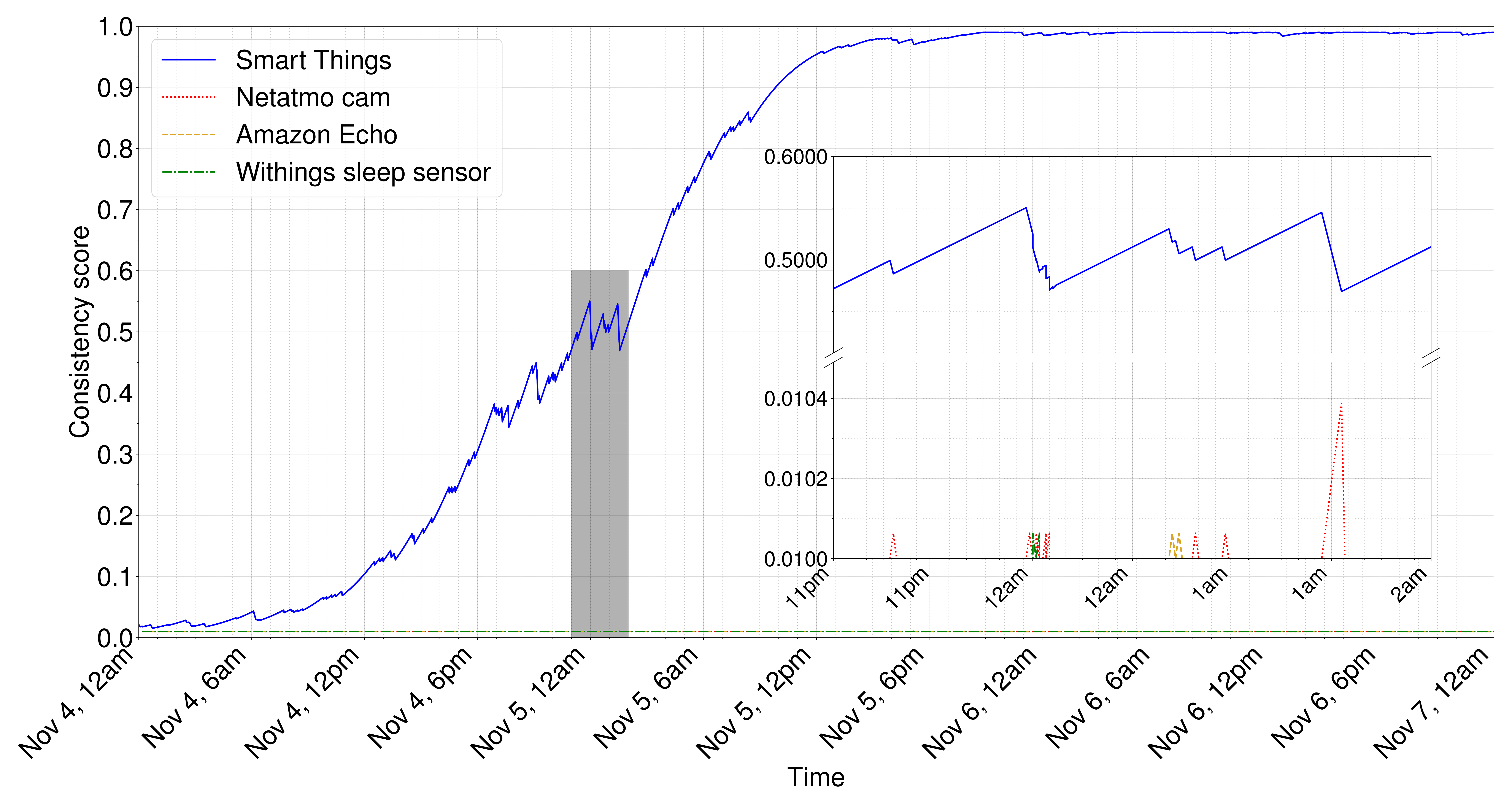}
				
				\caption{Dynamics of consistency score for Smart Things instances in real-time.}
				\label{fig:c5_consistency_demo}
			\end{centering}
		\end{figure}
	
		We update the consistency score with two rates: rising on positive output at rate $\lambda_r$ and falling on negative output at rate $\lambda_f$ -- these rates can be configured by network operators. To update the consistency score, we use sigmoid functions which are commonly used in processes like trust management \cite{Firdhous2012} as they exhibit a soft start and end, and are bounded within 0 and 1. The raw score is represented by sequence $\{S_t\}$ beginning at time $t=0$, is dynamically updated by:
		\begin{equation}
		S_t = \frac{S_{t-1} \times e^\lambda }{1+S_{t-1}\times(e^\lambda-1)}
		\label{eq:S_t}
		\end{equation}	
		where, $S_{t-1}$ is the previous value of the estimated score, and $\lambda$ is set dynamically depending on the latest output of the model (\ie $\lambda_r>0$ for positive output, and $\lambda_f<0$ for negative output). %The rising and falling rates are configurable parameters which can be decided by network operators. 
		Network operators can configure their $\lambda_r$ and $\lambda_f$ based on their preferred policy in terms of how quickly (or slowly) they want to rise/fall the score. Depending upon the time expected $T$ to reach to a ``target score'' $S^{*}$  (between 0 and 1) from the mid-level score 0.50, we derive the value $\lambda$ by: 
		\begin{equation}
		\lambda = \frac{\log(\frac{S^{*}}{1-S^{*}})}{T}
		\label{eq:lamda}
		\end{equation}
		\begin{figure}[t!]
			\begin{center}
				\includegraphics[width=0.6\textwidth]{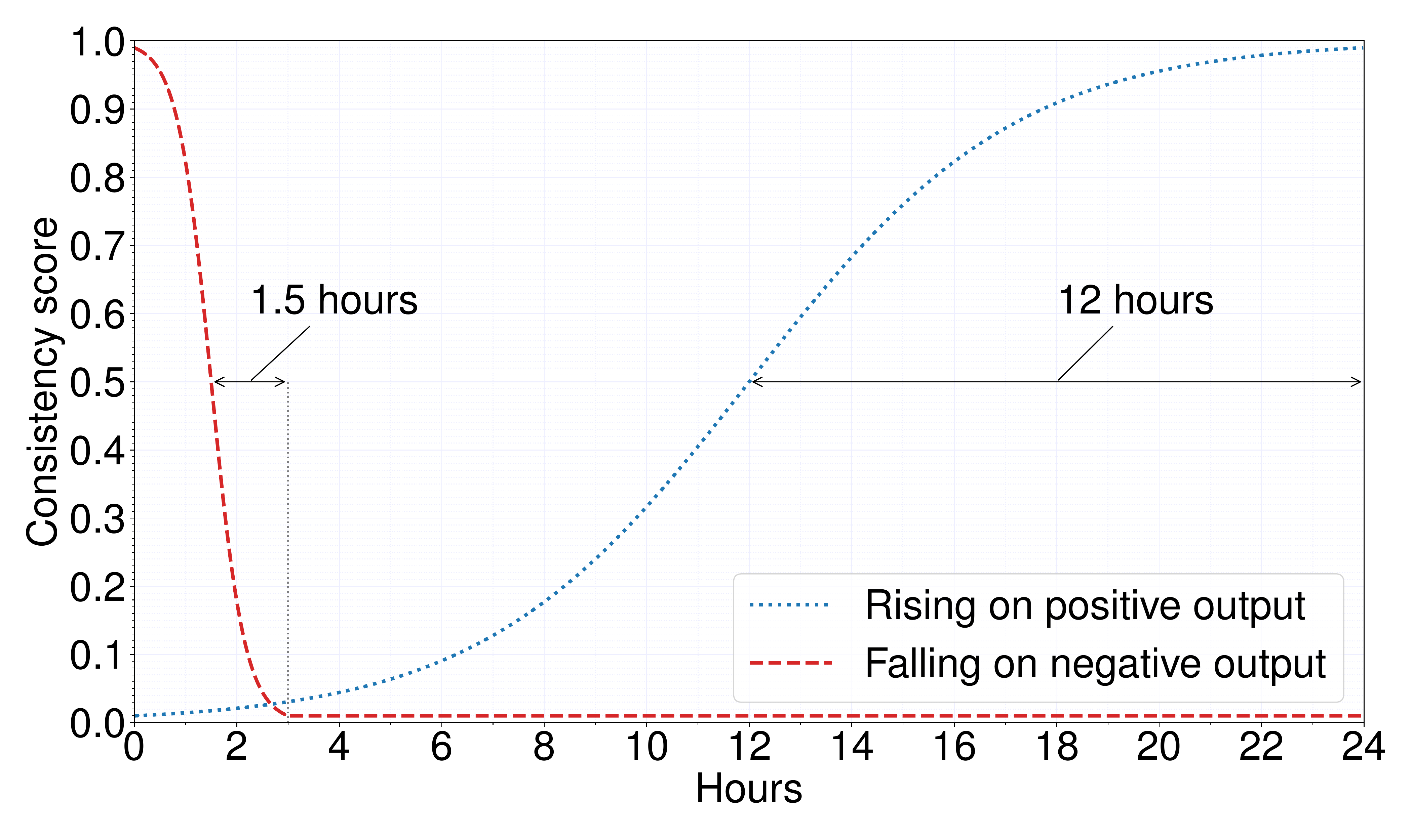}
				\caption{Real-time update rate of consistency score -- for a given model, it falls fast (from highest to lowest value in 3 hours) on continuous negative outputs, and rises slowly (from lowest to highest value in 24 hours) on continuous positive outputs.}
				\label{fig:c5_example_trust_curve}
			\end{center}
			
		\end{figure}

		In this chapter, we choose a conservative policy whereby the consistency score rises at slower rate than it falls. Our $\lambda$ values are the same for all models and configured in such a way that it will take 12 hours to reach to a very high score 0.99 from the score 0.50 ($\lambda_r=0.0064$) in case of successive positive outputs from the model, while  it will need only 1.5 hours to reach a very small score 0.01 from the score 0.50 ($\lambda_f=-0.0511$) in case of successive negative outputs. Fig.~\ref{fig:c5_example_trust_curve} shows two sample curves of consistency scores with our chosen $\lambda$ values, each is monotonically rising and falling on successive positive and negative outputs respectively. We note that both curves saturate (reaching to ultimate values 0 and 1) in infinite time, and change very slowly beyond certain levels, \ie above $0.99$ for the rising curve and below $0.01$ for the falling curve. In other words, entering into these regions can stifle agility of our real-time monitoring (specially for detecting attacks in real-time). For example, in order to fall from 0.999 to 0.99 it will take at least 45 minutes (half the time needed to fall from 0.99 to 0.50). Similarly, it needs 6 hours to rise from 0.001 to 0.01. Therefore, we cap the score at $0.99$ and $0.01$ as our saturation levels, and also initialize the score by $S_{0}=0.01$.

	\subsection{Monitoring Phases}
		For monitoring behavior of each IoT device we consider two phases: (1) initial phase, and (2) stable phase. Initial phase begins once a device connects to the network for the first time (discovered). During this phase, our inference engine (shown in Fig.~\ref{fig:c5_inferenceengine}) aims determine the device type (classification) by asking all existing models. To achieve this aim, every instance of the device traffic is fed to all models in real-time and their outputs are obtained. If multiple models give positive outputs, then our conflict resolution is applied to choose a winner model. During this phase, the consistency score of all winner models (giving positive output) is tracked till the score of one model reaches an acceptable level (\ie a threshold chosen by the network operator, say 0.90) whereby the device type is verified. At this point the device gets labeled by a known class, its intended model is determined, and its initial phase completes. Stable phase begins upon completion of the initial phase. In the stable phase, the inference engine  uses only the intended model to monitor the real-time behavior of the device. In the next section, we see how the consistency score of the intended model will be used to detect any change of behavior.

\section{Performance Evaluation}\label{sec:c5_operation}

	We now evaluate the efficacy of our inference engine. First, we evaluate the performance of one-class models and conflict resolution in selecting an intended model for a given device during its initial phase of monitoring. Once the device type is classified (with sufficiently high level of consistency), we next demonstrate behavioral changes using temporal consistency score of the intended model during its stable phase of monitoring. Finally, we show the efficacy of our inference engine. We also compare our one-class classification with a multi-class classification method. 

	\subsection{Device Classification}
		We begin by evaluating the performance of device classification using part of test instances from DATA1 (\ie only two weeks spanning from 1-Nov-2016 to 14-Nov-2016). We show in Fig.~\ref{fig:c5_confmap} the confusion matrix of classification before and after resolving conflicts. Every clustering model (listed in rows) is presented by test instances of IoT devices (listed in columns). For a given cell, the value indicates the percentage of instances (from the device in corresponding column) that receive positive output from the model in the corresponding row.
		
		\begin{figure}[t!]
			\begin{center}
				\subfloat[Raw output of clustering models.]{
						\includegraphics[width=0.6\textwidth]{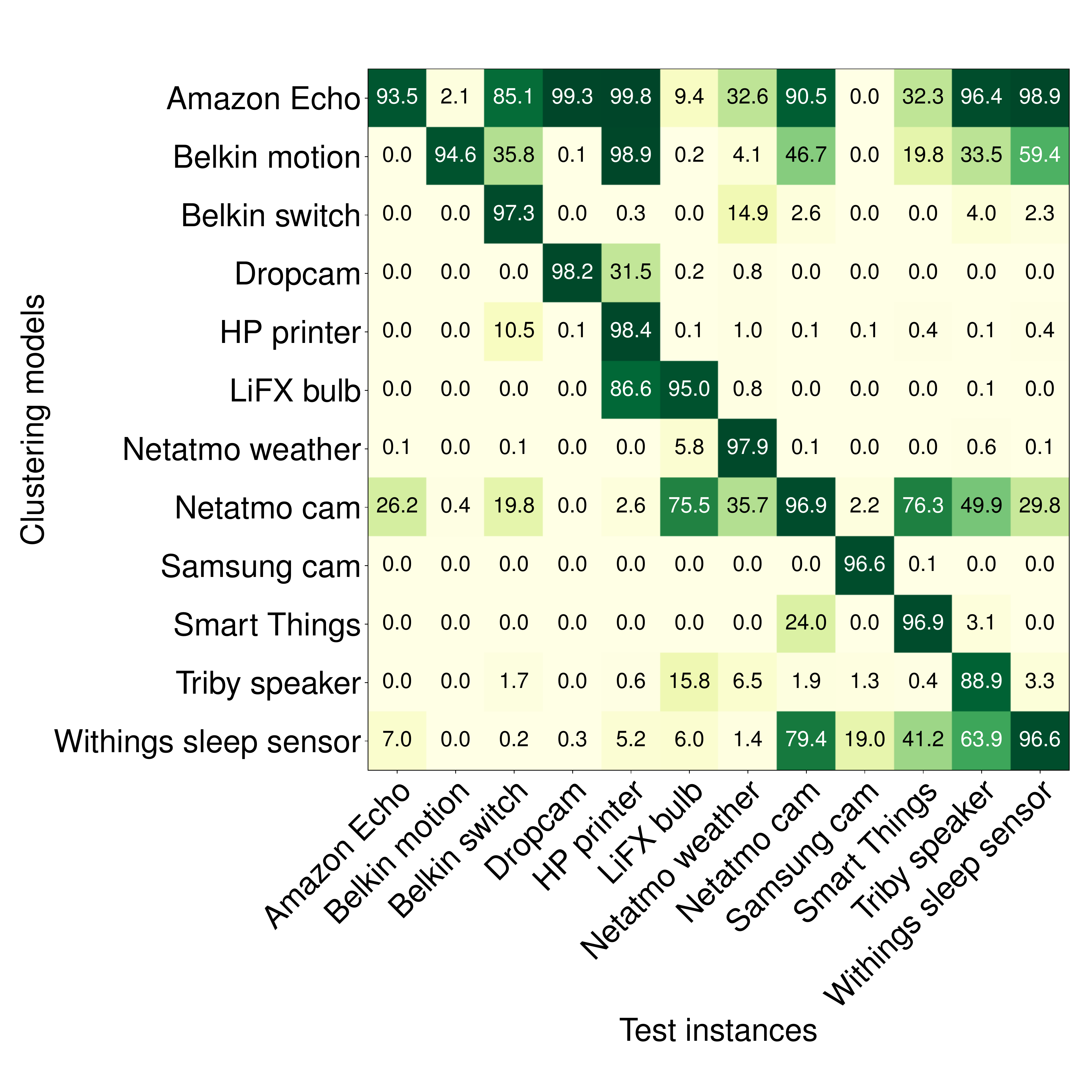}
						\label{fig:c5_conf_oneclass}
				}\qquad
				\subfloat[Refined output after conflict resolution.]{
						\includegraphics[width=0.6\textwidth]{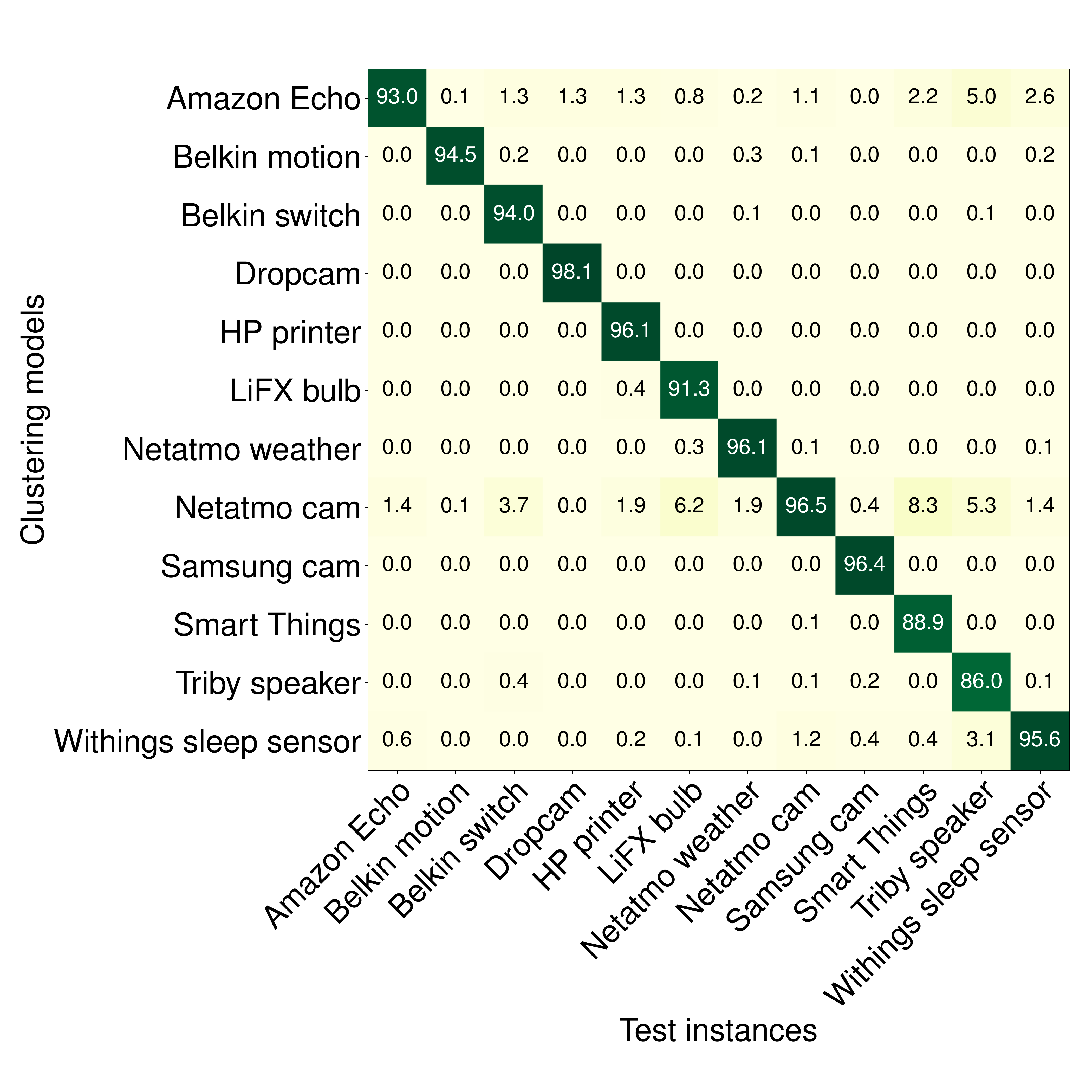}
						\label{fig:c5_conf_combine}
				}
				
				\caption{Confusion matrix of device classification: (a) raw output of clustering models; and (b) refined output after conflict resolution.}
				
				\label{fig:c5_confmap}
			\end{center}
		\end{figure}
			
		Starting from raw outputs in Fig.~\ref{fig:c5_conf_oneclass}, it can be seen that all models correctly detect majority of instances from their own class as shown by diagonal elements of the confusion matrix -- except Triby speaker with $88.9$\%, others display more than $93.5$\% of correct detection (\ie true positive). However, we observe that models incorrectly detect device instances from other classes (\ie false positive) as shown by non-diagonal elements of the confusion matrix. For example, models for Amazon Echo and Belkin motion incorrectly give positive output to $99.8$\% and $98.9$\% of instances from HP printer. Considering the raw outputs of various models,  we found $70$\% of test instances are detected by more than one model (in addition to their expected model), and $2$\% of test instances are not detected by any of the models. Next, we select the winner model for each test instance using model confidence-level (\S\ref{sec:c5_confidence}). 
	
		Fig.~\ref{fig:c5_conf_combine} shows the confusion map after conflict resolution. It clearly shows a significant enhancement in performance of our device classification by selecting the model with the highest confidence. Note that the average false positive rate has reduced to less than $0.4$\% while the average true positive rate is $93.9$\%. 
	
		Also, we observe that the conflict resolver has slightly reduced the rate of true positive for almost all models. Note that Smart Things is impacted more compared to other models by experiencing  a drop from $96.9$\% to $88.9$\% in its true positive rate largely because of the Netatmo camera model which gives positive output with high confidence for $8.3$\% of Smart Things instances. Focusing on the model of Netatmo camera, we found that its clusters overlap with a number of clusters of several devices such as Belkin switch, LiFX, Smart things, and Triby speaker, and hence results in false positives. This is mainly because of the aperiodic behavior of the Netatmo camera which is event triggered -- camera transmits video to its cloud server whenever it recognizes a human face or detects a motion. As a result, it displays a wider range of activity pattern at longer time scales, overlapping with traffic patterns of other devices. For example, the average byte rate of incoming NTP traffic of Netatmo camera at 8-min timescale can take a value from $0$ to $700$ bytes-per-min. Such a wide range overlaps with the value range of the same attribute for LiFX bulb (varying between $8 - 12$ bytes-per-min) and Smart Things (varying between $15 - 25$ bytes-per-min).

	\vspace{-1em}
	\subsection{Detecting Behavioral Change}
	\vspace{-1em}
		\begin{figure}[t]
			\begin{center}
				\subfloat[Belkin switch.]{
					\includegraphics[width=1\textwidth]{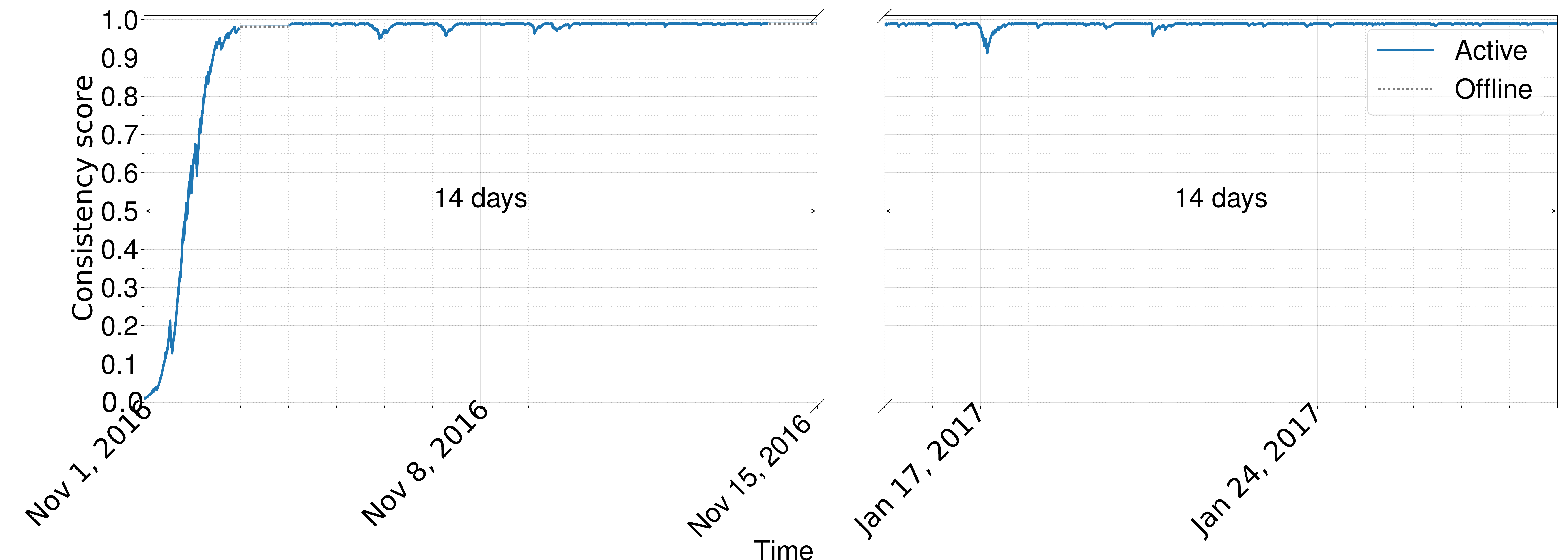}\quad
					\label{fig:c5_consistency_belkin_switch}
				}
				\qquad
				\subfloat[Triby speaker.]{
					\includegraphics[width=1\textwidth]{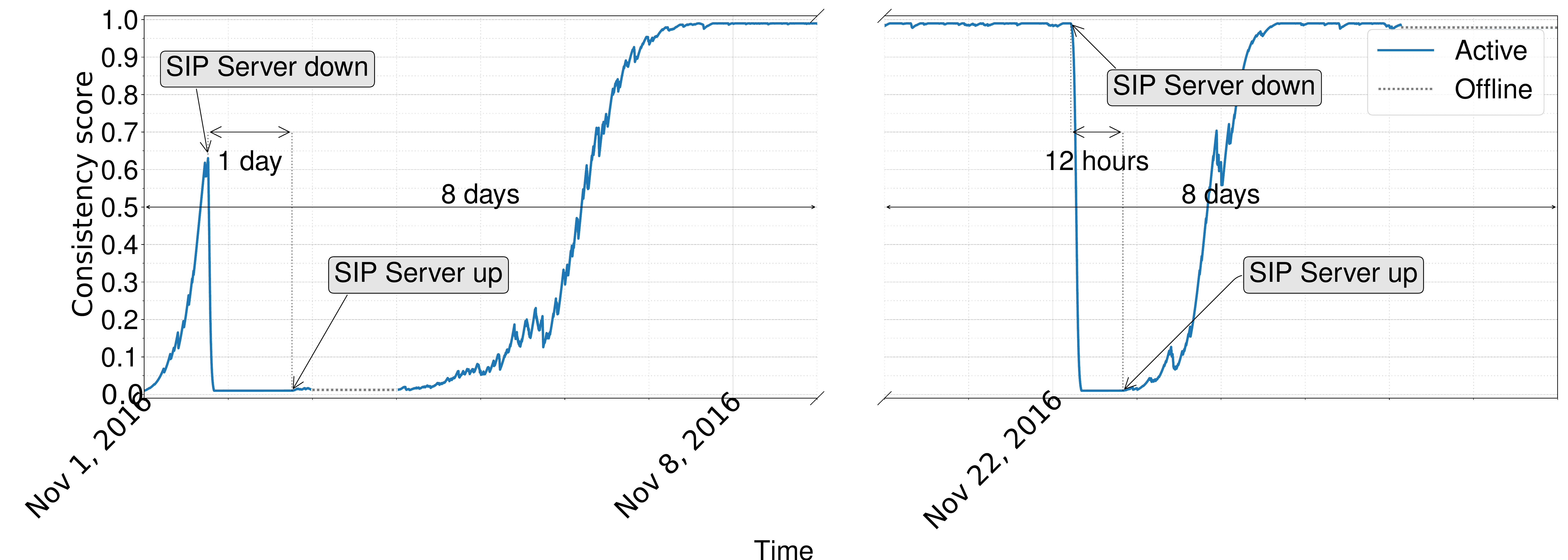}\quad
					\label{fig:c5_consistency_triby}
				}
				
				\caption{Time-trace of consistency score for normal behavior in: (a) Belkin switch; and (b) Triby speaker.}
				
				\label{fig:c5_consistency_curves}
			\end{center}
			\vspace{-1em}
		\end{figure}
		In the previous subsection, we showed the efficacy of our system in classifying individual device instances using an array of models. Once classified, we monitor activity of each IoT device in real-time using its intended model. We now check how our models highlight behavioral changes by tracking dynamics of their consistency scores (\S\ref{sec:c5_consistency}). For this evaluation, we use a longer portion of DATA1 spanning from 01-Nov-2016 to 31-Mar-2017.
		
		Fig~\ref{fig:c5_consistency_belkin_switch} shows the consistency score of our inference engine, for traffic instances of Belkin switch over a period between Nov 1, 2016 and Jan 28, 2017. It is seen that the score ramps up to $99$\% within the first 48 hours --  the device then goes offline for a day as shown by dashed gray lines, and comes back online on Nov 4. After that device instances are consistently detected by the intended model, and hence the score remains high with minor changes over this long period.

		Fig~\ref{fig:c5_consistency_triby} illustrates a scenario where consistency score drops for a relatively short period of time (due to temporary change of behavior in traffic of Triby speaker), and rises afterwards. We manually inspected packet traces corresponding to these temporary drops of the score. We found that a remote SIP server ({\myverb{sip.invoxia.com}}), with which Triby speaker keeps a continuous TCP connection, was responding with ACK/RST packets during those periods, as highlighted by red rows in Fig.~\ref{fig:c5_tribyWireshark}, indicating the expected SIP service was not operational. Note that other services for Triby speaker were functional normally.
		\begin{figure}[t]
			\begin{center}
				\includegraphics[width=1\textwidth]{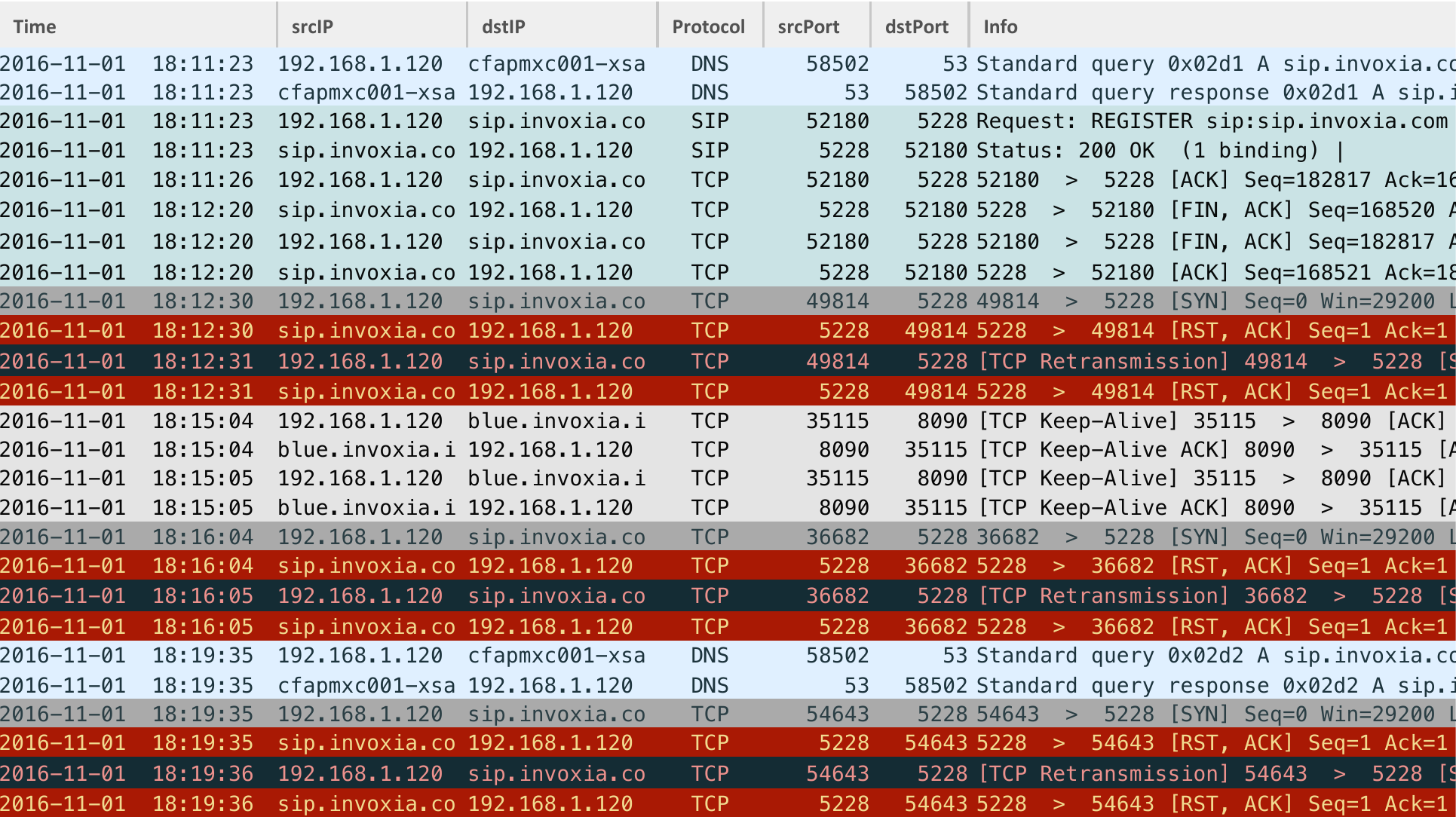}\quad
				\caption{Wireshark capture of Triby speaker packets showing outage of SIP server.}
				
				\label{fig:c5_tribyWireshark}
			\end{center}
		\vspace{-1em}
		\end{figure}

		\begin{figure}[t]
			\begin{center}
				
				\subfloat[Original model (Dropcam).]{
					\includegraphics[width=1\textwidth]{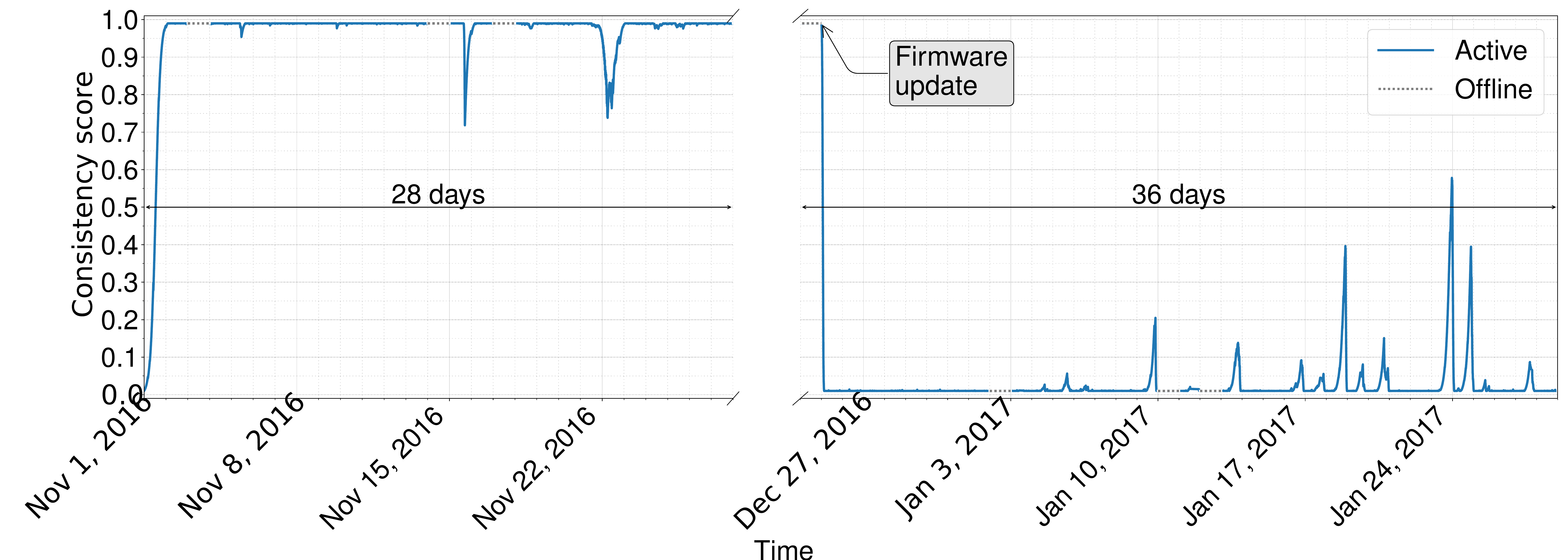}
					\label{fig:c5_consistency_dropcam_change}
				}
				\quad
				\subfloat[Re-trained model (Dropcam).]{
					\includegraphics[width=1\textwidth]{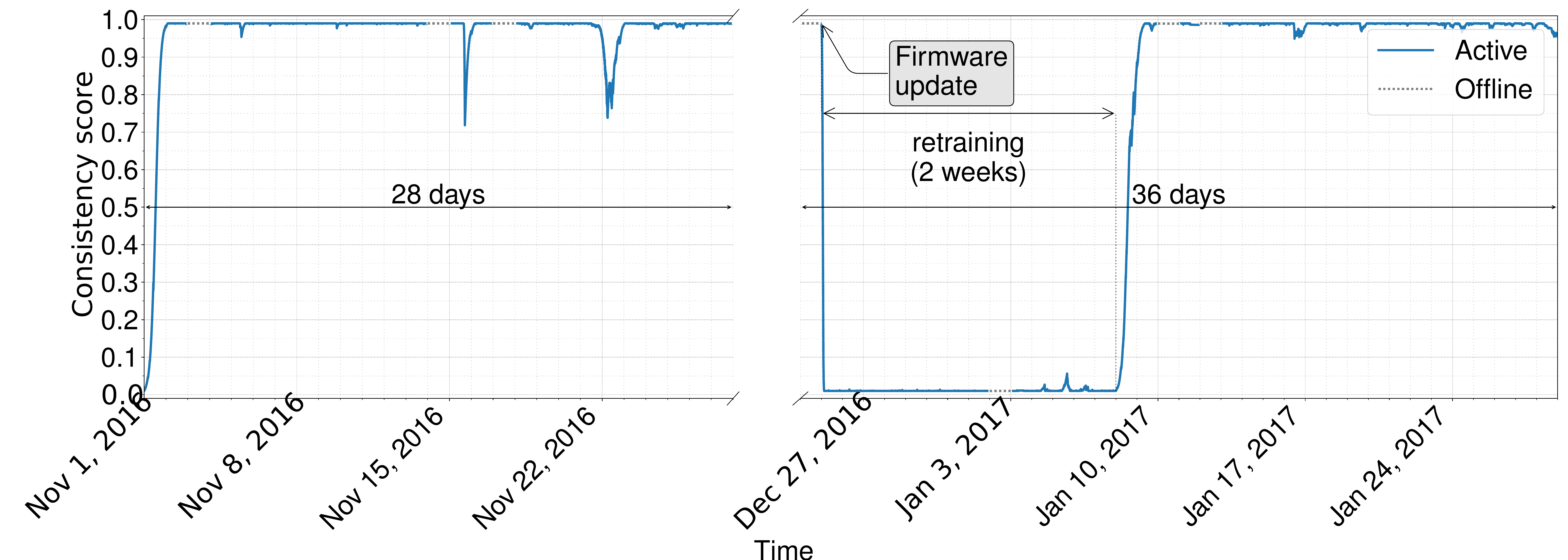}\quad
					\label{fig:c5_consistency_dropcam_fixed}
				}

				\caption{Time-trace of consistency score due to firmware upgrade in Dropcam traffic: (a) original model; and (b) re-trained model.}
				
				\label{fig:c5_consistency_dropcam}
			\end{center}
		\vspace{-1em}
		\end{figure}

		We also note that the behavior of a device may change permanently due to firmware upgrade. Fig~\ref{fig:c5_consistency_dropcam_change} illustrates this scenario for Dropcam. We can see that the consistency score of Dropcam model remains high throughout Nov 2016 until Dec 5, 2016 when the device goes off-line, as shown by dotted gray lines  --  a couple of slight drops are observed on Nov 15 and Nov 22 which get restored fairly quickly (infrequent mis-classification is not surprising due to minor overlaps between clusters of various models). However, once Dropcam comes back online on Dec 25, the score steeply drops to its lowest possible value $0.01$ and stays at that level during the whole January. It is seen that the score sometimes jumps up to $0.20$ or even $0.30$, but it quickly drops back to its minimum value. We again manually inspect the packet traces of Dropcam and found its behavior permanently changed. Note that the Dropcam network activity is dominated by a single  TLS connection which the device establishes with its cloud server ({\myverb{nexus-us1.dropcam.com}}) \cite{TMC18}, typically sending packets of size $156$ bytes and receiving packets of size $66$ bytes. Manual inspections revealed that the rate of packets for this flow changed in both directions (while packet sizes remained unchanged), resulting in a decrease of upstream bitrate from $1896$ bps to $1120$ bps and downstream bitrate from $584$ bps to $424$ bps. It is also important to note that the firmware upgrade for Dropcam is done automatically when it reboots.
		{\rev
			Confirming firmware updates (after they occur) requires manual inspection only when our inference engine flags anomalies. Note that automatic detecting of firmware updates requires labeled instances which are not available to us -- each manufacturer uses a different method to update device firmware remotely. Therefore, we are not able comment on the ability of our attributes in detecting firmware updates.
		}
		Once it is confirmed, we re-train the Dropcam model using an additional two weeks' worth of data (between Dec 25, 2016 and Jan 07, 2017) after this firmware upgrade -- adding new instances to training dataset resulted in an increase in the number of PCA components (from 9 to 11) for Dropcam, while the number of clusters remained the same. Fig.~\ref{fig:c5_consistency_dropcam_fixed} shows how consistency score returns back to its perfect level after augmenting the Dropcam model with attributes of the firmware upgrade -- the score is shown by dashed gray lines during the two weeks of re-training period.
		
	\subsection{Detecting Attacks}\label{sec:c5_attack}
	
		We now evaluate the performance of our inference engine against attack traffic. For this evaluation, we use our second dataset DATA2. It consists of well-annotated attack and benign traffic of ten real IoT devices namely Amazon Echo, TPlink switch, Belkin motion sensor, Belkin switch, LiFX bulb, Netatmo camera, Hue bulbs, iHome switch, Samsung Smart camera, and Google Chromecast. These attacks on IoT devices are in various types including directly targeted attacks such as ARP spoofing, TCP SYN flooding, Fraggle (UDP flooding), and Ping of Death, and also reflection attacks such as SNMP, SSDP, TCP SYN, and Smurf. Each type is at three different rates (\ie low: 1 packet-per-second, medium: 10 pps, and high: 100 pps). Additionally, attacks are diversified in terms of the location of attacker being remote or local to victim/reflector IoT devices. In total, DATA2 contains 200 attack sessions, and each lasts for around 10 minutes.
		
		\begin{table}[t]
			\centering
			\caption{Summary of DATA2: device instances and clustering parameters -- benign traces for training, and mix of benign and attack traces for testing).}
			\label{tab:c5_attack_classifier_parm}
			
			\begin{adjustbox}{width=0.7\textwidth}
				\def\arraystretch{1.1}
				\begin{tabular}{|l|cc|cc|}\hline
					& \multicolumn{2}{c|}{\textbf{Instance count}} & \multicolumn{2}{c|}{\shortstack{\\\textbf{Unsupervised classifier}\\ \textbf{parameters}}}                                                                                                                                  \\\hline
					Device        &  \shortstack{\\Training \\(4-week)}             & \shortstack{Testing \\(4-week)}            & \shortstack{\\\# Principal\\components} &  \shortstack{\#\\clusters} \\\hline
					Amazon Echo   &  27102 &  27510 &                    20 &              256 \\
					Belkin motion &  38229 &  37216 &                    13 &              256 \\
					Belkin switch &  21038 &  12689 &                    17 &              256 \\
					Chromecast   &  17396 &  24316 &                    17 &              512 \\
					Hue bulb      &  17329 &  25830 &                    19 &              512 \\
					LiFX bulb         &  25903 &  26181 &                    15 &              256 \\
					Netatmo cam   &  13529 &  10639 &                    16 &              256 \\
					Samsung cam   &  38227 &  36747 &                    15 &              256 \\
					TPlink switch &  38211 &  35205 &                    14 &              128 \\
					iHome         &  37866 &  35761 &                    16 &              128 \\ \hline   
				\end{tabular}
			\end{adjustbox}
		\end{table}		
		
		Note that DATA2 was collected from a different IoT environment, and therefore we need to regenerate our clustering models using data of IoT behaviors specific to that environment. From DATA2 traces, we choose four weeks' worth of data (\ie May 28-31, Jun 8-19, Oct 9-19) containing pure benign traffic for training, and the remaining four weeks (Jun 1-8, Jun 19-20, Oct 19-Nov 10) containing a mix of benign and attack traffic for testing. Table~\ref{tab:c5_attack_classifier_parm} shows the number of instances (training and testing) per each device as well as parameters of individual clustering models.

		Let us now evaluate the efficacy of individual models against traffic mix of attack and benign instances. We measure four metrics: fraction of attack instances getting negative output (TN: true negative), fraction of benign instances getting negative output (FN: false negative), fraction of benign instances getting positive output (TP: true positive), and fraction of attack instances getting positive output (FP: false positive). On average, our models yield acceptable performance metrics -- TN, FN, TP, and FP equals to $92.0$\%, $6.1$\%, $93.9$\%, and $8.0$\%, respectively. 
		
		Focusing on attacks, Table~\ref{tab:c5_attack_types_direct} and \ref{tab:c5_attack_types_Reflection} show the detection rate of our models for direct and reflection attacks. Each attack type-location scenario in columns (\eg ARP Spoofing R$\rightarrow$D: remote attacker launching direct spoofing attack to device) is repeated three times at rates 1, 10, and 100 pps.
						
		\begin{table}[b]
			\centering
			\caption{Detection rate (\%) of direct attacks: per model (in rows) and per attack-type (in columns).}
			\label{tab:c5_attack_types_direct}
			
			\begin{adjustbox}{width=0.8\textwidth}
				\def\arraystretch{1.1}
				\begin{tabular}{|l|lll|lll|lll|lll|lll|lll|}
					\hline
					\textbf{Attack} & \multicolumn{3}{c|}{\textbf{ARP Spoofing}} & \multicolumn{3}{c|}{\textbf{Ping of Death}} & \multicolumn{6}{c|}{\textbf{TCP SYN}} & \multicolumn{6}{c|}{\textbf{Fraggle}} \\
					\hline
					\textbf{Attacker} & \multicolumn{3}{c|}{\textbf{L$\rightarrow$D}} & \multicolumn{3}{c|}{\textbf{L$\rightarrow$D}} & \multicolumn{3}{c|}{\textbf{L$\rightarrow$D}} & \multicolumn{3}{c|}{\textbf{R$\rightarrow$D}} & \multicolumn{3}{c|}{\textbf{L$\rightarrow$D}} & \multicolumn{3}{c|}{\textbf{R$\rightarrow$D}} \\
					
					\textbf{Rate} &               \textbf{1} &   \textbf{10} &  \textbf{100} &               \textbf{1} &  \textbf{ 10} &  \textbf{100} &              \textbf{ 1} &   \textbf{10} &  \textbf{100} &               \textbf{1} &   \textbf{10} &  \textbf{100} &               \textbf{1} &   \textbf{10} &  \textbf{100} &               \textbf{1} &   \textbf{10} &  \textbf{100} \\
					\hline
					Amazon Echo   &             $100$ &   $75$ &   $88$ &                 &      &      &                 &      &      &                 &      &      &             $100$ &  $100$ &  $100$ &             $100$ &  $100$ &  $100$ \\
					Belkin motion &              $80$ &   $70$ &   $80$ &               \hl{$70$} &    \hl{$50$} &    \hl{$10$} &                \hl{$10$}&     \hl{$30$} &     \hl{$0$} &              $75$ &  $100$ &   $95$ &                 \hl{$0$} &      \hl{$0$} &    \hl{$10$} &                 &      &      \\
					Belkin switch &             $100$ &   $90$ &  $100$ &             $100$ &  $100$ &  $100$ &             $100$ &  $100$ &  $100$ &             $100$ &  $100$ &  $100$ &                 &      &      &                 &      &      \\
					Chromecast   &              $80$ &   $80$ &   $90$ &                 &      &      &             $100$ &  $100$ &  $100$ &              $50$ &   $70$ &   $90$ &                 &      &      &                 &      &      \\
					Hue bulb      &              $70$ &   $90$ &   $90$ &              $90$ &   $90$ &  $100$ &             $100$ &  $100$ &  $100$ &             $100$ &  $100$ &  $100$ &                 &      &      &                 &      &      \\
					LiFX bulb         &              $88$ &  $100$ &  $100$ &             $100$ &  $100$ &  $100$ &                 &      &      &                 &      &      &             $100$ &  $100$ &  $100$ &             $100$ &  $100$ &  $100$ \\
					Netatmo cam   &              $25$ &   $88$ &   $75$ &                 &      &      &             $100$ &  $100$ &  $100$ &              $80$ &  $100$ &  $100$ &                 &      &      &                 &      &      \\
					Samsung cam   &             $100$ &   $90$ &   $90$ &             $100$ &  $100$ &  $100$ &             $100$ &  $100$ &  $100$ &             $100$ &  $100$ &  $100$ &             $100$ &  $100$ &  $100$ &             $100$ &  $100$ &  $100$ \\
					TPlink switch &              $60$ &   $70$ &   $60$ &             $100$ &  $100$ &  $100$ &             $100$ &  $100$ &  $100$ &             $100$ &  $100$ &  $100$ &                 &      &      &                 &      &      \\
					iHome         &             $100$ &  $100$ &  $100$ &                 &      &      &                 &      &      &                 &      &      &                 &      &      &                 &      &      \\			\hline
				\end{tabular}
				
			\end{adjustbox}
		\end{table}
		
		\begin{table}[t]
			\centering
			\caption{Detection rate (\%) of reflection attacks: per model (in rows) and per attack-type (in columns).}
			\label{tab:c5_attack_types_Reflection}
			
			\begin{adjustbox}{width=0.99\textwidth}
				\def\arraystretch{1.1}
				\begin{tabular}{|l|rrr|rrr|rrr|rrr|rrr|rrr|rrr|rrr|rrr|}
					\hline
					\textbf{Attack} & \multicolumn{3}{c|}{\textbf{Smurf}} & \multicolumn{9}{c|}{\textbf{SNMP}} & \multicolumn{9}{c|}{\textbf{SSDP}} & \multicolumn{6}{c|}{\textbf{TcpSynReflection}} \\
					\hline
					\shortstack{\\ \textbf{Attacker \&}\\ \hfill \textbf{Victim}} & \multicolumn{3}{c|}{\textbf{L$\rightarrow$D$\rightarrow$L}} & \multicolumn{3}{c|}{\textbf{L$\rightarrow$D$\rightarrow$L}} & \multicolumn{3}{c|}{\textbf{L$\rightarrow$D$\rightarrow$R}} & \multicolumn{3}{c|}{\textbf{R$\rightarrow$D$\rightarrow$R}} & \multicolumn{3}{c|}{\textbf{L$\rightarrow$D$\rightarrow$L}} & \multicolumn{3}{c|}{\textbf{L$\rightarrow$D$\rightarrow$R}} & \multicolumn{3}{c|}{\textbf{R$\rightarrow$D$\rightarrow$R}} & \multicolumn{3}{c|}{\textbf{L$\rightarrow$D$\rightarrow$L}} & \multicolumn{3}{c|}{\textbf{R$\rightarrow$D$\rightarrow$R}} \\ 
					\textbf{Rate} &                             \textbf{1} &   \textbf{10} &  \textbf{100} &                             \textbf{1} & \textbf{10} & \textbf{100} &                             \textbf{1} &   \textbf{10} &  \textbf{100} &                             \textbf{1} &   \textbf{10} &  \textbf{100} &                             \textbf{1} &  \textbf{10} &  \textbf{100} &                             \textbf{1} &   \textbf{10} &  \textbf{100} &                             \textbf{1} &   \textbf{10} &  \textbf{100} &                             \textbf{1} &   \textbf{10} &  \textbf{100} &                             \textbf{1} &   \textbf{10} &  \textbf{100} \\
					\hline
					Amazon Echo   &                               &      &      &                               &    &     &                               &      &      &                               &      &      &                               &     &      &                               &      &      &                               &      &      &                               &      &      &                               &      &      \\
					Belkin motion &                               &      &      &                               &    &     &                               &      &      &                               &      &      &                           $100$ &  $90$ &   $80$ &                           $100$ &  $100$ &  $100$ &                           $100$ &  $100$ &  $100$ &                            \hl{$30$} &   \hl{$30$} &    \hl{$0$} &                            $85$ &   $95$ &  $100$ \\
					Belkin switch &                               &      &      &                               &    &     &                               &      &      &                               &      &      &                               &     &      &                               &      &      &                               &      &      &                           $100$ &  $100$ &  $100$ &                           $100$ &   $95$ &  $100$ \\
					Chromecast   &                               &      &      &                               &    &     &                               &      &      &                               &      &      &                             \hl{$0$} &  \hl{$30$} &    \hl{$0$} &                           $100$ &  $100$ &  $100$ &                           $100$ &  $100$ &  $100$ &                           $100$ &  $100$ &  $100$ &                            $40$ &   $90$ &  $100$ \\
					Hue bulb      &                            $90$ &  $100$ &  $100$ &                               &    &     &                               &      &      &                               &      &      &                           $100$ &  $90$ &  $100$ &                           $100$ &  $100$ &  $100$ &                           $100$ &  $100$ &  $100$ &                           $100$ &  $100$ &  $100$ &                           $100$ &  $100$ &  $100$ \\
					LiFX bulb         &                           $100$ &  $100$ &  $100$ &                               &    &     &                               &      &      &                               &      &      &                               &     &      &                               &      &      &                               &      &      &                               &      &      &                               &      &      \\
					Netatmo cam   &                               &      &      &                               &    &     &                               &      &      &                               &      &      &                               &     &      &                               &      &      &                               &      &      &                           $100$ &  $100$ &  $100$ &                           $100$ &  $100$ &  $100$ \\
					Samsung cam   &                           $100$ &  $100$ &  $100$ &                             \hl{$0$} &  \hl{$0$} &   \hl{$0$} &                            $30$ &  $100$ &  $100$ &                           $100$ &  $100$ &  $100$ &                               &     &      &                               &      &      &                               &      &      &                           $100$ &  $100$ &  $100$ &                           $100$ &  $100$ &  $100$ \\
					TPlink switch &                           $100$ &  $100$ &  $100$ &                               &    &     &                               &      &      &                               &      &      &                               &     &      &                               &      &      &                               &      &      &                           $100$ &  $100$ &  $100$ &                           $100$ &  $100$ &  $100$ \\
					iHome         &                               &      &      &                               &    &     &                               &      &      &                               &      &      &                               &     &      &                               &      &      &                               &      &      &                               &      &      &                               &      &      \\
					\hline
				\end{tabular}
				
			\end{adjustbox}
			
		\end{table}	 
		
		Starting from Table~\ref{tab:c5_attack_types_direct} corresponding to direct attacks, it can be seen that the average detection rate for ARP Spoofing, Ping of Death, TCP SYN flooding, and Fraggle is $84.3$\%, $89.4$\%, $91.3$\%, and $86.2$\%, respectively. However, we observe that the Belkin motion model displays a poor performance in detecting attacks launched from local attackers (highlighted cells). {For example, the detection rates of Ping of Death, TCP SYN flooding, and Fraggle are  43.3\%, 13.3\%, and 3.0\% respectively.}  This is mainly because Belkin motion typically communicates with its mobile App locally by UPnP messages reporting current state of the sensor, and hence local attacks are not seen as so abnormal by the corresponding model -- soon we will further investigate and address this issue. 
		\vspace{-0.7em} 
		
		{\rev Moving to Table~\ref{tab:c5_attack_types_Reflection}} to check the performance of models against reflection attacks, we see the rate of detection for Smurf, SNMP, SSDP and TCP SYN reflection attacks on average is $99.1$\%, $58.8$\%, $88.5$\%, and $92.0$\%, respectively. Again, we observe that some of broadcast attacks (\ie SSDP reflection attack on Chromecast) and local attacks (\ie TCPsyn on Belkin motion and SNMP on Samsung cam) are missed. This is primarily because we only monitor local traffic targeted to IoT devices (\S\ref{sec:c5_telemtry}), and hence broadcast traffic and reflected outgoing local traffic gets missed. It is important to note that models of Belkin motion and Hue bulb are detecting local SSDP reflection attacks (L$\rightarrow$D$\rightarrow$L) only because these two devices have limited processing power, and hence under local SSDP attacks their normal operation (activity pattern of other flows) gets impacted leading to abnormal behavior.
		
		\textbf{Enhancing Detection Rate}: As discussed earlier in this section, models in general perform well for a mix of benign and attack traffic except in certain situations. Among all models, we found that the Belkin motion model does not perform well especially for attack instances, and results in a relatively high FP $32.1$\%. To further investigate such performance, we look at its confidence-level. Fig.~\ref{fig:c5_confidence_level_belkinmotion} shows the CDF of the Belkin model confidence for incorrectly classified attack instances (FP) as well as correctly classified benign instances (TP). We note that the model gives a very low confidence-level (less than 2.5\%) for a majority  ($61$\%) of the FP instances while such low confidence is seen for a tiny fraction ($3$\%) of the TP instances. 
		Again the acceptable confidence-level will be chosen by the network operator depending on their desired sensitivity. In our case, choosing confidence threshold 2.5\%, the performance metrics is significantly enhanced for Belkin model -- FP is improved down to $12.5$\% while TP is slightly degraded (from $95.5$\% to $92.6$\%).  
		
		\begin{figure}[t]
			\begin{center}
				\includegraphics[width=0.6\textwidth]{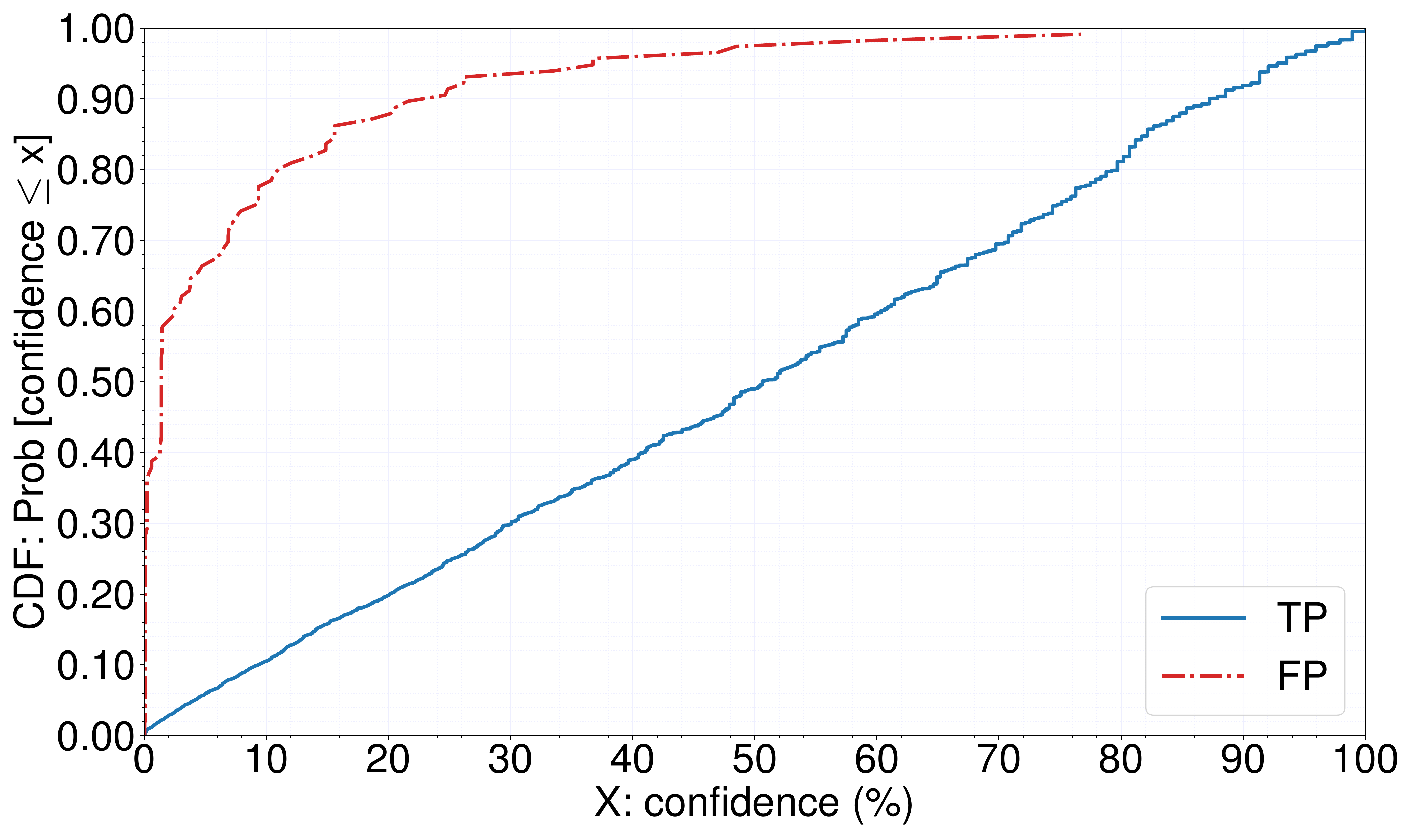}\quad
				\caption{CDF: distribution of confidence-level for Belkin motion instances.}
				\label{fig:c5_confidence_level_belkinmotion}
			\end{center}	
		\end{figure}
		
		\begin{table}[b]
			\centering
			\caption{Performance of one-class classifiers for mix of attack and benign traffic.}
			\label{tab:c5_attack_performance}
			
			\begin{adjustbox}{width=0.5\columnwidth}
				\def\arraystretch{1.1}
				\begin{tabular}{lrr|rr}
					\toprule
					{}&\multicolumn{4}{c}{Detected as} \\
					{}&\multicolumn{2}{c|}{attack} & \multicolumn{2}{c}{benign} \\
					{} &
					\shortstack{TN (\%)}&
					\shortstack{FN (\%)}&
					\shortstack{TP (\%)}&
					\shortstack{FP (\%)}\\
					\midrule
					Amazon Echo   &   $98.8$ &   $6.0$ &  $94.0$ &   $1.2$ \\
					TPlink switch &   $95.9$ &   $5.8$ &  $94.2$ &   $4.1$ \\
					Belkin motion &   $87.5$ &   $7.4$ &  $92.6$ &  $12.5$ \\
					Belkin switch &   $99.2$ &   $8.1$ &  $91.9$ &   $0.8$ \\
					LiFX bulb     &   $99.3$ &   $7.3$ &  $92.7$ &   $0.7$ \\
					Netatmo cam   &   $94.6$ &   $6.0$ &  $94.0$ &   $5.4$ \\
					Hue bulb      &   $97.3$ &  $15.7$ &  $84.3$ &   $2.7$ \\
					iHome         &  $100.0$ &   $7.1$ &  $92.9$ &   $0.0$ \\
					Samsung cam   &   $92.4$ &   $6.7$ &  $93.3$ &   $7.6$ \\
					Chromecast   &   $81.9$ &  $10.2$ &  $89.8$ &  $18.1$ \\
					\bottomrule
				\end{tabular}
			\end{adjustbox}
			
		\end{table}
		
		Such enhancement is observed across all models after filtering model outputs with confidence less than 2.5\%, and thus overall TN, FN, TP, and FP reaches to $94.7$\%, $9.03$\%, $92.0$\%, and $5.3$\%, respectively across all models. We show in Table~\ref{tab:c5_attack_performance} the performance of individual models after this enhancement. We can see that every model now displays acceptable value in performance metrics (high rate of true alarms and low rate of false alarms).

		\textbf{Performance Comparison of One-Class vs. Multi-Class:} Lastly, we compare the performance of our one-class classifier scheme versus previously studied multi-class classifiers (including \hyperref[chap:characterization]{Chapter~\ref*{chap:characterization}}). For our comparison, we use Random Forest algorithm (based on decision trees) to generate and tune a multi-class model using the training instances (same as for our one-class models) from DATA2.
		
		Before comparing the two schemes, we need our devices to operate in their stable phase of monitoring. Note that, the first two days of testing data contains pure benign traffic from all of the ten devices. This amount of data is sufficient for all of intended one-class models to be selected (consistency score of winner models exceeds our chosen threshold 0.90). In other words, every device passes its initial phase and enters into the stable phase. In the stable phase, the inference engine is expected to give negative output whenever attack traffic instances are present and generate positive output for pure benign traffic.

		Fig.~\ref{fig:c5_oneClassMulti} illustrates the consistency of the two schemes for a sample of traffic from Samsung smart cam during a week period with 380 instances of attack traffic -- each instance is worth a minute of traffic. In Fig.~\ref{fig:c5_oneClass} we plot the real-time consistency score for Samsung smart cam during attack periods -- attack instances are marked by red `$\times$'. It can be seen that the model correctly detects attack traffic instances by giving them negative outputs, causing a drop in the consistency score. We note that sometimes the consistency score keeps falling down even when attack finishes. This is because the impact of some attacks persist in attributes of a few following instances (up to 8 minutes). We can see that during intense attack periods (Jun 2 and Jun 3), the consistency score of the one-class model of Samsung camera drops to its lowest level, well highlighting a significant change of behavior in device traffic. 
		
		\begin{figure}[t]
			\begin{center}
				\mbox{
					\subfloat[one-class clustering model.]{
						\includegraphics[width=0.45\textwidth]{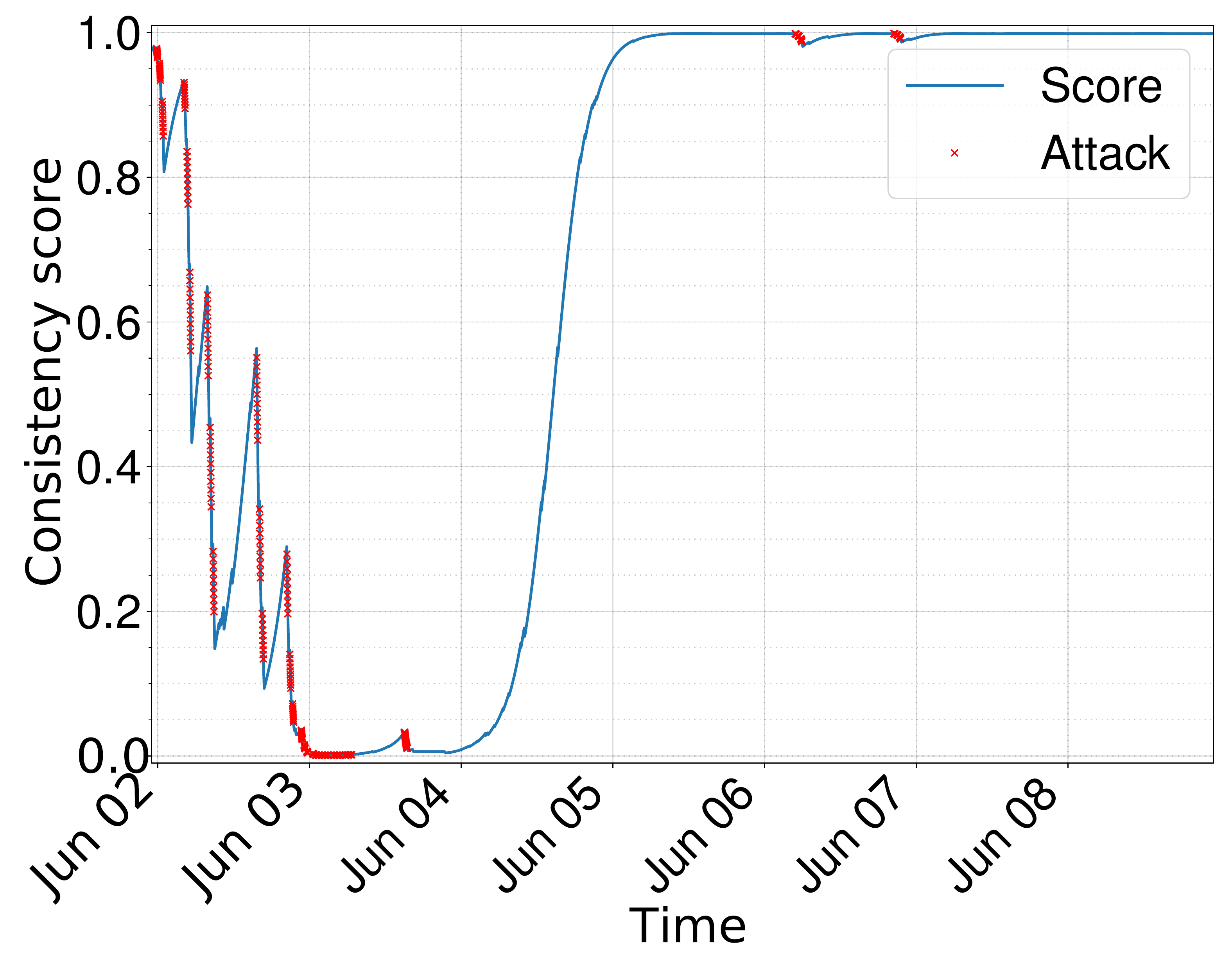}
						\label{fig:c5_oneClass}
					}
				}
				\mbox{
					\subfloat[multi-class decision-tree model.]{
						\includegraphics[width=0.45\textwidth]{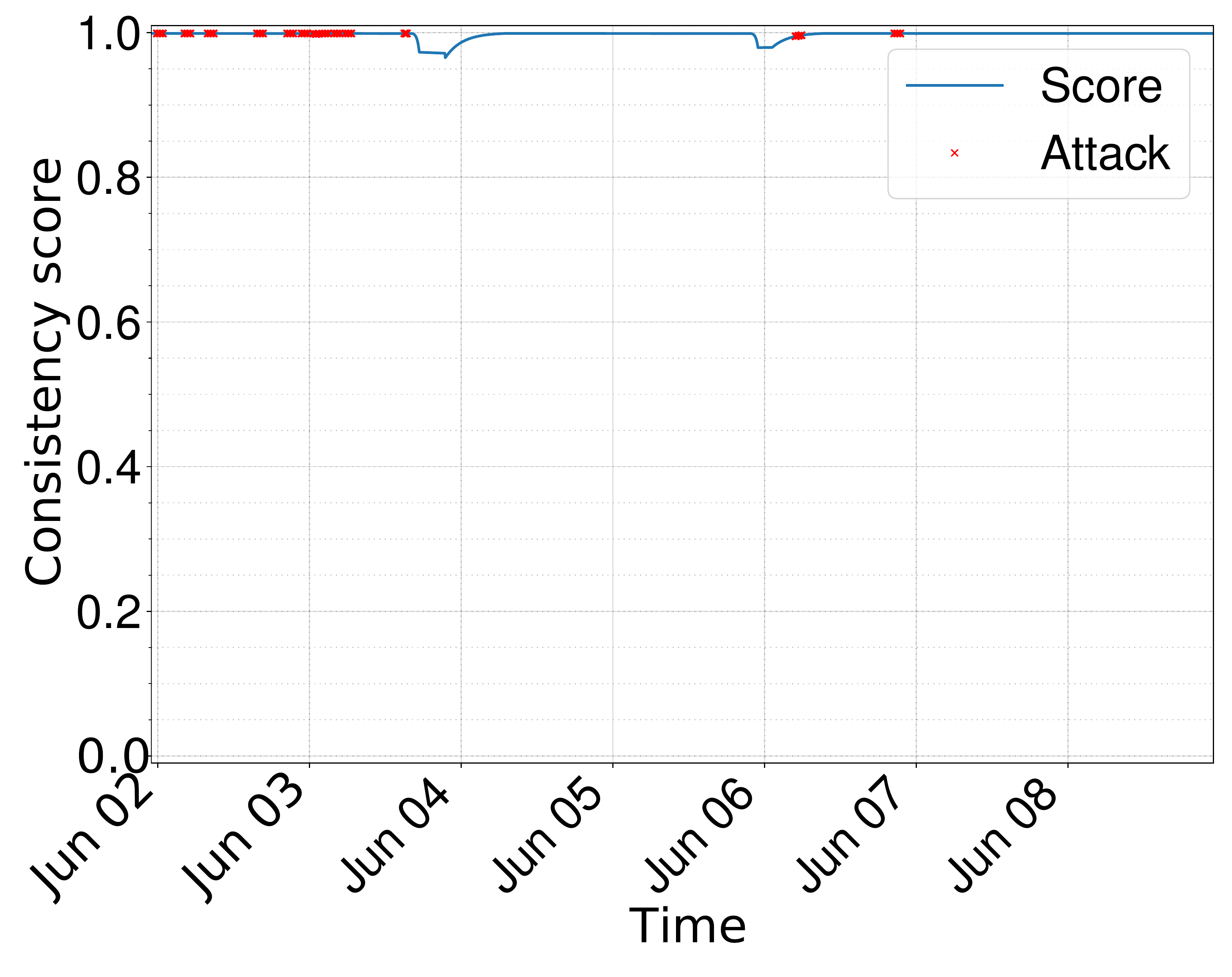}
						\label{fig:c5_multiClass}
					}
					
				}
				
				\caption{Performance comparison for traffic of Samsung cam during attack: (a) one-class model; and (b) multi-class model.}
				
				\label{fig:c5_oneClassMulti}
			\end{center}
		\vspace{-1em}
		\end{figure} 
		
		On the other hand, it is seen in Fig.~\ref{fig:c5_multiClass} that the multi-class model is insensitive to attacks. The consistency score of the model remains high during the whole week, and does not noticeably get affected by attack instances, as shown in Fig.~\ref{fig:c5_multiClass}. Even though the Random Forest model gives negative output to some attack instances ($34.6$\%), but each negative output is immediately followed by a sequence of positive outputs keeping the real-time consistency score at a very high value -- incorrectly suggesting that the device behaves normally.  
		
		We observe that the one-class clustering model has, by far, more ability to highlight (detect) anomalies in device behavior compared to the multi-class decision-tree model -- detection rate of $92.6$\% compared to $34.6$\%. 
		%({\color{red}do you consider confidence of random-forest too?})
		We also note that multi-class model correctly classifies $98.0$\% of benign instances (TP) while this metric is slightly lower ($94.7$\%) for one-class model.

		{\rev
			Although the performance of one-class classifier models is compared with a multi-class classifier for a limited number of devices, we have found that one-class classification is more scalable compared to multi-class classification. Note that adding more devices would exponentially increase the complexity of the multi-class model. For one-class models, instead, our conflict resolution method ensures each classifier works independently even for a large number of devices. We note that the cost of one-class classification is expected to increase linearly with the number of devices.
		}  
		
		Note that these two approaches are fundamentally different in their way of modeling: one-class models are generative (learn distribution of each class) while multi-class models are discriminative (learn decision boundary between various classes). As a result, one-class models become sensitive to changes in any attribute while multi-class models become sensitive to changes in only discriminative attributes.

\vspace{-1em}
\section{Conclusion}\label{sec:c5_con}
	\vspace{-1em}
    Real-time traffic monitoring is of paramount importance for network operators who manage a diverse set of IoT devices. In this chapter, we developed a modular classifier to identify IoT devices from their network behavior using a set of clustering models. We further fine-tuned our classification models to not only identify IoT devices but also detect the cyber-attacks from network traffic. We augmented our machine learning-based system of classifiers with a conflict resolver and a model consistency mechanism to track the behavioral changes of devices. Finally, we evaluated the efficacy of our system by applying it to traffic traces from 12 IoT devices, and demonstrate its ability to detect behavioral changes and cyber-attacks with an overall accuracy of more than 94\%. There are many interesting aspects of this work that warrant further study. We outline some directions in the next chapter.

\chapter{Conclusions and Future Work}
	\label{chap:conclusion}
	\vspace{-5mm}
	\minitoc
	
	\section{Conclusions}
		The Internet of Things has become a natural extension to the physical world and its influence is found in every aspect of our lives. Although they have endless potential to offer immense experiences and benefits to the users, the rapid innovation and proliferation make them vulnerable to security and privacy breaches. The recent attacks on IoT networks clearly indicate the possible catastrophic consequences if we fail to give adequate attention to IoT security. On the other hand, traditional IT security mechanisms fail to protect the IoT networks due to the large-scale deployments and heterogeneous behavior of the devices.
		
		Smart environment operators still fail to recognize the IoT assets and detect cyber-attacks or compromised behaviors in real-time due to the lack of tools to enable visibility into the IoT network. This thesis is an attempt to develop novel IoT behavioral monitoring mechanisms using network analytics to automate the IoT device identification and classification. It also deduces their operating context and detects anomalous behavior indicative of IoT cyber-attacks.
		
		This thesis opened by highlighting the ecosystem of IoT, especially in the perspective of security and privacy, by considering the market segments, security risks, challenges in protection, role of the stakeholders and vulnerability assessment methods. We have also carried out a comprehensive survey on existing IoT security solutions and behavioral monitoring methods. Following this, three key contributions to the field of IoT security have been presented by providing better visibility into IoT network:
		\begin{itemize}
			\item We captured and synthesized the IoT network traffic traces from a testbed which is equipped with 28 consumer IoT devices. These traces were used to profile the behavioral characteristics of IoT devices based on the activity and signalling patterns. We developed a machine learning based classification framework using the attributes obtained from the traffic characteristics to classify the IoT devices. The proposed classification architecture was trained and tested with six months worth of data collected from an IoT testbed that showed more than 99\% accuracy in classifying the IoT device types.
			
			\item We reduced the cost of attribute extraction by proposing flow level programmable telemetry at multiple timescales. We proposed a real-time IoT behavioral monitoring solution that can recognize the IoT devices and their states of operations. Furthermore, we showed how an operator can further optimize the cost of telemetry and how our inference engine can be used to identify behavioral changes including cyber-attacks. 
			
			\item We developed a modularized inference engine using an unsupervised machine learning algorithm, which allows changes to be accommodated in the device behaviors without system wide retraining. Additionally, we proved that unsupervised machine learning models are very sensitive with regards to detecting the behavioral changes including low-rate cyber-attacks which showed 94\% detection rate.
		\end{itemize}
	
	\section{Future Work}
		This work makes significant contribution towards IoT behavioral monitoring with better visibility into the IoT network using network analytics. The novel mechanisms for IoT behavioral monitoring outlined in this thesis can be improved and refined based on real-world deployment scenarios. Some of the key refinements are outlined below:
		
		\begin{itemize}
			\item The inference engine developed in this thesis has the ability to detect behavioral changes by monitoring for anomalous network traffic patterns. Techniques to identify the exact traffic flow that is causing the anomalous activity should be further researched and developed in the future.  Identifying and quarantining the anomalous flows during the attack, and performing forensics on compromised devices will be a sensible future requirement and can be a step forward for this work~\cite{Sommer2010}.
			
			\item We proposed a classification framework to deduce the operating context by identifying states of the devices. However, we have not investigated the probability of each state and transitional probabilities among them as a part of this work. It will also be useful to detect the attacks that exploit the legitimate network activities of devices, which will be a challenge to the current inference engine. For example, if a device is configured to work dependent on a web service, an attacker may compromise the web service to exploit the device using its dependency~\cite{Surbatovich2017}. During this attack the traffic between the web service and the device may follow a legitimate pattern and then it is difficult to trace any anomalies. Monitoring the state transitioning pattern and frequency may be a way forward to detect the attacks.
			
			\item The flow level telemetry is purely used in this thesis to monitor the IoT traffic activities. Although it is difficult to spoof the flow patterns of a device, there are possibilities an attacker could carefully mimic the flow patterns from a compromised device. This issue can be mitigated by including some packet level attributes from sampled traffic which fingerprint the payloads (\eg entropy of packets) in addition to flow level telemetry. %Stealth 
		\end{itemize}
	
	We hope other researchers will explore the future directions identified above.

%------------------------------------------------------------------
%	BIBLIOGRAPHY
%------------------------------------------------------------------
\backmatter
\renewcommand{\refname}{References} % Change the default bibliography title

% Run biber command for first time
%-------- IMP: DO NOT UPDATE THE LIBRARY DIRECTLY -----
%--------- IT IS LINKED TO ZOTERO----
%\bibliography{References} % Input your bibliography file
\appto{\bibsetup}{\sloppy}
\apptocmd{\sloppy}{\hbadness 10000\relax}{}{}
\begin{singlespace}
\setlength\bibitemsep{10pt}   % length between two different entries
\printbibliography[heading=bibintoc,title={References}]
\end{singlespace}

%------------------------------------------------------------------

% \input{./tex/appendix_1.tex}

\end{document}